\documentclass[11pt,a4paper]{article}

\usepackage[utf8]{inputenc}
\usepackage[T1]{fontenc}
\usepackage{lmodern}
\usepackage{geometry}
\usepackage{setspace}
\usepackage{booktabs}
\usepackage{longtable}
\usepackage{comment}
\usepackage{array}
\usepackage{threeparttable}
\usepackage{graphicx}
\usepackage{amsmath, amssymb}
\usepackage{hyperref}
\usepackage{xcolor}
\usepackage{enumitem}
\usepackage{natbib}
\usepackage{placeins}

\title{
Open Bitcoin Metrics: Verifiable Full-Node-Derived Bitcoin Time Series for Economic Research
}

\author{
Diego R. Llanos\textsuperscript{1,*}
\\
\textsuperscript{1}Department of Computer Science, University of Valladolid, Spain
\\
\textsuperscript{*}Corresponding author: diego.llanos@uva.es
}

\date{}

\begin{document}

\maketitle

\begin{abstract}
Bitcoin research increasingly relies on on-chain indicators to study network activity, monetary issuance, transaction demand, miner incentives, coin-age behavior, and long-run monetary dynamics. However, many commonly used Bitcoin metrics are dispersed across commercial platforms, subject to heterogeneous definitions, or not fully reproducible from primary blockchain data. This manuscript introduces Open Bitcoin Metrics (OBM), a reproducible, full-node-derived dataset and reference guide for Bitcoin on-chain time series designed for economic and econometric research. The dataset provides documented daily series covering block production, block-space usage, transaction counts, supply, issuance, fees, miner revenue, mining difficulty, estimated hashrate, Bitcoin Days Destroyed, dormancy, liveliness, UTXO counts, spent output value, and related UTXO-age indicators. Metrics are reconstructed from a locally maintained Bitcoin Core full node, a persistent spent-output indexer, or deterministic transformations of previously generated OBM series. Each series is accompanied by open-source Python code, stable identifiers, explicit definitions, metadata, validation procedures, interpretive caveats, and comparisons with the closest publicly available metrics. The dataset is intended to support transparent empirical research, replication, teaching, and comparative analysis across monetary economics, financial economics, and blockchain studies.

The dataset, related software and documentation can be found at \url{https://github.com/diegorllanos/open-bitcoin-metrics}.
\end{abstract}

\newpage

\tableofcontents
\newpage

\section{Introduction}

Bitcoin has generated a growing body of empirical literature across economics, finance, monetary studies, and computer science. Many of these studies require specialized on-chain time series that are not directly available from standard macro-financial databases. These series are used to study network activity, monetary issuance, transaction demand, fee-market dynamics, miner incentives, coin-age behavior, UTXO-state evolution, and long-run monetary properties. However, many commonly used Bitcoin metrics are scattered across specialized providers, blockchain explorers, and commercial platforms. Even when the resulting charts or data downloads are publicly visible, their precise definitions, timestamp conventions, treatment of edge cases, and reconstruction algorithms are often only partially documented.

This creates a methodological problem for empirical research and a broader problem for the Bitcoin ecosystem. Bitcoin is built around the principle that critical claims should be independently verifiable rather than accepted on authority. The familiar maxim ``do not trust, verify'' is usually applied to monetary ownership, consensus validation, and the operation of full nodes, but the same principle is also relevant to the metrics used to describe Bitcoin itself. If researchers, educators, builders, market participants, or policy analysts rely on opaque indicators, then part of the interpretation of Bitcoin's monetary and technical behavior is delegated to unverifiable data pipelines. Auditable metrics help reduce this hidden trust layer by enabling inspection of how an indicator was defined, reconstructed, validated, and derived from primary blockchain data.

This paper describes Open Bitcoin Metrics (OBM), an open and reproducible dataset of Bitcoin on-chain metrics derived from a locally maintained Bitcoin Core full node, a persistent spent-output indexer, and deterministic transformations of previously generated OBM series. The project has three objectives. First, it provides econometric-ready Bitcoin time series with stable identifiers, explicit units, regular frequencies, and documented aggregation conventions. Second, it documents the algorithms used to reconstruct each metric from primary blockchain data or from OBM source series, allowing independent replication. Third, it facilitates comparison, extension, and auditability by making the source code, output files, metadata, validation procedures, and metric-level reference guide openly available.

\paragraph{Contributions.} This manuscript makes five contributions. First, it introduces OBM as a reproducible, full-node-derived dataset of Bitcoin on-chain time series with stable identifiers, explicit units, and documented aggregation conventions. Second, it provides a metric-level reference guide covering definitions, economic interpretation, data-source requirements, algorithms, validation procedures, output formats, limitations, and public comparators for each included series. Third, it documents a reusable spent-output indexing pipeline that supports metrics requiring previous-output reconstruction, including fees, issuance, miner revenue, spent value, Bitcoin Days Destroyed, dormancy, and UTXO-age indicators. Fourth, it offers a transparent baseline against which provider-specific Bitcoin metrics can be compared, especially when definitions, timestamp conventions, smoothing choices, or entity-adjustment heuristics differ across sources. Fifth, it extends Bitcoin's verification ethos to empirical research by making the code and data sufficiently open for researchers to inspect, rerun, audit, and modify the metric-generation process.

The dataset is designed around the following principles:

\begin{enumerate}
    \item \textit{Primary-source derivation}: metrics are reconstructed from Bitcoin blockchain data obtained through a full node whenever possible.
    \item \textit{Transparent definitions}: each variable is associated with an explicit mathematical definition and computational algorithm.
    \item \textit{Econometric usability}: series are provided at regular time intervals, with clear timestamp conventions, aggregation rules, units, and metadata.
    \item \textit{Versioned reproducibility}: scripts, outputs, documentation, and data releases are archived and versioned.
    \item \textit{Validation}: selected metrics are checked using internal identities, consistency tests, and diagnostic comparisons with independent public sources.
    \item \textit{Verification-oriented openness}: users are not required to treat OBM as an authority, but can inspect, rerun, compare, and modify the code that generates the metrics.
\end{enumerate}

The objective of OBM is not to replace commercial data providers, which often offer broad metric coverage, refined interfaces, higher-frequency data, entity-adjusted indicators, and market analytics. Rather, OBM provides a transparent baseline for Bitcoin research: selected on-chain time series reconstructed through documented procedures, distributed with stable names and explicit limitations. In this sense, OBM complements existing data services while extending Bitcoin's verification ethos to the empirical metrics used to study the system.

The dataset, related software and documentation can be found at \url{https://github.com/diegorllanos/open-bitcoin-metrics}.

\section{Related Work}

\label{sec:related_work}

\begin{table}[ht!]
\centering
\scriptsize
\begin{tabular}{p{0.12\textwidth}p{0.26\textwidth}p{0.22\textwidth}p{0.26\textwidth}}
\hline
\textbf{Source} & \textbf{Relevant content} & \textbf{Access model} & \textbf{Auditability} \\
\hline
\citet{bitcoincore2026} & Primary block and transaction data, full-node validation, RPC access to the locally verified chain. & Free and open source. & Authoritative source of data. Metrics can be generated by querying Bitcoin nodes running this software, but no algorithms are provided for any metric. \\
\hline
\citet{mempool2026api} & Open-source explorer, REST and WebSocket APIs, mempool, block, transaction, mining, and fee data. & Free public service, with enterprise options for production use. & Medium to high. The explorer is open source and useful for reconstruction, though not all OBM-style daily metrics are packaged directly. \\
\hline
\citet{coinmetrics2026community} & Network Data metrics, including block counts, supply, fees, miner revenue, transaction counts, and coin-age indicators. & Freemium. Community API access exists for selected data, while most systematic coverage requires an API key or commercial access. & Medium. Metric identifiers and definitions are documented, but the full reconstruction code is not generally open. \\
\hline
\citet{blockchair2026api} (API) and \citet{blockchair2026dumps} (database dumps) & Blockchain explorer, chart catalog, SQL-like API, and downloadable database dumps. & Freemium. Limited API testing is available, with paid plans for heavier API use; dumps are publicly advertised. & Medium. Raw data access and dumps support independent reconstruction, but chart definitions are often concise. \\
\hline
\citet{glassnode2026pricing} & Broad set of Bitcoin on-chain, market, derivatives, and entity-adjusted metrics. & Freemium. A free Studio tier exists, while advanced metrics, higher resolution, broader history, and API-oriented use require paid plans or add-ons. & Medium to low. Conceptual definitions are available for many metrics, but provider heuristics and low-level conventions are not fully reproducible. \\
\hline
\citet{cryptoquant2026pricing} & On-chain and market indicators, including UTXO, miner, exchange, network, and valuation metrics. & Freemium. Limited free access is available, while professional API access and broader historical coverage require paid plans. & Medium to low. Some formula-level documentation is clear, but exact implementation conventions are not fully auditable. \\
\hline
\citet{blockchaincom2026charts} & Public Bitcoin charts, explorer information, and chart endpoints for common network indicators. & Free for many charts and chart-style access. & Medium to low. Useful for validation, but the methodology is usually brief and not equivalent to reproducible code. \\
\hline
\citet{bitbo2026pricing} & Bitcoin charting platform with on-chain, market, and macro indicators. & Freemium. Public charts are available, while larger data download quotas, alerts, and private indicators require a subscription. & Medium to low. Useful chart reference, but not a full algorithmic specification. \\
\hline
\citet{newhedge2026api} & Bitcoin analytics platform with on-chain metrics, valuation indicators, and API access. & Freemium to subscription. Limited chart views are available with free accounts, while API and full data access require an Advanced subscription. & Medium to low. Useful comparator, but systematic data access and implementation details are commercial. \\
\hline
\citet{bitinfocharts2026bitcoin} & Public cryptocurrency statistics, including current and historical network indicators for Bitcoin and other assets. & Free public website. & Low. Useful for diagnostic comparison, but with limited methodological detail and limited programmatic reproducibility. \\
\hline
\citet{ycharts2026fees} & Financial charting platform that republishes selected Bitcoin statistics, often sourced from Blockchain.com. & Subscription-oriented, with public indicator previews and trials. & Low. Useful as a secondary source, but not a primary blockchain-data methodology. \\
\hline
\citet{bitcoinmagazinepro2026home} & Public and subscription Bitcoin charts, including several coin-age and valuation indicators. & Freemium. Free charts are available, while full chart access and private tools require a subscription. & Low. Useful interpretive source, but not a reproducible metric specification. \\
\hline
\citet{checkonchain2026charts} & Bitcoin on-chain charting suite focused on market-cycle, lifespan, supply, and valuation metrics. & Free public charting site, with associated research and newsletter products. & Low to medium. Useful as a charting reference, but not generally a source of open reconstruction code. \\
\hline
\citet{tradingdigits2026realizedprice} & Public Bitcoin market-cycle and on-chain charts, including realized-price and supply-adjusted indicators. & Free public charts. & Low. Useful for comparison and interpretation, but limited methodological disclosure. \\
\hline
\end{tabular}
\caption{Main public sources for Bitcoin-related time series.}
\label{tab:public_bitcoin_data_sources}
\end{table}

Empirical research on Bitcoin increasingly depends on time series that summarize information embedded in the blockchain. These series are used to study network activity, transaction demand, monetary issuance, miner incentives, fee-market dynamics, coin-age behavior, and long-run monetary properties. Unlike conventional macroeconomic and financial variables, however, many Bitcoin on-chain indicators are not distributed by public statistical agencies. They are commonly reconstructed by specialized data providers, blockchain explorers, or individual researchers from primary ledger data. This creates a recurrent reproducibility problem: two series with similar labels may differ because of distinct timestamp conventions, treatment of chain reorganizations, entity-adjustment heuristics, address clustering rules, smoothing choices, or definitions of transferred value.

Public and auditable Bitcoin time series are therefore important for at least three reasons. First, they make empirical results easier to replicate because the underlying variables can be recomputed from a documented procedure rather than obtained from an opaque interface. Second, they improve comparability across studies by attaching stable names, units, frequencies, and aggregation rules to each variable. Third, they support methodological scrutiny because researchers can inspect whether a metric is derived directly from consensus data, from explorer-level transaction data, from provider-specific heuristics, or from market-price sources. This is especially relevant for economic research, where small definitional differences can materially affect econometric results, particularly in early Bitcoin history, around daily boundaries, or for UTXO-age metrics such as Coin Days Destroyed.

The primary source of auditable on-chain Bitcoin information is the Bitcoin blockchain itself, accessed via a fully synchronized Bitcoin Core node \citep{nakamoto2008bitcoin,bitcoincore2026}. A full node allows researchers to verify the main chain, retrieve block and transaction data, reconstruct previous outputs when required, and apply explicit timestamp and aggregation conventions. This approach maximizes auditability but imposes computational costs and requires metric-specific software. Consequently, most researchers rely on public data providers, charting services, or blockchain explorers. These sources are useful for discovery, validation, and cross-checking, but they differ considerably in openness, methodological detail, and access conditions.

Table~\ref{tab:public_bitcoin_data_sources} summarizes the main public sources that are most relevant for the metrics considered in this paper. The classification is deliberately practical. ``Free'' means that at least some charts or API endpoints can be accessed without a paid subscription. ``Freemium'' means that a limited public tier exists, but historical depth, download volume, higher-resolution data, API access, or advanced metrics require a paid plan. ``Subscription'' means that the source is primarily commercial for systematic use, even if selected pages or previews are publicly visible. Access conditions change over time, so the table should be interpreted as a snapshot of the services at the time of writing.

This landscape suggests a useful distinction between accessibility and auditability. A metric can be publicly visible while still being difficult to audit if the provider does not disclose the complete reconstruction procedure. Conversely, raw block and transaction data can be fully auditable but inconvenient for economists unless transformed into regular, documented, econometric-ready time series. All time series sources listed in Table~\ref{tab:public_bitcoin_data_sources} provide useful information. However, most of them are best viewed as comparison sources rather than fully reproducible research pipelines.

Open Bitcoin Metrics' contribution addresses this gap. The objective is not to replace commercial providers, which often offer broad coverage, refined interfaces, higher-frequency data, entity-adjusted indicators, and market analytics. Rather, the objective is to provide a transparent baseline for economic research: selected Bitcoin on-chain time series reconstructed from a full node, distributed with explicit definitions, stable series identifiers, regular frequencies, open-source scripts, validation checks, and documented limitations. This makes OBM complementary to existing providers. Their series remain valuable external benchmarks, while OBM emphasizes auditability, reproducibility, and direct derivation from primary blockchain data.

\FloatBarrier

\section{Methods}

\subsection{Overview of the Data Generation Pipeline}

The OBM data-generation pipeline is designed to transform primary Bitcoin blockchain data into regular, documented, econometric-ready time series. The dataset contains two broad types of time series: primary and derived.

Primary series are reconstructed from Bitcoin blockchain data obtained from a locally maintained Bitcoin Core full node. Depending on the metric, the required information may be purely block-level, transaction-level, coinbase-level, or spent-output-level. Simple block-level metrics, such as daily block or transaction counts, can be computed directly by scanning blocks over the requested date interval. More demanding metrics, such as Bitcoin Days Destroyed, spent output value, transaction fees, issuance, and miner revenue, require resolving transaction inputs to the previous outputs they spend. For these metrics, OBM uses a persistent spent-output indexer, described below.

Derived series are deterministic transformations of already generated OBM series. Examples include ratios, cumulative sums, and normalized measures. For instance, the fee share of miner revenue is computed from daily transaction fees and daily realized issuance, and supply is computed as a cumulative sum of daily issuance. Derived series do not query Bitcoin Core directly. They read existing OBM-compatible CSV files or aggregate tables and apply the documented transformation.

The general workflow is therefore:

\begin{enumerate}
    \item synchronize and maintain a Bitcoin Core full node;
    \item extract block-level and transaction-level information from the validated main chain;
    \item reconstruct previous outputs where required;
    \item store reusable spent-output and block-level aggregates in a persistent indexer database;
    \item export metric-specific daily series from either the node scan, the indexer database, or existing OBM CSV files;
    \item produce monthly versions where appropriate, using metric-specific aggregation rules;
    \item validate outputs using internal identities and external diagnostic comparisons;
    \item publish updated files, scripts, metadata, and documentation.
\end{enumerate}

This architecture separates computationally expensive blockchain reconstruction from lightweight metric export. Metrics that require the same spent-output resolution pass are computed from a shared intermediate database rather than by independently scanning the full blockchain for each series.

\subsection{Bitcoin Node Configuration}

The dataset is reconstructed from a Bitcoin Core full node. The node configuration used for data extraction is documented to support reproducibility.

\begin{itemize}
    \item Bitcoin Core version: 30.2.
    \item Network: mainnet.
    \item Indexing mode: \texttt{txindex=1}.
    \item Pruning: disabled.
    \item Operating system: Ubuntu 22.04.5 LTS.
    \item Update cadence: daily.
\end{itemize}

A non-pruned node is strongly preferred for OBM production runs. Although several block-level metrics can be computed from ordinary block metadata, spent-output metrics require reconstructing previous outputs over the historical chain. Maintaining a non-pruned full node avoids dependence on third-party APIs and preserves the project's primary-source reproducibility principle.

Bitcoin Core is accessed through its JSON-RPC interface. Authentication is handled either through explicit RPC credentials or through Bitcoin Core's cookie-authentication mechanism. All scripts use the locally verified main chain reported by the node. The scripts do not rely on blockchain explorers or commercial data providers for the construction of the primary metric.

\subsection{Time Convention and Aggregation}

Unless otherwise stated, daily series are aggregated using Coordinated Universal Time (UTC). A block \(b\) is assigned to a calendar day according to

\[
d(b) = \mathrm{UTCDate}(t_b),
\]

where \(t_b\) is the timestamp associated with block \(b\), as returned by Bitcoin Core. Alternative timestamp conventions, such as median time past, are discussed in the limitations section when relevant.

Bitcoin block timestamps are not strictly monotonic with respect to block height. Therefore, scripts that locate a date interval by block timestamp first identify an approximate height interval and then expand it by a safety margin. This reduces the risk of missing boundary blocks whose timestamps fall inside the requested UTC date interval but whose heights lie slightly outside the approximate interval obtained by timestamp-based search.

Daily aggregates are computed as sums, averages, medians, ratios, or end-of-period observations depending on the economic meaning of each variable. Flow variables, such as fees, issuance, miner revenue, spent value, and Bitcoin Days Destroyed, are aggregated as daily sums. Stock variables, such as supply, are reported as end-of-period or cumulative quantities, depending on the specific series definition. Ratio variables, such as fees as a share of miner revenue, are computed from the relevant daily numerator and denominator, rather than by averaging lower-level ratios.

Monthly versions, when distributed, are computed from daily values using metric-specific aggregation rules. Monthly flows are generally sums of daily flows. Monthly ratios are preferably recomputed from monthly summed components rather than obtained as arithmetic averages of daily ratios, unless a different convention is explicitly documented.

\subsection{The Spent-Output Indexer}
\label{sec:The Spent-Output Indexer}

Several OBM metrics require information that is not directly available from a single block without reconstructing previous outputs. In Bitcoin's UTXO model, each non-coinbase transaction input spends a previous transaction output. To compute metrics such as transaction fees, spent value, Bitcoin Days Destroyed, dormancy, or long-term-holder spent value, the value and creation time of each spent previous output must be known.

For this reason, OBM uses a persistent spent-output indexer, implemented as a script called \texttt{obm\_spent\_output\_indexer.py}. The indexer scans the blockchain in height order, maintains a local SQLite representation of the live outpoint state, and stores daily aggregate quantities needed by several metrics. This avoids repeating the same expensive full-chain input-resolution procedure for each individual time series.

The indexer maintains a live outpoint table with primary key \((\texttt{txid}, \texttt{vout})\). For each unspent output, the table stores its value in satoshis, creation timestamp, and creation height. When a new transaction output is created, it is inserted into the table. When a later transaction input spends it, the indexer retrieves its value and creation time, updates the relevant daily aggregates, and deletes the spent outpoint from the live table. Thus, the table represents the current UTXO-like state required for future input resolution, rather than a full historical log of all spent outputs.

For each spent input \(i\), the indexer obtains the previous output value \(V_i\), its creation time \(t_i^{\mathrm{create}}\), and the spending block time \(t_i^{\mathrm{spend}}\). The spent-output age is computed as

\[
A_i =
\max\left(0,\frac{t_i^{\mathrm{spend}} - t_i^{\mathrm{create}}}{86400}\right),
\]

where age is measured in days. Negative apparent ages can occur because Bitcoin block timestamps are not strictly monotonic. OBM floors such cases to zero and records the event count in the indexer metadata.

The Coin Days Destroyed contribution of input \(i\) is $\mathrm{CDD}_i = V_i A_i$. Internally, the indexer stores this quantity as satoshi-days. Export scripts convert satoshi-days into BTC-days by dividing by \(100{,}000{,}000\).

The indexer also computes transaction fees. For a non-coinbase transaction \(j\), let \(V_j^{\mathrm{in}}\) denote the total value of the previous outputs spent by its inputs, and let \(V_j^{\mathrm{out}}\) denote the total value of its outputs. The transaction fee is $F_j = V_j^{\mathrm{in}} - V_j^{\mathrm{out}}$.

Since the indexer already resolves every input to its previous output, the input values required for this calculation are available during the same pass. The total fees of a block are then obtained by summing \(F_j\) over all non-coinbase transactions in that block.

For each block, the indexer also reads the total value of the coinbase transaction outputs. This value represents BTC-denominated miner revenue for the block, because the coinbase transaction pays both newly issued BTC and transaction fees. Realized issuance is then computed as

\[
\mathrm{Issuance}_b =
\mathrm{CoinbaseOutput}_b - \mathrm{Fees}_b.
\]

This definition distinguishes newly issued BTC from fees, which are transfers of already existing BTC from users to miners.

The main daily aggregate table stores, for each UTC date with processed block activity, the following quantities:

\begin{itemize}
    \item total spent output value;
    \item spent output count;
    \item Bitcoin Days Destroyed;
    \item sum of spent-output ages;
    \item spent value and CDD for outputs aged at least 365 days;
    \item spent value and CDD for outputs aged at least 155 days;
    \item spent value and CDD for outputs younger than 155 days;
    \item total transaction fees;
    \item total coinbase output value;
    \item realized issuance;
    \item count of negative-age outputs floored to zero.
\end{itemize}

The indexer also stores age-band aggregates. The current age bands are:

\[
\begin{aligned}
&0\text{d--}1\text{d},\quad
1\text{d--}1\text{w},\quad
1\text{w--}1\text{m},\quad
1\text{m--}3\text{m},\quad
3\text{m--}6\text{m},\quad
6\text{m--}1\text{y},\\
&1\text{y--}2\text{y},\quad
2\text{y--}3\text{y},\quad
3\text{y--}5\text{y},\quad
5\text{y--}7\text{y},\quad
7\text{y--}10\text{y},\quad
10\text{y+}.
\end{aligned}
\]

For each date and age band, the indexer stores the spent value, CDD, and the spent output count. These tables support later export of age-distribution metrics without requiring a second full-chain scan.

The indexer is persistent and resumable. It records the last processed block height, block hash, timestamp, and date in a metadata table. When restarted, it resumes from the next block after the last committed height. Before continuing, it checks that the stored last-processed block hash still matches the block hash reported by Bitcoin Core at that height. If the hashes differ, the script aborts rather than attempting an automatic rollback. This conservative policy avoids silently mixing data from inconsistent chain states.

The indexer periodically commits SQLite changes according to a configurable \texttt{--commit\_every} parameter. This allows long historical builds to be interrupted and resumed without restarting from genesis, while controlling the trade-off between speed and the amount of work lost after an interruption.

Historical duplicate coinbase transaction identifiers are handled explicitly. Early Bitcoin contains exceptional duplicate \((\texttt{txid}, \texttt{vout})\) pairs. When the indexer encounters such a duplicate live outpoint, it overwrites the earlier entry with the later one and records the event in metadata. The overwritten output is not counted as spent and therefore does not generate spent value or CDD. This convention is documented because it affects the treatment of rare historical edge cases.

\subsection{Metric Exporters and Derived Scripts}

OBM separates the shared indexing step from metric-specific export. The spent-output indexer produces an internal SQLite database. Separate exporter scripts, then read the aggregate tables and write standard OBM CSV files. For example, \texttt{obm\_cdd\_btcxdays\_daily} is exported from the daily CDD aggregate, while \texttt{obm\_issuance\_btc\_daily} is exported from the daily issuance aggregate. Similarly, daily transaction fees and BTC-denominated miner revenue can be exported from the stored fee and coinbase-output aggregates.

This architecture has three advantages. First, expensive spent-output reconstruction is performed only once. Second, metric-specific scripts remain simple and auditable. Third, the same indexer database can support additional future metrics, such as dormancy, binary CDD, spent-output count, long-term-holder spent value, short-term-holder spent value, and age-band CDD.

For some dates, especially in the earliest Bitcoin period, the indexer may not create a daily aggregate row because no block is assigned to that UTC date under the block-timestamp convention. Export scripts apply metric-specific missing-date rules. For flow variables such as issuance, fees, miner revenue, CDD, and spent value, a missing aggregate row is exported as zero. For ratio variables, such as dormancy or fee share of miner revenue, a zero denominator leads to a missing value rather than a zero value. This distinction avoids assigning economic meaning to undefined ratios.

Other OBM series are derived from published CSV files rather than directly from the indexer database. For example, the fee share of miner revenue is computed as

\[
\mathrm{FeeShare}_d =
\frac{
\mathrm{FeesBTC}_d
}{
\mathrm{Issuance}_d + \mathrm{FeesBTC}_d
},
\]

using the already exported daily fees and daily issuance series. This design makes derived metrics reproducible from public CSV files alone and keeps the indexer as an internal source for primitive or semi-primitive aggregates.

\subsection{Variable Naming Convention}

Each time series uses a stable identifier of the form:

\[
\texttt{obm\_<metric>\_<unit or variant>\_<frequency>}.
\]

The prefix \texttt{obm} identifies the Open Bitcoin Metrics dataset. Lowercase snake-case identifiers improve usability across Python, R, Stata, SQL, and CSV workflows. Examples include \texttt{obm\_tx\_count\_daily}, \texttt{obm\_cdd\_btcxdays\_daily}, and \texttt{obm\_fee\_share\_revenue\_ratio\_daily}. 

Units are included when they clarify the value's interpretation. For example, \texttt{btc} denotes BTC-denominated quantities, \texttt{btcxdays} denotes BTC-days, and \texttt{ratio} denotes a dimensionless ratio. The final component indicates the observation frequency, such as \texttt{daily}. This convention is intended to make identifiers explicit while preserving compatibility with common data-analysis environments.

\section{Description of OBM available metrics}
\label{Description of OBM available metrics}

This section provides a compact description of the OBM metrics included in the current release. The objective is not to repeat the full reference guide, but to give readers a concise interpretation of each series, clarify its economic meaning, and identify the closest publicly available comparators where appropriate. The metrics cover block production, block-space usage, coin-age behavior, mining difficulty and estimated hashrate, miner compensation, monetary issuance and supply, raw transaction and UTXO activity, and spent-value indicators by age. For each metric, the discussion emphasizes what the series measures, what it should not be interpreted as measuring, and how it can be used in empirical Bitcoin research. Full mathematical definitions, data-source requirements, algorithmic details, validation notes, output formats, and limitations are provided in Appendix~\ref{Reference guide of OBM available metrics}.

\subsection{\texttt{obm\_block\_count\_daily}: Daily Block Count}

The series \texttt{obm\_block\_count\_daily} reports the number of Bitcoin blocks assigned to each UTC calendar day. At first sight, this is one of the simplest OBM metrics, but it is also one of the most useful reference series in the dataset. Bitcoin targets an average inter-block interval of approximately ten minutes, so the long-run benchmark is about 144 blocks per day. The realized daily count, however, fluctuates because block discovery is probabilistic, mining difficulty adjusts only at discrete intervals, and block timestamps do not align mechanically with calendar-day boundaries.

Economically, this metric should be interpreted as a measure of realized block production rather than a direct indicator of transaction demand or economic activity. Its main value is contextual. It tells researchers how many settlement batches the Bitcoin network produced on a given day, and it helps interpret variation in other daily aggregates. For instance, a low daily transaction count, fee total, or block-weight total may partly reflect a day with fewer blocks rather than a pure change in user demand. The metric is also a natural normalization variable for block-level series such as transactions, fees, issuance, miner revenue, and block-space usage.

The closest public comparator identified in the appendix is Coin Metrics' \texttt{BlkCnt} series, which counts the main-chain blocks created within the interval. Conceptually, this is very close to \texttt{obm\_block\_count\_daily}, although the exact daily values may differ across providers due to timestamp conventions, finality policies, or chain reorganization handling. Further methodological details, validation notes, and limitations are provided in Appendix~\ref{sec:Daily Block Count}.

\subsection{\texttt{obm\_block\_weight\_wu\_daily}: Daily Block Weight in Weight Units}

The series \texttt{obm\_block\_weight\_wu\_daily} measures the total block weight of all Bitcoin blocks assigned to each UTC calendar day. Unlike block count, which counts how many blocks were produced, this metric measures the amount of consensus-weighted block space actually included in the chain during the day. Since the SegWit capacity rule is expressed in weight units rather than bytes, this is the appropriate block-space metric for modern Bitcoin analysis. The series is a daily total, not an average per block.

For economic research, daily block weight is a compact measure of realized block-space usage. It complements transaction count by showing whether blocks are comparatively empty or full, and it complements fee metrics by helping researchers interpret congestion and competition for inclusion. A day with high total block weight may reflect many blocks, fuller blocks, or both, so the metric is especially informative when used together with \texttt{obm\_block\_count\_daily}. It can also support derived indicators such as average block weight per block or fees per weight unit, both of which are useful for studying Bitcoin's fee market and the changing intensity of block-space demand.

The closest public comparator identified in the appendix is Coin Metrics' \texttt{BlkWghtTot}, which reports the sum of block weights over the interval. Coin Metrics' \texttt{BlkWghtMean}, Newhedge's Bitcoin Block Weight chart, and Bitcoin Visuals' block-weight chart are related but less direct comparators because they focus on mean or median block weight rather than the total daily sum. Bytes-based block-size metrics from other providers are useful for broader block-space analysis, but they are not direct substitutes for a weight-unit-denominated series. Further methodological details, validation notes, and limitations are provided in Appendix~\ref{Daily Block Weight in Weight Units}.

\subsection{\texttt{obm\_cdd\_age\_band\_btcxdays\_daily}: Daily Bitcoin Days Destroyed by Age Band}

The table \texttt{obm\_cdd\_age\_band\_btcxdays\_daily} decomposes daily Bitcoin Days Destroyed according to the age of the outputs that are spent. Instead of reporting a single scalar value, it provides a wide daily table in which each column corresponds to an output-age band, such as \texttt{0d\_1d}, \texttt{1d\_1w}, \texttt{1y\_2y}, or \texttt{10y+}. Each entry reports the BTC-days destroyed by outputs in that age cohort on the corresponding UTC date. In this sense, the metric asks not only how much coin age was destroyed, but also where in the age distribution that destruction came from.

Economically, this metric is useful because aggregate CDD can hide very different underlying behaviors. A large CDD observation may be caused by a substantial movement of moderately aged coins or by a smaller movement of very old coins that had remained inactive for years. The age-band decomposition makes this distinction visible. It is therefore especially useful for studying dormant-supply activation, long-term holder behavior, distribution episodes, wallet consolidation, and the age profile of on-chain movements within each age band. It also complements the corresponding spent-value age-band table: spent value shows how many BTC moved by cohort, while CDD by age band shows how much accumulated coin age was destroyed by cohort.

The closest public age-distribution comparator is CryptoQuant's Spent Output Age Bands metric, which also groups spent outputs by age. However, the public metric reports the BTC value spent by age band rather than the BTC-days destroyed by age band. Glassnode's SOAB and SVAB indicators, Checkonchain's spent-volume age-band charts, and scalar CDD series from Glassnode, CryptoQuant, Bitbo, and Coin Metrics are also useful comparators, but none of them appears to provide a named daily wide table exactly equivalent to \texttt{obm\_cdd\_age\_band\_btcxdays\_daily}. Further methodological details, validation notes, and limitations are provided in Appendix~\ref{Daily Bitcoin Days Destroyed by Age Band}.

\subsection{\texttt{obm\_cdd\_btcxdays\_daily}: Bitcoin Days Destroyed}

The series \texttt{obm\_cdd\_btcxdays\_daily} reports daily Bitcoin Days Destroyed, a classic coin-age metric that combines value and dormancy. For each spent output, OBM multiplies the BTC value of the output by the number of days elapsed between the block in which the output was created and the block in which it was spent. The daily series is obtained by summing these value-age contributions across all non-coinbase inputs included in blocks assigned to the same UTC calendar day. Its unit is BTC-days.

The economic intuition is simple but powerful: moving one BTC that has been idle for 100 days destroys 100 BTC-days, while moving one BTC created only a few hours ago destroys very little coin age. The metric, therefore, gives greater weight to the movement of older coins than to the rapid circulation of recently created outputs. For economic research, this makes CDD useful as an indicator of dormant-supply activation, long-term holder behavior, older-coin redistribution, and episodes in which coins with substantial accumulated age re-enter transactional activity. It should not be interpreted as payment volume, exchange volume, user count, or entity-adjusted economic settlement. It is a raw spent-output metric that may reflect self-transfers, wallet reorganization, custodial activity, or exchange operations, as well as economically meaningful transfers.

The closest named public comparator is Coin Metrics' \texttt{TxTfrValDayDst}, or Xfer'd Days Destroyed. This metric is highly relevant because it also measures transferred days destroyed, but exact equality should not be expected: Coin Metrics uses full-day accounting, whereas \texttt{obm\_cdd\_btcxdays\_daily} uses fractional days by dividing elapsed seconds by 86,400. Glassnode's CDD and CryptoQuant's CDD metrics are also close conceptual comparators, although their public documentation does not expose a fully reproducible full-node reconstruction procedure with all low-level conventions. Further methodological details, validation notes, and limitations are provided in Appendix~\ref{Bitcoin Days Destroyed}.

\subsection{\texttt{obm\_cdd\_per\_supply\_days\_daily}: Bitcoin Days Destroyed per Unit of Supply}

The series \texttt{obm\_cdd\_per\_supply\_days\_daily} normalizes daily Bitcoin Days Destroyed by the outstanding Bitcoin supply on the same UTC date. It is computed as a deterministic transformation of two OBM source series: \texttt{obm\_cdd\_btcxdays\_daily} and \texttt{obm\_supply\_btc\_daily}. Because the numerator is measured in BTC-days and the denominator in BTC, the result is in days. A value of 0.015, for example, means that the coin age destroyed on that date was equivalent to 0.015 days per bitcoin outstanding.

Economically, this metric is a supply-normalized version of CDD. Raw CDD tends to become less directly comparable across Bitcoin history because the outstanding supply changes dramatically over time. Dividing by supply helps separate two effects: a larger monetary base mechanically permits more BTC-days to be destroyed, while unusually intense movement of aged coins reflects a behavioral change in how dormant or older supply is being spent. The metric is therefore useful for comparing coin-age destruction across different monetary epochs, especially when studying long-term holder activity, dormant-supply movement, distribution episodes, and supply-adjusted on-chain behavior.

The closest public equivalents are Glassnode's Supply-Adjusted CDD and CryptoQuant's Supply-Adjusted CDD, both of which use the same central idea: dividing CDD by total or circulating supply. Bitcoin Magazine Pro, MacroMicro, TradingDigits, and Checkonchain also provide related charting references. The advantage of the OBM version is that the ratio is derived from two explicitly documented OBM source series, uses a stable identifier, reports the unit as days, and records the value as missing rather than zero when the supply denominator is zero. Further methodological details, validation notes, and limitations are provided in Appendix~\ref{Bitcoin Days Destroyed per Unit of Supply}.

\subsection{\texttt{obm\_difficulty\_eod\_daily}: End-of-Day Bitcoin Mining Difficulty}

The series \texttt{obm\_difficulty\_eod\_daily} reports the Bitcoin mining difficulty associated with the highest-height block assigned to each UTC calendar day. It is therefore an end-of-day observation of a protocol state variable, not a daily average. This distinction matters because mining difficulty changes only at retarget boundaries and remains constant within ordinary difficulty epochs. If no block is assigned to a given UTC date, the series records the value as missing rather than assigning a misleading zero.

Economically, mining difficulty captures how demanding it is, relative to Bitcoin's reference difficulty, to produce a valid proof-of-work block. It is not a direct measure of observed hashrate, mining profitability, transaction demand, or user activity. Its main role is to describe the protocol-level mining environment in which miners operate. For empirical work, it is useful for studying difficulty-adjustment cycles, mining-sector conditions, issuance timing, network security, and the interaction between miner incentives, fees, block production, and Bitcoin price. It also provides one of the main ingredients for estimated hashrate metrics, although hashrate itself requires an additional estimation convention.

The closest public comparator identified in the appendix is Coin Metrics' \texttt{DiffLast}, which reports the difficulty of the last block in the interval. This is closely aligned with the OBM end-of-day convention. Coin Metrics' \texttt{DiffMean}, BitInfoCharts' average daily difficulty, Blockchain.com's Network Difficulty chart, Glassnode's difficulty-related charts, and CoinWarz's difficulty chart are useful related references, but they may rely on average, sampled, or insufficiently documented aggregation conventions. Further methodological details, validation notes, and limitations are provided in Appendix~\ref{End-of-Day Bitcoin Mining Difficulty}.

\subsection{\texttt{obm\_dormancy\_days\_daily}: Daily Dormancy}

The series \texttt{obm\_dormancy\_days\_daily} measures the average age, in days, of the coins spent on each UTC date, weighted by spent BTC value. It is computed by dividing the daily Bitcoin Days Destroyed by the daily spent output value. Because the numerator is measured in BTC-days and the denominator in BTC, the result is in days. When no spent value is observed, the metric is undefined and is recorded as missing, since a day with no spent value does not imply zero dormancy.

Economically, dormancy turns the intuition behind Bitcoin Days Destroyed into an average spent-age indicator. A high value means that, on that day, the coins being spent had accumulated substantial age before moving. A low value means that the spent value came mostly from recently created or recently active outputs. This makes the metric useful for studying dormant-supply activation, long-term holder behavior, distribution episodes, UTXO turnover, and the relationship between coin-age dynamics and market conditions. It should not be interpreted as payment volume, user activity, transaction count, or entity-adjusted settlement value, since it is a raw spent-output measure and does not identify users, exchanges, custodians, self-transfers, or change outputs.

The closest public comparator identified in the appendix is Glassnode's Bitcoin Dormancy chart, exposed as \texttt{indicators.AverageDormancy}. CryptoQuant's Average Dormancy chart is also directly related. Both use the same broad idea of dividing coin days destroyed by the amount of coins moved, although exact equivalence depends on the provider's transfer-volume convention, timestamp assignment, fractional-day treatment, entity-adjustment policy, and reorganization handling. Coin Metrics does not appear to provide a named public dormancy metric, but a related proxy could be constructed from its days-destroyed and transferred-value components. Further methodological details, validation notes, and limitations are provided in Appendix~\ref{Daily Dormancy}.

\subsection{\texttt{obm\_est7d\_hashrate\_ehs\_daily}: Estimated 7-Day Network Hashrate in EH/s}

The series \texttt{obm\_est7d\_hashrate\_ehs\_daily} estimates the aggregate computational rate devoted to Bitcoin mining, expressed in exahashes per second. Unlike difficulty, hashrate is not directly observed on-chain. It must be inferred from the proof-of-work difficulty of observed blocks and from the elapsed time over which those blocks were produced. OBM uses a trailing 7-day UTC calendar window, which is why \texttt{est7d} is part of the series identifier. A 14-day version would not use the same metric; it would use a different smoothing convention with a different identifier.

Economically, this metric serves as a proxy for mining activity and network security. It complements difficulty by combining the protocol target with realized block production over a rolling window. The 7-day window reduces the noise generated by Bitcoin's stochastic block-discovery process, making the series more stable than a one-day estimate while still responsive to changes in mining conditions. In empirical research, it can be used alongside miner revenue, fees, issuance, and the Bitcoin price to study mining incentives. It is also relevant for energy-related research, since estimated hashrate is often used as an intermediate variable in models of Bitcoin mining electricity consumption. It should not be read as a direct observation of actual machines, mining pools, geographic locations, hardware efficiency, energy sources, or electricity use.

The closest public comparators identified in the appendix are Glassnode's \texttt{mining.HashRateMean} and Blockchain.com's Total Hash Rate chart, because both explicitly present the estimated network hashrate. BitInfoCharts, CoinWarz, Newhedge, MacroMicro, Bitbo, and mempool.space provide useful secondary references or reconstruction sources. However, hashrate estimates are especially convention-dependent: providers may differ in window length, smoothing, timestamp convention, elapsed-time definition, unit scale, and treatment of sparse early windows. Further methodological details, validation notes, and limitations are provided in Appendix~\ref{Estimated 7-Day Network Hashrate in EH/s}.

\subsection{\texttt{obm\_fee\_share\_revenue\_ratio\_daily}: Fees as Share of Miner Revenue}

The series \texttt{obm\_fee\_share\_revenue\_ratio\_daily} measures the fraction of BTC-denominated miner compensation that comes from transaction fees. It is a derived metric computed from two OBM source series: \texttt{obm\_fees\_btc\_daily} and \texttt{obm\_issuance\_btc\_daily}. Since miners are compensated through both newly issued BTC and transaction fees, the metric reports fees divided by issuance plus fees. OBM expresses this value as a unit ratio rather than as a percentage. Thus, a value of 0.025 means that transaction fees represented 2.5\% of BTC-denominated miner revenue on that date.

Economically, this metric is a compact indicator of the changing composition of Bitcoin miner compensation. Its importance grows from the fact that Bitcoin's block subsidy declines through successive halvings, while transaction fees are expected to play a larger role in the long-run security budget. The series is therefore useful for studying subsidy eras, halving episodes, fee-market maturation, and periods in which block-space demand becomes economically significant relative to issuance. It should not be interpreted as total miner revenue, miner profit, or fiat-denominated mining income, since it does not include Bitcoin's market price or any cost-side variable.

The closest public comparators identified in the appendix are Coin Metrics' \texttt{FeeRevPct} and Glassnode's \texttt{mining.RevenueFromFees}, because both define the metric as fees divided by total miner revenue. CryptoQuant's Fees to Reward Ratio, BitInfoCharts' Fee in Reward, Bitbo's Fees as Percent of Total Block Reward, Newhedge's Percent Miner Revenue from Bitcoin Fees, and Bitcoin Magazine Pro's Miner Revenue (Fees vs Rewards) are also close public analogs. The main contribution of the OBM version is that the ratio is transparently derived from two documented OBM source series, reported as a unit ratio, and assigned a clear missing-value convention when the denominator is zero. Further methodological details, validation notes, and limitations are provided in Appendix~\ref{Fees as Share of Miner Revenue}.

\subsection{\texttt{obm\_fees\_btc\_daily}: Daily Transaction Fees}

The series \texttt{obm\_fees\_btc\_daily} reports the total amount of BTC paid as transaction fees by non-coinbase transactions included in blocks assigned to each UTC calendar day. For each non-coinbase transaction, the fee is computed as the difference between the total value of the previous outputs spent by the transaction inputs and the total value of the transaction outputs. The daily value is then obtained by summing these transaction-level fees across all blocks assigned to the same UTC date.

Economically, this metric measures the BTC-denominated amount users paid for block-space inclusion. It is therefore one of the central variables for studying Bitcoin's fee market. High daily fees may reflect stronger competition for settlement, greater congestion, higher fee rates, or some combination of transaction demand and block-space scarcity. The series is also a key component of miner compensation, since miner revenue in BTC equals issuance plus fees. For this reason, it supports research on miner incentives, the effects of halvings, security budget sustainability, and the gradual transition from subsidy-dominated to fee-supported miner revenue. It should not be interpreted as total miner revenue, miner profit, or a fee-rate metric, since it reports total fees in BTC rather than fees per virtual byte or per transaction.

The closest public comparator identified in the appendix is Coin Metrics' \texttt{FeeTotNtv}, which reports total fees in native units. Glassnode's \texttt{fees.VolumeSum} and Blockchain.com's Total Transaction Fees (BTC) chart are also close conceptual comparators, while Blockchair, Newhedge, Bitbo, BitInfoCharts, and YCharts provide useful secondary references. Exact equality across providers should not be assumed, especially because timestamp conventions, previous-output reconstruction, chain-reorganization treatment, and smoothing or charting choices may differ. Further methodological details, validation notes, and limitations are provided in Appendix~\ref{Daily Transaction Fees}.

\subsection{\texttt{obm\_issuance\_btc\_daily}: Daily Bitcoin Issuance}

The series \texttt{obm\_issuance\_btc\_daily} measures the amount of new BTC actually created in blocks assigned to each UTC calendar day. OBM defines issuance as a realized flow rather than merely as the theoretical block subsidy implied by the protocol schedule. For each block, the total output value of the coinbase transaction includes both newly issued BTC and transaction fees collected by the miner. Since fees are transfers of already existing BTC, OBM subtracts transaction fees from the coinbase output value to isolate the newly created supply. The daily series is then obtained by summing realized issuance across all blocks assigned to the same UTC date.

Economically, this metric is a monetary-flow variable. It measures the daily increase in Bitcoin supply and is therefore central for studying scarcity, supply growth, halving regimes, issuance shocks, and the monetary side of miner compensation. It should be distinguished from three nearby concepts. It is not the theoretical maximum subsidy, because miners may underclaim the amount they are allowed to create. It is not miner revenue, because miner revenue also includes transaction fees. It is not the circulating supply, because supply is a cumulative stock, while issuance is a daily flow. The metric is especially informative when interpreted alongside \texttt{obm\_block\_count\_daily}, since daily issuance depends mechanically on the number of blocks assigned to each UTC date.

The closest public comparators identified in the appendix are Coin Metrics' \texttt{IssTotNtv} and Glassnode's \texttt{supply.Issued}, because both provide named issuance metrics in native units. Newhedge, MacroMicro, and Bitbo also provide useful daily-issuance references, while Blockchain.com, YCharts, and Blockchair are more appropriate as supply-stock comparators or as sources from which approximate issuance could be derived. The distinctive contribution of OBM is that it reconstructs realized issuance from full-node-derived data, separates newly created BTC from transaction fees, assigns blocks to UTC dates using an explicit convention, and avoids relying only on the theoretical reward schedule. Further methodological details, validation notes, and limitations are provided in Appendix~\ref{Daily Bitcoin Issuance}.

\subsection{\texttt{obm\_liveliness\_ratio\_daily}: Daily Liveliness Ratio}

The series \texttt{obm\_liveliness\_ratio\_daily} summarizes the cumulative balance between coin-age destruction and coin-age creation. It is computed from two OBM source series: \texttt{obm\_cdd\_btcxdays\_daily} and \texttt{obm\_supply\_btc\_daily}. The numerator is cumulative Bitcoin Days Destroyed, while the denominator is a daily-supply-based approximation to cumulative coin-days created. The resulting value is dimensionless and is reported as a ratio. In intuitive terms, liveliness asks how much of the coin age historically created by Bitcoin has later been destroyed through spending.

Economically, this metric is a long-horizon indicator of holding versus spending behavior. When old coins move and destroy substantial accumulated coin age, liveliness tends to rise. When coins remain inactive, and the supply continues to accumulate coin days, liveliness tends to stabilize or decline more slowly. This makes the series a useful cumulative counterpart to daily CDD and dormancy. It can help distinguish periods dominated by accumulation from periods characterized by substantial movement of older coins. It should not be read as a daily activity measure, because it is cumulative and slow-moving, nor as an entity-adjusted measure of economic spending, because it does not identify lost coins, custodial movements, exchange activity, self-transfers, or wallet reorganizations.

The closest public comparator identified in the appendix is Glassnode's \texttt{indicators.Liveliness}, because it explicitly defines liveliness as cumulative Coin Days Destroyed divided by cumulative coin-days created. ChainExposed and CoinGlass use similar conceptual frameworks, while Newhedge and BGeometrics provide useful visual references with less methodological disclosure. Coin Metrics and CryptoQuant provide relevant CDD or supply components, but not a directly named public liveliness ratio equivalent to \texttt{obm\_liveliness\_ratio\_daily}. Further methodological details, validation notes, and limitations are provided in Appendix~\ref{Daily Liveliness Ratio}.

\subsection{\texttt{obm\_miner\_revenue\_btc\_daily}: Daily Miner Revenue in BTC}

The series \texttt{obm\_miner\_revenue\_btc\_daily} reports the total BTC-denominated compensation paid to miners through coinbase transactions in blocks assigned to each UTC calendar day. At the block level, the coinbase transaction pays the miner both newly issued BTC and transaction fees. Accordingly, daily miner revenue can be read either as the sum of coinbase transaction outputs assigned to that date or, equivalently, as daily issuance plus daily fees.

Economically, this is one of the central mining-sector variables in the OBM dataset. It measures the BTC flow paid to miners for producing blocks, before any cost-side consideration. It is therefore useful for studying Bitcoin's security budget, miner incentives, halving events, subsidy eras, and the transition from subsidy-dominated compensation toward a larger fee component. It also provides the denominator for the fee-share-of-miner-revenue metric. The series should not be interpreted as miner profit, since it does not account for electricity costs, hardware depreciation, pool fees, taxes, financing costs, or other operating expenses. It is also not fiat-denominated miner income, as it reports revenue in BTC rather than USD, EUR, or other fiat units.

The closest public comparator identified in the appendix is Glassnode's \texttt{mining.RevenueSum}, which defines total miner revenue as fees plus newly minted coins and exposes the metric in BTC. A close Coin Metrics benchmark can also be constructed by adding \texttt{IssTotNtv} and \texttt{FeeTotNtv}, although that is a component-based construction rather than a single, directly named gross miner-revenue metric. Blockchain.com and YCharts mainly provide USD-denominated miner-revenue series, while Newhedge, BitInfoCharts, Bitcoin Magazine Pro, MacroMicro, Bitbo, and Blockchair provide useful secondary references. Further methodological details, validation notes, and limitations are provided in Appendix~\ref{Daily Miner Revenue in BTC}.

\subsection{\texttt{obm\_raw\_output\_value\_btc\_daily}: Daily Raw Output Value in BTC}

The series \texttt{obm\_raw\_output\_value\_btc\_daily} measures the total BTC value of outputs created by non-coinbase transactions in blocks assigned to each UTC calendar day. Coinbase outputs are excluded because they represent miner compensation and are already captured by issuance and miner-revenue metrics. The result is a raw output-side accounting measure: it sums what ordinary transactions create as outputs, without attempting to classify which outputs are payments, change, self-transfers, or internal reorganizations.

Economically, this metric is a transparent baseline for on-chain value activity. It complements transaction count by measuring value rather than frequency, and it complements spent value by looking at the output side of non-coinbase transactions rather than the input side. Its strength is precisely its simplicity: it avoids opaque entity-adjustment heuristics and gives researchers a reproducible lower-level series. Its limitation is equally important. It should not be interpreted as clean payment volume or economically meaningful transfer volume, because Bitcoin transactions often include change outputs and may reflect batching, self-transfers, exchange operations, custodial wallet management, and wallet consolidation.

The closest public comparator identified in the appendix is Blockchain.com's Output Value Per Day chart, because it reports the total value of transaction outputs and explicitly includes change. YCharts republishes this Blockchain.com series, while Blockchair's output-total-based on-chain volume chart is another close secondary comparator. Glassnode's \texttt{transactions.TransfersVolumeSum}, Coin Metrics' \texttt{TxTfrValNtv}, Newhedge's transaction-volume indicators, and BitInfoCharts' sent-value statistics are useful related transfer-volume references, but they are less direct because they may use transfer-value conventions, adjusted heuristics, or undocumented provider filters. Further methodological details, validation notes, and limitations are provided in Appendix~\ref{Daily Raw Output Value in BTC}.

\subsection{\texttt{obm\_spent\_output\_count\_daily}: Daily Spent Output Count}

The series \texttt{obm\_spent\_output\_count\_daily} reports the number of previous outputs consumed by non-coinbase transaction inputs in blocks assigned to each UTC calendar day. It is therefore an input-side UTXO activity metric: it counts spent outputs, not transactions, addresses, users, entities, or payments. A single Bitcoin transaction may use one or more inputs, so this metric captures an aspect of transaction structure that is not visible in the daily transaction count alone.

Economically, the daily spent output count is useful for studying UTXO-flow dynamics. It helps distinguish ordinary transaction-count variation from days in which transactions consume unusually many previous outputs. This makes the series particularly informative when interpreting spent value, CDD, dormancy, age-band metrics, and threshold-based old-coin indicators. Large values may reflect periods of wallet consolidation, exchange activity, batching, custodial reorganization, or other input-heavy transaction patterns. The metric should not be interpreted as a measure of users, payments, or economically meaningful transfers, because it deliberately remains a raw UTXO-level count.

The closest public comparators identified in the appendix are Glassnode's spent-output age-band count charts, because they count spent outputs within specified lifespan intervals. CryptoQuant's Spent Output Age Bands framework is also related, although it is generally value-based rather than a simple count of spent outputs. SOPR metrics from Glassnode and CryptoQuant use spent outputs as their unit of analysis, but they report profitability ratios rather than counts. No reviewed provider appears to publish a simple named daily total exactly equivalent to \texttt{obm\_spent\_output\_count\_daily}. Further methodological details, validation notes, and limitations are provided in Appendix~\ref{Daily Spent Output Count}.

\subsection{\texttt{obm\_spent\_value\_age\_band\_btc\_daily}: Daily Spent Output Value by Age Band in BTC}

The table \texttt{obm\_spent\_value\_age\_band\_btc\_daily} decomposes the daily spent output value according to the age of the outputs at the moment they are spent. It is a wide daily table rather than a scalar series: each row corresponds to one UTC date, and each age-band column reports the BTC value of spent outputs whose age falls within that band. The table therefore asks not only how much BTC value was consumed through transaction inputs, but also how old that spent value was.

Economically, this metric is a useful map of the age composition of on-chain turnover. It helps distinguish young-output churn from the movement of coins that had remained inactive for months or years. This is valuable when interpreting spikes in spent value, because a large daily value may be driven by recently created outputs, by wallet-management activity, by medium-aged outputs, or by genuinely old dormant supply. The table also provides the basis for age-band shares and supports consistency checks with total spent value and age-threshold metrics. It should not be interpreted as entity-adjusted payment volume or settlement value, since it does not identify users, exchanges, custodians, change outputs, or self-transfers.

The closest public comparators identified in the appendix are Glassnode's Spent Volume by Age, Glassnode's SVAB and SOAB families, CryptoQuant's Spent Output Age Bands, and Checkonchain's Spent Volume Age Bands. These metrics share the central idea of classifying spent coins by age, although providers differ in whether they report absolute BTC values, percentages, entity-adjusted variants, or provider-defined age bands. The distinctive OBM contribution is the publication of a reproducible, BTC-denominated, full-node-derived wide table with explicit age-band columns and raw spent-output accounting. Further methodological details, validation notes, and limitations are provided in Appendix~\ref{Daily Spent Output Value by Age Band in BTC}.

\subsection{\texttt{obm\_spent\_value\_btc\_daily}: Daily Spent Output Value in BTC}

The series \texttt{obm\_spent\_value\_btc\_daily} reports the total BTC value of previous outputs consumed by non-coinbase transaction inputs in blocks assigned to each UTC calendar day. It is a raw input-side value-flow metric. Each transaction input contributes the value of the previous output it spends, regardless of whether the transaction represents an ordinary payment, a self-transfer, wallet consolidation, batching, exchange activity, change management, or custodial reorganization.

Economically, this metric measures gross UTXO turnover in BTC. It complements transaction count by capturing value rather than frequency, and it complements raw output value by focusing on consumed previous outputs rather than newly created transaction outputs. It is also a core building block for other OBM indicators: dormancy divides CDD by spent value, age-band tables decompose it by output age, and threshold metrics isolate the older or younger parts of it. Its simplicity is useful, but also limiting. It should not be interpreted as adjusted transfer volume, economic payment volume, or exchange-adjusted settlement value.

The closest public conceptual analogs identified in the appendix are Glassnode's family of Spent Volume metrics and CryptoQuant's Spent Output Value Bands. Coin Metrics' \texttt{TxTfrValNtv} and adjusted transfer-value metrics are related, but they answer a transfer-volume question rather than exactly replicating OBM's raw spent-output-value definition. Blockchain.com's Output Value Per Day and the corresponding YCharts series are also related, but they sum newly created outputs rather than previously spent outputs. Further methodological details, validation notes, and limitations are provided in Appendix~\ref{Daily Spent Output Value in BTC}.

\subsection{\texttt{obm\_spent\_value\_ge155d\_btc\_daily}: Spent Output Value Whose Age Is at Least 155 Days in BTC}

The series \texttt{obm\_spent\_value\_ge155d\_btc\_daily} reports the BTC value of spent outputs that had remained unspent for at least 155 days before being consumed. The age threshold is used as a binary filter: outputs aged 155 days or more are included, and younger outputs are excluded. The unit is therefore BTC, not BTC-days. This is important because the series measures thresholded spent value rather than conventional Bitcoin Days Destroyed.

Economically, this metric provides a raw long-term holder spent-value indicator under the 155-day convention. It highlights days in which longer-held coins move on-chain and helps separate older-supply movement from the ordinary turnover of recently active outputs. It is useful for studying dormant-supply activation, distribution episodes, threshold-based UTXO flows, and the share of daily spent value associated with outputs above the 155-day cutoff. The metric should not be interpreted as entity-adjusted long-term holder selling, because it does not identify users, entities, exchanges, custodians, changes of ownership, or self-transfers.

The closest public comparator identified in the appendix is Glassnode's Spent Volume by LTH/STH chart, although Glassnode's LTH/STH methodology is entity-based and uses a smoothed transition around the 155-day boundary rather than OBM's hard spent-output threshold. Glassnode's Spent Volume by Age and CryptoQuant's Spent Output Age Bands are also close conceptual comparators, while LTH-SOPR and long-term-holder supply metrics are related but measure profitability ratios or stock variables rather than thresholded spent value. Further methodological details, validation notes, and limitations are provided in Appendix~\ref{Spent Output Value whose age is at least 155 Days in BTC}.

\subsection{\texttt{obm\_spent\_value\_ge365d\_btc\_daily}: Spent Output Value Whose Age Is at Least 365 Days in BTC}

The series \texttt{obm\_spent\_value\_ge365d\_btc\_daily} reports the BTC value of spent outputs that had remained unspent for at least one year before being consumed. Like the 155-day metric, it uses age as a threshold rather than as a multiplier. Outputs aged at least 365 days are included, and all younger outputs are excluded. The series is measured in BTC, not in BTC-days.

Economically, this metric isolates the movement of long-dormant value. It is more restrictive than the 155-day series and is especially useful for identifying days when older supply becomes active. It can help researchers examine dormant-supply activation, long-term holder behavior, old-coin turnover, and the share of daily spent value associated with outputs that had survived at least one full year without being spent. It should not be interpreted as conventional CDD or as entity-adjusted long-term-holder selling. It remains a raw spent-output threshold metric and can be affected by self-transfers, wallet consolidation, exchange operations, custodial wallet management, and batching.

The closest public comparators identified in the appendix are Glassnode's Spent Volume by Age and CryptoQuant's Spent Output Age Bands, because both classify spent value by output age. A rough public benchmark could be approximated by summing age bands of one year or older, provided the bands are available in compatible BTC units and use comparable conventions. Glassnode's SOAB, HODL Waves, and CryptoQuant's UTXO Age Bands are related, but SOAB may be reported as a relative distribution, and HODL-wave-style indicators are stock metrics rather than spent-value flows. Further methodological details, validation notes, and limitations are provided in Appendix~\ref{Spent Output Value whose age is at least 365 Days in BTC}.

\subsection{\texttt{obm\_spent\_value\_lt155d\_btc\_daily}: Spent Output Value Younger Than 155 Days in BTC}

The series \texttt{obm\_spent\_value\_lt155d\_btc\_daily} reports the BTC value of spent outputs whose age is strictly lower than 155 days. Unlike the older-than-threshold exporters, this metric is derived from two already generated OBM series: total daily spent value and spent value from outputs aged at least 155 days. It is therefore the young-output complement of \texttt{obm\_spent\_value\_ge155d\_btc\_daily}. Together, the two series decompose the total spent value into younger and older spent-output components.

Economically, this metric captures recently active coin turnover. It is useful for studying short-age UTXO flows, short-term-holder behavior under a raw age-threshold convention, and the changing composition of daily spent value. A high value indicates that the day's spent output value came substantially from outputs younger than 155 days. Used together with the \texttt{ge155d} counterpart, it allows researchers to compute young-output and old-output shares of daily spent value. The series should not be interpreted as entity-adjusted short-term holder selling, because it does not classify users or entities and does not filter for self-transfers, changes in outputs, exchange activity, or custodial wallet management.

The closest public comparator identified in the appendix is Glassnode's Spent Volume by LTH/STH chart, especially its short-term-holder component. Glassnode's Spent Volume by Age, Glassnode's SVAB, and CryptoQuant's Spent Output Age Bands are also close conceptual references, though their age bands, entity treatment, smoothing, and reporting formats may differ from OBM's hard-threshold construction. CryptoQuant's STH-SOPR and related short-term-holder supply or profitability metrics are useful contextual indicators, but they are not spent-value metrics. Further methodological details, validation notes, and limitations are provided in Appendix~\ref{Spent Output Value Younger Than 155 Days in BTC}.

\subsection{\texttt{obm\_supply\_btc\_daily}: Bitcoin Supply}

The series \texttt{obm\_supply\_btc\_daily} reports Bitcoin supply as the cumulative sum of realized daily issuance from the beginning of Bitcoin's calendar history. It is derived from \texttt{obm\_issuance\_btc\_daily}, starting on 2009-01-01, and is best interpreted as an end-of-day supply stock obtained by accumulating newly issued BTC. This construction links the monetary stock directly to the realized issuance flow, rather than relying only on the theoretical subsidy schedule.

Economically, this metric is one of the central monetary variables in OBM. It measures the amount of BTC created up to each date and is therefore essential for studying scarcity, monetary expansion, halving regimes, supply growth, and supply-normalized indicators. It is also useful as a bridge between daily issuance, which is a flow, and longer-run monetary quantities, which are stocks. The series should not be interpreted as a measure of liquid supply, active supply, lost coins, available float, or entity-adjusted circulating balances. It is a cumulative realized-issuance measure, not a behavioral or ownership-based supply metric.

The closest public comparator identified in the appendix is Coin Metrics' \texttt{SplyCur}, or Current Supply. Glassnode's \texttt{supply.Current} is also a close public reference, while Blockchain.com's Total Circulating Bitcoin chart, YCharts' Bitcoin Supply series, Newhedge's circulating-supply page, and Blockchair's circulation chart provide useful secondary comparators. The main distinction is that \texttt{obm\_supply\_btc\_daily} is transparently derived from the OBM realized issuance series, whereas some public references may rely on theoretical reward schedules, different timestamp conventions, or undocumented supply-reconstruction choices. Further methodological details, validation notes, and limitations are provided in Appendix~\ref{Bitcoin Supply}.

\subsection{\texttt{obm\_tx\_count\_daily}: Daily Transaction Count}

The series \texttt{obm\_tx\_count\_daily} reports the number of Bitcoin transactions confirmed in blocks assigned to each UTC calendar day. It is computed from block-level metadata by summing the transaction count of all blocks whose timestamps fall on the corresponding date. Coinbase transactions are included because they are valid transactions contained in blocks and are part of the block-level transaction total reported by Bitcoin Core.

Economically, the daily transaction count is a basic but important proxy for on-chain network activity. It captures how many transactions were settled on-chain each day, regardless of their value, fee, size, number of inputs, or number of outputs. Its usefulness increases when interpreted together with other OBM metrics: fees reveal the price paid for inclusion, block weight captures block-space usage, spent value captures input-side value turnover, and CDD captures the age-weighted movement of coins. The metric should not be interpreted as a direct count of users, payments, or economically distinct transfers, because transactions may contain batching, change outputs, self-transfers, exchange activity, custodial operations, and other non-payment uses.

The closest public comparator identified in the appendix is Coin Metrics' \texttt{TxCnt}, which reports the transaction count over the interval. Blockchain.com's Confirmed Transactions Per Day chart, Glassnode's raw transaction-count series, and Blockchair's transaction-count chart are also strong public comparators. Bitbo, Newhedge, BitInfoCharts, YCharts, and Token Terminal provide useful secondary references. The OBM contribution is a simple, reproducible full-node implementation that counts transactions from block metadata, includes coinbase transactions by definition, and assigns blocks to UTC days using an explicit timestamp convention. Further methodological details, validation notes, and limitations are provided in Appendix~\ref{Daily Transaction Count}.

\subsection{\texttt{obm\_utxo\_eod\_count\_daily}: End-of-Day UTXO Count}

The series \texttt{obm\_utxo\_eod\_count\_daily} reports the number of spendable unspent transaction outputs after processing the highest-height block assigned to each UTC calendar day. It is therefore an end-of-day protocol-state metric, not a daily flow. If no block is assigned to a given date under the selected timestamp convention, the value is recorded as missing rather than forcing a misleading zero or carrying forward an artificial state observation.

Economically, the UTXO count measures the size of Bitcoin's spendable output set, which is the ledger state that full nodes must maintain to validate future transactions. It is relevant for scalability, node resource requirements, and the long-run evolution of Bitcoin's state burden. A growing UTXO set may reflect output splitting, wallet fragmentation, many small outputs, or activity patterns that increase the number of spendable outputs. A declining UTXO count may indicate consolidation, where transactions consume more outputs than they create. The metric should not be interpreted as a count of users, wallets, addresses, entities, or BTC value held. It counts unspent outputs, not people or money balances.

The closest public comparator identified in the appendix is Blockchain.com's Unspent Transaction Outputs chart, which reports the number of valid UTXOs and explicitly excludes OP\_RETURN outputs. Glassnode's UTXO Set Growth chart and CryptoQuant's UTXO Count chart are also close conceptual comparators. Exact equivalence should not be assumed, because providers may differ in snapshot timing, treatment of provably unspendable outputs, handling of immature coinbase outputs, no-block dates, timestamp conventions, and chain-reorganization policy. Further methodological details, validation notes, and limitations are provided in Appendix~\ref{End-of-Day UTXO Count}.

\section{Data records}

The dataset is deposited in \url{https://github.com/diegorllanos/open-bitcoin-metrics/}. The repository contains processed time series, accompanying plots, and reproducibility scripts.

Each time series is distributed as a CSV file with the following minimum schema. See the documentation that accompanies each metric for additional details. 

\begin{center}
\begin{tabular}{llp{0.3\textwidth}}
\toprule
\textbf{Column} & \textbf{Type} & \textbf{Description} \\
\midrule
\texttt{date} & Date & Calendar date in UTC \\
\texttt{series\_id} & String & Stable OBM series identifier \\
\texttt{value} & Numeric & Observed value \\
\texttt{unit} & String & Measurement unit \\
\texttt{frequency} & String & Daily or monthly \\
\texttt{release\_version} & String & Dataset release version \\
\bottomrule
\end{tabular}
\end{center}

\section{Usage notes}

The OBM dataset is intended for researchers working on Bitcoin economics, monetary analysis, financial econometrics, blockchain analytics, and digital asset policy. Typical applications include:

\begin{itemize}
    \item Modeling Bitcoin demand using on-chain activity proxies;
    \item Studying long-term holder behavior through coin-age metrics;
    \item Examining fee-market development and miner incentives;
    \item Constructing scarcity-adjusted or supply-normalized indicators;
    \item Teaching reproducible blockchain data analysis.
\end{itemize}

The metric families included in OBM can support several common lines of empirical research in Bitcoin economics. Table~\ref{tab:obm-economic-use-cases} summarizes representative use cases and the OBM series that may be most directly relevant. The table is not intended as an exhaustive mapping, but as a guide for researchers who wish to identify suitable variables for monetary, financial, mining-sector, or on-chain activity studies.

\begin{table}[t]
\centering
\small
\begin{tabular}{p{0.28\textwidth} p{0.38\textwidth} p{0.26\textwidth}}
\hline
\textbf{Research question} & \textbf{Relevant OBM metrics} & \textbf{Economic interpretation} \\
\hline
Monetary issuance and scarcity &
\texttt{obm\_issuance\_btc\_daily}, \texttt{obm\_supply\_btc\_daily}, \texttt{obm\_block\_count\_daily} &
Daily monetary creation, cumulative supply, and realized block production across subsidy eras. \\
\hline
Miner incentives and security budget &
\texttt{obm\_fees\_btc\_daily}, \texttt{obm\_miner\_revenue\_btc\_daily}, \texttt{obm\_fee\_share\_revenue\_ratio\_daily}, \texttt{obm\_difficulty\_eod\_daily}, \texttt{obm\_est7d\_hashrate\_ehs\_daily} &
BTC-denominated miner compensation, fee dependence, mining difficulty, and estimated network hashrate. \\
\hline
Transaction demand and block-space pressure &
\texttt{obm\_tx\_count\_daily}, \texttt{obm\_block\_weight\_wu\_daily}, \texttt{obm\_fees\_btc\_daily}, \texttt{obm\_raw\_output\_value\_btc\_daily} &
Confirmed transaction activity, realized block-space usage, fee-market pressure, and raw output-side value activity. \\
\hline
Long-term holder and dormant-supply behavior &
\texttt{obm\_cdd\_btcxdays\_daily}, \texttt{obm\_cdd\_per\_supply\_days\_daily}, \texttt{obm\_dormancy\_days\_daily}, \texttt{obm\_liveliness\_ratio\_daily}, \texttt{obm\_spent\_value\_ge155d\_btc\_daily}, \texttt{obm\_spent\_value\_ge365d\_btc\_daily} &
Coin-age destruction, average spent age, cumulative holding versus spending behavior, and threshold-based movement of older outputs. \\
\hline
UTXO-state evolution and node burden &
\texttt{obm\_utxo\_eod\_count\_daily}, \texttt{obm\_spent\_output\_count\_daily}, \texttt{obm\_spent\_value\_age\_band\_btc\_daily}, \texttt{obm\_cdd\_age\_band\_btcxdays\_daily} &
Evolution of the spendable output set, input-side UTXO turnover, and age composition of spent value and destroyed coin age. \\
\hline
\end{tabular}
\caption{Illustrative economic use cases for OBM metric families.}
\label{tab:obm-economic-use-cases}
\end{table}

When using these time series, users should nonetheless consider the following caveats:

\begin{itemize}
    \item On-chain transactions do not map one-to-one to economic transactions;
    \item Address-based metrics do not identify individual users;
    \item Gross spent output value may include self-churn and change outputs;
    \item USD-denominated metrics depend on external price sources;
    \item Daily aggregation depends on the chosen timestamp convention.
\end{itemize}

\section{Code and data availability}

All data series, Python scripts, documentation files, plots, and reproducibility instructions associated with Open Bitcoin Metrics are available in the project repository:

\begin{center}
\url{https://github.com/diegorllanos/open-bitcoin-metrics/}
\end{center}

The repository contains the processed OBM time series, metric-specific Python scripts, README files documenting the definition and execution of each metric, validation notes, and the standard output schema used across the dataset. Each metric is associated with a specific script distributed in the same repository. Daily rolling updates are maintained in the GitHub repository, so the main branch should be understood as the live development and update location of the project.

For citation and reproducibility, the dataset version described in this manuscript is archived as \textbf{Open Bitcoin Metrics v0.1.0}. This version corresponds to the first public release of the OBM dataset and reference guide. The archived release provides a fixed snapshot of the code, documentation, and data files used to support the present manuscript. The release can be accessed at:

\begin{center}
\url{https://github.com/diegorllanos/open-bitcoin-metrics/releases/tag/v0.1.0}
\end{center}

A persistent archival copy of this release is available at:

\begin{center}
\url{https://doi.org/10.5281/zenodo.21156871}
\end{center}

Users who rely on the dataset for empirical work should cite both this manuscript and the archived dataset release, rather than only the live GitHub repository, because the latter may evolve as metrics are added, corrected, or extended. A \texttt{CITATION.cff} file is provided in the repository to facilitate citation through GitHub, Zenodo, Zotero, and other bibliographic tools.

The standard citation for the archived dataset release is:

\begin{quote}
Llanos, D. R. (2026). \textit{Open Bitcoin Metrics (OBM): Reproducible full-node Bitcoin on-chain time series, version 0.1.0}. Zenodo. \url{https://doi.org/10.5281/zenodo.21156871}
\end{quote}

The release identifier \texttt{OBM v0.1.0} is also recorded in the \texttt{release\_version} field of the distributed CSV files. This field allows users to track which dataset release produced a given observation and helps distinguish archived releases from later rolling updates.

The scripts are intended to be rerunnable by users who maintain a synchronized Bitcoin Core full node with the configuration described in Section~3.2. Some metrics can be computed directly from block-level data through Bitcoin Core RPC calls, while metrics involving transaction fees, spent value, Bitcoin Days Destroyed, dormancy, miner revenue, realized issuance, and UTXO-age indicators require the spent-output indexer described in Section~\ref{sec:The Spent-Output Indexer}. A minimal example of metric regeneration is:

\begin{verbatim}
python obm_tx_count_daily.py \
  --start_date 2024-01-01 \
  --end_date 2024-01-31 \
  --use_default_cookie \
  --release_version "OBM v0.1.0" \
  --output obm_tx_count_daily.csv
\end{verbatim}

The exact command-line options differ across metrics and are documented in the corresponding README files and in Appendix~A. The general reproducibility workflow is to synchronize a Bitcoin Core full node, run the required metric-specific script or build the spent-output indexer when previous-output reconstruction is needed, export the standardized CSV file, and compare the result with the archived \texttt{OBM v0.1.0} release or with the validation checks documented for each metric.

\section{Limitations}

The dataset has several limitations. First, the initial release focuses on metrics that can be reconstructed from full-node data with transparent assumptions. Second, some economically meaningful quantities, such as adjusted transaction volume or entity-level activity, require heuristics that may introduce classification errors. Third, comparison with commercial data providers is complicated by differences in definitions, timestamp conventions, and filtering rules. Fourth, daily updates may be revised in later releases if bugs, definitional refinements, or edge cases involving blockchain reorganization are identified.

\section{Acknowledgements}

Prof. Llanos acknowledges financial support by the Spanish Ministerio de Ciencia e Innovación and by the European Regional Development Fund (ERDF) program of the European Union, under Grant PID2022-142292NB-I00 (NATASHA Project); and by the Junta de Castilla y León Regional Goverment, Spain, under Grant VA127P25 (HUSAR Project).

\section{Author contributions}

D.R.L. designed the dataset, implemented the extraction and validation pipelines, wrote the software, curated the data records, and prepared the manuscript. ChatGPT was used as an editorial and software-development assistant. The author reviewed, verified, and takes responsibility for all code, data, references, and manuscript content.

\section{Declaration of Competing interests}

The author declares no competing interests.


\appendix

\section{Reference guide of OBM available metrics}
\label{Reference guide of OBM available metrics}

This section contains one subsection per metric. Each subsection includes its identifier, its mathematical definition, the economic interpretation of the metric, data source and input requirements, a sketch of the algorithm used to calculate it, a description of the metric-specific input parameters, the aggregation rule, the output format, the method used for technical validation, and a discussion of known limitations and usefulness of this metric. 

\subsection{\texttt{obm\_block\_count\_daily}: Daily Block Count}
\label{sec:Daily Block Count}


\paragraph{Definition.}
The daily block count measures the number of Bitcoin blocks assigned to a given UTC calendar day. Let \(B_d\) denote the set of blocks assigned to day \(d\). The daily block count is defined as:

\[
\mathrm{BlockCount}_d = |B_d|,
\]

where \(|B_d|\) denotes the cardinality of the set of blocks assigned to day \(d\).

A block \(b\) is assigned to day \(d\) according to the UTC calendar date derived from its block timestamp \(t_b\): $d(b) = \mathrm{UTCDate}(t_b)$. The resulting series therefore reports the total number of blocks included in the Bitcoin main chain per UTC day. Since Bitcoin targets an average inter-block interval of approximately ten minutes, the expected long-run average is approximately 144 blocks per day. However, the realized daily number of blocks fluctuates because block discovery is probabilistic, mining difficulty adjusts discretely, and block timestamps do not necessarily align with exact calendar-day boundaries.

\paragraph{Economic interpretation.}
The daily block count is a basic indicator of Bitcoin block production. It does not directly measure transaction demand, user activity, or settlement value. Rather, it captures the number of blocks available for transaction settlement on each UTC day.

For economic research, this metric is useful in three main ways. First, it provides a transparent measure of realized block production, allowing researchers to observe deviations from the expected average of approximately 144 blocks per day. Second, it is useful as a normalization variable for other block-level aggregates, such as transaction count, total fees, miner revenue, issuance, or block weight. Third, it helps interpret daily variation in other on-chain metrics. For example, a lower-than-usual daily transaction count may partly reflect fewer blocks produced that day rather than weaker transaction demand.

The metric should not be interpreted as a direct measure of network health. Short-run deviations from 144 blocks per day are normal under Bitcoin's probabilistic block-discovery process. Persistent deviations may instead reflect changes in aggregate hash rate, the timing of difficulty adjustments, or other mining-related dynamics.

\paragraph{Similar metrics publicly available}

The OBM series \texttt{obm\_block\_count\_daily} is directly comparable to several publicly available Bitcoin block-count metrics. The closest public equivalent is Coin Metrics' \emph{Block Cnt} metric, with MetricID \texttt{BlkCnt}.\footnote{\url{https://community-api.coinmetrics.io/v4/timeseries/asset-metrics?assets=btc\&metrics=BlkCnt}} Coin Metrics defines it as ``the sum count of blocks created that interval that were included in the main (base) chain,'' reports it in units of blocks, and makes it available at one-day and one-hour intervals through its asset-metrics API.

Coin Metrics also documents two important methodological choices: only main-chain, non-orphaned blocks are counted, and interval assignment is based on median time for chains that use median time, otherwise on block timestamps. For Bitcoin, this is therefore highly comparable to \texttt{obm\_block\_count\_daily}, although exact daily values could differ if Coin Metrics applies a different time convention, finality policy, or chain-reorganization handling.

Glassnode provides a closely related \emph{Bitcoin Blocks Mined} chart, internally exposed as \texttt{blockchain.BlockCount}.\footnote{\url{https://studio.glassnode.com/charts/blockchain.BlockCount?a=BTC}} Glassnode describes the metric as ``the number of blocks created and included in the main blockchain in that time period.'' The conceptual definition is therefore very close to \texttt{obm\_block\_count\_daily}. However, the public chart page does not provide a full reproducible algorithm comparable to the \texttt{obm\_block\_count\_daily} script. It indicates data access through CSV, JSON, API, Excel, and MCP interfaces, but the exact implementation details, including timestamp convention, confirmation policy, and reorganization treatment, are not fully specified on the public metric page.

Blockchain.com does not appear to expose a clearly named public \texttt{n-blocks} chart in the currently indexed chart list, but it does provide a Charts and Statistics API whose date parameters are documented as UTC and whose charts are programmatically accessible. Its block explorer also exposes current and historical block-level information. Consequently, Blockchain.com can be used to reconstruct a daily block-count series by grouping blocks by timestamp, but we have not found a currently documented, directly labeled daily block-count endpoint with an explicit algorithm equivalent to \texttt{obm\_block\_count\_daily}. Related Blockchain.com chart pages usually provide a short methodology statement for the specific chart, but we have not found such a methodology page for the daily block count itself.

Bitbo provides a \emph{Bitcoin Total Number of Blocks Daily} chart.\footnote{\url{https://charts.bitbo.io/blocks-daily/}} The page states that the chart represents the total number of blocks mined each day and explains the 144-block-per-day target. However, it also says that the chart attempts to report actual blocks per day ``while using a 10-day moving average,'' which makes it less directly comparable to \texttt{obm\_block\_count\_daily} if the plotted or exported series is smoothed rather than a raw daily count. The page provides an interpretation, but not a full reproducible algorithm.

Newhedge provides a \emph{Bitcoin Blocks per Day} chart.\footnote{\url{https://newhedge.io/bitcoin/blocks-per-day}} It defines the metric as the number of blocks mined on the Bitcoin blockchain within a 24-hour period and exposes a documented API endpoint\footnote{\url{https://newhedge.io/api/v2/metrics/blocks-per-day/blocks\_per\_day}}, but the page indicates that API access requires a paid plan. The public definition is sufficient to identify the series as a close comparator, but it does not disclose enough implementation detail to determine exact equivalence with \texttt{obm\_block\_count\_daily}.

BitInfoCharts reports current network statistics, including \emph{Blocks Count}, \emph{Blocks last 24h}, and \emph{Blocks avg. per hour}.\footnote{\url{https://bitinfocharts.com/}} This is useful as a current or recent diagnostic comparator, but it is not a clearly documented downloadable daily historical time series for \texttt{obm\_block\_count\_daily}. We have not found a public algorithm beyond the apparent interpretation that it counts blocks over the last 24 hours.

Finally, Blockchair and mempool.space provide block-level data access rather than a simple, pre-packaged daily block-count metric. Blockchair offers Bitcoin block data and a daily updated TSV database dumps. \footnote{\url{https://blockchair.com/bitcoin/blocks}}\footnote{\url{https://blockchair.com/dumps}}. 
mempool.space provides a public REST API for blocks and is itself an open-source Bitcoin explorer.\footnote{\url{https://mempool.space/docs/api/rest}}\footnote{\url{https://github.com/mempool/mempool}} Both sources could be used to reconstruct a series equivalent to \texttt{obm\_block\_count\_daily} by counting blocks whose timestamps fall within each UTC day. Nevertheless, they should be treated as reconstruction sources rather than as named daily block-count metrics unless the researcher explicitly documents the reconstructed grouping convention.

\paragraph{Data source and input requirements.}
The metric is obtained from a running Bitcoin Core full node through the JSON-RPC interface. For each block, the script queries Bitcoin Core to obtain the block hash and then retrieves the decoded block object. The block timestamp is read from the block metadata returned by Bitcoin Core and converted into a UTC calendar date.

This metric does not require reconstructing previous transaction outputs, accessing the UTXO set, extracting addresses, parsing transaction values, computing fees, or using external price data. Consequently, it is one of the simplest and most robust OBM metrics. It can be computed solely from block-level information.

\paragraph{Algorithm.}
The script \texttt{compute\_obm\_block\_count\_daily.py} implements the following procedure:

\begin{enumerate}
    \item Parse the user-provided date interval, \texttt{\symbol{45}\symbol{45}start\_date} and \texttt{\symbol{45}\symbol{45}end\_date}, using the format \texttt{YYYY-MM-DD}. Both dates are interpreted as UTC and included in the output.

    \item Convert the starting date into the timestamp corresponding to 00:00:00 UTC of that day, and the ending date into the timestamp corresponding to 23:59:59 UTC of that day.

    \item Query the local Bitcoin Core node using \texttt{getblockchaininfo} to determine the current best chain height.

    \item Use \texttt{getblockhash} and \texttt{getblock} to retrieve block timestamps and locate an approximate height interval covering the requested date range. This is done through a binary search over block heights.

    \item Expand the approximate height interval using the metric-specific safety parameter \texttt{\symbol{45}\symbol{45}height\_margin}. This margin is added before the estimated starting height and after the estimated ending height.

    \item Scan all blocks in the expanded height interval. For each height \(h\), the script:
    \begin{enumerate}
        \item obtains the corresponding block hash using \texttt{getblockhash};
        \item retrieves the decoded block object using \texttt{getblock};
        \item extracts the block timestamp \(t_b\);
        \item assigns the block to a UTC calendar date \(d(b)\);
        \item adds one unit to the daily count if \(d(b)\) lies between the requested starting and ending dates.
    \end{enumerate}

    \item Initialize all dates in the requested interval with a value of zero before scanning. This ensures that the output file contains one row for each calendar day in the requested interval, even if no block was assigned to that day.

    \item Write the resulting time series to a CSV file using the standardized OBM schema:
    \[
    \texttt{date},\quad
    \texttt{series\_id},\quad
    \texttt{value},\quad
    \texttt{unit},\quad
    \texttt{frequency},\quad
    \texttt{release\_version}.
    \]

    \item Optionally generate a plot of the resulting series when the plotting flag is activated. The plot title includes the series description and the selected date interval.
\end{enumerate}

\paragraph{Metric-specific input parameter.}
The only input parameter specific to this metric is:

\begin{itemize}
    \item \texttt{\symbol{45}\symbol{45}height\_margin}: number of extra blocks scanned before and after the approximate height interval associated with the requested date range.
\end{itemize}

This parameter is needed because Bitcoin block timestamps are not strictly monotonic with respect to block height. Although block heights are strictly ordered by chain position, miners' timestamps can occasionally shift slightly backward or forward relative to neighboring blocks. As a result, a binary search based only on block timestamps could identify a height interval that is too narrow and accidentally exclude blocks whose timestamps fall inside the requested UTC date interval. To avoid this problem, the script first obtains an approximate height interval and then scans an expanded interval $[h_{\min}^{\mathrm{scan}}, h_{\max}^{\mathrm{scan}}]$, where:

\[
h_{\min}^{\mathrm{scan}}
=
\max(0, h_{\min}^{\mathrm{approx}} - m),
\]
\[
h_{\max}^{\mathrm{scan}}
=
\min(h_{\mathrm{tip}}, h_{\max}^{\mathrm{approx}} + m),
\]

and \(m\) is the value of \texttt{\symbol{45}\symbol{45}height\_margin}. The default value is 288 blocks, approximately two days of Bitcoin block production at the expected rate of 144 blocks per day. This choice is deliberately conservative. The margin does not change the dates written to the output file; it only widens the internal block scan. During the scan, the script still counts only blocks whose UTC dates fall between \texttt{\symbol{45}\symbol{45}start\_date} and \texttt{\symbol{45}\symbol{45}end\_date}, inclusive.

A larger value of \texttt{\symbol{45}\symbol{45}height\_margin} increases robustness at the cost of scanning more blocks. A smaller value improves speed but increases the risk of missing boundary blocks in unusual timestamp configurations. For ordinary daily updates, the default value is sufficient. For full historical reconstruction or formal archival releases, a larger margin can be used as an additional safeguard. The OBM repository uses the default value.

\paragraph{Aggregation rule.}
The daily value is computed as the number of blocks assigned to the same UTC calendar date:
 
\[
\mathrm{BlockCount}_d =
\sum_{b: \mathrm{UTCDate}(t_b)=d} 1.
\]

The aggregation rule is therefore a daily count. Monthly versions of this metric, if distributed, should also be computed as sums of the corresponding daily values:

\[
\mathrm{BlockCount}_m =
\sum_{d \in m} \mathrm{BlockCount}_d.
\]

\paragraph{Output format.}
The output file contains one observation per UTC date. Each row has the following fields:

\begin{center}
\begin{tabular}{llp{0.45\textwidth}}
\toprule
\textbf{Column} & \textbf{Example} & \textbf{Description} \\
\midrule
\texttt{date} & \texttt{2024-01-01} & UTC calendar date \\
\texttt{series\_id} & \texttt{obm\_block\_count\_daily} & Stable OBM series identifier \\
\texttt{value} & \texttt{147} & Number of blocks assigned to the UTC date \\
\texttt{unit} & \texttt{blocks} & Measurement unit \\
\texttt{frequency} & \texttt{daily} & Observation frequency \\
\texttt{release\_version} & \texttt{OBM v0.1.0} & Dataset release version \\
\bottomrule
\end{tabular}
\end{center}

\paragraph{Technical validation.}
Several checks are either implemented by the script or recommended for validating this metric. First, the script verifies that the requested date range is valid and that the starting date is not later than the ending date. Second, it checks that the requested interval does not extend beyond the current chain tip. Third, the output is initialized with all calendar dates in the requested interval, which helps detect missing observations. Fourth, the daily values can be independently checked by counting the number of blocks whose timestamps fall within each UTC day. Finally, the number of scanned blocks and the effective height interval are printed during execution, allowing the user to verify that the requested period has been covered.

For additional validation, the total number of counted blocks over a period can be compared with the number of scanned blocks whose timestamps fall inside the requested date interval. Selected periods can also be compared with external Bitcoin data providers. Such comparisons should be interpreted as diagnostic rather than definitive, because providers may differ in their timestamp conventions, treatment of chain reorganizations, or daily boundary rules.

\paragraph{Known limitations.}
The daily block count is simple and reproducible, but it has several limitations. First, it measures blocks, not transactions, users, transaction demand, or economic activity. Second, it does not contain information on block size, block weight, transaction fees, transaction volume, or miner revenue. Third, short-run deviations from the expected value of approximately 144 blocks per day are normal under Bitcoin's probabilistic block-discovery process and should not be interpreted mechanically as changes in network health. Fourth, the assignment of blocks to days depends on the block timestamp convention. In OBM, the default convention is to use the UTC date derived from the block timestamp returned by Bitcoin Core. This convention is transparent and reproducible, but alternative conventions, such as median time past, could produce small differences near daily boundaries.

Despite these limitations, \texttt{obm\_block\_count\_daily} is a useful baseline series in the OBM dataset. It provides a transparent measure of realized block production and serves as a natural normalization variable for interpreting other daily metrics, such as transaction count, issuance, fees, miner revenue, and block-space utilization.
\subsection{\texttt{obm\_block\_weight\_wu\_daily}: Daily Block Weight in Weight Units}
\label{Daily Block Weight in Weight Units}

\paragraph{Definition.}
The daily block weight series measures the total block weight of all Bitcoin blocks assigned to a given UTC calendar day. Let \(B_d\) denote the set of blocks assigned to day \(d\). For each block \(b \in B_d\), let \(W_b\) denote the block weight, measured in Bitcoin weight units. The daily block weight series is defined as:
\[
\mathrm{BlockWeightWU}_d =
\sum_{b \in B_d}
W_b.
\]

A block \(b\) is assigned to day \(d\) according to the UTC calendar date derived from its block timestamp \(t_b\): $d(b)=\mathrm{UTCDate}(t_b)$. The resulting series therefore reports the total amount of block weight included in the Bitcoin main chain per UTC day. Its unit is \texttt{WU}, denoting Bitcoin weight units. This is a daily total, not an average per block.

\paragraph{Economic interpretation.}
The daily block weight series is a measure of Bitcoin's block space usage. It complements the daily block count and the daily transaction count by capturing the amount of weighted block data included in the chain on each UTC date. Since Bitcoin's post-SegWit block capacity constraint is expressed in weight units, block weight is the natural block-space metric for modern Bitcoin analysis.

For economic research, this metric is useful in several ways. First, it provides a transparent measure of realized block-space usage. Second, it helps interpret fee-market dynamics, because higher block-space usage may be associated with stronger competition for inclusion in blocks. Third, it can be combined with transaction fees to construct fee-per-weight-unit indicators. Fourth, it can be used in conjunction with block count to distinguish days with high total block weight, since many blocks were produced on days with high average block weight per block. Fifth, it provides a useful normalization variable for transaction count, raw output value, fees, and other activity measures.

The metric should not be interpreted as a direct measure of economic activity or transaction demand. Total daily block weight depends partly on the number of blocks assigned to the day, which fluctuates under Bitcoin's probabilistic block-discovery process. A day with unusually high block weight may reflect either more blocks, fuller blocks, or both. For this reason, the average block weight per block is a useful derived diagnostic:
\[
\mathrm{AvgBlockWeightWU}_d =
\frac{
\mathrm{BlockWeightWU}_d
}{
\mathrm{BlockCount}_d
},
\qquad
\mathrm{BlockCount}_d > 0.
\]

\paragraph{Similar metrics publicly available.}
The OBM series \texttt{obm\_block\_weight\_wu\_daily} is comparable to public metrics usually labeled \emph{block weight}, \emph{sum block weight}, \emph{mean block weight}, \emph{block size}, or \emph{block-space utilization}. In OBM, the metric is defined as the total block weight of all Bitcoin blocks assigned to a given UTC calendar day. Let \(B_d\) denote the set of blocks assigned to UTC date \(d\), and let \(\mathrm{Weight}_b\) denote the block weight of block \(b\), as returned by Bitcoin Core. The daily value is
\[
\mathrm{BlockWeightWU}_d
=
\sum_{b \in B_d} \mathrm{Weight}_b .
\]
The unit is weight units (WU). This is a total daily aggregate, not an average per block. Therefore, daily variation reflects both the number of blocks assigned to the UTC day and the amount of block weight each block uses. The average block weight can be derived separately as
\[
\frac{\texttt{obm\_block\_weight\_wu\_daily}}
{\texttt{obm\_block\_count\_daily}},
\]
for dates with a positive block count.

The closest public comparator is Coin Metrics' \emph{Sum Block Weight}, with MetricID \texttt{BlkWghtTot}\footnote{\url{https://community-api.coinmetrics.io/v4/timeseries/asset-metrics?assets=btc&metrics=BlkWghtTot}} (credentials required). Coin Metrics defines this metric as the sum weight of all blocks created during the interval and makes it available at one-day frequency. It also provides \emph{Mean Block Weight}, with MetricID \texttt{BlkWghtMean} \footnote{\url{https://community-api.coinmetrics.io/v4/timeseries/asset-metrics?assets=btc&metrics=BlkWghtMean}} (credentials required), defined as the mean weight of all blocks created during the interval. These metrics are highly relevant because they use the same consensus-based block weight rather than traditional block size in bytes. The main conceptual difference is that \texttt{obm\_block\_weight\_wu\_daily} is explicitly a total daily series, so the direct Coin Metrics comparator is \texttt{BlkWghtTot}, while \texttt{BlkWghtMean} corresponds more closely to a derived OBM average block weight. Exact equality should not be assumed without checking timestamp assignment, finality policy, chain-reorganization handling, and whether the provider uses block timestamps or another interval convention. 

Newhedge provides a \emph{Bitcoin Block Weight} chart.\footnote{\url{https://newhedge.io/bitcoin/block-weight}} The page defines Bitcoin Block Weight as the daily average block weight. This makes it related, but not identical, to \texttt{obm\_block\_weight\_wu\_daily}. Newhedge's metric appears to be an average per block, whereas OBM reports the total daily sum of block weights. A comparable OBM diagnostic would therefore be \texttt{obm\_block\_weight\_wu\_daily} divided by \texttt{obm\_block\_count\_daily}. The public Newhedge page is useful as a chart-level comparator, but it does not provide a full reproducible algorithm specifying timestamp convention, treatment of reorganizations, or whether the reported daily value is the arithmetic mean, median, or another aggregation. 

Bitcoin Visuals provides a \emph{Block Weight} chart.\footnote{\url{https://bitcoinvisuals.com/chain-block-weight}} The page describes block weight as a measure of Bitcoin block size that incorporates the SegWit witness discount, gives the formula $\mathrm{Weight} = 3 \times \mathrm{stripped\ size} + \mathrm{total\ size}$, and notes the consensus limit of 4 million weight units per block. However, the chart is described as ``daily median block weight'', not ``total daily block weight''. Therefore, it is useful for validating the block-weight concept and for comparing central tendencies in block-space usage, but it is not a direct equivalent of \texttt{obm\_block\_weight\_wu\_daily}. A direct comparison would require either deriving a median from OBM block-level data or, if available, summing Bitcoin Visuals block-level weights. 

Glassnode provides several related block-size metrics, including \emph{Bitcoin Block Size (Mean)}\footnote{\url{https://studio.glassnode.com/charts/blockchain.BlockSizeMean?a=BTC}} and \emph{Bitcoin Block Size (Total)}. \footnote{\url{https://studio.glassnode.com/charts/blockchain.BlockSizeSum?a=BTC}} Glassnode defines the mean block-size metric as the mean size of all blocks created within the time period, measured in bytes, and the total block-size metric as the total size of all blocks created within the time period, also measured in bytes. These are related block-space metrics, but they are not direct equivalents because they use bytes rather than weight units. Since SegWit, block weight is the consensus-relevant capacity measure, while byte size alone does not fully capture the witness discount. Glassnode's block-size series can therefore be useful secondary comparators for broad block-space usage, but they should not replace a WU-denominated series when the research question concerns post-SegWit block capacity. 

Blockchain.com provides an \emph{Average Block Size (MB)} chart.\footnote{\url{https://www.blockchain.com/explorer/charts/avg-block-size}} The page describes the metric as the average block size over the past 24 hours in megabytes. This is related to block-space usage, but it differs from \texttt{obm\_block\_weight\_wu\_daily} in three ways: it reports an average rather than a total, it uses megabytes rather than weight units, and it does not directly reflect the SegWit weight formula. Blockchain.com is therefore useful as a historical block-size comparator, but not as a direct benchmark for daily total block weight. 

Blockchair provides block-level explorer data and a Bitcoin charts catalog.\footnote{\url{https://blockchair.com/bitcoin/blocks}} Although Blockchair does not appear to provide a directly named daily total block-weight metric equivalent to \texttt{obm\_block\_weight\_wu\_daily}, it can serve as a reconstruction source if block-level weight fields are available through its explorer or API. A researcher could sum block weights by UTC date to build an external series. However, unless the provider publishes a named total daily block-weight chart with documented aggregation rules, Blockchair should be treated as a lower-level data source rather than as a direct public comparator.

Other public sources often report block size rather than block weight. These include blockchain-size charts, average block-size charts, and capacity-utilization indicators. Such metrics are related because they measure the amount of data or block space used by the network, but they are not equivalent to \texttt{obm\_block\_weight\_wu\_daily}. The distinction is especially important after SegWit, because weight units are the consensus measure that distinguishes between prices witnessed and non-witnessed data. A bytes-based metric can be informative for storage and bandwidth analysis, while a weight-based metric is more appropriate for block-capacity and fee-market analysis.

Overall, the strongest public comparator for \texttt{obm\_block\_weight\_wu\_daily} is Coin Metrics \texttt{BlkWghtTot}, because it explicitly reports the sum of block weights over the interval. Coin Metrics \texttt{BlkWghtMean}, Newhedge's \emph{Bitcoin Block Weight}, and Bitcoin Visuals' \emph{Block Weight} chart are useful related comparators, but they report average or median block weight rather than total daily block weight. Glassnode and Blockchain.com provide block-size metrics in bytes or megabytes, which are useful for broader block-space analysis but are not direct WU-denominated equivalents. The distinctive contribution of OBM is that it computes total daily block weight directly from the Bitcoin Core block-level \texttt{weight} field, assigns blocks to UTC calendar days using a documented timestamp convention, reports integer weight units, and provides a reproducible full-node-derived series suitable for fee-market and block-space utilization studies.

\paragraph{Data source and input requirements.}
The metric is obtained from a running Bitcoin Core full node through the JSON-RPC interface. For each block in the selected interval, the script retrieves the decoded block object using:
\[
\texttt{getblock <block\_hash> 1}.
\]
The block weight is read from the block-level \texttt{weight} field returned by Bitcoin Core.

This metric does not require reconstructing previous transaction outputs. It does not require the spent-output indexer database, transaction-level input resolution, address extraction, user clustering, entity identification, external price data, or third-party APIs. It also does not require \texttt{txindex=1}, because the required information is available from block metadata.

The metric requires access to a synchronized Bitcoin Core full node and JSON-RPC credentials or cookie authentication. The script uses the locally verified main chain reported by the node. As with other directly scanned OBM metrics, the block timestamp returned by Bitcoin Core is used to assign each block to a UTC calendar date.

\paragraph{Algorithm.}
The script \texttt{compute\_obm\_block\_weight\_wu\_daily.py} implements the following procedure:

\begin{enumerate}
    \item Parse the user-provided date interval, \texttt{\symbol{45}\symbol{45}start\_date} and \texttt{\symbol{45}\symbol{45}end\_date}, using the format \texttt{YYYY-MM-DD}. Both dates are interpreted as UTC dates and both are included in the output.

    \item Convert the starting date into the timestamp corresponding to 00:00:00 UTC of that day, and the ending date into the timestamp corresponding to 23:59:59 UTC of that day.

    \item Connect to the local Bitcoin Core node through JSON-RPC, using either explicit RPC credentials, environment variables, or cookie authentication.

    \item Query the local node using \texttt{getblockchaininfo} to determine the current best-chain height.

    \item Locate an approximate height interval covering the requested date range. This is done by using block timestamps and binary search over block heights.

    \item Expand the approximate height interval using the metric-specific safety parameter \texttt{\symbol{45}\symbol{45}height\_margin}. This reduces the risk of missing boundary blocks because Bitcoin block timestamps are not strictly monotonic with respect to height.

    \item Initialize all dates in the requested interval with value zero. This ensures that the output file contains one row for each selected UTC date.

    \item Scan every block in the expanded height interval. For each height \(h\), the script:
    \begin{enumerate}
        \item obtains the corresponding block hash using \texttt{getblockhash};
        \item retrieves the decoded block using \texttt{getblock} with verbosity level 1;
        \item extracts the block timestamp \(t_b\);
        \item assigns the block to a UTC date \(d(b)\);
        \item skips the block if \(d(b)\) falls outside the requested date interval;
        \item reads the block-level \texttt{weight} field;
        \item adds the block weight to the daily total for \(d(b)\).
    \end{enumerate}

    \item Write the resulting time series to a CSV file using the standardized OBM schema:
    \[
    \texttt{date},\quad
    \texttt{series\_id},\quad
    \texttt{value},\quad
    \texttt{unit},\quad
    \texttt{frequency},\quad
    \texttt{release\_version}.
    \]

    \item Optionally generate a plot of the resulting series when the plotting flag is activated.
\end{enumerate}

The script therefore computes the metric directly from decoded block metadata. It does not query the spent-output indexer, does not resolve transaction inputs, and does not reconstruct previous outputs.

\paragraph{Metric-specific input parameters.}
The metric-specific input parameter is:

\begin{itemize}
    \item \texttt{\symbol{45}\symbol{45}height\_margin}: number of extra blocks scanned before and after the approximate height interval associated with the requested date range.
\end{itemize}

This parameter is needed because Bitcoin block timestamps are not strictly monotonic with respect to block height. A timestamp-based binary search provides only an approximate height interval. To avoid excluding blocks whose timestamps fall inside the requested UTC interval but whose heights lie slightly outside the approximate interval, the script expands the scan interval by \(m\) blocks on both sides:
\[
h^{\mathrm{scan}}_{\min} =
\max(0, h^{\mathrm{approx}}_{\min} - m),
\]
\[
h^{\mathrm{scan}}_{\max} =
\min(h_{\mathrm{tip}}, h^{\mathrm{approx}}_{\max} + m).
\]

The default value is 288 blocks, approximately two days of expected Bitcoin block production. This is deliberately conservative. The margin only widens the internal block scan; the script still counts only blocks whose UTC dates fall inside the requested interval.

The script also accepts standard RPC, output, release-version, and plotting parameters:
\texttt{\symbol{45}\symbol{45}rpc\_url}, \texttt{\symbol{45}\symbol{45}rpc\_user}, \texttt{\symbol{45}\symbol{45}rpc\_password}, \texttt{\symbol{45}\symbol{45}cookie\_path}, \texttt{\symbol{45}\symbol{45}use\_default\_cookie}, \texttt{\symbol{45}\symbol{45}rpc\_timeout}, \texttt{\symbol{45}\symbol{45}output}, \texttt{\symbol{45}\symbol{45}release\_version}, \texttt{\symbol{45}\symbol{45}plot}, \texttt{\symbol{45}\symbol{45}plot\_output}, and \texttt{\symbol{45}\symbol{45}quiet}.

\paragraph{Aggregation rule.}
The daily value is computed as the sum of block weights for all blocks assigned to the same UTC calendar date:
\[
\mathrm{BlockWeightWU}_d =
\sum_{b: \mathrm{UTCDate}(t_b)=d}
W_b.
\]

The aggregation rule is therefore a daily sum. Monthly versions of this metric, if distributed, should also be computed as sums of the corresponding daily values:
\[
\mathrm{BlockWeightWU}_m =
\sum_{d \in m}
\mathrm{BlockWeightWU}_d.
\]

This convention preserves the interpretation of the metric as the total amount of block weight included over the corresponding period.

\paragraph{Output format.}
The output file contains one observation per UTC date. Each row has the following fields:

\begin{center}
\begin{tabular}{llp{0.40\textwidth}}
\toprule
\textbf{Column} & \textbf{Example} & \textbf{Description} \\
\midrule
\texttt{date} & \texttt{2024-01-01} & UTC calendar date \\
\texttt{series\_id} & \texttt{obm\_block\_weight\_wu\_daily} & Stable OBM series identifier \\
\texttt{value} & \texttt{568923456} & Total block weight assigned to the UTC date \\
\texttt{unit} & \texttt{WU} & Weight units \\
\texttt{frequency} & \texttt{daily} & Observation frequency \\
\texttt{release\_version} & \texttt{OBM v0.1.0} & Dataset release version \\
\bottomrule
\end{tabular}
\end{center}

Values are integer quantities measured in weight units. No floating-point arithmetic is required for the metric itself.

\paragraph{Technical validation.}
Several internal checks are used to validate this metric during execution. First, the script verifies that the requested date range is valid and that \texttt{\symbol{45}\symbol{45}start\_date} is not later than \texttt{\symbol{45}\symbol{45}end\_date}. Second, it verifies that \texttt{\symbol{45}\symbol{45}height\_margin} is non-negative. Third, it checks that RPC authentication is valid, including cookie-file existence and format when cookie authentication is used. Fourth, it connects to Bitcoin Core and retrieves the current chain tip using \texttt{getblockchaininfo}. Fifth, it locates and expands the block-height interval corresponding to the requested timestamp range. Sixth, it scans all blocks in the expanded interval and counts only blocks whose UTC date lies inside the requested interval. Seventh, it verifies that each counted block contains a \texttt{weight} field. Eighth, it rejects negative block-weight values. Ninth, it initializes all requested UTC dates to zero and writes one row per selected date.

Additional consistency checks can be performed using related OBM metrics. Since the total daily block weight depends partly on the number of blocks assigned to the day, the metric \texttt{obm\_block\_weight\_wu\_daily} should be interpreted together with \texttt{obm\_block\_count\_daily}. A useful diagnostic is:
\[
\mathrm{AvgBlockWeightWU}_d =
\frac{
\mathrm{BlockWeightWU}_d
}{
\mathrm{BlockCount}_d
},
\qquad
\mathrm{BlockCount}_d > 0.
\]
For post-SegWit blocks, this average should be consistent with Bitcoin's consensus block-weight limit. Unusually high or low daily totals should be checked against daily block count, because part of the variation may reflect the number of blocks assigned to the UTC date rather than block fullness alone.

The metric can also be combined with fee data. For example, daily fees per weight unit can be computed as:
\[
\mathrm{FeesPerWU}_d =
\frac{
\mathrm{FeesBTC}_d
}{
\mathrm{BlockWeightWU}_d
},
\qquad
\mathrm{BlockWeightWU}_d > 0.
\]
This derived quantity can be useful for fee-market analysis, although it is not part of the primary OBM series.

For external validation, selected periods can be compared with blockchain explorers or public block-space indicators. Such comparisons should be interpreted cautiously because providers may report average block weight, block size in bytes, stripped size, smoothed capacity utilization, or percentage of capacity used rather than total daily block weight.

\paragraph{Known limitations.}
The daily block weight series is transparent and reproducible, but it has several limitations. First, it measures total daily block weight, not average block weight per block. Therefore, it partly reflects variation in the number of blocks assigned to each UTC date. Second, it is a block-space usage metric, not a direct measure of transaction demand, user activity, economic value, or network health. Third, it depends on the block timestamp convention used to assign blocks to UTC dates. Fourth, it does not distinguish between witness and non-witness data beyond the consensus weight formula already reflected in the block-level \texttt{weight} field. Fifth, it may differ from public metrics reported as block size in bytes, average block size, average block weight, or capacity-utilization percentage. Sixth, it is directly computed from decoded block metadata and does not require the spent-output indexer.

Despite these limitations, \texttt{obm\_block\_weight\_wu\_daily} is a useful OBM block-space series. It provides a simple, full-node-derived measure of total daily block weight and complements block count, transaction count, transaction fees, raw output value, and other activity metrics. Its main value is to provide a reproducible block-space usage baseline for empirical studies of Bitcoin transaction demand and fee-market dynamics.          
\subsection{\texttt{obm\_cdd\_age\_band\_btcxdays\_daily}: Daily Bitcoin Days Destroyed by Age Band}
\label{Daily Bitcoin Days Destroyed by Age Band}


\paragraph{Definition.}
The daily Bitcoin-Days-Destroyed-by-age-band table decomposes total daily Bitcoin Days Destroyed according to the age of the outputs at the moment they are spent. Unlike scalar CDD metrics, this metric is a wide, vector-valued daily table: each row corresponds to one UTC calendar date, and each age-band column reports the BTC-days contributed by outputs spent on that date whose age falls within the corresponding band.

Let \(B_d\) denote the set of blocks assigned to day \(d\). For each non-coinbase transaction input \(i\) included in a transaction in block \(b \in B_d\), let \(v_i\) denote the BTC value of the previous output spent by that input, and let \(a_i\) denote the age of that previous output, measured in days. The age of the spent output is defined as:
\[
a_i =
\max\left(
0,
\frac{t^{\mathrm{spent}}_i - t^{\mathrm{created}}_i}{86400}
\right),
\]
where \(t^{\mathrm{created}}_i\) is the timestamp of the block in which the spent output was created, and \(t^{\mathrm{spent}}_i\) is the timestamp of the block in which it is spent.

For each age band \(k\), the daily age-band CDD component is defined as:
\[
\mathrm{CDDBand}_{d,k} =
\sum_{b \in B_d}
\sum_{i \in I_b}
v_i a_i \mathbf{1}\{a_i \in k\},
\]
where \(I_b\) denotes the set of non-coinbase transaction inputs included in block \(b\), and \(\mathbf{1}\{\cdot\}\) is an indicator function equal to one when the spent output age belongs to band \(k\), and zero otherwise.

A block \(b\) is assigned to day \(d\) according to the UTC calendar date derived from its block timestamp \(t_b\):
\[
d(b)=\mathrm{UTCDate}(t_b).
\]

The output table uses the following age bands: 0d--1d, 1d--1w, 1w--1m, 1m--3m, 3m--6m, 6m--1y, 1y--2y, 2y--3y, 3y--5y, 5y--7y, 7y--10y, and 10y+. The sum across all age-band columns should recover total daily Bitcoin Days Destroyed:
\[
\mathrm{CDD}_d =
\sum_k
\mathrm{CDDBand}_{d,k}.
\]

\paragraph{Economic interpretation.}
The daily Bitcoin-Days-Destroyed-by-age-band table describes the age composition of destroyed coin age. It indicates whether daily CDD is driven primarily by recently created outputs, medium-aged outputs, or outputs that had remained inactive for several years.

For economic research, this table is useful in several ways. First, it decomposes aggregate CDD by output age, providing richer information than a single scalar CDD series. Second, it helps distinguish CDD spikes generated by large movements of moderately aged outputs from spikes generated by smaller movements of very old outputs. Third, it complements the spent-value age-band table: the spent-value table shows how much BTC moved by age cohort, while the CDD age-band table shows how much coin age was destroyed by age cohort. Fourth, it supports validation and interpretation of aggregate \texttt{obm\_cdd\_btcxdays\_daily}. Fifth, it provides the basis for constructing CDD age-band shares, such as the fraction of daily CDD associated with outputs older than one year.

The metric should not be interpreted as spent value, transaction volume, user activity, or entity-adjusted settlement value. CDD multiplies spent value by output age, so a small amount of very old spent value can contribute more to the series than a larger amount of recently active spent value. The metric is also not entity-adjusted: it does not identify users, entities, exchanges, custodians, change outputs, or self-transfers. As with other raw spent-output metrics, wallet consolidation, batching, custodial reorganization, and internal transfers may affect the observed CDD age distribution.

\paragraph{Similar metrics publicly available.}
The OBM table \texttt{obm\_cdd\_age\_band\_btcxdays\_daily} is related to public metrics usually labeled \emph{Coin Days Destroyed}, \emph{CDD}, \emph{Spent Output Age Bands}, \emph{Spent Volume Age Bands}, or \emph{coin-age distribution of spent outputs}. Unlike scalar CDD series, this metric is a wide daily table: each row corresponds to one UTC calendar date, and each age-band column reports the BTC-days contributed by spent outputs assigned to that age band. Let \(S_d\) denote the set of previous outputs consumed by non-coinbase transaction inputs in blocks assigned to UTC day \(d\), let \(v_i\) denote the BTC value of spent output \(i\), and let \(a_i\) denote its age in days. For each age band \(k=[\ell_k,u_k)\), the table reports
\[
\mathrm{CDDAgeBandBTCxDays}_{d,k}
=
\sum_{i \in S_d}
v_i a_i \mathbf{1}\{\ell_k \leq a_i < u_k\}.
\]
The final band is open-ended. In OBM, the fixed age-band columns correspond to each age band. The metric is measured in BTC-days, not BTC. It therefore decomposes destroyed coin age by output-age cohort, rather than decomposing spent value by output-age cohort.

The closest public scalar metric is Glassnode's \emph{Bitcoin Coin Days Destroyed (CDD)}.\footnote{\url{https://studio.glassnode.com/charts/indicators.Cdd?a=BTC}} Glassnode defines CDD for a transaction as the number of coins in the transaction multiplied by the number of days since those coins were last spent. This is conceptually aligned with the aggregate version of the OBM CDD calculation. However, Glassnode's public CDD chart is a scalar series, not a wide table that decomposes CDD into age bands. It is therefore useful for validating the aggregate level of CDD, but not for validating the age-band distribution itself. An exact comparison would also require harmonizing timestamp conventions, age computation, fractional-day treatment, entity-adjustment status, and chain-reorganization policy. 

CryptoQuant also provides a scalar \emph{Coin Days Destroyed (CDD)} metric.\footnote{\url{https://cryptoquant.com/asset/btc/chart/network-indicator/coin-days-destroyed-cdd}} Its methodological documentation states that, when a UTXO is destroyed, CDD is calculated as the sum of the number of days between creation and spending multiplied by the UTXO amount. This is close to the OBM spent-output definition of CDD. Nevertheless, as with Glassnode, the public CryptoQuant CDD metric is an aggregate scalar series rather than a CDD-by-age-band table. It is therefore a relevant benchmark for total CDD, but not a direct public equivalent for \texttt{obm\_cdd\_age\_band\_btcxdays\_daily}. 

The closest public age-distribution comparator is CryptoQuant's \emph{Spent Output Age Bands} metric. \footnote{\url{https://cryptoquant.com/asset/btc/chart/network-indicator/spent-output-age-bands}} CryptoQuant describes this metric as a set of spent outputs grouped by the age band in which their lifespan falls, with each colored band representing the total value of spent outputs in the corresponding interval. This is structurally close to the OBM table because both classify spent outputs by age. However, CryptoQuant's public metric is value-based, not CDD-based: it reports the BTC value of spent outputs by age band, whereas \texttt{obm\_cdd\_age\_band\_btcxdays\_daily} multiplies each spent output value by its age and reports BTC-days. A CDD-by-age-band table could be derived from spent-output-level data, but it cannot be obtained from value bands alone unless the within-band age distribution is also known. 

Glassnode's \emph{Spent Output Age Bands} (SOAB) is also related.\footnote{\url{https://studio.glassnode.com/charts/indicators.Soab?a=BTC}} Glassnode describes SOAB as a bundle of all spent outputs created within specified age bands, with each line representing the percentage of spent outputs created within the time interval denoted in the legend. This is useful for comparing the age distribution of spent outputs, but it is not a direct equivalent of OBM's CDD age-band table. SOAB is typically expressed as a percentage distribution over spent outputs rather than as BTC-days contributed by each band. Thus, it captures age composition, but not the value-weighted and age-weighted magnitude of destroyed coin age. 

Glassnode's \emph{Spent Volume Age Bands} (SVAB)\footnote{\url{https://studio.glassnode.com/charts/indicators.Svab?a=BTC}} is another relevant comparator. Glassnode defines SVAB as a separation of on-chain transfer volume based on coin age, where each band represents the percentage of spent volume associated with coins last moved within the corresponding age interval. This is closer to OBM's spent-value age-band table than to the CDD age-band table. It can help assess whether value movement is concentrated in young or old cohorts, but it does not directly show how much CDD each cohort contributes, because CDD also multiplies value by age. 

Checkonchain provides \emph{Spent Volume Age Bands} and related revived-supply tools.\footnote{\url{https://charts.checkonchain.com/}} Its chart catalog describes Spent Volume Age Bands as an on-chain spent-volume breakdown by coin age, with absolute and relative versions. This is useful for comparing OBM's spent-value age-band table and for interpreting old-coin spending episodes. However, it is not a direct CDD-by-age-band comparator unless the provider also reports, or allows reconstruction of, the age-weighted BTC-days contribution of each cohort. Public documentation does not provide a full reproducible CDD age-band algorithm. 

Bitbo provides a scalar \emph{CDD Bitcoin Chart}.\footnote{\url{https://charts.bitbo.io/cdd/}} The page explains CDD as the product of coins spent and the number of days accumulated while the coins remained unmoved, and offers chart access options such as API, CSV, XLSX, and JSON. This is useful as an aggregate CDD comparator, but it is not an age-band decomposition. It does not indicate how total CDD is distributed across output-age cohorts. 

Coin Metrics provides a related scalar metric, \texttt{TxTfrValDayDst}\footnote{\url{https://community-api.coinmetrics.io/v4/timeseries/asset-metrics?assets=btc&metrics=TxTfrValDayDst}}, which measures transferred days destroyed. It can be used as an aggregate CDD-like comparator, and it is especially useful because Coin Metrics documents the metric identifier and interval availability. However, Coin Metrics does not appear to expose a publicly wide table that decomposes transferred days destroyed by output-age band. Therefore, it is a useful aggregate benchmark, but not a direct public equivalent of \texttt{obm\_cdd\_age\_band\_btcxdays\_daily}.

Blockchair provides block-level, transaction-level, input-level, and output-level explorer data.\footnote{\url{https://blockchair.com/bitcoin/outputs}} This makes Blockchair useful as a reconstruction source. A researcher could resolve spent inputs to previous outputs, compute each output's age, multiply each value by age, assign the contribution to an age band, and aggregate by UTC date. However, Blockchair does not appear to publish a named daily wide table of CDD by age band. It should therefore be treated as a source of lower-level data rather than as a direct comparator.

HODL Waves, UTXO Age Bands, and current-supply age-distribution charts should also be distinguished from this metric. For example, Glassnode's HODL Waves and CryptoQuant's UTXO Age Bands describe the age distribution of currently unspent supply, not the age distribution of destroyed coin age.\footnote{\url{https://docs.glassnode.com/guides-and-tutorials/metric-guides/age-distribution/hodl-waves} and \url{https://cryptoquant.com/asset/btc/chart/network-indicator/utxo-age-bands}} These are stock metrics, whereas \texttt{obm\_cdd\_age\_band\_btcxdays\_daily} is a daily flow table. They are useful for contextualizing the stock of old coins that could potentially be spent, but they do not measure the CDD actually generated by each cohort on a given day. 

Overall, no reviewed public provider appears to expose a named daily wide table exactly equivalent to \texttt{obm\_cdd\_age\_band\_btcxdays\_daily}. The closest aggregate CDD comparators are Glassnode's \texttt{indicators.Cdd}, CryptoQuant's CDD metric, Bitbo's CDD chart, and Coin Metrics' \texttt{TxTfrValDayDst}. The closest age-distribution comparators are Glassnode's SOAB and SVAB, CryptoQuant's Spent Output Age Bands, and Checkonchain's Spent Volume Age Bands. These sources are useful for external interpretation, but they either report aggregate CDD, value by age band, percentage age distributions, or spent-volume age bands rather than BTC-days by age band. The distinctive contribution of OBM is that it publishes a reproducible, full-node-derived, BTC-days-denominated wide daily table with explicit CDD age-band columns, raw spent-output accounting, documented UTC timestamp convention, optional row-sum validation against aggregate \texttt{obm\_cdd\_btcxdays\_daily}, and direct comparability with \texttt{obm\_spent\_value\_age\_band\_btc\_daily}.

\paragraph{Data source and input requirements.}
The metric is exported from the persistent OBM spent-output indexer database, rather than computed by an independent metric-specific blockchain scan. The indexer, described in Sect.~\ref{sec:The Spent-Output Indexer}, is built from a running Bitcoin Core full node and stores reusable daily age-band aggregates in a SQLite database. For \texttt{obm\_cdd\_age\_band\_btcxdays\_daily}, the relevant source table is \texttt{daily\_age\_band\_aggregates}, and the relevant fields are \texttt{date}, \texttt{age\_band}, and \texttt{cdd\_sats\_days}.

The indexer computes this table during its sequential scan of the blockchain. For each non-coinbase transaction input, it resolves the input to the previous output it spends, retrieves the previous output value and creation timestamp from the live outpoint state, computes the elapsed age in days, assigns the spent output to one of the fixed age bands, and accumulates the corresponding value-age product in the daily age-band aggregate. Internally, the indexer stores:
\[
\mathrm{cdd\_sats\_days}_{d,k} =
\sum_{b \in B_d}
\sum_{i \in I_b}
v^{\mathrm{sats}}_i a_i \mathbf{1}\{a_i \in k\}.
\]

The exporter converts each stored satoshi-day aggregate into BTC-days by dividing by \(100{,}000{,}000\):
\[
\mathrm{CDDBand}_{d,k} =
\frac{\mathrm{cdd\_sats\_days}_{d,k}}{100{,}000{,}000}.
\]

The exporter script does not query Bitcoin Core directly. It requires an existing SQLite database generated by \texttt{obm\_spent\_output\_indexer.py}, with metadata identifying it as an OBM spent-output indexer database and with processed-date metadata covering the requested output interval. The script checks that the requested ending date is not beyond the indexer's maximum date. If a requested date-band pair has no row in \texttt{daily\_age\_band\_aggregates}, the exporter writes zero for that age-band column. This convention is appropriate because each age-band CDD component is a flow variable: if no CDD is assigned to a given age band on a UTC date, the daily value for that band is zero.

The OBM spent-output indexer is designed to be built from the Bitcoin genesis block onward. Accordingly, the exporter assumes that the indexer database contains a complete historical pass from block height 0 through the maximum processed height recorded in the database metadata. Under this invariant, a missing date-band pair in \texttt{daily\_age\_band\_aggregates} does not indicate an unprocessed observation; it indicates that no CDD contribution was assigned to that age band and UTC date under the indexer's timestamp and age-band conventions.

This metric does not require address extraction, user clustering, entity identification, external price data, third-party APIs, or a direct Bitcoin Core connection at export time. However, it depends on the prior successful execution of the spent-output indexer, which itself requires a synchronized, non-pruned Bitcoin Core full node and access to transaction-level data sufficient to reconstruct previous-output values and creation timestamps. The exported table, therefore, inherits the indexer's UTC block-timestamp convention, previous-output reconstruction procedure, age-band assignment convention, fractional-day age convention, treatment of negative apparent ages, chain-consistency checks, release metadata, and handling of historical duplicate outpoint edge cases.

\paragraph{Algorithm.}
The script \texttt{export\_obm\_cdd\_age\_band\_btcxdays\_daily.py} implements the following procedure:

\begin{enumerate}
    \item Parse the user-provided date interval, \texttt{\symbol{45}\symbol{45}start\_date} and \texttt{\symbol{45}\symbol{45}end\_date}, using the format \texttt{YYYY-MM-DD}. Both dates are interpreted as UTC dates, and both are included in the output.

    \item Open the persistent SQLite database generated by \texttt{obm\_spent\_output\_indexer.py}. The path to this database is provided through the \texttt{\symbol{45}\symbol{45}state\_db} argument.

    \item Validate that the SQLite database corresponds to the OBM spent-output indexer. The script checks the indexer metadata, including the expected \texttt{indexer\_id} and the presence of processing metadata such as \texttt{last\_processed\_height}, \texttt{last\_processed\_date}, and \texttt{max\_processed\_date}.

    \item Verify that the database contains the table \texttt{daily\_age\_band\_aggregates} and that this table contains the required fields \texttt{date}, \texttt{age\_band}, and \texttt{cdd\_sats\_days}.

    \item Verify that the requested \texttt{\symbol{45}\symbol{45}end\_date} is not later than the maximum UTC date processed by the indexer. If the requested interval extends beyond the indexed range, the script aborts and requires the indexer to be run further before export.

    \item Infer the dataset release version from the indexer metadata. If the database does not contain a release-version field, the script uses the fallback value provided through the \texttt{\symbol{45}\symbol{45}release\_version} argument.

    \item Initialize one complete output row for each UTC date in the requested interval, setting all CDD age-band columns to zero.

    \item Query \texttt{daily\_age\_band\_aggregates} for all date-band observations in the requested interval.

    \item For each retrieved row, verify that \texttt{age\_band} belongs to the expected set of age-band labels. Unexpected labels cause the script to abort.

    \item For each retrieved row, verify that the corresponding \texttt{cdd\_sats\_days} value is non-negative and that there is no duplicate date-band aggregate.

    \item Convert each value from satoshi-days to BTC-days:
    \[
    \mathrm{CDDBand}_{d,k} =
    \frac{\mathrm{cdd\_sats\_days}_{d,k}}{100{,}000{,}000}.
    \]

    \item Write the resulting wide table to a CSV file. Each row corresponds to one UTC date, and each age-band column reports BTC-day-denominated CDD for that band.

    \item Optionally validate the row sums when the \texttt{\symbol{45}\symbol{45}validate\_row\_sums} flag is activated. In that case, the script checks that the sum across all age-band columns equals the value of \texttt{daily\_aggregates.cdd\_sats\_days}, converted to BTC-days, within the configured tolerance.

    \item Optionally generate a stacked area plot when the \texttt{\symbol{45}\symbol{45}plot} flag is activated. The stacked areas represent the contribution of each age band to total daily Bitcoin Days Destroyed.
\end{enumerate}

This exporter does not query Bitcoin Core, does not maintain an outpoint state, and does not reconstruct spent outputs at export time. The computationally expensive previous-output resolution, age calculation, CDD accumulation, and age-band assignment have already been performed by the spent-output indexer. The export script is therefore a lightweight, deterministic transformation from the indexed database to a wide OBM CSV table.

\paragraph{Metric-specific input parameters.}
The input parameters specific to this exporter are:

\begin{itemize}
    \item \texttt{\symbol{45}\symbol{45}state\_db}: path to the persistent SQLite database generated by the Python script that indexes the blockchain, \texttt{obm\_spent\_output\_indexer.py}. This database must contain the \texttt{daily\_age\_band\_aggregates} table and the metadata required to verify that it corresponds to the OBM spent-output indexer.

    \item \texttt{\symbol{45}\symbol{45}start\_date}: starting date of the selected interval, inclusive, in \texttt{YYYY-MM-DD} format. The date is interpreted as a UTC calendar date.

    \item \texttt{\symbol{45}\symbol{45}end\_date}: ending date of the selected interval, inclusive, in \texttt{YYYY-MM-DD} format. The script verifies that this date is not later than the maximum date processed by the indexer.

    \item \texttt{\symbol{45}\symbol{45}release\_version}: fallback dataset release version used only if the indexer database does not contain a \texttt{release\_version} metadata field.

    \item \texttt{\symbol{45}\symbol{45}output}: path of the output CSV file to be written.

    \item \texttt{\symbol{45}\symbol{45}validate\_row\_sums}: optional flag that instructs the script to verify that the sum across all CDD age-band columns equals total CDD from \texttt{daily\_aggregates.cdd\_sats\_days}, when that aggregate is available.

    \item \texttt{\symbol{45}\symbol{45}row\_sum\_tolerance}: tolerance, in BTC-days, used by \texttt{\symbol{45}\symbol{45}validate\_row\_sums}. The default is \(0.00000001\) BTC-days.

    \item \texttt{\symbol{45}\symbol{45}plot}: optional flag that instructs the script to generate a stacked area plot of all CDD age-band time series.

    \item \texttt{\symbol{45}\symbol{45}plot\_output}: optional path for the generated plot. If this argument is omitted while \texttt{\symbol{45}\symbol{45}plot} is used, the plot is saved next to the CSV file using the same base name and a \texttt{.png} extension.
\end{itemize}

The exporter does not accept Bitcoin Core RPC parameters, \texttt{\symbol{45}\symbol{45}height\_margin}, \texttt{\symbol{45}\symbol{45}commit\_every}, \texttt{\symbol{45}\symbol{45}min\_confirmations}, or \texttt{\symbol{45}\symbol{45}reset\_state\_db}. These parameters belong to the spent-output indexer, not to the CDD age-band exporter. The exporter assumes that the indexer database has already been built and updated through the requested ending date. Its role is only to read \texttt{daily\_age\_band\_aggregates}, pivot it into a wide daily table, convert satoshi-day values into BTC-days, and write the standardized output.

\paragraph{Aggregation rule.}
For each UTC date \(d\), the value in each age-band column is computed as the sum of the BTC-day contributions of all previous outputs spent on that date whose age belongs to the corresponding band:
\[
\mathrm{CDDBand}_{d,k} =
\sum_{b \in B_d}
\sum_{i \in I_b}
v_i a_i \mathbf{1}\{a_i \in k\}.
\]

The row sum across all age bands recovers total daily Bitcoin Days Destroyed:
\[
\mathrm{CDD}_d =
\sum_k
\mathrm{CDDBand}_{d,k}.
\]

Monthly versions of this table, if distributed, should be computed by summing each age-band column over the corresponding daily observations:
\[
\mathrm{CDDBand}_{m,k} =
\sum_{d \in m}
\mathrm{CDDBand}_{d,k}.
\]

This convention preserves the interpretation of each column as a BTC-day-denominated CDD flow associated with a specific age band.

\paragraph{Output format.}
The output file contains one observation per UTC date and one column per age band. It therefore departs from the scalar OBM schema with a single \texttt{value} column, but retains the usual metadata fields.

\begin{center}
\begin{tabular}{lp{0.25\textwidth}p{0.25\textwidth}}
\toprule
\textbf{Column} & \textbf{Example} & \textbf{Description} \\
\midrule
\texttt{date} & \texttt{2024-01-01} & UTC calendar date \\
\texttt{series\_id} & \texttt{obm\_cdd\_age\_band \_btcxdays\_daily} & Stable OBM table identifier \\
\texttt{cdd\_0d\_1d\_btcxdays} & \texttt{2500.123456789012} & CDD from outputs aged 0d to 1d \\
\texttt{cdd\_1d\_1w\_btcxdays} & \texttt{119000.000000000000} & CDD from outputs aged 1d to 1w \\
\texttt{cdd\_1w\_1m\_btcxdays} & \texttt{802000.250000000000} & CDD from outputs aged 1w to 1m \\
\texttt{cdd\_1m\_3m\_btcxdays} & \texttt{3210000.000000000000} & CDD from outputs aged 1m to 3m \\
\texttt{cdd\_3m\_6m\_btcxdays} & \texttt{5190000.750000000000} & CDD from outputs aged 3m to 6m \\
\texttt{cdd\_6m\_1y\_btcxdays} & \texttt{7330000.000000000000} & CDD from outputs aged 6m to 1y \\
\texttt{cdd\_1y\_2y\_btcxdays} & \texttt{15400000.000000000000} & CDD from outputs aged 1y to 2y \\
\texttt{cdd\_2y\_3y\_btcxdays} & \texttt{20100000.000000000000} & CDD from outputs aged 2y to 3y \\
\texttt{cdd\_3y\_5y\_btcxdays} & \texttt{28900000.000000000000} & CDD from outputs aged 3y to 5y \\
\texttt{cdd\_5y\_7y\_btcxdays} & \texttt{33100000.000000000000} & CDD from outputs aged 5y to 7y \\
\texttt{cdd\_7y\_10y\_btcxdays} & \texttt{21000000.000000000000} & CDD from outputs aged 7y to 10y \\
\texttt{cdd\_10y\_plus\_btcxdays} & \texttt{18000000.000000000000} & CDD from outputs aged 10y or more \\
\texttt{unit} & \texttt{BTC-days} & Shared unit for all age-band columns \\
\texttt{frequency} & \texttt{daily} & Observation frequency \\
\texttt{release\_version} & \texttt{OBM v0.1.0} & Dataset release version \\
\bottomrule
\end{tabular}
\end{center}

All CDD age-band columns are measured in BTC-days and are written with twelve decimal places. The additional decimal precision is useful because the indexer uses fractional-day ages.

\paragraph{Technical validation.}
Several internal checks are used to validate this metric at export time. First, the script verifies that the requested date range is valid and that \texttt{\symbol{45}\symbol{45}start\_date} is not later than \texttt{\symbol{45}\symbol{45}end\_date}. Second, it checks that the SQLite state database exists, opens it, and sets the connection to query-only mode. Third, it verifies that the database metadata identify it as an OBM spent-output indexer database, by checking the expected \texttt{indexer\_id}. Fourth, it checks that the indexer database contains processing metadata, including \texttt{last\_processed\_height}, \texttt{last\_processed\_date}, and \texttt{max\_processed\_date}. Fifth, the script verifies that the table \texttt{daily\_age\_band\_aggregates} is present in the database and that it contains the fields \texttt{date}, \texttt{age\_band}, and \texttt{cdd\_sats\_days}. Sixth, it verifies that the requested ending date is not later than the maximum date processed by the indexer. If the requested interval extends beyond the indexed range, the script aborts and requires the indexer to be updated before the export is attempted.

For each requested date, the exporter initializes all expected age-band columns to zero. For each row retrieved from \texttt{daily\_age\_band\_aggregates}, the script checks that the age-band label belongs to the expected set of labels, that the corresponding value is non-negative, and that the date-band pair is not duplicated. Each retrieved value is converted from satoshi-days to BTC-days by dividing by \(100{,}000{,}000\). The resulting file contains one complete wide row per calendar date in the requested interval.

When the optional \texttt{\symbol{45}\symbol{45}validate\_row\_sums} flag is used, the script also compares the sum across all CDD age-band columns with total CDD from \texttt{daily\_aggregates.cdd\_sats\_days}, converted to BTC-days. The validation identity is:
\[
\sum_k
\mathrm{CDDBand}_{d,k}
=
\mathrm{CDD}_d.
\]
The check is performed within the configured \texttt{\symbol{45}\symbol{45}row\_sum\_tolerance}. This is a strong internal consistency check when both \texttt{daily\_age\_band\_aggregates} and \texttt{daily\_aggregates} are generated by the same indexer run.

Additional consistency checks can be performed using related OBM series. In particular, the row sum across all CDD age-band columns should equal \texttt{obm\_cdd\_btcxdays\_daily} for every date when both outputs are generated from the same indexer database and release version. The table can also be compared with \texttt{obm\_spent\_value\_age\_band\_btc\_daily}: the spent-value table shows how much BTC moved by age cohort, whereas the CDD table shows how much coin age was destroyed by age cohort. Comparisons with external CDD age-distribution metrics are useful as diagnostics but should not be interpreted as strict equality tests, because providers may differ in timestamp conventions, age-band boundaries, fractional-day treatment, entity adjustment, change-output treatment, transfer-value definition, and historical edge-case handling.

\paragraph{Known limitations.}
The daily Bitcoin-Days-Destroyed-by-age-band table is useful but definition-sensitive. First, \texttt{obm\_cdd\_age\_band\_btcxdays\_daily} is a raw spent-output coin-age table, not an entity-adjusted measure. It does not identify users, entities, custodians, exchanges, self-transfers, or change outputs. Second, the table depends on the block timestamp convention used both to compute output age and to assign spending blocks to calendar days. Third, the age-band boundaries are fixed, so small differences around a boundary can move an output from one column to another. Fourth, the table inherits the fractional-day age convention used for CDD. Fifth, the table is wide and vector-valued rather than a scalar \texttt{value} series, so software expecting the standard scalar OBM schema must treat it separately. Sixth, the metric can be affected by self-transfers, wallet consolidation, batching, exchange operations, and custodial wallet management. Seventh, it reports BTC-days, not BTC value or fiat-denominated value. Eighth, it requires the persistent spent-output indexer database and therefore depends on the correctness and completeness of the indexer run. Ninth, the simple reorganization policy of the indexer detects inconsistencies and aborts, but it does not automatically roll back the state database.

Despite these limitations, \texttt{obm\_cdd\_age\_band\_btcxdays\_daily} is a useful OBM coin-age distribution table. It provides a transparent decomposition of daily Bitcoin Days Destroyed by output age and complements scalar CDD, spent-value, dormancy, threshold-based CDD, and UTXO-flow metrics.    
\subsection{\texttt{obm\_cdd\_btcxdays\_daily}: Bitcoin Days Destroyed}
\label{Bitcoin Days Destroyed}


\paragraph{Definition.}
The Bitcoin Days Destroyed series measures the amount of accumulated coin age destroyed by outputs spent in blocks assigned to a given UTC calendar day. Let \(S_d\) denote the set of transaction outputs spent in blocks assigned to day \(d\). For each spent output \(i \in S_d\), let \(v_i\) denote its value in BTC, let \(t_i^{\mathrm{created}}\) denote the timestamp of the block in which the output was created, and let \(t_i^{\mathrm{spent}}\) denote the timestamp of the block in which the output is spent. The age of the spent output is defined as:

\[
a_i =
\max\left(
0,
\frac{
t_i^{\mathrm{spent}} - t_i^{\mathrm{created}}
}{86400}
\right),
\]

where age is measured in days. The contribution of spent output \(i\) to Bitcoin Days Destroyed is $\mathrm{CDD}_i = v_i a_i$. The daily Bitcoin Days Destroyed series is then defined as:

\[
\mathrm{CDD}_d =
\sum_{i \in S_d} v_i a_i.
\]

A block \(b\) is assigned to day \(d\) according to the UTC calendar date derived from its block timestamp \(t_b\): $d(b) = \mathrm{UTCDate}(t_b)$.

The resulting series therefore reports the total number of BTC-days destroyed by outputs spent in blocks assigned to each UTC day. The unit is BTC-days, because the metric multiplies a BTC-denominated value by the number of days that value remained unspent. The term ``destroyed'' does not mean that bitcoins are destroyed. It means that the accumulated age of a spent output is reset when that output is consumed by a transaction input.

\paragraph{Economic interpretation.}
Bitcoin Days Destroyed is a coin-age metric. It gives more weight to the movement of older coins than to the movement of recently active coins. For example, spending \(1\) BTC that has remained unmoved for \(1\) day destroys \(1\) BTC-day, whereas spending \(1\) BTC that has remained unmoved for \(100\) days destroys \(100\) BTC-days. Similarly, spending \(10\) BTC that have remained unmoved for \(100\) days destroys \(1000\) BTC-days.

For economic research, this metric is useful because it distinguishes ordinary transaction activity from the movement of aged supply. A high transaction count may reflect recent turnover, exchange operations, batching, or repeated movement of recently spent coins. By contrast, a spike in Bitcoin Days Destroyed indicates that either a large amount of BTC has moved, old coins have moved, or both. The series is therefore relevant for studying long-term holder activity, dormant-supply movement, coin-age dynamics, distribution episodes, and the relationship between on-chain activity and market conditions.

The metric should not be interpreted as a direct measure of transaction demand, user count, or payment volume. It is a measure of the destroyed coin age. A large value may arise from a small number of very old outputs, a large number of moderately old outputs, or large BTC amounts that have been inactive for some time. It is therefore best interpreted jointly with other OBM series, such as \texttt{obm\_tx\_count\_daily}, \texttt{obm\_block\_count\_daily}, \texttt{obm\_spent\_value\_btc\_daily}, and future dormancy-related indicators.

\paragraph{Similar metrics publicly available.}
The OBM series \texttt{obm\_cdd\_btcxdays\_daily} is comparable to a family of public Bitcoin Coin Days Destroyed or Bitcoin Days Destroyed metrics. As we stated above, our metric is defined as a raw spent-output measure. For every previous output consumed by a transaction input, the value of the spent output, measured in BTC, is multiplied by the number of days between the block timestamp at which the output was created and the block timestamp at which it was spent. The daily value is then obtained by summing all such BTC-day contributions for outputs spent in blocks assigned to the same UTC calendar day. This definition is deliberately mechanical: it does not attempt to identify entities, exchanges, self-transfers, change outputs, or economically adjusted transfers. It is therefore best described as a reproducible raw spent-output Bitcoin Days Destroyed series.

The closest named public comparator is Coin Metrics' \emph{Xfer'd Days Destroyed}, with MetricID \texttt{TxTfrValDayDst}\footnote{\url{https://community-api.coinmetrics.io/v4/timeseries/asset-metrics?assets=btc\&metrics=TxTfrValDayDst}}, but this query requires supplying credentials. Coin Metrics documents it as the sum of native units transferred during an interval multiplied by the number of days since those native units were last transferred, with availability at one-day and one-block intervals. Its documentation further states that, when an unspent output is spent, the time it remained unspent is multiplied by its value and summed for the day. A relevant difference is that Coin Metrics explicitly uses full days only, so, for example, 2.5 days count as 2 days. By contrast, \texttt{obm\_cdd\_btcxdays\_daily} uses fractional days by dividing elapsed seconds by 86400. Therefore, \texttt{TxTfrValDayDst} is a highly relevant benchmark, but exact equality should not be expected. The corresponding API endpoint can be queried.

Coin Metrics provides a public definition and relevant implementation conventions, but not a fully open script that reconstructs the metric from raw Bitcoin Core data. \texttt{TxTfrValDayDst} may differ because they use transfer-level rather than spent-output-level accounting, different daily-interval labeling conventions, alternative block timestamp conventions, or special handling of early coinbase-origin outputs. These differences are especially visible in the earliest Bitcoin period, when the number of non-coinbase transactions was very small, and a single definitional choice could dominate the daily value.

Empirically, comparisons with \texttt{TxTfrValDayDst} show larger discrepancies during the first days of Bitcoin transaction history, while the OBM fractional spent-output version becomes increasingly close to the provider series after the earliest observations. For this reason, OBM does not claim that \texttt{obm\_cdd\_btcxdays\_daily} replicates \texttt{TxTfrValDayDst}. Instead, it provides a transparent, reproducible, raw spent-output CDD series, with its construction fully documented.

When validating this series against external providers, comparisons should therefore be interpreted as diagnostic rather than definitive. A useful validation practice is to report comparisons both over the full sample and after excluding the earliest Bitcoin period, such as the first 30 or 90 days. The full-sample comparison documents early-chain convention differences, whereas the restricted comparison is more informative about the metric's general behavior.

Glassnode provides a \emph{Bitcoin Coin Days Destroyed (CDD)} metric, exposed in Studio as \texttt{indicators.Cdd}\footnote{\url{https://studio.glassnode.com/charts/indicators.Cdd?a=BTC}} (credentials required). Glassnode defines CDD for a transaction as the number of coins in the transaction multiplied by the number of days since those coins were last spent. This is conceptually close to \texttt{obm\_cdd\_btcxdays\_daily}, and the page indicates access through CSV, JSON, API, Excel, and MCP interfaces. However, the public chart page does not disclose a full reproducible algorithm. In particular, it does not specify all details that matter for exact replication, such as whether fractional days are retained or floored, how transaction-level aggregation is mapped to daily intervals, how change or transfer semantics are treated, and how reorganizations or early-chain edge cases are handled.

CryptoQuant provides a \emph{Coin Days Destroyed (CDD)} metric\footnote{\url{https://cryptoquant.com/asset/btc/chart/network-indicator/coin-days-destroyed-cdd}} (credentials required). It also offers methodological documentation.\footnote{\url{https://userguide.cryptoquant.com/cryptoquant-metrics/utxo/coin-days-destroyed-cdd}} Its documentation states that, when a UTXO is destroyed, CDD is calculated as the sum over spent outputs of the output lifespan multiplied by the UTXO amount. This is very close to the 	\texttt{obm\_cdd\_btcxdays\_daily} spent-output definition. CryptoQuant therefore provides one of the clearest public formula-level descriptions among commercial data providers. Nevertheless, the public documentation does not appear to provide open-source reconstruction code or all low-level conventions needed for exact replication, such as timestamp convention, fractional-day handling, and chain-reorganization policy.

Bitbo provides a \emph{CDD Bitcoin Chart}.\footnote{\url{https://charts.bitbo.io/cdd/}} The page explains that each coin accumulates days while it remains unmoved and that, when the coin is spent, the accumulated days are multiplied by the number of coins spent. Bitbo also exposes chart access options labelled API, CSV, XLSX, and JSON. The definition is economically equivalent to the usual CDD interpretation and useful for visual comparison. However, the public page is not an algorithmic specification. It does not document all choices required for strict reproducibility from a Bitcoin node.

Newhedge provides a \emph{Bitcoin Coin Days Destroyed (CDD)} chart.\footnote{\url{https://newhedge.io/bitcoin/coin-days-destroyed}} and exposes an API endpoint\footnote{\url{https://newhedge.io/api/v2/metrics/coin-days-destroyed/coin\_days\_destroyed}} (credentials required). The page defines the metric as the number of bitcoins transacted multiplied by the number of days since those bitcoins were last moved. It also gives the formula
\[
\mathrm{CDD}=\mathrm{Number\ of\ Coins}\times\mathrm{Days\ Since\ Last\ Move}
\]
The definition is useful, but the public documentation is not detailed enough to establish exact equivalence with 	\texttt{obm\_cdd\_btcxdays\_daily}.

Blockchair provides a \emph{Bitcoin Coin-Days Destroyed} chart.\footnote{\url{https://blockchair.com/bitcoin/charts/coindays-destroyed}} The chart description states that it shows how many Bitcoin coin-days were destroyed by transacting BTC. Blockchair is also useful because it provides block and transaction-level explorer data and downloadable datasets. However, for this metric, the public chart page provides only a high-level definition, not a full algorithm specifying spent-output reconstruction, timestamp treatment, fractional-day convention, and aggregation rules. It should therefore be treated as a useful external comparison source rather than as a fully reproducible methodological benchmark.

Bitcoin Magazine Pro provides a \emph{Bitcoin: Coin Days Destroyed} chart.\footnote{\url{https://www.bitcoinmagazinepro.com/charts/coin-days-destroyed-cdd/}} It defines CDD as the number of coins moved on-chain multiplied by the number of days since those coins were last moved, and notes that Coin Days Destroyed is also referred to as Bitcoin Days Destroyed. The page is useful for conceptual exposition and market interpretation, but it is not a reproducible data specification. Moreover, the free tier only shows the 90-day moving average for this metric. It does not provide enough information to determine whether the underlying data are raw, smoothed, supply-adjusted, entity-adjusted, or computed with specific timestamp conventions. Related metrics on the same site, such as Supply Adjusted CDD and Value Days Destroyed, should be treated as derivatives rather than direct equivalents of \texttt{obm\_cdd\_btcxdays\_daily}.

BitInfoCharts reports \emph{Days Destroyed last 24h / Total Bitcoins} and also displays block-level pages with large days-destroyed observations. This is useful as a current diagnostic source and as an exploratory block-level reference, but it is not a clean downloadable daily historical series equivalent to OBM's raw CDD. Moreover, the published indicator is normalized by total bitcoins, so it is closer to a supply-adjusted or per-supply version than to raw \texttt{obm\_cdd\_btcxdays\_daily}. 

Overall, the strongest public comparators for \texttt{obm\_cdd\_btcxdays\_daily} are Coin Metrics \texttt{TxTfrValDayDst}, Glassnode \texttt{indicators.Cdd}, and CryptoQuant's CDD metric. Coin Metrics is especially valuable because it provides a named metric identifier and explicitly states the one-day interval and full-day convention. CryptoQuant is valuable because it states the spent-output formula directly. Glassnode is widely used and conceptually aligned, but its public method description is thinner. Bitbo, Newhedge, Blockchair, Bitcoin Magazine Pro, and BitInfoCharts provide useful public charts or secondary benchmarks, but they do not disclose enough implementation detail to make their algorithms fully auditable. This comparison supports the contribution of \texttt{obm\_cdd\_btcxdays\_daily}: it provides not only a CDD-like time series, but also an explicit full-node reconstruction procedure, a stable series identifier, and a transparent statement of conventions, including fractional output age, UTC block-date assignment, and the absence of entity or transfer adjustment.

\paragraph{Data source and input requirements.}
The metric is exported from the persistent OBM spent-output indexer database, rather than computed by an independent metric-specific blockchain scan. The indexer, described in Sect.~\ref{sec:The Spent-Output Indexer}, is built from a running Bitcoin Core full node and stores reusable daily spent-output aggregates in a SQLite database. For \texttt{obm\_cdd\_btcxdays\_daily}, the relevant source table is \texttt{daily\_aggregates}, and the relevant field is \texttt{cdd\_sats\_days}.

The indexer computes CDD during its sequential scan of the blockchain. For each non-coinbase transaction input, it resolves the input to the previous output it spends, retrieves the previous output value and creation timestamp from the live outpoint state, computes the elapsed age in days, and adds the resulting value-age product to the daily CDD aggregate. Internally, this quantity is stored as satoshi-days:
\[
\mathrm{cdd\_sats\_days}_d
=
\sum_{i \in S_d}
\mathrm{value\_sats}_i
\max\left(
0,
\frac{t^{\mathrm{spent}}_i - t^{\mathrm{created}}_i}{86400}
\right),
\]
where \(S_d\) is the set of outputs spent in blocks assigned to UTC date \(d\). The exporter converts the stored satoshi-day aggregate into BTC-days by dividing by \(100{,}000{,}000\):
\[
\mathrm{CDD}_d =
\frac{\mathrm{cdd\_sats\_days}_d}{100{,}000{,}000}.
\]

The exported \texttt{value} field is written with eight decimal places, matching the BTC-denominated precision convention used by other OBM series.

The exporter script does not query Bitcoin Core directly. It requires an existing SQLite database generated by \texttt{obm\_spent\_output\_indexer.py}, with metadata identifying it as an OBM spent-output indexer database and with processed-date metadata covering the requested output interval. The script checks that the requested ending date is not beyond the indexer's maximum date. If a requested date has no row in \texttt{daily\_aggregates}, the exporter writes a zero value for that date. This convention is appropriate for \texttt{obm\_cdd\_btcxdays\_daily} when the indexer database fully covers the requested interval: because CDD is a flow variable, if no CDD contribution is assigned to a UTC date under the indexer's timestamp convention, the daily value is set to zero rather than missing.

This metric does not require address extraction, user clustering, entity identification, external price data, third-party APIs, or a direct Bitcoin Core connection at export time. However, it depends on the prior successful execution of the spent-output indexer, which itself requires a synchronized, non-pruned Bitcoin Core full node and access to transaction-level data sufficient to reconstruct previous-output values and creation timestamps. The exported series, therefore, inherits the indexer's UTC block-timestamp convention, previous-output reconstruction procedure, fractional-day age convention, treatment of negative apparent ages, and handling of historical duplicate outpoint edge cases.

\paragraph{Algorithm.}
The script \texttt{export\_obm\_cdd\_btcxdays\_daily.py} implements the following procedure:

\begin{enumerate}
    \item Parse the user-provided date interval, \texttt{\symbol{45}\symbol{45}start\_date} and \texttt{\symbol{45}\symbol{45}end\_date}, using the format \texttt{YYYY-MM-DD}. Both dates are interpreted as UTC dates, and both are included in the output.

    \item Open the persistent SQLite database generated by \texttt{obm\_spent\_output\_indexer.py}. The path to this database is provided through the \texttt{\symbol{45}\symbol{45}state\_db} argument.

    \item Validate that the SQLite database corresponds to the OBM spent-output indexer. The script checks the indexer metadata, including the expected \texttt{indexer\_id} and the presence of processing metadata such as \texttt{last\_processed\_height}, \texttt{last\_processed\_date}, and \texttt{max\_processed\_date}.

    \item Verify that the requested \texttt{\symbol{45}\symbol{45}end\_date} is not later than the maximum UTC date processed by the indexer. If the requested interval extends beyond the indexed range, the script aborts and requires the indexer to be run further before export.

    \item Infer the dataset release version from the indexer metadata. If the database does not contain a release-version field, the script uses the fallback value provided through the \texttt{\symbol{45}\symbol{45}release\_version} argument.

    \item For each UTC date \(d\) in the requested interval, query the \texttt{daily\_aggregates} table for the field \texttt{cdd\_sats\_days}. If a row exists for date \(d\), the script reads the stored satoshi-day CDD value. If no row exists, the script assigns zero CDD to that date.

    \item Convert each daily value from satoshi-days to BTC-days:
    \[
    \mathrm{CDD}_d =
    \frac{\mathrm{cdd\_sats\_days}_d}{100{,}000{,}000}.
    \]

    \item Write the resulting time series to a CSV file using the standardized OBM schema:
    \[
    \texttt{date},\quad
    \texttt{series\_id},\quad
    \texttt{value},\quad
    \texttt{unit},\quad
    \texttt{frequency},\quad
    \texttt{release\_version}.
    \]

    \item Optionally generate a plot of the exported series when the plotting flag is activated. If \texttt{\symbol{45}\symbol{45}plot\_output} is not provided, the plot is saved next to the CSV file with a \texttt{.png} extension. The plot title includes the series description and the selected date interval.
\end{enumerate}

This exporter does not query Bitcoin Core, does not maintain an outpoint state, and does not reconstruct spent outputs at export time. The computationally expensive previous-output resolution, age computation, and daily CDD accumulation have already been performed by the spent-output indexer, which stores the daily aggregate in \texttt{daily\_aggregates}. The export script is therefore a lightweight, deterministic transformation from the indexed database to the OBM CSV format.

\paragraph{Metric-specific input parameters.}
The input parameters specific to this exporter are:

\begin{itemize}
    \item \texttt{\symbol{45}\symbol{45}state\_db}: path to the persistent SQLite database generated by the indexer Python script \texttt{obm\_spent\_output\_indexer.py}. This database must contain the \texttt{daily\_aggregates} table and the metadata required to verify that it corresponds to the OBM spent-output indexer.

    \item \texttt{\symbol{45}\symbol{45}start\_date}: starting date of the selected interval, inclusive, in \texttt{YYYY-MM-DD} format. The date is interpreted as a UTC calendar date.

    \item \texttt{\symbol{45}\symbol{45}end\_date}: ending date of the selected interval, inclusive, in \texttt{YYYY-MM-DD} format. The script verifies that this date is not later than the maximum date processed by the indexer.

    \item \texttt{\symbol{45}\symbol{45}release\_version}: fallback dataset release version used only if the indexer database does not contain a \texttt{release\_version} metadata field.

    \item \texttt{\symbol{45}\symbol{45}output}: path of the output CSV file to be written.

    \item \texttt{\symbol{45}\symbol{45}plot}: optional flag that instructs the script to generate a plot of the exported series.

    \item \texttt{\symbol{45}\symbol{45}plot\_output}: optional path for the generated plot. If this argument is omitted while \texttt{\symbol{45}\symbol{45}plot} is used, the plot is saved next to the CSV file using the same base name and a \texttt{.png} extension.
\end{itemize}

Unlike the previous metric-specific CDD reconstruction script, the current exporter does not accept Bitcoin Core RPC parameters, \texttt{\symbol{45}\symbol{45}height\_margin}, \texttt{\symbol{45}\symbol{45}min\_confirmations}, \texttt{\symbol{45}\symbol{45}commit\_every}, or \texttt{\symbol{45}\symbol{45}reset\_state\_db}. These parameters now belong to the spent-output indexer, not to the CDD exporter. The exporter assumes that the indexer database has already been built and updated through the requested ending date. Its role is only to read \texttt{cdd\_sats\_days} from \texttt{daily\_aggregates}, convert the stored satoshi-day values into BTC-days, and write the standardized OBM output file.

\paragraph{Aggregation rule.}
The daily value is computed as the sum of CDD contributions for all outputs spent in blocks assigned to the same UTC calendar date:

\[
\mathrm{CDD}_d =
\sum_{i \in S_d}
v_i
\max\left(
0,
\frac{
t_i^{\mathrm{spent}} - t_i^{\mathrm{created}}
}{86400}
\right).
\]

The aggregation rule is therefore a daily sum of spent-output BTC-days. Monthly versions of this metric, if distributed, should also be computed as sums of the corresponding daily values:

\[
\mathrm{CDD}_m =
\sum_{d \in m} \mathrm{CDD}_d.
\]

\paragraph{Output format.}
The output file contains one observation per UTC date. Each row has the following fields:

\begin{center}
\begin{tabular}{llp{0.3\textwidth}}
\toprule
\textbf{Column} & \textbf{Example} & \textbf{Description} \\
\midrule
\texttt{date} & \texttt{2024-01-01} & UTC calendar date \\
\texttt{series\_id} & \texttt{obm\_cdd\_btcxdays\_daily} & Stable OBM series identifier \\
\texttt{value} & \texttt{123456789.12345678} & Daily Coin Days Destroyed \\
\texttt{unit} & \texttt{BTC-days} & Measurement unit \\
\texttt{frequency} & \texttt{daily} & Observation frequency \\
\texttt{release\_version} & \texttt{OBM v0.1.0} & Dataset release version \\
\bottomrule
\end{tabular}
\end{center}

\paragraph{Technical validation.}
Several internal checks are used to validate this metric at export time. First, the script verifies that the requested date range is valid and that \texttt{\symbol{45}\symbol{45}start\_date} is not later than \texttt{\symbol{45}\symbol{45}end\_date}. Second, it checks that the SQLite state database exists, opens it, and sets the connection to query-only mode. Third, it verifies that the database metadata identifies it as an OBM spent-output indexer database by checking the expected \texttt{indexer\_id}. Fourth, it checks that the indexer database contains processing metadata, including \texttt{last\_processed\_height}, \texttt{last\_processed\_date}, and \texttt{max\_processed\_date}. Fifth, it verifies that the requested ending date is not later than the maximum date processed by the indexer. If the requested interval extends beyond the indexed range, the script aborts and requires the indexer to be updated before the export is attempted.

For each date in the requested interval, the exporter reads \texttt{cdd\_sats\_days} from \texttt{daily\_aggregates}. If the date is absent from the aggregate table, the script exports zero CDD for that date. This convention is appropriate for \texttt{obm\_cdd\_btcxdays\_daily} because CDD is a flow variable: if no CDD contribution is assigned to a UTC date under the indexer's timestamp convention, the daily value is zero rather than missing. The OBM spent-output indexer is designed to be built from the Bitcoin genesis block onward. Accordingly, the exporter assumes that the indexer database contains a complete historical pass from block height 0 through the maximum processed height recorded in the database metadata. Under this invariant, a missing row in \texttt{daily\_aggregates} for a requested UTC date does not indicate an unprocessed date. It indicates that no spent-output aggregate was assigned to that date under the indexer's timestamp convention. For \texttt{obm\_cdd\_btcxdays\_daily}, such dates are therefore exported as zero.

Each retrieved value is converted from satoshi-days to BTC-days by dividing by \(100{,}000{,}000\), and the resulting file is written using the standard OBM schema:
\[
\texttt{date},\quad
\texttt{series\_id},\quad
\texttt{value},\quad
\texttt{unit},\quad
\texttt{frequency},\quad
\texttt{release\_version}.
\]

The script also propagates the release version stored in the indexer metadata. If the metadata field is absent, it uses the fallback value supplied through \texttt{\symbol{45}\symbol{45}release\_version}. The number of rows written must equal the number of calendar days in the requested interval, because the exporter iterates explicitly over all dates between \texttt{\symbol{45}\symbol{45}start\_date} and \texttt{\symbol{45}\symbol{45}end\_date}, inclusive. Exported values should be non-negative, since the indexer floors negative apparent output ages to zero before accumulating CDD.

The exporter also relies on the OBM invariant that the spent-output indexer is built from the Bitcoin genesis block onward; users employing non-standard databases should verify this condition before interpreting missing aggregate rows as zero values.

Additional consistency checks can be performed using related OBM series exported from the same indexer database. In particular, \texttt{obm\_dormancy\_days\_daily} should satisfy
\[
\mathrm{Dormancy}_d =
\frac{\mathrm{CDD}_d}{\mathrm{SpentValueBTC}_d}
\]
for every date \(d\) with positive spent value. Similarly, normalized versions such as the OBM metric \texttt{obm\_cdd\_per\_supply\_days\_daily} should reproduce the corresponding ratio between \texttt{obm\_cdd\_btcxdays\_daily} and \texttt{obm\_supply\_btc\_daily}. Comparisons with external CDD-like series, such as Coin Metrics' \texttt{TxTfrValDayDst}, Glassnode's CDD metric, or CryptoQuant's CDD metric, are useful as diagnostics but should not be interpreted as strict equality tests. Differences may arise from timestamp conventions, fractional-day versus full-day age treatment, transfer-level versus spent-output-level definitions, historical duplicate-transaction handling, or early-chain edge cases.

\paragraph{Known limitations.}
Bitcoin Days Destroyed is informative but definition-sensitive. First, \texttt{obm\_cdd\_btcxdays\_daily} is a raw spent-output metric, not an entity-adjusted or transfer-adjusted metric. It does not identify users, entities, custodians, exchanges, self-transfers, or change outputs. Second, the metric depends on the block timestamp convention used both to compute output age and to assign spending blocks to calendar days. Third, early Bitcoin data are especially sensitive to conventions regarding coinbase-origin outputs, daily-interval labeling, and the first non-coinbase transactions. Fourth, the metric requires a persistent local state database, which becomes part of the reproducible computation. Fifth, the simple reorganization policy detects inconsistencies and aborts, but it does not automatically roll back the state database. Finally, it is important to note that the reconstruction explicitly handles the two historical duplicate coinbase transaction pairs that existed before BIP30 enforcement by overwriting the earlier txid:vout entry in the local outpoint state and recording the event in metadata.

Despite these limitations, \texttt{obm\_cdd\_btcxdays\_daily} is one of the central coin-age metrics in the OBM dataset. It provides a transparent measure of destroyed coin age and serves as a foundation for studying dormant supply, long-term holder activity, dormancy, supply-adjusted CDD, and the interaction between on-chain behavior and market conditions.
\subsection{\texttt{obm\_cdd\_per\_supply\_days\_daily}: Bitcoin Days Destroyed per Unit of Supply}
\label{Bitcoin Days Destroyed per Unit of Supply}


\paragraph{Definition.} The Bitcoin-Days-Destroyed-per-unit-of-supply series measures daily Bitcoin Days Destroyed normalized by the outstanding Bitcoin supply. It is a derived metric computed from the daily Bitcoin Days Destroyed and Bitcoin supply series. Let $CDD_d$ denote the total number of BTC-days destroyed on UTC calendar day $d$, as reported by \texttt{obm\_cdd\_btcxdays\_daily}. Let $\mathrm{SupplyBTC}_d$ denote the outstanding Bitcoin supply on the same day, as reported by \texttt{obm\_supply\_btc\_daily}. The supply-normalized CDD metric is defined as:
\[
\mathrm{CDDPerSupply}_d =
\frac{\mathrm{CDD}_d}{\mathrm{SupplyBTC}_d}.
\]

The numerator is measured in BTC-days, and the denominator is measured in BTC. Therefore, the resulting unit is days.

The resulting series reports the amount of destroyed coin age per bitcoin outstanding. For example, a value of 0.015 means that the coin age destroyed on that date was equivalent to 0.015 days per bitcoin of outstanding supply.

When the denominator ($\mathrm{SupplyBTC}_d$) is zero, the ratio is undefined. This situation can occur only at the very beginning of Bitcoin history, before positive supply has been created under the selected timestamp convention. In such cases, OBM records the value as missing rather than zero. This convention avoids the incorrect implication that no coin age was destroyed relative to a positive monetary base. Instead, no positive supply exists from which a per-supply value can be computed.

\paragraph{Economic interpretation.} The CDD-per-supply series measures destroyed coin age relative to the size of the Bitcoin monetary base. It is therefore a supply-normalized version of Bitcoin Days Destroyed.

For economic research, this metric is useful in several ways. First, it makes CDD values more comparable across periods with very different levels of outstanding supply. Second, it helps distinguish growth in raw CDD caused by a larger monetary base from growth in CDD caused by unusually intense movement of aged coins. Third, it provides a normalized coin-age indicator for studying long-term holder activity, dormant-supply movement, distribution episodes, and supply-adjusted on-chain behavior. Fourth, it can be used together with raw CDD, transaction count, spent value, and dormancy-related indicators to separate activity intensity from the age profile of spent coins.

The metric should not be interpreted as a direct measure of transaction demand, payment volume, user activity, or miner revenue. It measures destroyed coin age per unit of outstanding supply. A high value indicates that the daily destruction of accumulated coin age was large relative to the monetary base, but it does not identify the entities involved, the economic purpose of the transactions, or whether the observed movements correspond to payments, self-transfers, exchange operations, custodial reorganization, or long-term holder distribution. 

\paragraph{Similar metrics publicly available.}
The OBM series \texttt{obm\_cdd\_per\_supply\_days\_daily} is comparable to the family of public metrics usually labelled \emph{Supply-Adjusted Coin Days Destroyed} or \emph{Supply-Adjusted CDD}. \texttt{obm\_cdd\_per\_supply\_days\_daily} is defined as a derived daily series obtained by dividing raw Bitcoin Days Destroyed by the outstanding Bitcoin supply on the same UTC calendar day. The numerator is measured in BTC-days and the denominator in BTC, so the resulting unit is days. This interpretation is important: the series measures destroyed coin age per bitcoin outstanding, not a dimensionless ratio. When the supply denominator is zero, \texttt{obm\_cdd\_per\_supply\_days\_daily} records the value as missing rather than zero.

The closest public equivalent is Glassnode's \emph{Bitcoin Supply-Adjusted CDD} metric, exposed as \texttt{indicators.CddSupplyAdjusted}\footnote{\url{https://studio.glassnode.com/charts/indicators.CddSupplyAdjusted?a=BTC}} (Advanced Plan required). Glassnode's methodological documentation states that Supply-Adjusted CDD ``simply divides CDD by the circulating supply'', described as the total amount of coins issued:
\[
\mathrm{Supply\mbox{-}Adjusted\ CDD}
=
\frac{\mathrm{CDD}}{\mathrm{Total\ Supply}}.
\]
This is very close to the \texttt{obm\_cdd\_per\_supply\_days\_daily} definition. The main difference is that Glassnode's public documentation does not fully specify all low-level conventions inherited from its underlying CDD and supply series, including fractional-day treatment, timestamp convention, entity adjustment status, treatment of early-chain edge cases, and reorganization handling. The Glassnode page offers CSV, JSON, API, Excel, and MCP access, but the public documentation is not equivalent to an open full-node reconstruction script.

CryptoQuant also provides a directly comparable \emph{Supply-Adjusted CDD} metric\footnote{\url{https://cryptoquant.com/asset/btc/chart/network-indicator/supply-adjusted-cdd}} (credentials required). CryptoQuant defines it as CDD divided by total supply. Its API documentation identifies the corresponding field as \texttt{sa\_cdd}, states that CDD is the sum of spent-output lifespan multiplied by value, and specifies that \texttt{sa\_cdd} is calculated as CDD over \texttt{supply\_total}. CryptoQuant, therefore, provides a clear formula-level definition. However, the public documentation does not provide a fully reproducible open-source algorithm for reconstructing the metric from a Bitcoin Core node, nor does it expose all implementation details needed to guarantee exact equivalence with \texttt{obm\_cdd\_per\_supply\_days\_daily}.

Bitcoin Magazine Pro provides a \emph{Bitcoin: Supply Adjusted Coin Days Destroyed} chart.\footnote{\url{https://www.bitcoinmagazinepro.com/charts/supply-adjusted-coin-days-destroyed-cdd/}} The page describes the metric as a variation of Coin Days Destroyed that divides CDD by the total number of bitcoins issued into the market. This matches the \texttt{obm\_cdd\_per\_supply\_days\_daily} concept. The page also provides a useful economic interpretation, emphasizing the comparability of the Bitcoin supply over time as it grows. Nevertheless, it is primarily a chart and explanatory page, not a reproducible metric specification. It does not disclose all calculation conventions, source data, timestamp rules, smoothing defaults, or chain-handling procedures. Moreover, only the 90-day median average is publicly available. 

MacroMicro provides a \emph{Bitcoin - Supply-Adjusted Coin Days Destroyed (CDD)} chart \footnote{\texttt{https://en.macromicro.me/charts/143317/bitcoin-supply-adjusted-token-burn-day}} (subscription required). The page identifies the metric as Supply-Adjusted CDD and describes it as an indicator of long-term holder behavior. It is useful as an additional public charting reference, but it does not provide a full algorithmic definition comparable to our metric \texttt{obm\_cdd\_per\_supply\_days\_daily}. In particular, the public page does not specify whether the underlying CDD is raw, entity-adjusted, smoothed, fractional-day-based, or computed under a specific timestamp convention.

TradingDigits provides a \emph{Bitcoin Supply Adjusted Coin Days Destroyed} chart.\footnote{\url{https://www.tradingdigits.io/supply-adjusted-cdd}} The page explains that Supply-Adjusted CDD refines standard CDD by dividing it by circulating supply. This is conceptually aligned with \texttt{obm\_cdd\_per\_supply\_days\_daily}. However, the publicly visible information is interpretive rather than methodological. It does not disclose a reproducible algorithm, the precise source of its CDD, supply inputs, or the treatment of low-level blockchain conventions.

Checkonchain provides CDD-related charts, including a supply-adjusted 7-day EMA CDD view.\footnote{\url{https://charts.checkonchain.com/btconchain/lifespan/cdd/cdd\_light.html}} This is related but not identical to \texttt{obm\_cdd\_per\_supply\_days\_daily}, because the public chart explicitly refers to a 7-day EMA and therefore appears to display a smoothed supply-adjusted variant rather than the raw daily ratio. It is useful for qualitative comparison, but not as a direct benchmark for the unsmoothed \texttt{obm\_cdd\_per\_supply\_days\_daily} series.

Coin Metrics does not appear to expose a named public metric that is exactly equivalent to \texttt{obm\_cdd\_per\_supply\_days\_daily}. However, it provides the raw CDD-like metric \texttt{TxTfrValDayDst}, and Supply metrics that could be combined externally to construct a similar ratio. Such a derived comparison would still inherit Coin Metrics' own conventions, including its full-day convention for transferred days destroyed. Therefore, a Coin Metrics-based ratio would be useful as a secondary benchmark, but it should not be described as a directly published equivalent unless the derived construction is explicitly documented.

Overall, the strongest public comparators for the \texttt{obm\_cdd\_per\_supply\_days\_daily} metric are Glassnode's \texttt{indicators.CddSupplyAdjusted} and CryptoQuant's \texttt{sa\_cdd}. Both state the central formula, CDD divided by supply. Bitcoin Magazine Pro, MacroMicro, TradingDigits, and Checkonchain provide useful public charting references, but their algorithmic documentation is thinner. The distinctive contribution of \texttt{obm\_cdd\_per\_supply\_days\_daily} is that the ratio is derived from two explicitly documented OBM source series, \texttt{obm\_cdd\_btcxdays\_daily} and \texttt{obm\_supply\_btc\_daily}, with a stable identifier, explicit unit in days, a missing-value convention when supply is zero, and validation identities that make the series auditable from its source files.

\paragraph{Data source and input requirements.} The metric is derived from two existing OBM CSV files, namely \texttt{obm\_cdd\_btcxdays\_daily.csv} and \texttt{obm\_supply\_btc\_daily.csv}. Unlike \texttt{obm\_cdd\_btcxdays\_daily}, this metric does not query Bitcoin Core and does not reconstruct previous transaction outputs directly. It is a deterministic transformation of two previously generated OBM time series.

The input files must follow the standard OBM schema:
\[
\texttt{date, series\_id, value, unit, frequency, release\_version}.
\]
The script verifies that the first input series is \texttt{obm\_cdd\_btcxdays\_daily}, that the second input series is \texttt{obm\_supply\_btc\_daily}, that the CDD series uses unit \texttt{BTC-days}, that the supply series uses unit \texttt{BTC}, and that both series have daily frequency. It also verifies that both input files contain all the required dates for the selected interval.

\paragraph{Algorithm.} The script \texttt{compute\_obm\_cdd\_per\_supply\_days\_daily.py} implements the following procedure:

\begin{enumerate}
    \item Read the input CSV file containing \texttt{obm\_cdd\_btcxdays\_daily}. This file is provided as the first mandatory positional argument.

    \item Read the input CSV file containing     \texttt{obm\_supply\_btc\_daily}. This file is provided as the second mandatory positional argument.

    \item Validate the schema of both input files. The script checks that each file contains the fields:
    \[
    \texttt{date, series\_id, value, unit, frequency, release\_version}.
    \]

    \item Validate that the input series identifiers are \texttt{obm\_cdd\_btcxdays\_daily} and \texttt{obm\_supply\_btc\_daily}, respectively.

    \item Validate that the CDD input series uses unit \texttt{BTC-days}, that the supply input series uses unit \texttt{BTC}, and that both input series have frequency \texttt{daily}.

    \item Determine the date interval. If \texttt{\symbol{45}\symbol{45}start\_date} and \texttt{\symbol{45}\symbol{45}end\_date} are provided, the script uses that interval. If one or both are omitted, the script uses the corresponding bounds of the common overlapping interval covered by both input files.

    \item Verify that both source files contain one and only one observation for every date in the selected interval. If any required date is missing, the script aborts.

    \item For each date $d$, read $CDD_d$ and $\mathrm{SupplyBTC}_d$. The script checks that both values are non-negative.

    \item If $\mathrm{SupplyBTC}_d > 0$, compute:
    \[
    CDDPerSupply_d =
    \frac{CDD_d}{SupplyBTC_d}.
    \]

    \item If $\mathrm{SupplyBTC}_d = 0$, record the observation as missing rather than zero.

    \item Infer the \texttt{release\_version} from the source observations in the selected interval. If multiple release versions are present, the script aborts to avoid mixing releases.

    \item Write the resulting time series to a CSV file using the standardized OBM schema:
    \[
    \texttt{date, series\_id, value, unit, frequency, release\_version}.
    \]

    \item Optionally generate a plot of the resulting series when the plotting flag is activated. Dates with missing values are skipped in the plot.
\end{enumerate}

\paragraph{Metric-specific input parameters.} The input parameters specific to this metric are:

\begin{itemize}
    \item \texttt{cdd\_csv}: mandatory positional argument containing the path to the \texttt{obm\_cdd\_btcxdays\_daily} CSV file.

    \item \texttt{supply\_csv}: mandatory positional argument containing the path to the \texttt{obm\_supply\_btc\_daily} CSV file.

    \item \texttt{\symbol{45}\symbol{45}start\_date}: starting date of the selected interval, inclusive. If omitted, the first common date available in both input files is used.

    \item \texttt{\symbol{45}\symbol{45}end\_date}: ending date of the selected interval, inclusive. If omitted, the last common date available in both input files is used.
\end{itemize}

The script also accepts \texttt{\symbol{45}\symbol{45}output}, \texttt{\symbol{45}\symbol{45}plot}, and \texttt{\symbol{45}\symbol{45}plot\_output}, following the same output and plotting conventions used by the other OBM scripts.

No Bitcoin Core RPC parameters are required, because this metric is not reconstructed directly from the blockchain. It is a deterministic transformation of already generated OBM time series.

\paragraph{Aggregation rule.} This metric is not aggregated directly from blocks or transactions. Instead, it is computed pointwise from two daily OBM series. For each UTC date $d$, the daily value is:
\[
CDDPerSupply_d =
\frac{CDD_d}{SupplyBTC_d}, \qquad SupplyBTC_d > 0.
\]

If $SupplyBTC_d = 0$, then $CDDPerSupply_d$ is undefined and is recorded as missing.

Monthly versions of this metric should not be computed as arithmetic averages of daily ratios without explicit justification. A more interpretable monthly version is obtained by first summing monthly CDD and using the appropriate monthly supply denominator, preferably the end-of-month supply stock:
\[
CDDPerSupply_m =
\frac{CDD_m}{SupplyBTC_m^{EOM}},
\]
where:
\[
CDD_m =
\sum_{d \in m} CDD_d.
\]
This convention preserves the interpretation of the metric as monthly destroyed coin age normalized by the outstanding Bitcoin supply at the end of the month. If an alternative denominator is used, such as average monthly supply, the choice should be documented explicitly.

\paragraph{Relationship with \texttt{obm\_cdd\_btcxdays\_daily} and \texttt{obm\_supply\_btc\_daily}.} Our series is directly derived from \texttt{obm\_cdd\_btcxdays\_daily} and \texttt{obm\_supply\_btc\_daily}. The former measures the daily flow of destroyed coin age, expressed in BTC-days, whereas the latter measures the outstanding Bitcoin supply, expressed in BTC. Dividing the former by the latter converts raw destroyed coin age into a supply-normalized value expressed in days.

For dates with positive supply, the defining identity is:
\[
CDDPerSupply_d =
\frac{CDD_d}{SupplyBTC_d}.
\]

This relationship provides a useful validation identity. Differences between the raw CDD series and the product of \texttt{obm\_cdd\_per\_supply\_days\_daily} and \texttt{obm\_supply\_btc\_daily} would indicate a timestamp mismatch, source-file inconsistency, implementation error, rounding issue, or release-version mismatch.

\paragraph{Output format.} The output file contains one observation per UTC date in the selected interval. Each row has the following fields:

\begin{center}
\begin{tabular}{llp{0.3\textwidth}}
\hline
Column & Example & Description \\
\hline
\texttt{date} & \texttt{2024-01-01} & UTC calendar date \\
\texttt{series\_id} & \texttt{obm\_cdd\_per\_supply\_days\_daily} &
Stable OBM series identifier \\
\texttt{value} & \texttt{0.012345678901} &
CDD divided by outstanding supply \\
\texttt{unit} & \texttt{days} & Measurement unit \\
\texttt{frequency} & \texttt{daily} & Observation frequency \\
\texttt{release\_version} & \texttt{OBM v0.1.0} &
Dataset release version inherited from the source interval \\
\hline
\end{tabular}
\end{center}

When the ratio is undefined because the supply denominator is zero, the \texttt{value} field is left empty. This indicates a missing value, not a zero value.

\paragraph{Technical validation.} Several internal checks are used to validate this metric. First, the script verifies that both source files conform to the OBM schema. Second, it checks that the source series identifiers are \texttt{obm\_cdd\_btcxdays\_daily} and \texttt{obm\_supply\_btc\_daily}. Third, it verifies that the CDD input uses unit \texttt{BTC-days}, that the supply input uses unit \texttt{BTC}, and that both inputs use frequency \texttt{daily}. Fourth, it checks that every date in the selected interval is present exactly once in both source files. Fifth, it verifies that source values are non-negative. Sixth, it records missing values only for dates where $\mathrm{SupplyBTC}_d = 0$. Seventh, it checks that the selected interval does not mix different release versions.

The primary validation identity is:
\[
CDDPerSupply_d =
\frac{CDD_d}{SupplyBTC_d}, \qquad SupplyBTC_d > 0.
\]
A secondary validation identity is:
\[
CDD_d =
CDDPerSupply_d \times SupplyBTC_d.
\]
These identities make the metric easy to audit and reproduce from the two source series.

\paragraph{Known limitations.} The CDD-per-supply series is simple and reproducible, but it has several limitations. First, it is a derived series and not an independent full-node reconstruction. Second, it inherits all definitional choices, timestamp conventions, and possible revisions of \texttt{obm\_cdd\_btcxdays\_daily} and \texttt{obm\_supply\_btc\_daily}. Third, it is undefined when the supply denominator equals zero, and OBM records such observations as missing rather than zero. Fourth, it is not entity-adjusted or transfer-adjusted, because the underlying CDD series is a raw spent-output metric. Fifth, it does not identify users, entities, exchanges, custodians, self-transfers, or change outputs. Sixth, it should not be interpreted as transaction demand, payment volume, or economic value transferred.

Despite these limitations, \texttt{obm\_cdd\_per\_supply\_days\_daily} is a useful derived OBM series. It provides a transparent supply-normalized measure of destroyed coin age and supports research on dormant-supply movement, long-term holder activity, coin-age dynamics, and comparisons of on-chain behavior across periods with different levels of outstanding Bitcoin supply.
\subsection{\texttt{obm\_difficulty\_eod\_daily}: End-of-Day Bitcoin Mining Difficulty}
\label{End-of-Day Bitcoin Mining Difficulty}

\paragraph{Definition.}
The end-of-day Bitcoin mining difficulty series reports the mining difficulty of the last block assigned to a given UTC calendar day. Difficulty is a protocol state variable, not a daily flow. It is therefore reported as an end-of-day observation rather than as a sum or average. The unit field is recorded as ``difficulty''. The unit is not repeated in the series identifier because difficulty is itself the named Bitcoin protocol state variable being measured.

Let \(B_d\) denote the set of blocks assigned to day \(d\). For each block \(b \in B_d\), let \(h_b\) denote its height and let \(D_b\) denote the mining difficulty reported by Bitcoin Core for that block. The end-of-day difficulty is defined as $\mathrm{DifficultyEOD}_d =
D_{b^\ast}$, where
\[
b^\ast =
\arg\max_{b \in B_d} h_b.
\]
That is, \(b^\ast\) is the highest-height block among all blocks assigned to UTC date \(d\).

A block \(b\) is assigned to day \(d\) according to the UTC calendar date derived from its block timestamp \(t_b\): $d(b)=\mathrm{UTCDate}(t_b)$. If no block is assigned to date \(d\), the series is undefined for that date and the output value is recorded as \texttt{NaN}. This convention avoids assigning a misleading value of zero to a protocol state variable.

\paragraph{Economic interpretation.}
Bitcoin mining difficulty measures how difficult it is to produce a valid proof-of-work block relative to Bitcoin's reference difficulty. It adjusts periodically so that the protocol continues to target an average inter-block interval of approximately ten minutes.

For economic research, this metric is useful in several ways. First, it captures the protocol-level mining condition faced by miners at the end of each UTC day. Second, it is useful for studying difficulty-adjustment cycles and mining dynamics. Third, it can be combined with block count, miner revenue, and transaction fees to analyze mining incentives. Fourth, it provides a natural input for estimated hashrate measures, although hashrate itself is not directly observed on-chain and requires an additional estimation convention. Fifth, it can be used as a control variable in empirical studies where mining conditions may affect network security, issuance timing, miner behavior, or Bitcoin market dynamics.

The metric should not be interpreted as observed hashrate, mining profitability, transaction demand, user activity, or network usage. Difficulty is a protocol state variable. It changes only at retarget boundaries and remains constant within ordinary difficulty epochs. Therefore, it is typically slow-moving and stepwise, unlike transaction count, fees, or block weight.

\paragraph{Similar metrics publicly available.}
The OBM series \texttt{obm\_difficulty\_eod\_daily} is comparable to public metrics usually labelled \emph{Bitcoin difficulty}, \emph{mining difficulty}, \emph{last difficulty}, \emph{average difficulty}, or \emph{network difficulty}. In OBM, the metric is defined as the mining difficulty of the highest-height block assigned to a given UTC calendar day. Let \(B_d\) denote the set of blocks assigned to UTC date \(d\). For each block \(b\), let \(h_b\) denote its height and let \(\mathrm{Difficulty}_b\) denote the difficulty value returned by Bitcoin Core. OBM defines
\[
b^\ast = \arg\max_{b \in B_d} h_b,
\]
and
\[
\mathrm{DifficultyEOD}_d = \mathrm{Difficulty}_{b^\ast}.
\]
If no block is assigned to date \(d\), the value is undefined and is written as \texttt{NaN}. This end-of-day convention treats difficulty as a protocol state variable, analogous to an end-of-period observation in financial or macroeconomic time series. It should therefore be distinguished from average daily difficulty and from forward-filled difficulty series.

The closest public comparator is Coin Metrics' \emph{Difficulty}, with MetricID \texttt{DiffLast}.\footnote{\url{https://api.coinmetrics.io/v4/timeseries/asset-metrics?assets=btc\&metrics=DiffLast}} Coin Metrics defines this metric as the difficulty of the last block in the considered time period. This is highly aligned with \texttt{obm\_difficulty\_eod\_daily}, because both use a last-observation convention over a daily interval. Coin Metrics also provides \emph{Mean Difficulty}, with MetricID \texttt{DiffMean}, defined as the mean difficulty of finding a valid block during the interval. \texttt{DiffMean} is related, but less directly comparable, because OBM deliberately avoids averaging a stepwise protocol state variable. 

Exact equality between \texttt{DiffLast} and \texttt{obm\_difficulty\_eod\_daily} should not be assumed without checking interval boundary conventions, timestamp assignment, chain-reorganization policy, and how dates with no blocks are treated. Coin Metrics' documentation also makes clear that the protocol adjusts difficulty periodically as a function of the hashing power deployed by miners. 

Glassnode provides Bitcoin mining difficulty charts and educational material on proof-of-work mining. Its documentation describes mining difficulty as the complexity of the puzzle miners must solve to find the next valid block, and explains that the difficulty adjustment occurs every 2016 blocks to regulate block production around the target interval. Glassnode, therefore, provides a useful conceptual and charting comparator for Bitcoin difficulty. However, the public documentation is less explicit than Coin Metrics about whether the daily series is last-observation, mean-observation, or forward-filled. Thus, Glassnode's difficulty-related series are useful for external comparison, but the aggregation convention should be verified before treating them as exact equivalents of \texttt{obm\_difficulty\_eod\_daily}. 

Blockchain.com provides a \emph{Network Difficulty} chart.\footnote{\url{https://www.blockchain.com/charts/difficulty}} The page defines difficulty as a measure of how difficult it is to mine a Bitcoin block, or technically, to find a hash below a given target. This is a close conceptual comparator for the OBM series. However, the public chart page does not specify whether daily observations correspond to end-of-day difficulty, average daily difficulty, a forward-filled state value, or another convention. It also does not fully document how no-block dates, timestamp boundaries, and historical revisions are handled. Blockchain.com should therefore be treated as a useful public charting reference rather than as a fully auditable methodological benchmark. 

BitInfoCharts provides a \emph{Bitcoin Difficulty} historical chart.\footnote{\url{https://bitinfocharts.com/comparison/bitcoin-difficulty.html}} The page explicitly labels the series as \emph{Average mining difficulty per day}. This makes it related but not identical to \texttt{obm\_difficulty\_eod\_daily}. Since Bitcoin difficulty is constant during a difficulty epoch and changes only at retarget boundaries, average and end-of-day values will often coincide on ordinary days but may differ on retarget days or days affected by boundary conventions. BitInfoCharts also reports current difficulty and next-retarget information on its Bitcoin statistics page, which is useful for diagnostics, but the public documentation does not provide a full, reproducible algorithm. 

CoinWarz provides a \emph{Bitcoin Difficulty Chart}.\footnote{\url{https://www.coinwarz.com/mining/bitcoin/difficulty-chart}} The page reports current and historical Bitcoin mining difficulty and is useful as a mining-focused public reference. However, it is primarily a mining calculator and charting source. Its public page does not provide enough methodological detail to determine whether historical daily values are last-observation, average, forward-filled, or sampled at a specific time. It should therefore be used as a secondary comparator rather than as a strict validation benchmark.

Difficulty-related hashrate charts should also be distinguished from \texttt{obm\_difficulty\_eod\_daily}. For example, Blockchain.com provides a total hash-rate chart and notes that the exact Bitcoin hashing power is unknown and must be estimated from the number of blocks being mined and the current block difficulty.\footnote{\url{https://www.blockchain.com/charts/hash-rate}} Hashrate metrics are derived estimates that combine difficulty with block-production assumptions over a selected window. OBM difficulty is instead a directly observed protocol state variable extracted from block metadata. Therefore, hashrate charts are useful companions, but not direct comparators. 

Block explorers and data APIs such as Blockchair can also support independent reconstruction. Block-level data generally includes the difficulty or compact target fields from which difficulty can be read or computed. A researcher could reconstruct \texttt{obm\_difficulty\_eod\_daily} by grouping blocks by UTC date and selecting the difficulty of the highest-height block in each group. However, unless a provider publishes a named end-of-day difficulty series with documented aggregation rules, such sources should be treated as lower-level reconstruction inputs rather than direct public comparators.

Overall, the strongest public comparator for \texttt{obm\_difficulty\_eod\_daily} is Coin Metrics \texttt{DiffLast}, because it explicitly reports the difficulty of the last block in the interval, matching the OBM end-of-day convention. Coin Metrics \texttt{DiffMean}, BitInfoCharts' average daily difficulty, Blockchain.com's network difficulty chart, Glassnode's difficulty-related charts, and CoinWarz's difficulty chart are useful related references, but their aggregation conventions may differ or may not be fully documented. The distinctive contribution of OBM is that it defines the daily observation precisely as the difficulty of the highest-height block assigned to the UTC date, writes \texttt{NaN} when no block is assigned to that date, uses Bitcoin Core's block-level \texttt{difficulty} field, and documents the block-timestamp convention and height-margin scan needed for reproducibility.

\paragraph{Data source and input requirements.}
The metric is obtained from a running Bitcoin Core full node through the JSON-RPC interface. For each block in the selected interval, the script retrieves the decoded block object using:
\[
\texttt{getblock <block\_hash> 1}.
\]
The mining difficulty is read from the block-level \texttt{difficulty} field returned by Bitcoin Core.

This metric does not require reconstructing previous transaction outputs. It does not require the spent-output indexer database, transaction-level input resolution, address extraction, user clustering, entity identification, external price data, or third-party APIs. It also does not require \texttt{txindex=1}, because the required information is available from block metadata.

The metric requires access to a synchronized Bitcoin Core full node and JSON-RPC credentials or cookie authentication. The script uses the locally verified main chain reported by the node. As with other directly scanned OBM metrics, the block timestamp returned by Bitcoin Core is used to assign each block to a UTC calendar date.

\paragraph{Algorithm.}
The script \texttt{compute\_obm\_difficulty\_eod\_daily.py} implements the following procedure:

\begin{enumerate}
    \item Parse the user-provided date interval, \texttt{\symbol{45}\symbol{45}start\_date} and \texttt{\symbol{45}\symbol{45}end\_date}, using the format \texttt{YYYY-MM-DD}. Both dates are interpreted as UTC dates and both are included in the output.

    \item Convert the starting date into the timestamp corresponding to 00:00:00 UTC of that day, and the ending date into the timestamp corresponding to 23:59:59 UTC of that day.

    \item Connect to the local Bitcoin Core node through JSON-RPC, using either explicit RPC credentials, environment variables, or cookie authentication.

    \item Query the local node using \texttt{getblockchaininfo} to determine the current best-chain height.

    \item Locate an approximate height interval covering the requested date range. This is done by using block timestamps and binary search over block heights.

    \item Expand the approximate height interval using the metric-specific safety parameter called \texttt{\symbol{45}\symbol{45}height\_margin}. This reduces the risk of missing boundary blocks because Bitcoin block timestamps are not strictly monotonic with respect to height.

    \item Initialize all dates in the requested interval with missing values. This ensures that the output file contains one row for each selected UTC date, including dates with no assigned blocks.

    \item Scan every block in the expanded height interval. For each height \(h\), the script:
    \begin{enumerate}
        \item obtains the corresponding block hash using \texttt{getblockhash};
        \item retrieves the decoded block using \texttt{getblock} with verbosity level 1;
        \item extracts the block timestamp \(t_b\);
        \item assigns the block to a UTC date \(d(b)\);
        \item skips the block if \(d(b)\) falls outside the requested date interval;
        \item reads the block-level \texttt{difficulty} field;
        \item compares the block height with the currently stored last-block height for that date;
        \item stores the block's difficulty if it is the highest-height block observed for that date.
    \end{enumerate}

    \item For each selected UTC date, write the stored difficulty value if at least one block was assigned to that date. Otherwise, write \texttt{NaN}.

    \item Write the resulting time series to a CSV file using the standardized OBM schema:
    \[
    \texttt{date},\quad
    \texttt{series\_id},\quad
    \texttt{value},\quad
    \texttt{unit},\quad
    \texttt{frequency},\quad
    \texttt{release\_version}.
    \]

    \item Optionally generate a plot of the resulting series when the plotting flag is activated.
\end{enumerate}

The script therefore computes the metric directly from decoded block metadata. It does not query the spent-output indexer, does not resolve transaction inputs, and does not reconstruct previous outputs.

\paragraph{Metric-specific input parameters.}
The metric-specific input parameter is:

\begin{itemize}
    \item \texttt{\symbol{45}\symbol{45}height\_margin}: number of extra blocks scanned before and after the approximate height interval associated with the requested date range.
\end{itemize}

This parameter is needed because Bitcoin block timestamps are not strictly monotonic with respect to block height. A timestamp-based binary search provides only an approximate height interval. To avoid excluding blocks whose timestamps fall inside the requested UTC interval but whose heights lie slightly outside the approximate interval, the script expands the scan interval by \(m\) blocks on both sides:
\[
h^{\mathrm{scan}}_{\min} =
\max(0, h^{\mathrm{approx}}_{\min} - m),
\]
\[
h^{\mathrm{scan}}_{\max} =
\min(h_{\mathrm{tip}}, h^{\mathrm{approx}}_{\max} + m).
\]

The default value is 288 blocks, approximately two days of expected Bitcoin block production. This is deliberately conservative. The margin only widens the internal block scan; the script still considers only blocks whose UTC dates fall inside the requested interval.

The script also accepts standard RPC, output, release-version, and plotting parameters:
\texttt{\symbol{45}\symbol{45}rpc\_url}, \texttt{\symbol{45}\symbol{45}rpc\_user}, \texttt{\symbol{45}\symbol{45}rpc\_password}, \texttt{\symbol{45}\symbol{45}cookie\_path}, \texttt{\symbol{45}\symbol{45}use\_default\_cookie}, \texttt{\symbol{45}\symbol{45}rpc\_timeout}, \texttt{\symbol{45}\symbol{45}output}, \texttt{\symbol{45}\symbol{45}release\_version}, \texttt{\symbol{45}\symbol{45}plot}, \texttt{\symbol{45}\symbol{45}plot\_output}, and \texttt{\symbol{45}\symbol{45}quiet}.

\paragraph{Aggregation rule.}
Difficulty is not aggregated as a flow. For each UTC date \(d\), the daily value is the difficulty of the highest-height block assigned to that date:
\[
\mathrm{DifficultyEOD}_d =
D_{b^\ast},
\qquad
b^\ast =
\arg\max_{b \in B_d} h_b.
\]

If no block is assigned to date \(d\), then \(B_d = \emptyset\), and \(\mathrm{DifficultyEOD}_d\) is undefined. In the output file, this case is recorded as \texttt{NaN}.

Monthly versions of this metric, if distributed, should not be computed as sums or arithmetic averages of daily difficulty values. Since difficulty is a protocol state variable, a monthly version should preferably be computed as an end-of-month observation:
\[
\mathrm{DifficultyEOM}_m =
\mathrm{DifficultyEOD}_{d^\ast},
\]
where \(d^\ast\) is the last date in month \(m\) for which a difficulty observation is defined. Alternatively, a monthly average could be constructed for specific empirical purposes, but it should be documented as a separate convention.

\paragraph{Output format.}
The output file contains one observation per UTC date. Each row has the following fields:

\begin{center}
\begin{tabular}{llp{0.45\textwidth}}

\toprule
\textbf{Column} & \textbf{Example} & \textbf{Description} \\
\midrule
\texttt{date} & \texttt{2024-01-01} & UTC calendar date \\
\texttt{series\_id} & \texttt{obm\_difficulty\_eod\_daily} & Stable OBM series identifier \\
\texttt{value} & \texttt{73197634206448.906250000000} & Difficulty of the last block assigned to the UTC date \\
\texttt{unit} & \texttt{difficulty} & Bitcoin mining difficulty \\
\texttt{frequency} & \texttt{daily} & Observation frequency \\
\texttt{release\_version} & \texttt{OBM v0.1.0} & Dataset release version \\
\bottomrule
\end{tabular}
\end{center}

Defined values are written with twelve decimal places. Dates with no assigned blocks are written as \texttt{NaN}. The unit field is recorded as \texttt{difficulty}; although difficulty is dimensionless, this label is more informative for users than a generic dimensionless unit.

\paragraph{Technical validation.}
Several internal checks are used to validate this metric during execution. First, the script verifies that the requested date range is valid and that \texttt{\symbol{45}\symbol{45}start\_date} is not later than \texttt{\symbol{45}\symbol{45}end\_date}. Second, it verifies that \texttt{\symbol{45}\symbol{45}height\_margin} is non-negative. Third, it checks that RPC authentication is valid, including cookie-file existence and format when cookie authentication is used. Fourth, it connects to Bitcoin Core and retrieves the current chain tip using \texttt{getblockchaininfo}. Fifth, it locates and expands the block-height interval corresponding to the requested timestamp range. Sixth, it scans all blocks in the expanded interval and considers only blocks whose UTC date lies inside the requested interval. Seventh, it verifies that each counted block contains a \texttt{difficulty} field. Eighth, it rejects negative difficulty values. Ninth, it selects the highest-height block assigned to each UTC date. Tenth, it writes \texttt{NaN} for selected dates with no assigned block.

Additional consistency checks can be performed using related OBM metrics. If the metric \texttt{obm\_block\_count\_daily} is positive for date \(d\), then \texttt{obm\_difficulty\_eod\_daily} should normally be defined for that date. Conversely, if block count is zero under the same timestamp convention, the end-of-day difficulty observation should be \texttt{NaN}. Difficulty should remain constant within ordinary retarget periods and change at difficulty-adjustment boundaries. Large changes should therefore coincide with retarget heights rather than arbitrary dates.

For external validation, selected periods can be compared with public difficulty series from blockchain explorers or data providers. Such comparisons should be interpreted cautiously because external sources may report end-of-day difficulty, average difficulty, block-level sampled difficulty, or forward-filled daily difficulty. They may also use different timestamp conventions or daily boundary rules.

\paragraph{Known limitations.}
The end-of-day difficulty series is transparent and reproducible, but it has several limitations. First, it reports end-of-day difficulty, not average daily difficulty. Second, it is defined using the highest-height block assigned to each UTC date, so it depends on the block timestamp convention used to assign blocks to days. Third, dates with no assigned blocks are recorded as \texttt{NaN}, not forward-filled. Users who need a fully filled state series may forward-fill the output as a downstream transformation, but that is not part of the primary OBM definition. Fourth, difficulty is not observed hashrate. Hashrate must be estimated from difficulty and realized block production over a chosen window. Fifth, the metric does not directly measure mining profitability, transaction demand, network usage, or economic activity. Sixth, it is computed directly from decoded block metadata and does not require the spent-output indexer.

Despite these limitations, \texttt{obm\_difficulty\_eod\_daily} is a useful OBM protocol-state series. It provides a simple, full-node-derived measure of Bitcoin mining difficulty and complements block count, block weight, fees, miner revenue, issuance, and future estimated-hashrate metrics.           
\subsection{\texttt{obm\_dormancy\_days\_daily}: Daily Dormancy}
\label{Daily Dormancy}


\paragraph{Definition.}
The daily dormancy series measures the value-weighted average age, in days, of outputs spent in blocks assigned to a given UTC calendar day. Let \(B_d\) denote the set of blocks assigned to day \(d\). For each non-coinbase transaction input \(i\) included in a transaction in block \(b \in B_d\), let \(v_i\) denote the value, in BTC, of the previous output spent by that input, and let \(a_i\) denote the age of that previous output, measured in days. The age of the spent output is defined as:
\[
a_i =
\max\left(
0,
\frac{t^{\mathrm{spent}}_i - t^{\mathrm{created}}_i}{86400}
\right),
\]
where \(t^{\mathrm{created}}_i\) is the timestamp of the block in which the spent output was created, and \(t^{\mathrm{spent}}_i\) is the timestamp of the block in which it is spent.

Daily Bitcoin Days Destroyed is:
\[
\mathrm{CDD}_d =
\sum_{b \in B_d}
\sum_{i \in I_b}
v_i a_i,
\]
where \(I_b\) denotes the set of non-coinbase transaction inputs included in block \(b\). Daily spent output value is:
\[
\mathrm{SpentValueBTC}_d =
\sum_{b \in B_d}
\sum_{i \in I_b}
v_i.
\]

Daily dormancy is then defined as:
\[
\mathrm{Dormancy}_d =
\frac{\mathrm{CDD}_d}{\mathrm{SpentValueBTC}_d},
\qquad
\mathrm{SpentValueBTC}_d > 0.
\]

The resulting unit is days, because BTC-days divided by BTC leaves days. If \(\mathrm{SpentValueBTC}_d = 0\), dormancy is undefined and is recorded as missing.

A block \(b\) is assigned to day \(d\) according to the UTC calendar date derived from its block timestamp \(t_b\):
\[
d(b)=\mathrm{UTCDate}(t_b).
\]

\paragraph{Economic interpretation.}
Daily dormancy measures the average age of spent coins, weighted by the BTC value of the outputs being spent. It can be interpreted as the average number of days that the BTC value spent on a given day had remained inactive before being consumed by transaction inputs.

For example, if the only outputs spent on a given day are 1 BTC aged 10 days and 3 BTC aged 20 days, then:
\[
\mathrm{CDD}_d = 1 \times 10 + 3 \times 20 = 70
\]
BTC-days, while:
\[
\mathrm{SpentValueBTC}_d = 1 + 3 = 4.
\]
The dormancy value is therefore:
\[
\mathrm{Dormancy}_d = \frac{70}{4} = 17.5
\]
days.

For economic research, this metric is useful because it separates the amount of value spent from the age profile of that value. A high spent value can correspond to recently active coins, while a high dormancy value indicates that the value being spent had, on average, remained inactive for a longer period. The series is therefore relevant for studying dormant-supply activation, long-term-holder behavior, distribution episodes, UTXO turnover, and the relationship between coin-age dynamics and market conditions.

The metric should not be interpreted as transaction count, payment volume, user activity, or entity-adjusted economic settlement value. It is a raw spent-output age metric. It does not identify users, entities, exchanges, custodians, change outputs, or self-transfers.

\paragraph{Similar metrics publicly available.}
The OBM series \texttt{obm\_dormancy\_days\_daily} is comparable to public metrics usually labeled \emph{Average Dormancy}, \emph{Average Coin Dormancy}, or simply \emph{Dormancy}. In OBM, the metric is defined as a derived daily series obtained by dividing daily Bitcoin Days Destroyed by daily spent output value:
\[
\mathrm{Dormancy}_d =
\frac{\mathrm{CDD}_d}{\mathrm{SpentValueBTC}_d},
\qquad
\mathrm{SpentValueBTC}_d > 0.
\]
Since \(\mathrm{CDD}_d\) is measured in BTC-days and \(\mathrm{SpentValueBTC}_d\) is measured in BTC, the resulting unit is days. The series can therefore be interpreted as the average age, in days, of the coins spent on day \(d\), weighted by the spent BTC value. When the denominator is zero, the value is undefined and should be recorded as missing rather than as zero. This distinction is important because a day with no spent value does not imply zero dormancy; it implies that no meaningful average spent age can be computed.

The closest public comparator is Glassnode's \emph{Bitcoin Dormancy} chart, exposed as \texttt{indicators.AverageDormancy.}\footnote{\url{https://studio.glassnode.com/charts/indicators.AverageDormancy?a=BTC}} Glassnode defines Dormancy as the average number of days destroyed per coin transacted, calculated as the ratio between coin days destroyed and total transfer volume. Its methodological guide further describes Average Coin Dormancy as the average number of days that each spent coin had remained dormant before it was moved, and states that the metric divides total coin-days destroyed by total coin volume transacted. This is conceptually very close to \texttt{obm\_dormancy\_days\_daily}. The main difference is that OBM uses its own raw spent-output value in the denominator, whereas Glassnode refers to total transfer volume. Exact equivalence, therefore, depends on whether the provider's transfer-volume convention coincides with OBM's raw spent-output-value convention, and on low-level choices such as timestamp assignment, fractional-day treatment, entity adjustment status, and reorganization handling. The public Glassnode page offers CSV, JSON, API, Excel, and MCP access, but it does not provide a full open-source reconstruction script. 

CryptoQuant provides a directly comparable \emph{Bitcoin: Average Dormancy} chart.\footnote{\url{https://cryptoquant.com/asset/btc/chart/network-indicator/average-dormancy}} CryptoQuant describes Average Dormancy as the average number of destroyed days of moved coins, calculated by dividing CDD by the total amount of movement of coins. Its API documentation identifies the field \texttt{average\_dormancy} and states that it is the average number of days destroyed per coin transacted. It also exposes \texttt{sa\_average\_dormancy}, a supply-adjusted version normalized by total supply.\footnote{\url{https://api.cryptoquant.com/v1/btc/network-indicator/dormancy}}. The formula-level description is close to OBM's definition. However, the public documentation does not provide a full node-level reconstruction algorithm, nor does it specify all conventions needed for exact replication, such as whether CDD uses fractional or integer days, whether transfer volume is raw or adjusted, and how daily boundaries and reorganizations are handled.

Coin Metrics does not appear to expose a named public metric exactly equivalent to \texttt{obm\_dormancy\_days\_daily}. However, it provides the components needed to construct a related dormancy measure. Its \texttt{TxTfrValDayDst} metric\footnote{\url{https://api.coinmetrics.io/v4/timeseries/asset-metrics?assets=btc\&metrics=TxTfrValDayDst}} measures the number of days destroyed by transferred units, while \texttt{TxTfrValNtv}\footnote{\url{https://api.coinmetrics.io/v4/timeseries/asset-metrics?assets=btc\&metrics=TxTfrValNtv}} measures the number of native units transferred during the interval. A Coin Metrics-based dormancy proxy could therefore be constructed as
\[
\frac{\texttt{TxTfrValDayDst}}{\texttt{TxTfrValNtv}}.
\]

This derived series would be conceptually similar to dormancy, but it would inherit Coin Metrics' own transfer-value and days-destroyed conventions. In particular, Coin Metrics documents transferred days destroyed as a transfer-based metric and uses full-day conventions for that metric. Therefore, this construction should be treated as a useful component-based comparator rather than as a direct replication of OBM's raw spent-output dormancy.

Glassnode also provides several related, but not identical, dormancy variants. Its \emph{Supply-Adjusted Dormancy} metric divides Average Coin Dormancy by circulating supply, while its point-in-time and entity-adjusted dormancy variants discard transactions between addresses attributed to the same entity. These variants are analytically useful, but they should not be treated as direct comparators for \texttt{obm\_dormancy\_days\_daily}. OBM's metric is raw, spent-output-based, and not entity-adjusted. Consequently, entity-adjusted or supply-adjusted dormancy series answer different research questions and rely on additional provider-specific assumptions.

CryptoQuant similarly provides \texttt{sa\_average\_dormancy}, a supply-adjusted dormancy series, alongside \texttt{average\_dormancy}. The supply-adjusted version is related but not equivalent, because it normalizes average dormancy by total supply. For comparison with OBM, the relevant CryptoQuant series is \texttt{average\_dormancy}, not \texttt{sa\_average\_dormancy}.

Several public charting sites provide dormancy-adjacent metrics rather than direct equivalents. Bitbo provides charts on dormant coins.\footnote{\url{https://charts.bitbo.io/dormant-coins/}} This chart tracks bitcoins that have not moved for more than one year, and is therefore a dormant-supply stock metric rather than a daily average spent-age flow. Checkonchain also provides a broad suite of Bitcoin on-chain charts, including lifespan and coin-age-related indicators.\footnote{\url{https://charts.checkonchain.com/}} Such charts are useful for qualitative comparison of long-term-holder behavior, but they should not be interpreted as direct implementations of \texttt{obm\_dormancy\_days\_daily} unless the specific denominator, numerator, and smoothing convention are documented.

Dormancy Flow metrics should also be distinguished from daily dormancy. For example, Glassnode defines Dormancy Flow as a ratio of holdings to the rolling annual USD value of coin-day destruction, rather than CDD divided by spent BTC value. Thus, Dormancy Flow is a valuation or cycle indicator derived from dormancy-related concepts, not a direct equivalent of the OBM daily dormancy series.

Overall, the strongest public comparators for \texttt{obm\_dormancy\_days\_daily} are Glassnode \texttt{indicators.AverageDormancy} and CryptoQuant \texttt{average\_dormancy}, because both define the metric as coin days destroyed divided by transacted or moved coin volume. A related Coin Metrics comparator can be constructed from \texttt{TxTfrValDayDst} divided by \texttt{TxTfrValNtv}, but this is a derived construction and not a directly named public dormancy metric. Supply-adjusted dormancy, entity-adjusted dormancy, Dormancy Flow, and dormant-supply charts are useful related indicators, but they are not direct equivalents. The distinctive contribution of OBM is that it defines daily dormancy explicitly from two auditable OBM source series, \texttt{obm\_cdd\_btcxdays\_daily} and \texttt{obm\_spent\_value\_btc\_daily}, reports the result in days, uses a documented missing-value convention when spent value is zero, and avoids proprietary entity-adjustment heuristics.

\paragraph{Data source and input requirements.}
The metric is exported from the persistent OBM spent-output indexer database, rather than computed by an independent metric-specific blockchain scan. The indexer, described in Sect.~\ref{sec:The Spent-Output Indexer}, is built from a running Bitcoin Core full node and stores reusable daily aggregates in a SQLite database. For \texttt{obm\_dormancy\_days\_daily}, the relevant source table is \texttt{daily\_aggregates}, and the relevant fields are \texttt{cdd\_sats\_days} and \texttt{spent\_value\_sats}.

The indexer computes both quantities during its sequential scan of the blockchain. For each non-coinbase transaction input, it resolves the input to the previous output it spends, retrieves the previous output value and creation timestamp from the live outpoint state, computes the elapsed age in days, and accumulates both the spent value and the value-age product. Internally, the indexer stores:
\[
\mathrm{cdd\_sats\_days}_d =
\sum_{b \in B_d}
\sum_{i \in I_b}
v^{\mathrm{sats}}_i
\max\left(
0,
\frac{t^{\mathrm{spent}}_i - t^{\mathrm{created}}_i}{86400}
\right),
\]
and:
\[
\mathrm{spent\_value\_sats}_d =
\sum_{b \in B_d}
\sum_{i \in I_b}
v^{\mathrm{sats}}_i.
\]

The exporter computes dormancy directly as:
\[
\mathrm{Dormancy}_d =
\frac{\mathrm{cdd\_sats\_days}_d}{\mathrm{spent\_value\_sats}_d},
\qquad
\mathrm{spent\_value\_sats}_d > 0.
\]

No satoshi-to-BTC conversion factor is needed in this ratio, because both numerator and denominator use satoshis as the value unit. The resulting unit is days.

The exporter script does not query Bitcoin Core directly. It requires an existing SQLite database generated by \texttt{obm\_spent\_output\_indexer.py}, with metadata identifying it as an OBM spent-output indexer database and with processed-date metadata covering the requested output interval. The script checks that the requested ending date is not beyond the maximum date processed by the indexer. If a requested date has no row in \texttt{daily\_aggregates}, or if \texttt{spent\_value\_sats} is zero, the exporter writes \texttt{NaN} for that date. This convention is appropriate because dormancy is undefined when no spent value exists.

The OBM spent-output indexer is designed to be built from the Bitcoin genesis block onward. Accordingly, the exporter assumes that the indexer database contains a complete historical pass from block height 0 through the maximum processed height recorded in the database metadata. Under this invariant, a missing row in \texttt{daily\_aggregates} for a requested UTC date does not indicate an unprocessed date; it indicates that no spent-output aggregate was assigned to that date under the indexer's timestamp convention. For \texttt{obm\_dormancy\_days\_daily}, such dates are recorded as \texttt{NaN}, because the spent-value denominator is zero and the ratio is undefined.

This metric does not require address extraction, user clustering, entity identification, external price data, third-party APIs, or a direct Bitcoin Core connection at export time. However, it depends on the prior successful execution of the spent-output indexer, which itself requires a synchronized, non-pruned Bitcoin Core full node and access to transaction-level data sufficient to reconstruct previous-output values and creation timestamps. The exported series therefore inherits the indexer's UTC block-timestamp convention, previous-output reconstruction procedure, fractional-day age convention, treatment of negative apparent ages, chain-consistency checks, release metadata, and handling of historical duplicate outpoint edge cases.

\paragraph{Algorithm.}
The script \texttt{export\_obm\_dormancy\_days\_daily.py} implements the following procedure:

\begin{enumerate}
    \item Parse the user-provided date interval, \texttt{\symbol{45}\symbol{45}start\_date} and \texttt{\symbol{45}\symbol{45}end\_date}, using the format \texttt{YYYY-MM-DD}. Both dates are interpreted as UTC dates and both are included in the output.

    \item Open the persistent SQLite database generated by \texttt{obm\_spent\_output\_indexer.py}. The path to this database is provided through the \texttt{\symbol{45}\symbol{45}state\_db} argument.

    \item Validate that the SQLite database corresponds to the OBM spent-output indexer. The script checks the indexer metadata, including the expected \texttt{indexer\_id} and the presence of processing metadata such as \texttt{last\_processed\_height}, \texttt{last\_processed\_date}, and \texttt{max\_processed\_date}.

    \item Verify that the requested \texttt{\symbol{45}\symbol{45}end\_date} is not later than the maximum UTC date processed by the indexer. If the requested interval extends beyond the indexed range, the script aborts and requires the indexer to be run further before export.

    \item Infer the dataset release version from the indexer metadata. If the database does not contain a release-version field, the script uses the fallback value provided through the \texttt{\symbol{45}\symbol{45}release\_version} argument.

    \item For each UTC date \(d\) in the requested interval, query the \texttt{daily\_aggregates} table for the fields \texttt{cdd\_sats\_days} and \texttt{spent\_value\_sats}. If a row exists for date \(d\), the script reads both values. If no row exists, the script records dormancy as undefined.

    \item If \(\mathrm{spent\_value\_sats}_d > 0\), compute:
    \[
    \mathrm{Dormancy}_d =
    \frac{\mathrm{cdd\_sats\_days}_d}{\mathrm{spent\_value\_sats}_d}.
    \]

    \item If \(\mathrm{spent\_value\_sats}_d = 0\), or if the date has no row in \texttt{daily\_aggregates}, write \texttt{NaN} in the value field. This indicates that dormancy is undefined, not zero.

    \item Write the resulting time series to a CSV file using the standardized OBM schema:
    \[
    \texttt{date},\quad
    \texttt{series\_id},\quad
    \texttt{value},\quad
    \texttt{unit},\quad
    \texttt{frequency},\quad
    \texttt{release\_version}.
    \]

    \item Optionally generate a plot of the exported series when the plotting flag is activated. Undefined observations are skipped in the plot. If \texttt{\symbol{45}\symbol{45}plot\_output} is not provided, the plot is saved next to the CSV file with a \texttt{.png} extension.
\end{enumerate}

This exporter does not query Bitcoin Core, does not maintain an outpoint state, and does not reconstruct spent outputs at export time. The computationally expensive previous-output resolution, spent-value aggregation, and CDD accumulation have already been performed by the spent-output indexer. The export script is therefore a lightweight, deterministic transformation from the indexed database to the OBM CSV format.

\paragraph{Metric-specific input parameters.}
The input parameters specific to this exporter are:

\begin{itemize}
    \item \texttt{\symbol{45}\symbol{45}state\_db}: path to the persistent SQLite database generated by the Python script \texttt{obm\_spent\_output\_indexer.py}. This database must contain the \texttt{daily\_aggregates} table and the metadata required to verify that it corresponds to the OBM spent-output indexer.

    \item \texttt{\symbol{45}\symbol{45}start\_date}: starting date of the selected interval, inclusive, in \texttt{YYYY-MM-DD} format. The date is interpreted as a UTC calendar date.

    \item \texttt{\symbol{45}\symbol{45}end\_date}: ending date of the selected interval, inclusive, in \texttt{YYYY-MM-DD} format. The script verifies that this date is not later than the maximum date processed by the indexer.

    \item \texttt{\symbol{45}\symbol{45}release\_version}: fallback dataset release version used only if the indexer database does not contain a \texttt{release\_version} metadata field.

    \item \texttt{\symbol{45}\symbol{45}output}: path of the output CSV file to be written.

    \item \texttt{\symbol{45}\symbol{45}plot}: optional flag that instructs the script to generate a plot of the exported series. Undefined observations are skipped.

    \item \texttt{\symbol{45}\symbol{45}plot\_output}: optional path for the generated plot. If this argument is omitted while \texttt{\symbol{45}\symbol{45}plot} is used, the plot is saved next to the CSV file using the same base name and a \texttt{.png} extension.
\end{itemize}

The exporter does not accept Bitcoin Core RPC parameters, \texttt{\symbol{45}\symbol{45}height\_margin},  \texttt{\symbol{45}\symbol{45}commit\_every}, \texttt{\symbol{45}\symbol{45}min\_confirmations}, or \texttt{\symbol{45}\symbol{45}reset\_state\_db}. These parameters belong to the spent-output indexer, not to the dormancy exporter. The exporter assumes that the indexer database has already been built and updated through the requested ending date. Its role is only to read \texttt{cdd\_sats\_days} and \texttt{spent\_value\_sats} from \texttt{daily\_aggregates}, compute the ratio when the denominator is positive, and write the standardized OBM output file.

\paragraph{Aggregation rule.}
This metric is not aggregated directly as a simple sum. It is a daily ratio computed from two daily aggregates stored by the spent-output indexer. For each UTC date \(d\), the daily value is:
\[
\mathrm{Dormancy}_d =
\frac{\mathrm{CDD}_d}{\mathrm{SpentValueBTC}_d},
\qquad
\mathrm{SpentValueBTC}_d > 0.
\]

Equivalently, in the indexer units:
\[
\mathrm{Dormancy}_d =
\frac{\mathrm{cdd\_sats\_days}_d}{\mathrm{spent\_value\_sats}_d},
\qquad
\mathrm{spent\_value\_sats}_d > 0.
\]

If \(\mathrm{SpentValueBTC}_d = 0\), then \(\mathrm{Dormancy}_d\) is undefined and is recorded as \texttt{NaN}. It is not recorded as zero, because a zero dormancy value would imply that positive spent value had an average age of zero days.

Monthly versions of this metric should not be computed as arithmetic averages of daily dormancy values without explicit justification. A more interpretable monthly version is obtained by first summing monthly CDD and monthly spent value, and then computing the ratio:
\[
\mathrm{Dormancy}_m =
\frac{\mathrm{CDD}_m}{\mathrm{SpentValueBTC}_m},
\qquad
\mathrm{SpentValueBTC}_m > 0,
\]
where:
\[
\mathrm{CDD}_m =
\sum_{d \in m}
\mathrm{CDD}_d,
\qquad
\mathrm{SpentValueBTC}_m =
\sum_{d \in m}
\mathrm{SpentValueBTC}_d.
\]

This convention preserves the interpretation of dormancy as the value-weighted average age of spent coins over the corresponding period.

\paragraph{Output format.}
The output file contains one observation per UTC date. Each row has the following fields:

\begin{center}
\begin{tabular}{llp{0.45\textwidth}}
\toprule
\textbf{Column} & \textbf{Example} & \textbf{Description} \\
\midrule
\texttt{date} & \texttt{2024-01-01} & UTC calendar date \\
\texttt{series\_id} & \texttt{obm\_dormancy\_days\_daily} & Stable OBM series identifier \\
\texttt{value} & \texttt{8.239184726415} & Value-weighted average age of spent outputs \\
\texttt{unit} & \texttt{days} & Measurement unit \\
\texttt{frequency} & \texttt{daily} & Observation frequency \\
\texttt{release\_version} & \texttt{OBM v0.1.0} & Dataset release version \\
\bottomrule
\end{tabular}
\end{center}

When dormancy is undefined because \(\mathrm{SpentValueBTC}_d = 0\), the \texttt{value} field is written as \texttt{NaN}. This indicates a missing ratio, not a zero value.

\paragraph{Technical validation.}
Several checks are implemented by the exporter, and additional consistency checks are recommended for validating this metric. First, the script verifies that the requested date range is valid and that \texttt{\symbol{45}\symbol{45}start\_date} is not later than \texttt{\symbol{45}\symbol{45}end\_date}. Second, it checks that the SQLite state database exists, opens it, and sets the connection to query-only mode. Third, it verifies that the database metadata identify it as an OBM spent-output indexer database, by checking the expected \texttt{indexer\_id}. Fourth, it checks that the indexer database contains processing metadata, including \texttt{last\_processed\_height}, \texttt{last\_processed\_date}, and \texttt{max\_processed\_date}. Fifth, it verifies that the table \texttt{daily\_aggregates} is present in the database. Sixth, it verifies that the requested ending date is not later than the maximum date processed by the indexer. If the requested interval extends beyond the indexed range, the script aborts and requires the indexer to be updated before the export is attempted.

Additional consistency checks can be performed using related OBM series exported from the same indexer database. For each date in the requested interval, the exporter reads \texttt{cdd\_sats\_days} and \texttt{spent\_value\_sats} from \texttt{daily\_aggregates}. If \texttt{spent\_value\_sats} is positive, the script computes dormancy as their ratio. If the date is absent from the aggregate table, or if \texttt{spent\_value\_sats} equals zero, the script records the value as \texttt{NaN}. This convention is appropriate because dormancy is a ratio whose denominator is spent value. When no value is spent, the value-weighted average age of spent coins is undefined.

The script also propagates the release version stored in the indexer metadata. If the metadata field is absent, it uses the fallback value supplied through \texttt{\symbol{45}\symbol{45}release\_version}. The number of rows written must equal the number of calendar days in the requested interval, because the exporter iterates explicitly over all dates between \texttt{\symbol{45}\symbol{45}start\_date} and \texttt{\symbol{45}\symbol{45}end\_date}, inclusive. Defined values should be non-negative, because the indexer floors negative apparent output ages to zero before accumulating CDD.

The primary validation identity is:
\[
\mathrm{Dormancy}_d =
\frac{\mathrm{CDD}_d}{\mathrm{SpentValueBTC}_d},
\qquad
\mathrm{SpentValueBTC}_d > 0.
\]
Equivalently:
\[
\mathrm{CDD}_d =
\mathrm{Dormancy}_d
\times
\mathrm{SpentValueBTC}_d,
\]
for all dates with defined dormancy. These identities should hold when \texttt{obm\_dormancy\_days\_daily}, \texttt{obm\_cdd\_btcxdays\_daily}, and \texttt{obm\_spent\_value\_btc\_daily} are generated from the same indexer release. Comparisons with external dormancy or coin-age series are useful as diagnostics but should not be interpreted as strict equality tests, because providers may differ in timestamp conventions, transferred-value definitions, spent-output definitions, fractional-day treatment, entity adjustment, change-output treatment, and historical edge-case handling.

\paragraph{Known limitations.}
The daily dormancy series is useful but definition-sensitive. First, \texttt{obm\_dormancy\_days\_daily} is a raw spent-output metric, not an entity-adjusted or transfer-adjusted metric. It inherits the limitations of both the \texttt{obm\_cdd\_btcxdays\_daily} metric and the \texttt{obm\_spent\_value\_btc\_daily} metric. Second, the metric depends on the block timestamp convention used both to compute output age and to assign spending blocks to calendar days. Third, the series is undefined on dates with zero spent value, and OBM records those observations as \texttt{NaN} rather than zero. Fourth, the metric can be affected by self-transfers, change-related activity, batching, consolidation transactions, exchange operations, and custodial wallet management. Fifth, it requires the persistent spent-output indexer database and therefore depends on the correctness and completeness of the indexer run. Sixth, the simple reorganization policy of the indexer detects inconsistencies and aborts, but it does not automatically roll back the state database.

Despite these limitations, \texttt{obm\_dormancy\_days\_daily} is a useful OBM coin-age series. It provides a transparent value-weighted measure of the average age of spent coins and complements raw CDD and spent-value metrics in the study of dormant supply, long-term-holder behavior, UTXO turnover, and the interaction between on-chain activity and market conditions.            
\subsection{\texttt{obm\_est7d\_hashrate\_ehs\_daily}: Estimated 7-Day Network Hashrate in EH/s}
\label{Estimated 7-Day Network Hashrate in EH/s}

\paragraph{Definition.}
The estimated 7-day network hashrate series reports an estimate of the aggregate computational rate devoted to Bitcoin mining, expressed in exahashes per second. Unlike difficulty, hashrate is not directly observed on-chain. It must be inferred from the proof-of-work difficulty of observed blocks and the elapsed time over which those blocks were produced.

The canonical OBM series uses a trailing 7-day rolling block-timestamp window. For each UTC date \(d\), let \(W_d\) denote the set of blocks whose timestamps fall in the interval:
\[
[d-6\ \mathrm{days}\ 00{:}00{:}00\ \mathrm{UTC},\ d\ 23{:}59{:}59\ \mathrm{UTC}].
\]
For each block \(b \in W_d\), let \(D_b\) denote the Bitcoin mining difficulty reported by Bitcoin Core for that block, and let \(t_b\) denote its block timestamp. The estimated hashrate in hashes per second is:
\[
\widehat{H}^{\mathrm{H/s}}_d =
\frac{
\sum_{b \in W_d} D_b 2^{32}
}{
\max_{b \in W_d}(t_b) - \min_{b \in W_d}(t_b)
}.
\]
The value is then converted to exahashes per second:
\[
\widehat{H}^{\mathrm{EH/s}}_d =
\frac{
\widehat{H}^{\mathrm{H/s}}_d
}{
10^{18}
}.
\]

The resulting OBM metric is:
\[
\texttt{obm\_est7d\_hashrate\_ehs\_daily}_d
=
\widehat{H}^{\mathrm{EH/s}}_d.
\]

If the rolling window contains fewer than the minimum required number of blocks, or if the elapsed time between the first and last block in the window is not positive, the estimate is undefined and the output value is recorded as \texttt{NaN}. The default minimum number of blocks is two.

The component \texttt{est7d} is part of the series identifier because the rolling-window length is part of the metric definition. A 7-day estimate and a 14-day estimate are not the same metric. If a different window is used, the script automatically adapts the series identifier, for example:
\[
\texttt{obm\_est14d\_hashrate\_ehs\_daily}
\]
for a 14-day rolling-window estimate.

\paragraph{Economic interpretation.}
The estimated 7-day network hashrate series is a mining-activity indicator. It approximates the aggregate rate at which miners are performing proof-of-work computations over the trailing 7-day window. It is therefore useful as a proxy for mining-sector activity and, more indirectly, for the computational security allocated to the Bitcoin network.

For economic research, the metric is useful in several ways. First, it complements mining difficulty by combining the protocol target with realized block production. Second, it provides a smoother mining-activity measure than a one-day estimate, because Bitcoin block discovery is stochastic and daily block counts fluctuate even when underlying hashrate is stable. Third, it can be used together with miner revenue, fees, issuance, and Bitcoin price to study miner incentives. Fourth, it is relevant for energy-related research, because estimated hashrate is often used as an intermediate variable in mining electricity-consumption models. Fifth, it can be used as a control variable in econometric studies where changes in mining conditions or network security may be relevant.

The metric should not be interpreted as directly observed hashrate. It is an estimate based on block difficulty and realized block timestamps. It does not identify individual miners, mining pools, geographic mining locations, energy sources, mining hardware, or electricity consumption. It also does not measure mining profitability directly. Its interpretation depends on the rolling-window convention, timestamp convention, and minimum-block rule.

\paragraph{Similar metrics publicly available.}
The OBM series \texttt{obm\_est7d\_hashrate\_ehs\_daily} is comparable to public metrics usually labelled \emph{estimated hashrate}, \emph{mean hash rate}, \emph{total hash rate}, \emph{network hashrate}, or \emph{Bitcoin hashrate}. In OBM, the metric is explicitly defined as a 7-day trailing-window estimate inferred from block-level difficulty and realized block production. For each UTC date \(d\), the canonical window contains blocks with timestamps in the trailing 7-day interval ending at 23:59:59 UTC on date \(d\). If \(W_d\) denotes this set of blocks, the estimate is
\[
\mathrm{EstimatedHashrateHPS}_d
=
\frac{\sum_{b \in W_d} \mathrm{Difficulty}_b 2^{32}}
{\max_{b \in W_d}(t_b)-\min_{b \in W_d}(t_b)},
\]
and the value is converted to exahashes per second as
\[
\mathrm{EstimatedHashrateEHS}_d
=
\frac{\mathrm{EstimatedHashrateHPS}_d}{10^{18}}.
\]
The metric is therefore not directly observed on-chain. It is an inference from the expected amount of proof-of-work represented by the blocks found in the rolling window and the elapsed time between the first and last block timestamps in that window. The 7-day window is part of the metric definition, because alternative windows, such as 14 or 30 days, produce distinct estimates with different smoothness and responsiveness.

The closest public comparator is Glassnode's \emph{Bitcoin Hash Rate} (\texttt{mining.HashRateMean}).\footnote{\url{https://studio.glassnode.com/charts/mining.HashRateMean?a=BTC}} Glassnode describes this metric as the average estimated number of hashes per second produced by miners in the network. This is conceptually close to \texttt{obm\_est7d\_hashrate\_ehs\_daily}, because both are estimated network hashrate series rather than directly observed quantities. Glassnode also allows moving-average displays, including 7-day views in its interface, which makes it useful for visual comparison with the OBM canonical 7-day estimate. However, the public page does not fully specify the underlying formula, whether the estimate is based on a 24-hour, 7-day, difficulty-epoch, or other estimation window, nor whether elapsed time is measured using first and last block timestamps, calendar-window duration, or expected block time. Exact equivalence should therefore not be assumed. 

Blockchain.com provides a \emph{Total Hash Rate (TH/s)} chart.\footnote{\url{https://www.blockchain.com/charts/hash-rate}} The page defines the metric as the estimated number of terahashes per second the Bitcoin network is performing in the last 24 hours. This is a relevant public comparator because it is also an inferred hashrate series based on network mining activity. Nevertheless, it is not directly equivalent to OBM's canonical metric, because Blockchain.com describes a last-24-hours estimate, whereas OBM uses a trailing 7-day window. The unit also differs: Blockchain.com reports TH/s, while OBM reports EH/s. The Blockchain.com chart is therefore useful for broad comparison, but a direct numerical comparison requires converting units and accounting for the shorter estimation window. 

Coin Metrics provides mining-related network data, including difficulty- and hashrate-type metrics, and its documentation distinguishes between last- and mean-difficulty metrics. Where a Coin Metrics hashrate series is used, it should be treated as an estimated network hashrate whose exact comparability depends on the provider's estimation window and timestamp convention. Coin Metrics' difficulty metrics are especially relevant because OBM's hashrate estimate is constructed from block difficulties and realized block timing. However, unless the precise Coin Metrics hashrate formula and window are matched, a Coin Metrics hashrate series should be treated as a related benchmark rather than as a direct replication of \texttt{obm\_est7d\_hashrate\_ehs\_daily}. A component-based comparison could also be built from Coin Metrics difficulty and block-count data, but this would still require documenting the chosen window and elapsed-time convention.

BitInfoCharts provides a \emph{Bitcoin Hashrate} historical chart.\footnote{\url{https://bitinfocharts.com/comparison/bitcoin-hashrate.html}} The page labels the series as average hashrate per day, expressed in hashes per second. This is related to OBM's estimate, but it is not identical: OBM uses a 7-day rolling block-timestamp window, while BitInfoCharts describes a daily average. The public page does not provide a full, reproducible algorithm, including the exact values for difficulty, block count, elapsed time, or smoothing. It should therefore be used as a secondary diagnostic comparator rather than as a strict methodological benchmark. 

CoinWarz provides Bitcoin hashrate charts, including current and historical global network hashrate.\footnote{\url{https://www.coinwarz.com/mining/bitcoin/hashrate-chart}} The page presents the current BTC hashrate and historical hashrate in graph format, with access to historical views back to 2009. CoinWarz is useful as a mining-focused public reference, but the public chart page does not disclose enough methodological detail to determine exact equivalence with OBM. In particular, it does not specify whether the historical series is based on a 24-hour window, a rolling multi-day window, a difficulty-period estimate, or another smoothing convention. 

Newhedge provides Bitcoin hashrate-related charts and mining indicators. These are relevant public comparators, but the exact aggregation and smoothing conventions should be verified before comparing them numerically with OBM. Hashrate estimates can differ materially depending on whether the provider uses daily block production, a 7-day rolling window, a 14-day or 30-day smoothing window, difficulty epochs, or expected block time. Newhedge should therefore be treated as a useful charting comparator rather than as a fully auditable benchmark unless its formula and window are documented for the selected series.

MacroMicro provides a Bitcoin hash rate chart in its crypto indicators.\footnote{\url{https://en.macromicro.me/charts/29046/bitcoin-hash-rate}} The page describes hash rate as the total computational power used to mine Bitcoin. This is useful as a public macro-style charting reference, but the public documentation is explanatory rather than algorithmic. It does not specify the exact estimation formula, the smoothing window, the timestamp convention, or how to handle early periods with few blocks. 

Bitbo provides a \emph{Bitcoin Hashrate \& Price} chart.\footnote{\url{https://charts.bitbo.io/hashrate/}} The page overlays Bitcoin price and mining hashrate, making it useful for visual inspection of long-run hashrate dynamics. However, the public page does not provide a full formula for the hashrate estimate or specify whether the displayed hashrate is daily, 7-day smoothed, 30-day smoothed, or based on another provider's convention. It should therefore be treated as a secondary visual comparator. 

mempool.space provides mining-related API endpoints and pool hash rate data. Its API documentation includes mining-pool hashrate endpoints rather than a simple named total-network 7-day hashrate series equivalent to OBM.\footnote{\url{https://mempool.space/docs/api/rest}} mempool.space is therefore useful for pool-level and lower-level mining analysis, and its block APIs could support independent reconstruction of a network's hash rate estimate. However, unless a total-network series is reconstructed using the same 7-day rolling-window formula as OBM, it should not be treated as a direct equivalent. 

Hashrate estimates should also be distinguished from difficulty series. Difficulty is directly available as a protocol state variable, while hashrate is inferred from difficulty and block production over a chosen time window. For example, Blockchain.com explicitly notes that exact hashing power is unknown and that total hash rate must be estimated from the number of blocks mined and current block difficulty. This distinction is central to OBM: \texttt{obm\_difficulty\_eod\_daily} records a protocol state, whereas \texttt{obm\_est7d\_hashrate\_ehs\_daily} estimates computational activity from a rolling block sample. 

Overall, the strongest public comparators for \texttt{obm\_est7d\_hashrate\_ehs\_daily} are Glassnode's \texttt{mining.HashRateMean} and Blockchain.com's \emph{Total Hash Rate}, because both explicitly present estimated network hashrate. BitInfoCharts, CoinWarz, Newhedge, MacroMicro, Bitbo, and mempool.space provide useful secondary references or reconstruction sources. The main limitation of external comparison is that hash rate estimates are highly convention-dependent: providers may differ in window length, smoothing, timestamp convention, elapsed-time definition, block inclusion rules, unit scale, and treatment of sparse early windows. The distinctive contribution of OBM is that it makes the estimation rule explicit: it uses a trailing 7-day UTC calendar window, sums \(\mathrm{Difficulty}_b 2^{32}\) over blocks in the window, divides by the elapsed time between the first and last block timestamps, reports the result in EH/s, writes \texttt{NaN} when the rolling window is insufficient, and encodes the window length directly in the series identifier.

\paragraph{Data source and input requirements.}
The metric is obtained from a running Bitcoin Core full node through the JSON-RPC interface. For each block in the expanded scan interval, the script retrieves decoded block metadata using:
\[
\texttt{getblock <block\_hash> 1}.
\]
The relevant block-level fields are:
\[
\texttt{time},\quad
\texttt{difficulty},\quad
\texttt{height}.
\]

This metric does not require reconstructing previous transaction outputs. It does not require the spent-output indexer database, transaction-level input resolution, address extraction, user clustering, entity identification, external price data, or third-party APIs. It also does not require \texttt{txindex=1}, because the required information is available from block metadata.

The metric requires access to a synchronized Bitcoin Core full node and JSON-RPC credentials or cookie authentication. The script uses the locally verified main chain reported by the node. As with other directly scanned OBM metrics, block timestamps returned by Bitcoin Core are used to construct the rolling windows.

\paragraph{Algorithm.}
The script \texttt{compute\_obm\_est7d\_hashrate\_ehs\_daily.py} implements the following procedure:

\begin{enumerate}
    \item Parse the user-provided output date interval, \texttt{\symbol{45}\symbol{45}start\_date} and \texttt{\symbol{45}\symbol{45}end\_date}, using the format \texttt{YYYY-MM-DD}. Both dates are interpreted as UTC dates and both are included in the output.

    \item Read the rolling-window length from \texttt{\symbol{45}\symbol{45}window\_days}. The default is seven calendar days. The canonical series identifier is therefore \texttt{obm\_est7d\_hashrate\_ehs\_daily}.

    \item Derive the output \texttt{series\_id} from the selected window length. For example, \texttt{\symbol{45}\symbol{45}window\_days 7} produces \texttt{obm\_est7d\_hashrate\_ehs\_daily}, whereas \texttt{\symbol{45}\symbol{45}window\_days 14} produces a 14-days window version, namely \texttt{obm\_est14d\_hashrate\_ehs\_daily}.

    \item Expand the scan date range backward by \(\texttt{window\_days}-1\) days. This ensures that the first requested output date has enough preceding block metadata to construct its trailing window.

    \item Connect to the local Bitcoin Core node through JSON-RPC, using either explicit RPC credentials, environment variables, or cookie authentication.

    \item Query the local node using \texttt{getblockchaininfo} to determine the current best-chain height.

    \item Locate an approximate height interval covering the expanded scan date range. This is done by using block timestamps and binary search over block heights.

    \item Expand the approximate height interval using the metric-specific safety parameter called \texttt{\symbol{45}\symbol{45}height\_margin}. The default is 2016 blocks. This reduces the risk of missing boundary blocks because Bitcoin block timestamps are not strictly monotonic with respect to height.

    \item Scan every block in the expanded height interval. For each height \(h\), the script:
    \begin{enumerate}
        \item obtains the corresponding block hash using \texttt{getblockhash};
        \item retrieves the decoded block using \texttt{getblock} with verbosity level 1;
        \item extracts the block timestamp \(t_b\);
        \item assigns the block to a UTC date;
        \item keeps the block if its UTC date lies inside the expanded scan date range;
        \item reads the block-level \texttt{difficulty} field;
        \item stores the block height, timestamp, date, and difficulty.
    \end{enumerate}

    \item Sort the retained blocks by timestamp, using height as a tie-breaker.

    \item For each output date \(d\), define the trailing rolling window ending at 23:59:59 UTC on date \(d\).

    \item Select all blocks whose timestamps fall inside that rolling window.

    \item If the number of blocks in the window is smaller than \texttt{\symbol{45}\symbol{45}min\_blocks}, write \texttt{NaN} for that date.

    \item Compute the elapsed time between the latest and earliest block timestamp in the window. If the elapsed time is not positive, write \texttt{NaN} for that date.

    \item Compute the estimated hashrate:
    \[
    \widehat{H}^{\mathrm{EH/s}}_d =
    \frac{
    \sum_{b \in W_d} D_b 2^{32}
    }{
    \left(\max_{b \in W_d}(t_b) - \min_{b \in W_d}(t_b)\right)10^{18}
    }.
    \]

    \item Write the resulting time series to a CSV file using the standardized OBM schema:
    \[
    \texttt{date},\quad
    \texttt{series\_id},\quad
    \texttt{value},\quad
    \texttt{unit},\quad
    \texttt{frequency},\quad
    \texttt{release\_version}.
    \]

    \item If \texttt{\symbol{45}\symbol{45}output} is omitted, derive the default output filename from the selected window length. For example, the default 7-day run writes \texttt{obm\_est7d\_hashrate\_ehs\_daily.csv}.

    \item Optionally generate a plot of the resulting series when the plotting flag is activated.
\end{enumerate}

The script therefore computes the metric directly from decoded block metadata. It does not query the spent-output indexer, does not resolve transaction inputs, and does not reconstruct previous outputs.

\paragraph{Metric-specific input parameters.}
The metric-specific input parameters are:

\begin{itemize}
    \item \texttt{\symbol{45}\symbol{45}window\_days}: length of the trailing rolling calendar-day window. The default is 7. This parameter is part of the metric definition and determines the output \texttt{series\_id}.

    \item \texttt{\symbol{45}\symbol{45}min\_blocks}: minimum number of blocks required in a rolling window for the estimate to be defined. The default is 2. If the number of blocks in a window is lower than this threshold, the output value is \texttt{NaN}.

    \item \texttt{\symbol{45}\symbol{45}height\_margin}: number of extra blocks scanned before and after the approximate height interval associated with the expanded scan date range. The default is 2016 blocks.
\end{itemize}

The \texttt{\symbol{45}\symbol{45}height\_margin} parameter is needed because Bitcoin block timestamps are not strictly monotonic with respect to block height. A timestamp-based binary search provides only an approximate height interval. To avoid excluding blocks whose timestamps fall inside a rolling window but whose heights lie slightly outside the approximate interval, the script expands the scan interval by \(m\) blocks on both sides:
\[
h^{\mathrm{scan}}_{\min} =
\max(0, h^{\mathrm{approx}}_{\min} - m),
\]
\[
h^{\mathrm{scan}}_{\max} =
\min(h_{\mathrm{tip}}, h^{\mathrm{approx}}_{\max} + m).
\]

The script also accepts standard RPC, output, release-version, and plotting parameters:
\texttt{\symbol{45}\symbol{45}rpc\_url}, \texttt{\symbol{45}\symbol{45}rpc\_user}, \texttt{\symbol{45}\symbol{45}rpc\_password}, \texttt{\symbol{45}\symbol{45}cookie\_path}, \texttt{\symbol{45}\symbol{45}use\_default\_cookie}, \texttt{\symbol{45}\symbol{45}rpc\_timeout}, \texttt{\symbol{45}\symbol{45}output}, \texttt{\symbol{45}\symbol{45}release\_version}, \texttt{\symbol{45}\symbol{45}plot}, \texttt{\symbol{45}\symbol{45}plot\_output}, and \texttt{\symbol{45}\symbol{45}quiet}.

\paragraph{Aggregation rule.}
This metric is not a simple daily sum. It is a rolling-window estimate. For each output date \(d\), the estimate is computed from all blocks in the trailing \(w\)-day window:
\[
W_d^{(w)} =
\{b:\ t_b \in [d-(w-1)\ \mathrm{days}\ 00{:}00{:}00,\ d\ 23{:}59{:}59]\}.
\]
For the canonical OBM series, \(w=7\). The estimate is:
\[
\widehat{H}^{\mathrm{EH/s}}_d =
\frac{
\sum_{b \in W_d^{(7)}} D_b 2^{32}
}{
\left(\max_{b \in W_d^{(7)}}(t_b)-\min_{b \in W_d^{(7)}}(t_b)\right)10^{18}
}.
\]

Monthly versions of this metric, if distributed, should not be computed by summing daily hashrate estimates. Since hashrate is a rate, lower-frequency versions should be computed either as period averages of the daily estimates, or by recomputing the same formula over a clearly specified monthly or rolling-window convention.

\paragraph{Output format.}
The output file contains one observation per UTC date. Each row has the following fields:

\begin{center}
\begin{tabular}{llp{0.35\textwidth}}
\toprule
\textbf{Column} & \textbf{Example} & \textbf{Description} \\
\midrule
\texttt{date} & \texttt{2024-01-01} & UTC calendar date \\
\texttt{series\_id} & \texttt{obm\_est7d\_hashrate\_ehs\_daily} & Window-specific OBM series identifier \\
\texttt{value} & \texttt{512.345678901234} & Estimated network hashrate \\
\texttt{unit} & \texttt{EH/s} & Exahashes per second \\
\texttt{frequency} & \texttt{daily} & Observation frequency \\
\texttt{release\_version} & \texttt{OBM v0.1.0} & Dataset release version \\
\bottomrule
\end{tabular}
\end{center}

Defined values are written with twelve decimal places. Undefined estimates are written as \texttt{NaN}. When a non-default window is selected, the \texttt{series\_id} field changes accordingly. For example, a 14-day run writes \texttt{obm\_est14d\_hashrate\_ehs\_daily}.

\paragraph{Technical validation.}
Several internal checks are used to validate this metric during execution. First, the script verifies that the requested output date range is valid and that \texttt{\symbol{45}\symbol{45}start\_date} is not later than \texttt{\symbol{45}\symbol{45}end\_date}. Second, it verifies that \texttt{\symbol{45}\symbol{45}window\_days} is at least one. Third, it verifies that \texttt{\symbol{45}\symbol{45}min\_blocks} is at least two. Fourth, it verifies that \texttt{\symbol{45}\symbol{45}height\_margin} is non-negative. Fifth, it checks that RPC authentication is valid, including cookie-file existence and format when cookie authentication is used. Sixth, it connects to Bitcoin Core and retrieves the current chain tip using \texttt{getblockchaininfo}. Seventh, it expands the scan date range backward by \(\texttt{window\_days}-1\) days. Eighth, it locates and expands the block-height interval corresponding to the expanded timestamp range. Ninth, it scans all blocks in the expanded interval and retains only blocks whose timestamps fall inside the expanded scan date range. Tenth, it verifies that each retained block contains a \texttt{difficulty} field. Eleventh, it rejects negative difficulty values. Twelfth, it sorts blocks by timestamp and uses height as a tie-breaker. Thirteenth, it counts the number of blocks in each rolling window. Fourteenth, it writes \texttt{NaN} when the window has fewer than the required number of blocks or when the elapsed time between the first and last block in the window is not positive.

Additional consistency checks can be performed using related OBM metrics. The rolling-window block counts should be consistent with \texttt{obm\_block\_count\_daily} over the same timestamp convention. Large movements in estimated hashrate should be checked against changes in block count, difficulty, or both. The series should also be compared with \texttt{obm\_difficulty\_eod\_daily}; difficulty gives the protocol target, whereas estimated hashrate combines that target with realized block timing.

For external validation, selected periods can be compared with public hashrate estimates from blockchain explorers, mining dashboards, or on-chain analytics providers. Such comparisons should be interpreted cautiously because provider estimates may use different windows, smoothing procedures, timestamp conventions, difficulty conventions, or formulas. The selected window length should always be documented when the series is used in empirical work.

\paragraph{Known limitations.}
The estimated 7-day network hashrate series is useful but definition-sensitive. First, hashrate is not directly observed on-chain; it is inferred from difficulty and realized block timing. Second, the estimate depends on the chosen rolling-window length. The canonical OBM series uses seven trailing calendar days, but other choices produce different metrics. Third, the estimate is affected by stochastic variation in block discovery, especially when short windows are used. Fourth, the metric depends on the block timestamp convention used to construct rolling windows. Fifth, it uses elapsed time between the first and last block timestamps in the window, so early windows or windows with few blocks may be noisy or undefined. Sixth, it does not identify miners, pools, hardware, electricity use, geographic distribution, or mining profitability. Seventh, it is computed directly from block metadata and does not require the spent-output indexer. Eighth, it may differ from provider series that use different smoothing windows or formulas.

Despite these limitations, \texttt{obm\_est7d\_hashrate\_ehs\_daily} is a useful OBM mining-activity series. It provides a transparent, full-node-derived estimate of network hashrate and complements difficulty, block count, block weight, fees, miner revenue, issuance, and energy-related empirical analyses.
\subsection{\texttt{obm\_fee\_share\_revenue\_ratio\_daily}: Fees as Share of Miner Revenue}
\label{Fees as Share of Miner Revenue}


\paragraph{Definition.}
The fees-as-share-of-miner-revenue series measures the fraction of BTC-denominated miner compensation accounted for by transaction fees. It is a derived metric computed from the daily issuance and daily transaction-fee series. Let \(\mathrm{FeesBTC}_d\) denote the total transaction fees paid in BTC on UTC calendar day \(d\), as reported by \texttt{obm\_fees\_btc\_daily}. Let \(\mathrm{Issuance}_d\) denote realized Bitcoin issuance on the same day, as reported by \texttt{obm\_issuance\_btc\_daily}. Since BTC-denominated miner revenue is the sum of issuance and fees, the fee-share ratio is defined as:

\[
\mathrm{FeeShare}_d
=
\frac{
\mathrm{FeesBTC}_d
}{
\mathrm{Issuance}_d + \mathrm{FeesBTC}_d
}.
\]

The resulting series is dimensionless. OBM reports it as a ratio rather than as a percentage. For example, a value of \(0.025\) means that transaction fees represented \(2.5\%\) of BTC-denominated miner revenue on that date.

When the denominator is zero $(\mathrm{Issuance}_d + \mathrm{FeesBTC}_d = 0)$, the ratio is undefined. This can occur in the earliest Bitcoin period under the UTC timestamp convention, when no blocks, no issuance, and no fees are assigned to a particular calendar day. In such cases, OBM records the value as missing rather than zero. This convention avoids the incorrect implication that fees represented \(0\%\) of positive miner revenue. Instead, there is no miner revenue from which a share can be computed.

\paragraph{Economic interpretation.}
The fee-share ratio measures the relative importance of transaction fees in BTC-denominated miner compensation. It is therefore a direct indicator of the transition from subsidy-dominated miner revenue toward fee-supported miner revenue.

For economic research, this metric is useful in several ways. First, it allows researchers to study how the composition of miner compensation changes across subsidy eras. Second, it provides a concise measure of the role of transaction fees in Bitcoin's long-run security budget. Third, it helps identify periods in which block-space demand becomes economically significant relative to issuance. Fourth, it is useful around halving events, where issuance mechanically falls, and the relative contribution of fees may increase. Fifth, it provides a normalized measure that is easier to compare across periods with different subsidy levels than raw fees alone.

The metric should not be interpreted as total miner revenue, miner profit, or fiat-denominated miner income. It is a relative composition measure in BTC terms. It does not account for Bitcoin's market price, mining costs, electricity costs, hardware depreciation, pool fees, financing costs, or operating margins.

\paragraph{Similar metrics publicly available.}
The OBM series \texttt{obm\_fee\_share\_revenue\_ratio\_daily} is comparable to a set of public metrics usually labeled \emph{revenue from fees}, \emph{fee-to-reward ratio}, \emph{fee percentage of total block reward}, or \emph{percent miner revenue from fees}. In OBM, the metric is defined as a derived daily ratio computed from two BTC-denominated source series: daily transaction fees and daily realized issuance. The resulting series is dimensionless and is reported by OBM as a ratio rather than as a percentage. When the denominator is zero, the value is recorded as missing rather than zero. This convention avoids implying that fees represented 0

The closest public comparator is Coin Metrics' \emph{Revenue from Fees (\%)}, with MetricID \texttt{FeeRevPct}\footnote{\url{https://community-api.coinmetrics.io/v4/timeseries/asset-metrics?assets=btc\&metrics=FeeRevPct}}, but this query requires supplying credentials. Coin Metrics defines this metric as the percentage of revenue derived from fees during the interval, equal to fees divided by revenue. It reports the unit as dimensionless and makes the metric available at one-day frequency. 

This is conceptually almost identical to \texttt{obm\_fee\_share\_revenue\_ratio\_daily}, provided that revenue is defined as fees plus newly issued coins. However, exact equality should not be assumed without checking the underlying conventions used for the fee and revenue components, including timestamp assignment, treatment of reorganizations, handling of underclaimed issuance, and whether the reported value is scaled as a percentage or as a unit ratio.

Glassnode provides a directly comparable \emph{Bitcoin Miner Revenue (Fees)} metric, exposed as \texttt{mining.RevenueFromFees}\footnote{\url{https://studio.glassnode.com/charts/mining.RevenueFromFees?a=BTC}} (Advanced Plan required). Glassnode defines the metric as the percentage of miner revenue derived from fees, that is, fees divided by fees plus minted coins. This is the same economic ratio as the \texttt{obm\_fee\_share\_revenue\_ratio\_daily} definition. The public chart page also indicates access through CSV, JSON, API, Excel, and MCP interfaces. However, the publicly available documentation does not provide a fully reproducible algorithm for reconstructing the two components from raw Bitcoin Core data, nor does it fully specify all low-level conventions, such as timestamp assignment, finality policy, early-chain handling, and the treatment of exceptional coinbase outputs.

CryptoQuant provides a \emph{Bitcoin: Fees to Reward Ratio} chart\footnote{\url{https://cryptoquant.com/asset/btc/chart/fees-and-revenue/fees-to-reward-ratio}} (registration required). The public chart description states that the metric is the percentage of fees in total block reward and that values lie between 0 and 1. This is a close match to \texttt{obm\_fee\_share\_revenue\_ratio\_daily}, especially because our metric also reports the series as a ratio rather than as a percentage. CryptoQuant also provides related fee and reward series under its fees-and-revenue category. Nevertheless, the public chart page does not disclose a full node-level reconstruction algorithm for the ratio, and the precise conventions inherited from the underlying fee and reward series are not fully visible from the chart description.

BitInfoCharts provides a \emph{Bitcoin Fee in Reward} historical chart.\footnote{\url{https://bitinfocharts.com/comparison/bitcoin-fee\_to\_reward.html}} The page describes the series as the average fee percentage in total block reward. This is closely related to our metric, because the total block reward can be interpreted as block subsidy plus transaction fees. BitInfoCharts also reports current network statistics, including \emph{Fee in Reward}. However, the public documentation is limited to a high-level chart label and does not provide a reproducible algorithm specifying the exact aggregation rule, timestamp convention, treatment of blocks with no fees, or whether the daily value is computed as a ratio of daily totals or as an average of block-level percentages.

Bitbo provides a \emph{Bitcoin Fees as Percent of Total Block Reward} chart.\footnote{\url{https://charts.bitbo.io/fees-percent-of-reward/}} The page explains that the chart shows what percentage of the block reward is composed of fees rather than block subsidy, and interprets higher values as indicating a larger role for fees in miner revenue. This is a strong conceptual match to \texttt{obm\_fee\_share\_revenue\_ratio\_daily}. However, Bitbo's public page is primarily explanatory and chart-oriented. It does not provide a complete reconstruction algorithm, and related Bitbo charts sometimes display moving-average versions by default. Therefore, it is useful as a public comparison source, but not as a fully auditable methodological benchmark.

Newhedge provides a \emph{Percent Miner Revenue from Bitcoin Fees} chart.\footnote{\url{https://newhedge.io/bitcoin/percent-miner-revenue-from-fees}} The page defines the metric as the percentage of a miner's total revenue that comes from transaction fees rather than block rewards. This is directly aligned with the \texttt{obm\_fee\_share\_revenue\_ratio\_daily} interpretation. The site also exposes image and API access, but full data access appears to require registration or a paid plan. The public page, therefore, identifies a close comparator, but does not provide enough methodological detail to determine exact equivalence.

Bitcoin Magazine Pro provides a \emph{Bitcoin: Miner Revenue (Fees vs Rewards)} chart.\footnote{\url{https://www.bitcoinmagazinepro.com/charts/bitcoin-miner-revenue-fees-vs-rewards/}} The page states that miners generate revenue from transaction fees and mining rewards, and that the chart shows the percentage of total revenue earned from each component. This is very close to the \texttt{obm\_fee\_share\_revenue\_ratio\_daily} decomposition of BTC-denominated miner compensation into fees and issuance. However, the page is designed for interpretation rather than reproducibility. It does not disclose the precise data source, the aggregation convention, the timestamp convention, or the algorithm used to compute the two shares.

Blockchain.com provides related fee and miner-revenue charts, including \emph{Total Transaction Fees (BTC)}, \emph{Total Transaction Fees (USD)}, and \emph{Miners Revenue (USD)}. It also provides a \emph{Cost \% of Transaction Volume} chart, but this is not equivalent to \texttt{obm\_fee\_share\_revenue\_ratio\_daily}, because it compares miner revenue with transaction volume rather than comparing fees with miner revenue. Blockchain.com can therefore be used to reconstruct a fee-share-like measure from component series, but the site does not appear to publish a directly named BTC-denominated fee share of miner revenue that is as close as Coin Metrics \texttt{FeeRevPct} or Glassnode \texttt{mining.RevenueFromFees}.

Overall, the strongest public comparators for \texttt{obm\_fee\_share\_revenue\_ratio\_daily} are Coin Metrics \texttt{FeeRevPct} and Glassnode \texttt{mining.RevenueFromFees}, because both define the metric as fees divided by total miner revenue. CryptoQuant's \emph{Fees to Reward Ratio}, BitInfoCharts' \emph{Fee in Reward}, Bitbo's \emph{Fees as Percent of Total Block Reward}, Newhedge's \emph{Percent Miner Revenue from Bitcoin Fees}, and Bitcoin Magazine Pro's \emph{Miner Revenue (Fees vs Rewards)} are also close public analogs. The main distinction is that \texttt{obm\_fee\_share\_revenue\_ratio\_daily} derives the ratio transparently from \texttt{obm\_fees\_btc\_daily} and \texttt{obm\_issuance\_btc\_daily}, reports it as a unit ratio, specifies the missing-value convention when the denominator is zero, and provides validation identities that make the calculation auditable from the underlying source series.

\paragraph{Data source and input requirements.}
The metric is derived from two existing OBM CSV files: \texttt{obm\_issuance\_btc\_daily.csv} and \texttt{obm\_fees\_btc\_daily.csv}. Unlike them, this metric does not query Bitcoin Core or reconstruct information directly from blocks or transactions. It is a deterministic transformation of both previously generated OBM time series.

The input files must follow the standard OBM schema:
\[
\texttt{date},\quad
\texttt{series\_id},\quad
\texttt{value},\quad
\texttt{unit},\quad
\texttt{frequency},\quad
\texttt{release\_version}.
\]

The script verifies that the first input series is \texttt{obm\_issuance\_btc\_daily}, that the second input series is \texttt{obm\_fees\_btc\_daily}, that both series use unit \texttt{BTC}, and that both series have a daily frequency. It also verifies that both input files contain all the required dates for the selected interval.

\paragraph{Algorithm.}
The script \texttt{compute\_obm\_fee\_share\_revenue\_ratio\_daily.py} implements the following procedure:

\begin{enumerate}
    \item Read the input CSV file containing \texttt{obm\_issuance\_btc\_daily}. This file is provided as the first mandatory positional argument.

    \item Read the input CSV file containing \texttt{obm\_fees\_btc\_daily}. This file is provided as the second mandatory positional argument.

    \item Validate the schema of both input files. The script checks that each file contains the fields:
    \[
    \texttt{date},\quad
    \texttt{series\_id},\quad
    \texttt{value},\quad
    \texttt{unit},\quad
    \texttt{frequency},\quad
    \texttt{release\_version}.
    \]

    \item Validate that the input series identifiers are \texttt{obm\_issuance\_btc\_daily} and \texttt{obm\_fees\_btc\_daily}, respectively.

    \item Validate that both input series use unit \texttt{BTC} and frequency \texttt{daily}.

    \item Determine the date interval. If \texttt{\symbol{45}\symbol{45}start\_date} and \texttt{\symbol{45}\symbol{45}end\_date} are provided, the script uses that interval. If one or both are omitted, the script uses the corresponding bounds of the common overlapping interval covered by both input files.

    \item Verify that both source files contain one and only one observation for every date in the selected interval. If any required date is missing, the script aborts.

    \item For each date \(d\), read \(\mathrm{Issuance}_d\) and \(\mathrm{FeesBTC}_d\). The script checks that both values are non-negative.

    \item Compute the denominator: $\mathrm{MinerRevenueBTC}_d = \mathrm{Issuance}_d + \mathrm{FeesBTC}_d$.

    \item If \(\mathrm{MinerRevenueBTC}_d > 0\), compute:
    \[
    \mathrm{FeeShare}_d =
    \frac{\mathrm{FeesBTC}_d}{\mathrm{MinerRevenueBTC}_d}.
    \]

    \item If \(\mathrm{MinerRevenueBTC}_d = 0\), record the observation as missing rather than zero.

    \item Infer the \texttt{release\_version} from the source observations in the selected interval. If multiple release versions are present, the script aborts to avoid mixing releases.

    \item Write the resulting time series to a CSV file using the standardized OBM schema:
    \[
    \texttt{date},\quad
    \texttt{series\_id},\quad
    \texttt{value},\quad
    \texttt{unit},\quad
    \texttt{frequency},\quad
    \texttt{release\_version}.
    \]

    \item Optionally generate a plot of the resulting series when the plotting flag is activated. Dates with missing values are skipped in the plot.
\end{enumerate}

\paragraph{Metric-specific input parameters.}
The input parameters specific to this metric are:

\begin{itemize}
    \item \texttt{issuance\_csv}: mandatory positional argument containing the path to the \texttt{obm\_issuance\_btc\_daily} CSV file.
    \item \texttt{fees\_csv}: mandatory positional argument containing the path to the \texttt{obm\_fees\_btc\_daily} CSV file.
    \item \texttt{\symbol{45}\symbol{45}start\_date}: starting date of the selected interval, inclusive. If omitted, the first common date available in both input files is used.
    \item \texttt{\symbol{45}\symbol{45}end\_date}: ending date of the selected interval, inclusive. If omitted, the last common date available in both input files is used.
\end{itemize}

The script also accepts \texttt{\symbol{45}\symbol{45}output}, \texttt{\symbol{45}\symbol{45}plot}, and \texttt{\symbol{45}\symbol{45}plot\_output}, following the same output and plotting conventions used by the other OBM scripts.

No Bitcoin Core RPC parameters are required, because this metric is not reconstructed directly from the blockchain. It is a deterministic transformation of already generated OBM time series.

\paragraph{Aggregation rule.}
This metric is not aggregated directly from blocks or transactions. Instead, it is computed pointwise from two daily OBM series. For each UTC date \(d\), the daily ratio is:

\[
\mathrm{FeeShare}_d
=
\frac{
\mathrm{FeesBTC}_d
}{
\mathrm{Issuance}_d + \mathrm{FeesBTC}_d
},
\qquad
\mathrm{Issuance}_d + \mathrm{FeesBTC}_d > 0.
\]

If $\mathrm{Issuance}_d + \mathrm{FeesBTC}_d = 0$, then \(\mathrm{FeeShare}_d\) is undefined and is recorded as missing.

Monthly versions of this metric should not be computed as arithmetic averages of daily ratios without explicit justification. A more interpretable monthly version is obtained by first summing monthly fees and monthly issuance and then computing:

\[
\mathrm{FeeShare}_m
=
\frac{
\mathrm{FeesBTC}_m
}{
\mathrm{Issuance}_m + \mathrm{FeesBTC}_m
},
\]

\noindent where:

\[
\mathrm{FeesBTC}_m =
\sum_{d \in m} \mathrm{FeesBTC}_d,
\qquad
\mathrm{Issuance}_m =
\sum_{d \in m} \mathrm{Issuance}_d.
\]

This convention weights days by their BTC-denominated miner revenue and preserves the metric's interpretation as the fee share of total miner compensation over the month.

\paragraph{Relationship with other OBM metrics.}
The series \texttt{obm\_fee\_share\_revenue\_ratio\_daily} is directly derived from \texttt{obm\_fees\_btc\_daily} and \texttt{obm\_issuance\_btc\_daily}. It can also be expressed using \texttt{obm\_miner\_revenue\_btc\_daily}, because $\mathrm{MinerRevenueBTC}_d = \mathrm{Issuance}_d + \mathrm{FeesBTC}_d$. Therefore, for dates with positive miner revenue:

\[
\mathrm{FeeShare}_d
=
\frac{\mathrm{FeesBTC}_d}
{\mathrm{MinerRevenueBTC}_d}.
\]

This relationship provides a useful validation identity. When the miner-revenue series is available, the denominator implied by \(\mathrm{Issuance}_d + \mathrm{FeesBTC}_d\) should match \texttt{obm\_miner\_revenue\_btc\_daily}. Differences would indicate a timestamp mismatch, source-file inconsistency, implementation error, or release-version mismatch.

\paragraph{Output format.}
The output file contains one observation per UTC date in the selected interval. Each row has the following fields:

\begin{center}
\begin{tabular}{llp{0.3\textwidth}}
\toprule
\textbf{Column} & \textbf{Example} & \textbf{Description} \\
\midrule
\texttt{date} & \texttt{2024-01-01} & UTC calendar date \\
\texttt{series\_id} & \texttt{obm\_fee\_share\_revenue\_ratio\_daily} & Stable OBM series identifier \\
\texttt{value} & \texttt{0.007216384512} & Fees divided by issuance plus fees \\
\texttt{unit} & \texttt{ratio} & Dimensionless ratio \\
\texttt{frequency} & \texttt{daily} & Observation frequency \\
\texttt{release\_version} & \texttt{OBM v0.1.0} & Dataset release version inherited from the source interval \\
\bottomrule
\end{tabular}
\end{center}

When the ratio is undefined because the denominator is zero, the \texttt{value} field is left empty. This indicates a missing value, not a zero ratio.

\paragraph{Technical validation.}
Several internal checks are used to validate this metric. First, the script verifies that both source files conform to the OBM schema. Second, it checks that the source series identifiers are \texttt{obm\_issuance\_btc\_daily} and \texttt{obm\_fees\_btc\_daily}. Third, it verifies that both inputs use unit \texttt{BTC} and frequency \texttt{daily}. Fourth, it checks that every date in the selected interval is present exactly once in both source files. Fifth, it verifies that source values are non-negative. Sixth, because issuance and fees are required to be non-negative, any defined ratio is constrained to lie between 0 and 1. Seventh, it records missing values only for dates where \(\mathrm{Issuance}_d + \mathrm{FeesBTC}_d = 0\). Eighth, it checks that the selected interval does not mix different release versions.

The primary validation identity is:

\[
\mathrm{FeeShare}_d
=
\frac{
\mathrm{FeesBTC}_d
}{
\mathrm{Issuance}_d + \mathrm{FeesBTC}_d
},
\qquad
\mathrm{Issuance}_d + \mathrm{FeesBTC}_d > 0.
\]

A secondary validation identity, when \texttt{obm\_miner\_revenue\_btc\_daily} is available, is:

\[
\mathrm{FeeShare}_d
=
\frac{
\mathrm{FeesBTC}_d
}{
\mathrm{MinerRevenueBTC}_d
}.
\]

These identities make the metric particularly easy to audit and reproduce.

\paragraph{Known limitations.}
The fee-share ratio is simple and reproducible, but it has several limitations. First, it is a derived series and not an independent full-node reconstruction. Second, it inherits all definitional choices, timestamp conventions, and possible revisions of our primary metrics, namely \texttt{obm\_issuance\_btc\_daily} and \texttt{obm\_fees\_btc\_daily}. Third, it is undefined when issuance plus fees equals zero, and OBM records such observations as missing rather than zero. Fourth, it reports a ratio, not a percentage. Fifth, it measures BTC-denominated revenue composition, not fiat-denominated miner income. Sixth, it does not measure miner profit, mining costs, energy costs, hardware costs, or operating margins.

Despite these limitations, \texttt{obm\_fee\_share\_revenue\_ratio\_daily} is a useful derived OBM series. It provides a transparent measure of the relative importance of transaction fees in miner compensation and supports research on Bitcoin's security budget, fee-market development, halving dynamics, and the long-run transition from subsidy-based to fee-supported miner revenue.
\subsection{\texttt{obm\_fees\_btc\_daily}: Daily Transaction Fees}
\label{Daily Transaction Fees}


\paragraph{Definition.}
The daily transaction-fee series measures the total amount of BTC paid as transaction fees by non-coinbase transactions included in blocks assigned to a given UTC calendar day. Let \(B_d\) denote the set of blocks assigned to day \(d\). For each non-coinbase transaction \(j\) included in block \(b \in B_d\), let \(V^{\mathrm{in}}_j\) denote the total value of the previous outputs spent by the transaction inputs, and let \(V^{\mathrm{out}}_j\) denote the total value of the transaction outputs. The transaction fee paid by transaction \(j\) is defined as $F_j = V^{\mathrm{in}}_j - V^{\mathrm{out}}_j$. The daily transaction-fee series is then defined as:
\[
\mathrm{FeesBTC}_d =
\sum_{b \in B_d}
\sum_{j \in T_b}
F_j,
\]
where \(T_b\) denotes the set of non-coinbase transactions included in block \(b\).

A block \(b\) is assigned to day \(d\) according to the UTC calendar date derived from its block timestamp \(t_b\): $d(b)=\mathrm{UTCDate}(t_b)$.

The resulting series therefore reports the total amount of BTC paid as transaction fees in the Bitcoin main chain per UTC day. Transaction fees are transfers of already existing bitcoins from transaction senders to miners. They are therefore distinct from newly issued BTC, although both components are paid to miners through the coinbase transaction.

\paragraph{Economic interpretation.}
Daily transaction fees are a BTC-denominated flow variable that measures the amount paid by users for block-space inclusion. The series is therefore a direct indicator of the fee component of Bitcoin transaction demand and miner compensation.

For economic research, this metric is useful in several ways. First, it helps study the development of Bitcoin's fee market. Second, it provides a measure of congestion and block-space demand, especially when interpreted jointly with transaction count, block count, block weight, or fee-rate metrics. Third, it is a key component of miner revenue, since miners receive both newly issued BTC and transaction fees. Fourth, it is needed to compute the fee share of miner revenue, a central metric for analyzing the long-run transition from subsidy-dominated to fee-supported miner compensation. Fifth, it can be used in empirical work on transaction demand, network usage, security-budget sustainability, and the effect of halvings on miner incentives.

The metric should not be interpreted as total miner revenue. Miner revenue is the sum of realized issuance and transaction fees:
\[
\mathrm{MinerRevenueBTC}_d =
\mathrm{IssuanceBTC}_d + \mathrm{FeesBTC}_d.
\]
Nor should it be interpreted as miner profit, because it does not account for electricity costs, hardware depreciation, pool fees, financing costs, or other operating expenses. It is also not a fee-rate metric, since it reports total fees in BTC rather than fees per virtual byte or per transaction.

\paragraph{Similar metrics publicly available.} The OBM series \texttt{obm\_fees\_btc\_daily} is comparable to several publicly available Bitcoin fee metrics that report the total amount of transaction fees paid in native units over daily intervals. In OBM, the metric is defined as the daily sum of fees paid by all non-coinbase transactions included in blocks assigned to a given UTC calendar day. For each non-coinbase transaction \(j\), the fee is computed as the difference between total input value and total output value, $F_j = V^{\mathrm{in}}_j - V^{\mathrm{out}}_j$, 
and the daily aggregate is
\[
\mathrm{FeesBTC}_d =
\sum_{b:\mathrm{UTCDate}(t_b)=d}
\sum_{j \in T^{\mathrm{noncb}}_b}
\left(V^{\mathrm{in}}_j - V^{\mathrm{out}}_j\right).
\]
Coinbase transactions are excluded from transaction-level fee computation because they collect both newly issued coins and the transaction fees paid by non-coinbase transactions. The resulting series measures BTC-denominated transaction fees, not fiat-denominated fees, and not total miner revenue.

The closest public comparator is Coin Metrics' \emph{Total Fees (native units)}, with MetricID \texttt{FeeTotNtv}. Coin Metrics defines this metric as the sum of all fees paid to miners, transaction validators, stakers, or block producers during the interval. It reports the metric in native units and makes it available at one-day and one-hour frequencies. The corresponding API endpoint can be queried\footnote{\url{https://api.coinmetrics.io/v4/timeseries/asset-metrics?assets=btc\&metrics=FeeTotNtv}} (authorization required). This is a strong benchmark for \texttt{obm\_fees\_btc\_daily}. However, exact equality should not be assumed without checking timestamp conventions. Coin Metrics' documentation notes that differences between its \texttt{FeeTotNtv} metric for BTC and other providers may arise because Coin Metrics uses the median block timestamp for BTC, whereas many other providers use the miner timestamp. This distinction is directly relevant because OBM assigns blocks to UTC days using the block timestamp returned by Bitcoin Core.

Glassnode provides a closely related \emph{Bitcoin Fees (Total)} metric, exposed as \texttt{fees.VolumeSum}.\footnote{\url{https://studio.glassnode.com/charts/fees.VolumeSum?a=BTC}} Glassnode describes the metric as the total amount of fees paid to miners, excluding issued or minted coins. This definition is conceptually very close to \texttt{obm\_fees\_btc\_daily}, because both isolate transaction fees from the block subsidy. The public page also indicates availability through CSV, JSON, API, Excel, and MCP interfaces. However, the publicly visible documentation is concise and does not fully specify the reconstruction algorithm, including the previous-output resolution, timestamp convention, confirmation policy, chain reorganization handling, or historical edge-case treatment.

Blockchain.com provides a \emph{Total Transaction Fees (BTC)} chart.\footnote{\url{https://www.blockchain.com/charts/transaction-fees}} The page defines the series as the total BTC value of all transaction fees paid to miners and explicitly states that coinbase block rewards are excluded. This is a close conceptual match to \texttt{obm\_fees\_btc\_daily}. Blockchain.com charts can also be accessed programmatically through its chart interface. Nevertheless, the public page does not provide a full reproducible algorithm. In particular, it does not specify all low-level choices needed to determine exact equivalence with OBM, such as daily boundary rules, reorganization handling, historical fee reconstruction, or whether the aggregation uses miner timestamps or another timing convention.

Blockchair provides a \emph{Total transaction fees (BTC)} chart within its Bitcoin charts catalog.\footnote{\url{https://blockchair.com/bitcoin/charts}} The chart catalog also includes related fee metrics, such as average transaction fee, median transaction fee, total transaction fees in USD, and fee-to-reward ratios. Blockchair also exposes block-level and transaction-level data, so, in principle, daily fees could be reconstructed independently by resolving transaction inputs and aggregating transaction-level fees by day. However, the public chart catalog gives only a high-level metric label for total fees and does not provide a complete algorithmic specification. It should therefore be treated as a useful comparison source rather than as a fully auditable methodological benchmark.

Newhedge provides a \emph{Total Bitcoin Transaction Fees per Day (BTC)} chart.\footnote{\url{https://newhedge.io/bitcoin/total-transaction-fees-btc}} The page defines the metric as the total BTC value of all transaction fees paid to miners on the Bitcoin network. This is directly aligned with the economic object measured by \texttt{obm\_fees\_btc\_daily}. However, the page does not disclose enough implementation detail to determine exact equivalence with OBM, and access to full data appears to require registration or a paid plan.

Bitbo provides a \emph{Bitcoin Fees Per Day} chart.\footnote{\url{https://charts.bitbo.io/fees-per-day/}} The page shows Bitcoin fees per day and includes both BTC-denominated and USD-denominated fee series. However, the default visualization emphasizes a 30-day moving average, while the raw daily series must be distinguished from smoothed views. This makes Bitbo useful for visual comparison and broad validation, but less suitable as a primary benchmark unless the raw daily BTC series is explicitly selected. The public page does not disclose a full fee-reconstruction algorithm.

BitInfoCharts reports several Bitcoin fee indicators.\footnote{\url{https://bitinfocharts.com/bitcoin/}} These include average transaction fee, median transaction fee, and fee-to-reward measures. These indicators are related to our metric, but most are not direct equivalents. Average and median transaction fees describe typical transaction-level costs rather than total daily fees, while fee-to-reward indicators are ratios rather than native-unit fee totals. BitInfoCharts is therefore useful as a secondary diagnostic source, but not as the main public analog for this OBM series.

YCharts republishes a \emph{Bitcoin Total Transaction Fees Per Day} series sourced from the Blockchain.com website.\footnote{\url{https://ycharts.com/indicators/bitcoin\_total\_transaction\_fees\_per\_day\_btc}} This is useful as an accessible financial-data reference for daily BTC transaction fees. However, it should be treated as a secondary source because it republishes an underlying provider series and does not provide an independent blockchain-level reconstruction algorithm.

Overall, the strongest public comparator for \texttt{obm\_fees\_btc\_daily} is Coin Metrics \texttt{FeeTotNtv}, because it provides a named metric identifier, native-unit reporting, one-day availability, API access, and explicit documentation of timestamp-related differences. Glassnode \texttt{fees.VolumeSum} and Blockchain.com's \emph{Total Transaction Fees (BTC)} are also close conceptual comparators, although their public methodology is less detailed. Blockchair, Newhedge, Bitbo, BitInfoCharts, and YCharts are useful secondary sources for cross-checking. The distinctive contribution of OBM is that it reconstructs BTC-denominated daily transaction fees from full-node data, excludes coinbase transactions from transaction-level fee computation, assigns blocks to UTC dates using an explicit timestamp convention, and documents the aggregation and validation rules needed for auditability.

\paragraph{Data source and input requirements.}
The metric is exported from the persistent OBM spent-output indexer database, rather than computed by an independent metric-specific blockchain scan. The indexer, described in Sect.~\ref{sec:The Spent-Output Indexer}, is built from a running Bitcoin Core full node and stores reusable daily aggregates in a SQLite database. For \texttt{obm\_fees\_btc\_daily}, the relevant source table is \texttt{daily\_aggregates}, and the relevant field is \texttt{fees\_sats}.

The indexer computes transaction fees during its sequential scan of the blockchain. For each non-coinbase transaction, it resolves every input to the previous output it spends, retrieves the value of that previous output from the live outpoint state, sums all input values, sums all output values, and computes the transaction fee as the difference between both quantities:
\[
F_j = V^{\mathrm{in}}_j - V^{\mathrm{out}}_j.
\]
The indexer then accumulates these transaction-level fees by UTC block date. Internally, this quantity is stored in satoshis:
\[
Da\mathrm{fees\_sats}_d =
\sum_{b \in B_d}
\sum_{j \in T_b}
\left(
V^{\mathrm{in,sats}}_j -
V^{\mathrm{out,sats}}_j
\right).
\]

The exporter converts the stored satoshi-denominated aggregate into BTC by dividing by \(100{,}000{,}000\):
\[
\mathrm{FeesBTC}_d =
\frac{\mathrm{fees\_sats}_d}{100{,}000{,}000}.
\]

The exporter script does not query Bitcoin Core directly. It requires an existing SQLite database generated by \texttt{obm\_spent\_output\_indexer.py}, with metadata identifying it as an OBM spent-output indexer database and with processed-date metadata covering the requested output interval. The script checks that the requested ending date is not beyond the maximum date processed by the indexer. If a requested date has no row in \texttt{daily\_aggregates}, the exporter writes a zero value for that date. This convention is appropriate for transaction fees because an absent aggregate row indicates that no fee contribution was assigned to that UTC date under the indexer's timestamp convention.

This metric does not require address extraction, user clustering, entity identification, external price data, third-party APIs, or a direct Bitcoin Core connection at export time. However, it depends on the prior successful execution of the spent-output indexer, which itself requires a synchronized, non-pruned Bitcoin Core full node and access to transaction-level data sufficient to reconstruct previous-output values. The exported series therefore inherits the indexer's UTC block-timestamp convention, previous-output reconstruction procedure, chain-consistency checks, and handling of historical duplicate outpoint edge cases.

\paragraph{Algorithm.}
The script \texttt{export\_obm\_fees\_btc\_daily.py} implements the following procedure:

\begin{enumerate}
    \item Parse the user-provided date interval, \texttt{\symbol{45}\symbol{45}start\_date} and \texttt{\symbol{45}\symbol{45}end\_date}, using the format \texttt{YYYY-MM-DD}. Both dates are interpreted as UTC dates and both are included in the output.

    \item Open the persistent SQLite database generated by \texttt{obm\_spent\_output\_indexer.py}. The path to this database is provided through the \texttt{\symbol{45}\symbol{45}state\_db} argument.

    \item Validate that the SQLite database corresponds to the OBM spent-output indexer. The script checks the indexer metadata, including the expected \texttt{indexer\_id} and the presence of processing metadata such as \texttt{last\_processed\_height}, \texttt{last\_processed\_date}, and \texttt{max\_processed\_date}.

    \item Verify that the requested \texttt{\symbol{45}\symbol{45}end\_date} is not later than the maximum UTC date processed by the indexer. If the requested interval extends beyond the indexed range, the script aborts and requires the indexer to be run further before export.

    \item Infer the dataset release version from the indexer metadata. If the database does not contain a release-version field, the script uses the fallback value provided through the \texttt{\symbol{45}\symbol{45}release\_version} argument.

    \item For each UTC date \(d\) in the requested interval, query the \texttt{daily\_aggregates} table for the field \texttt{fees\_sats}. If a row exists for date \(d\), the script reads the stored satoshi-denominated fee value. If no row exists, the script assigns zero fees to that date.

    \item Convert each daily value from satoshis to BTC:
    \[
    \mathrm{FeesBTC}_d =
    \frac{\mathrm{fees\_sats}_d}{100{,}000{,}000}.
    \]

    \item Write the resulting time series to a CSV file using the standardized OBM schema:
    \[
    \texttt{date},\quad
    \texttt{series\_id},\quad
    \texttt{value},\quad
    \texttt{unit},\quad
    \texttt{frequency},\quad
    \texttt{release\_version}.
    \]

    \item Optionally generate a plot of the exported series when the plotting flag is activated. If \texttt{\symbol{45}\symbol{45}plot\_output} is not provided, the plot is saved next to the CSV file with a \texttt{.png} extension. The plot title includes the series description and the selected date interval.
\end{enumerate}

This exporter does not query Bitcoin Core, does not maintain an outpoint state, and does not reconstruct transaction fees at export time. The computationally expensive previous-output resolution and fee reconstruction have already been performed by the spent-output indexer, which stores the daily aggregate in \texttt{daily\_aggregates}. The export script is therefore a lightweight, deterministic transformation from the indexed database to the OBM CSV format.

\paragraph{Metric-specific input parameters.}
The input parameters specific to this exporter are:

\begin{itemize}
    \item \texttt{\symbol{45}\symbol{45}state\_db}: path to the persistent SQLite database generated by \texttt{obm\_spent\_output\_indexer.py}. This database must contain the \texttt{daily\_aggregates} table and the metadata required to verify that it corresponds to the OBM spent-output indexer.

    \item \texttt{\symbol{45}\symbol{45}start\_date}: starting date of the selected interval, inclusive, in \texttt{YYYY-MM-DD} format. The date is interpreted as a UTC calendar date.

    \item \texttt{\symbol{45}\symbol{45}end\_date}: ending date of the selected interval, inclusive, in \texttt{YYYY-MM-DD} format. The script verifies that this date is not later than the maximum date processed by the indexer.

    \item \texttt{\symbol{45}\symbol{45}release\_version}: fallback dataset release version used only if the indexer database does not contain a \texttt{release\_version} metadata field.

    \item \texttt{\symbol{45}\symbol{45}output}: path of the output CSV file to be written.

    \item \texttt{\symbol{45}\symbol{45}plot}: optional flag that instructs the script to generate a plot of the exported series.

    \item \texttt{\symbol{45}\symbol{45}plot\_output}: optional path for the generated plot. If this argument is omitted while \texttt{\symbol{45}\symbol{45}plot} is used, the plot is saved next to the CSV file using the same base name and a \texttt{.png} extension.
\end{itemize}

The exporter does not accept Bitcoin Core RPC parameters, \texttt{\symbol{45}\symbol{45}height\_margin}, \texttt{\symbol{45}\symbol{45}min\_confirmations}, \texttt{\symbol{45}\symbol{45}commit\_every}, or \texttt{\symbol{45}\symbol{45}reset\_state\_db}. These parameters belong to the spent-output indexer, not to the fee exporter. The exporter assumes that the indexer database has already been built and updated through the requested ending date. Its role is only to read \texttt{fees\_sats} from \texttt{daily\_aggregates}, convert the stored satoshi values into BTC, and write the standardized OBM output file.

\paragraph{Aggregation rule.}
The daily value is computed as the sum of transaction fees paid by all non-coinbase transactions included in blocks assigned to the same UTC calendar date:
\[
\mathrm{FeesBTC}_d =
\sum_{b \in B_d}
\sum_{j \in T_b}
\left(
V^{\mathrm{in}}_j -
V^{\mathrm{out}}_j
\right).
\]

The aggregation rule is therefore a daily sum of transaction fees. Monthly versions of this metric, if distributed, should also be computed as sums of the corresponding daily values:
\[
\mathrm{FeesBTC}_m =
\sum_{d \in m}
\mathrm{FeesBTC}_d.
\]

This convention preserves the interpretation of the metric as the total amount of BTC paid as transaction fees over the corresponding period.

\paragraph{Output format.}
The output file contains one observation per UTC date. Each row has the following fields:

\begin{center}
\begin{tabular}{llp{0.45\textwidth}}
\toprule
\textbf{Column} & \textbf{Example} & \textbf{Description} \\
\midrule
\texttt{date} & \texttt{2024-01-01} & UTC calendar date \\
\texttt{series\_id} & \texttt{obm\_fees\_btc\_daily} & Stable OBM series identifier \\
\texttt{value} & \texttt{12.38492017} & Total transaction fees paid on that date \\
\texttt{unit} & \texttt{BTC} & Measurement unit \\
\texttt{frequency} & \texttt{daily} & Observation frequency \\
\texttt{release\_version} & \texttt{OBM v0.1.0} & Dataset release version \\
\bottomrule
\end{tabular}
\end{center}

\paragraph{Technical validation.}
Several internal checks are used to validate this metric at export time. First, the script verifies that the requested date range is valid and that \texttt{\symbol{45}\symbol{45}start\_date} is not later than \texttt{\symbol{45}\symbol{45}end\_date}. Second, it checks that the SQLite state database exists, opens it, and sets the connection to query-only mode. Third, it verifies that the database metadata identify it as an OBM spent-output indexer database, by checking the expected \texttt{indexer\_id}. Fourth, it checks that the indexer database contains processing metadata, including \texttt{last\_processed\_height}, \texttt{last\_processed\_date}, and \texttt{max\_processed\_date}. Fifth, it verifies that the table \texttt{daily\_aggregates} is present in the database. Sixth, it verifies that the requested ending date is not later than the maximum date processed by the indexer. If the requested interval extends beyond the indexed range, the script aborts and requires the indexer to be updated before the export is attempted.

For each date in the requested interval, the exporter reads \texttt{fees\_sats} from \texttt{daily\_aggregates}. If the date is absent from the aggregate table, the script exports zero fees for that date. This convention is appropriate for \texttt{obm\_fees\_btc\_daily} because transaction fees are a flow variable: if no fee contribution is assigned to a UTC date under the indexer's timestamp convention, the daily value is zero rather than missing. Each retrieved value is converted from satoshis to BTC by dividing by \(100{,}000{,}000\), and the resulting file is written using the standard OBM schema:
\[
\texttt{date},\quad
\texttt{series\_id},\quad
\texttt{value},\quad
\texttt{unit},\quad
\texttt{frequency},\quad
\texttt{release\_version}.
\]

The script also propagates the release version stored in the indexer metadata. If the metadata field is absent, it uses the fallback value supplied through \texttt{\symbol{45}\symbol{45}release\_version}. The number of rows written must equal the number of calendar days in the requested interval, because the exporter iterates explicitly over all dates between \texttt{\symbol{45}\symbol{45}start\_date} and \texttt{\symbol{45}\symbol{45}end\_date}, inclusive. Exported values should be non-negative, since valid Bitcoin transaction fees are non-negative under the reconstruction procedure used by the indexer.

Additional consistency checks can be performed using related OBM series exported from the same indexer database. In particular,
\[
\mathrm{MinerRevenueBTC}_d =
\mathrm{IssuanceBTC}_d + \mathrm{FeesBTC}_d
\]
should hold for every date \(d\), up to decimal representation. Equivalently,
\[
\mathrm{FeesBTC}_d =
\mathrm{MinerRevenueBTC}_d -
\mathrm{IssuanceBTC}_d.
\]
These identities should hold when the metrics \texttt{obm\_fees\_btc\_daily}, \texttt{obm\_issuance\_btc\_daily}, and \texttt{obm\_miner\_revenue\_btc\_daily} are exported from the same indexer database using the same timestamp convention and release version. Comparisons with external fee series are useful as diagnostics but should not be interpreted as strict equality tests, because providers may differ in timestamp conventions, confirmation policies, reorganization handling, and historical edge-case treatment.

\paragraph{Known limitations.}
The daily transaction-fee series is conceptually simple, but it inherits several limitations from the spent-output reconstruction pipeline. First, \texttt{obm\_fees\_btc\_daily} is exported from the OBM spent-output indexer database and therefore depends on the correctness and completeness of the indexer run. Second, the metric depends on the block timestamp convention used to assign blocks to calendar days. Third, the series reports fees in BTC, not in fiat currency, so it does not capture the fiat-denominated cost paid by users or fiat-denominated miner income. Fourth, it does not measure fee rates, such as satoshis per virtual byte, and therefore does not directly measure the price of block space per unit of transaction weight. Fifth, it is not entity-adjusted and does not identify users, exchanges, custodians, self-transfers, or transaction purpose. Sixth, the simple reorganization policy of the indexer detects inconsistencies and aborts, but it does not automatically roll back the state database.

Despite these limitations, \texttt{obm\_fees\_btc\_daily} is a core OBM series. It provides a transparent BTC-denominated measure of the daily fee component of Bitcoin transaction activity and serves as a foundation for studying fee markets, block-space demand, miner incentives, fee share of miner revenue, and Bitcoin's long-run security budget.

\subsection{\texttt{obm\_issuance\_btc\_daily}: Daily Bitcoin Issuance}
\label{Daily Bitcoin Issuance}


\paragraph{Definition.}
The daily Bitcoin issuance series measures the number of new bitcoins actually created in blocks assigned to a given UTC calendar day. Let \(B_d\) denote the set of blocks assigned to day \(d\). For each block \(b \in B_d\), let \(C_b\) denote the total output value of the coinbase transaction, and let \(F_b\) denote the total transaction fees paid by the non-coinbase transactions included in the same block. The realized issuance of block \(b\) is defined as:

\[
I_b = C_b - F_b.
\]

The daily Bitcoin issuance is then defined as:

\[
\mathrm{Issuance}_d =
\sum_{b \in B_d} I_b.
\]

A block \(b\) is assigned to day \(d\) according to the UTC calendar date derived from its block timestamp \(t_b\): $d(b) = \mathrm{UTCDate}(t_b)$. The resulting series therefore reports the total amount of newly created BTC assigned to each UTC day. This definition distinguishes realized issuance from the theoretical maximum subsidy schedule. The coinbase transaction is the only transaction in a block that can create new bitcoins. However, the total output value of the coinbase transaction generally includes two components: newly issued BTC and transaction fees collected from transactions in the block. Since transaction fees are transfers of already existing bitcoins, they must be subtracted from the total coinbase output value to isolate the newly created supply.

\paragraph{Economic interpretation.}
Daily Bitcoin issuance is a monetary-flow variable. It measures the realized daily increase in Bitcoin supply, expressed in BTC. For economic research, this metric is central to studying Bitcoin's monetary schedule, halving periods, supply growth, scarcity dynamics, and the transition from subsidy-based to fee-based miner compensation.

This metric should be distinguished from three related quantities. First, it is not the theoretical block subsidy implied by the protocol schedule, because miners may claim less than the maximum allowed amount in the coinbase transaction. Second, it is not miner revenue, because miner revenue includes both new issuance and transaction fees. Third, it is not the circulating supply, because the circulating supply is a cumulative stock, whereas the daily issuance is a flow.

The metric is especially useful when combined with \texttt{obm\_block\_count\_daily}, since daily issuance mechanically depends on the number of blocks assigned to a UTC day. Around halving events, the series also allows researchers to identify the realized transition from one subsidy regime to the next using block-level data rather than relying only on calendar approximations.

\paragraph{Similar metrics publicly available.}
The OBM series \texttt{obm\_issuance\_btc\_daily} is comparable to public metrics usually labeled \emph{Bitcoin issuance}, \emph{total issuance}, \emph{new coins issued}, \emph{newly minted coins}, or \emph{daily supply increase}. \texttt{obm\_issuance\_btc\_daily} is defined as the realized number of new bitcoins created in blocks assigned to a given UTC calendar day. This definition distinguishes realized issuance from the theoretical maximum subsidy schedule. The coinbase transaction collects both newly issued BTC and transaction fees, but only the former represents new monetary creation. Therefore, transaction fees must be subtracted from the total coinbase output value to isolate the amount of newly issued BTC.

The closest public comparator is Coin Metrics' \emph{Total Issuance (native units)}, with MetricID \texttt{IssTotNtv}.\footnote{\url{https://community-api.coinmetrics.io/v4/timeseries/asset-metrics?assets=btc\&metrics=IssTotNtv}} Coin Metrics defines this metric as the sum of all new supply units issued that day, reports it in native units, and makes it available at one-day and one-hour frequencies. 

Coin Metrics also distinguishes this metric from \texttt{IssContNtv}, which measures issuance from protocol-mandated continuous emission schedules. For Bitcoin, \texttt{IssTotNtv} is the most relevant Coin Metrics benchmark for \texttt{obm\_issuance\_btc\_daily}. However, exact equality should not be assumed without checking the underlying conventions used for time assignment, chain reorganizations, underclaimed coinbase rewards, and the treatment of historical edge cases. The distinction between theoretical issuance and realized issuance is especially important: recall that \texttt{obm\_issuance\_btc\_daily} computes realized issuance from coinbase outputs after subtracting fees, whereas some public supply charts may rely on the theoretical reward schedule.

Glassnode provides a directly related \emph{Bitcoin Issuance} chart\texttt{supply.Issued}\footnote{\url{https://studio.glassnode.com/charts/supply.Issued?a=BTC}} (Advanced Plan required). Glassnode describes this metric as the total number of new coins added to the current supply, that is, minted or released to the network. This is conceptually close to \texttt{obm\_issuance\_btc\_daily}. Glassnode also describes issuance in its supply-dynamics material as the number of newly minted BTC coins mined each day, with visible variation due to the number of blocks mined per day and halving events. The public chart page indicates access through CSV, JSON, API, Excel, and MCP interfaces. However, the publicly visible documentation does not provide a full reproducible algorithm specifying whether issuance is reconstructed from coinbase outputs, inferred from the subsidy schedule, adjusted for underclaimed rewards, or assigned to dates using a particular timestamp convention.

Blockchain.com provides a \emph{Total Circulating Bitcoin} chart\footnote{\url{https://www.blockchain.com/charts/total-bitcoins}} This is not a daily issuance-flow series, but it is closely related because daily issuance can be approximated as the first difference of the circulating supply series. Blockchain.com explains that the chart shows how many bitcoins have been mined or put into circulation and states in its methodology that the number of bitcoins in circulation is calculated from the theoretical reward defined by the Bitcoin protocol. This makes Blockchain.com useful for broad supply-schedule validation, but it is not equivalent to \texttt{obm\_issuance\_btc\_daily}. In particular, a first-difference version of this chart would reflect theoretical issuance under the provider's convention, rather than necessarily realized issuance computed from actual coinbase outputs and fees.

YCharts republishes a \emph{Bitcoin Supply} series sourced from Blockchain.com.\footnote{\url{https://ycharts.com/indicators/bitcoin\_supply}} As it happens with Blockchain.com, this is a supply-stock series rather than a named daily issuance-flow series. It can be used to compute a daily supply increment, but the resulting increment should be treated as a secondary comparator rather than a direct public equivalent. Its usefulness depends on the same methodological limitation: the underlying Blockchain.com methodology is based on the theoretical reward defined by the Bitcoin protocol.

Newhedge provides a \emph{Bitcoin Circulating Supply} page.\footnote{\url{https://newhedge.io/bitcoin/circulating-supply}} The page reports several related quantities, including circulating supply, percent issued, left to mine, total mined blocks, and \emph{Daily Issuance (BTC)}. This makes Newhedge a useful public reference for daily issuance. It also exposes a documented API endpoint for its circulating-supply metric.\footnote{\url{https://newhedge.io/api/v2/metrics/circulating-supply/total\_circulating\_bitcoin}} However, API access appears to require an Advanced plan, and the public page does not disclose a full algorithm for the daily issuance figure. It is therefore useful as a public charting and diagnostic source, but not as a fully auditable benchmark for \texttt{obm\_issuance\_btc\_daily}.

MacroMicro provides a \emph{Bitcoin - Issuance by Day} chart.\footnote{\url{https://en.macromicro.me/charts/29069/bitcoin-issuance}} The page reports a Bitcoin daily issuance series and relates it to Bitcoin market capitalization. A related MacroMicro page explains the issuance logic using the approximately ten-minute block interval and the halving of block rewards every four years. This is a relevant public comparator for daily issuance, but the public documentation appears to be explanatory rather than algorithmic. It does not specify whether the series is computed from actual block-level coinbase outputs, from block count multiplied by the applicable subsidy, or from a theoretical issuance schedule.

Bitbo provides a related chart, \emph{Bitcoin Ancient Supply vs Daily Issuance}.\footnote{\texttt{https://charts.bitbo.io/ancient-supply-daily-issuance/}} The page states that daily issuance is the number of new bitcoins created and added to circulation each day through mining, and its chart description identifies a daily issuance line and a raw daily issuance line. It also offers API, CSV, XLSX, and JSON access options. This is useful for visual comparison and may be valuable for cross-checking daily variation. However, the public page does not provide a full reconstruction algorithm, nor does it specify whether the raw daily issuance line is based on actual realized coinbase outputs, daily block counts multiplied by subsidy, or another provider-specific convention.

Blockchair provides a \emph{Bitcoin circulation} chart.\footnote{\url{https://blockchair.com/bitcoin/charts/circulation}} This is again a supply-stock series rather than a direct issuance-flow series. Blockchair also provides block-level and transaction-level data, so daily realized issuance could in principle be reconstructed by reading coinbase transactions and subtracting transaction fees, following a procedure similar to \texttt{obm\_issuance\_btc\_daily}. Nevertheless, the public circulation chart should be treated as a related supply comparator rather than as a named equivalent of \texttt{obm\_issuance\_btc\_daily}.

Overall, the strongest public comparators for \texttt{obm\_issuance\_btc\_daily} are Coin Metrics \texttt{IssTotNtv} and Glassnode \texttt{supply.Issued}, because both provide named issuance metrics in native units. Newhedge, MacroMicro, and Bitbo provide useful chart-level references for daily issuance, but their public methodology is less detailed. Blockchain.com, YCharts, and Blockchair are useful for supply-stock validation or for deriving approximate issuance from changes in circulating supply, but they should not be treated as direct equivalents unless the derivation is explicitly documented. The main contribution of \texttt{obm\_issuance\_btc\_daily} is that it defines issuance as a realized full-node-derived flow, separates newly created BTC from transaction fees in the coinbase transaction, assigns blocks to UTC calendar days using an explicit timestamp convention, and provides a reproducible algorithm rather than relying only on the theoretical reward schedule.

\paragraph{Data source and input requirements.}
The metric is exported from the persistent OBM spent-output indexer database, rather than computed by an independent block-scanning script. The indexer (see Sect.~\ref{sec:The Spent-Output Indexer}) is built from a running Bitcoin Core full node and stores reusable daily aggregates in a SQLite database. For \texttt{obm\_issuance\_btc\_daily}, the relevant source table is \texttt{daily\_aggregates}, and the relevant field is \texttt{issuance\_sats}.

The indexer computes realized issuance at the block level as the difference between the total output value of the coinbase transaction and the total transaction fees paid by the non-coinbase transactions included in the same block. Therefore, the exported daily issuance series inherits the indexer's UTC block-timestamp convention, its treatment of previous-output reconstruction, and its handling of historical edge cases. The exporter converts the stored satoshi-denominated daily aggregate into BTC by dividing by \(100{,}000{,}000\):

\[
\mathrm{IssuanceBTC}_d =
\frac{\mathrm{issuance\_sats}_d}{100{,}000{,}000}.
\]

The exporter script does not query Bitcoin Core directly. It requires an existing SQLite database generated by \texttt{obm\_spent\_output\_indexer.py}, with metadata identifying it as an OBM spent-output indexer database and with processed-date metadata covering the requested output interval. The script checks that the requested ending date is not beyond the maximum date processed by the indexer. If a requested date has no row in \texttt{daily\_aggregates}, the exporter writes a zero value for that date. This convention is appropriate for issuance because an absent aggregate row indicates that no block-level aggregate activity was assigned to that UTC date under the indexer's timestamp convention.

The OBM spent-output indexer is designed to be built from the Bitcoin genesis block onward. Accordingly, the exporter assumes that the indexer database contains a complete historical pass from block height 0 through the maximum processed height recorded in the database metadata. Under this invariant, a missing row in \texttt{daily\_aggregates} for a requested UTC date does not indicate an unprocessed date; it indicates that no block-level aggregate activity was assigned to that date under the indexer's timestamp convention.

This metric does not require address extraction, user clustering, entity identification, external price data, third-party APIs, or a direct Bitcoin Core connection at export time. However, it does depend on the prior successful execution of the spent-output indexer, which itself requires a synchronized, non-pruned Bitcoin Core full node and access to transaction-level data sufficient to reconstruct fees and coinbase-output aggregates.

\paragraph{Algorithm.}
The script \texttt{export\_obm\_issuance\_btc\_daily.py} implements the following procedure:

\begin{enumerate}
    \item Parse the user-provided date interval, \texttt{\symbol{45}\symbol{45}start\_date} and \texttt{\symbol{45}\symbol{45}end\_date}, using the format \texttt{YYYY-MM-DD}. Both dates are interpreted as UTC dates and both are included in the output.

    \item Open the persistent SQLite database generated by \texttt{obm\_spent\_output\_indexer.py}. The path to this database is provided through the \texttt{\symbol{45}\symbol{45}state\_db} argument.

    \item Validate that the SQLite database corresponds to the OBM spent-output indexer. The script checks the indexer metadata, including the indexer identifier and the presence of processing metadata such as the last processed height and the maximum processed date.

    \item Verify that the requested \texttt{\symbol{45}\symbol{45}end\_date} is not later than the maximum UTC date processed by the indexer. If the requested interval extends beyond the indexed range, the script aborts and requires the indexer to be run further before export.

    \item Infer the dataset release version from the indexer metadata. If the database does not contain a release-version field, the script uses the fallback value provided through the \texttt{\symbol{45}\symbol{45}release\_version} argument.

    \item For each UTC date \(d\) in the requested interval, query the \texttt{daily\_aggregates} table for the field \texttt{issuance\_sats}. If a row exists for date \(d\), the script reads the stored satoshi-denominated issuance value. If no row exists, the script assigns zero issuance to that date.

    \item Convert each daily value from satoshis to BTC:
    \[
    \mathrm{IssuanceBTC}_d =
    \frac{\mathrm{issuance\_sats}_d}{100{,}000{,}000}.
    \]

    \item Write the resulting time series to a CSV file using the standardized OBM schema:
    \[
    \texttt{date},\quad
    \texttt{series\_id},\quad
    \texttt{value},\quad
    \texttt{unit},\quad
    \texttt{frequency},\quad
    \texttt{release\_version}.
    \]

    \item Optionally generate a plot of the exported series when the plotting flag is activated. The plot title includes the series description and the selected date interval.
\end{enumerate}

Unlike the earlier direct block-scanning implementation, this exporter does not query Bitcoin Core and does not reconstruct transaction fees at export time. The computationally expensive reconstruction has already been performed by the spent-output indexer, which stores the daily issuance aggregate in \texttt{daily\_aggregates}. The export script is therefore a lightweight, deterministic transformation from the indexed database to the OBM CSV format.

\paragraph{Metric-specific input parameter.}
The only input parameter specific to this metric is:

\begin{itemize}
    \item \texttt{\symbol{45}\symbol{45}height\_margin}: number of extra blocks scanned before and after the approximate height interval associated with the requested date range.
\end{itemize}

This parameter is needed because Bitcoin block timestamps are not strictly monotonic with respect to block height. Although block heights are strictly ordered by chain position, miners' timestamps can occasionally shift slightly backward or forward relative to neighboring blocks. As a result, a binary search based only on block timestamps could identify a height interval that is too narrow and accidentally exclude blocks whose timestamps fall inside the requested UTC date interval.

To avoid this problem, the script first obtains an approximate height interval and then scans an expanded interval $[h_{\min}^{\mathrm{scan}}, h_{\max}^{\mathrm{scan}}]$, where 

\[
h_{\min}^{\mathrm{scan}}
=
\max(0, h_{\min}^{\mathrm{approx}} - m),
\]
\[
h_{\max}^{\mathrm{scan}}
=
\min(h_{\mathrm{tip}}, h_{\max}^{\mathrm{approx}} + m),
\]

and \(m\) is the value of \texttt{\symbol{45}\symbol{45}height\_margin}. The default value is 288 blocks, approximately two days of Bitcoin block production at the expected rate of 144 blocks per day. This choice is deliberately conservative. The margin does not change the dates written to the output file; it only widens the internal block scan. During the scan, the script still counts only blocks whose UTC dates fall between \texttt{\symbol{45}\symbol{45}start\_date} and \texttt{\symbol{45}\symbol{45}end\_date}, inclusive.

A larger value of \texttt{\symbol{45}\symbol{45}height\_margin} increases robustness at the cost of scanning more blocks. A smaller value improves speed but increases the risk of missing boundary blocks in unusual timestamp configurations. For ordinary daily updates, the default value is sufficient. For full historical reconstruction or formal archival releases, a larger margin can be used as an additional safeguard. The OBM repository uses the default value.

\paragraph{Aggregation rule.}
The daily value is computed as the sum of realized block issuance for all blocks assigned to the same UTC calendar date:

\[
\mathrm{Issuance}_d =
\sum_{b: \mathrm{UTCDate}(t_b)=d} I_b,
\]

\noindent where $I_b = C_b - F_b$. The aggregation rule is therefore a daily sum. Monthly versions of this metric, if distributed, should also be computed as sums of the corresponding daily values:

\[
\mathrm{Issuance}_m =
\sum_{d \in m} \mathrm{Issuance}_d.
\]

\paragraph{Output format.}
The output file contains one observation per UTC date. Each row has the following fields:

\begin{center}
\begin{tabular}{llp{0.35\textwidth}}
\toprule
\textbf{Column} & \textbf{Example} & \textbf{Description} \\
\midrule
\texttt{date} & \texttt{2024-01-01} & UTC calendar date \\
\texttt{series\_id} & \texttt{obm\_issuance\_btc\_daily} & Stable OBM series identifier \\
\texttt{value} & \texttt{918.75000000} & Realized daily Bitcoin issuance \\
\texttt{unit} & \texttt{BTC} & Measurement unit \\
\texttt{frequency} & \texttt{daily} & Observation frequency \\
\texttt{release\_version} & \texttt{OBM v0.1.0} & Dataset release version \\
\bottomrule
\end{tabular}
\end{center}

\paragraph{Technical validation.}
Several internal checks are used to validate this metric at export time. First, the script verifies that the requested date range is valid and that \texttt{\symbol{45}\symbol{45}start\_date} is not later than \texttt{\symbol{45}\symbol{45}end\_date}. Second, it checks that the SQLite state database exists, opens it, and sets the connection to query-only mode. Third, it verifies that the database metadata identify it as an OBM spent-output indexer database, by checking the expected \texttt{indexer\_id}. Fourth, it checks that the indexer database contains processing metadata, including \texttt{last\_processed\_height}, \texttt{last\_processed\_date}, and \texttt{max\_processed\_date}. Fifth, it verifies that the table \texttt{daily\_aggregates} is present in the database. Sixth, it checks that the requested ending date is not later than the maximum date processed by the indexer. If the requested interval extends beyond the indexed range, the script aborts and requires the indexer to be updated before the export is attempted.

For each date in the requested interval, the exporter reads \texttt{issuance\_sats} from \texttt{daily\_aggregates}. If the date is absent from the aggregate table, the script exports zero issuance for that date. This convention is appropriate for issuance because an absent daily aggregate row indicates that no block-level aggregate activity was assigned to that UTC date under the indexer's timestamp convention. Each retrieved value is converted from satoshis to BTC by dividing by \(100{,}000{,}000\), and the resulting file is written using the standard OBM schema:
\[
\texttt{date},\quad
\texttt{series\_id},\quad
\texttt{value},\quad
\texttt{unit},\quad
\texttt{frequency},\quad
\texttt{release\_version}.
\]

The script also propagates the release version stored in the indexer metadata. If the metadata field is absent, it uses the fallback value supplied through \texttt{\symbol{45}\symbol{45}release\_version}. The number of rows written must equal the number of calendar days in the requested interval, because the exporter iterates explicitly over all dates between \texttt{\symbol{45}\symbol{45}start\_date} and \texttt{\symbol{45}\symbol{45}end\_date}, inclusive.

Additional consistency checks can be performed using related OBM series exported from the same indexer database. In particular, for every date \(d\),
\[
\mathrm{MinerRevenueBTC}_d
=
\mathrm{IssuanceBTC}_d + \mathrm{FeesBTC}_d,
\]

This identity should hold up to decimal representation when \texttt{obm\_issuance\_btc\_daily}, \texttt{obm\_fees\_btc\_daily}, and \texttt{obm\_miner\_revenue\_btc\_daily} are exported from the same indexer database, using the same timestamp convention and release version.

\paragraph{Known limitations.}
Daily Bitcoin issuance is conceptually simple, but its exact reconstruction requires careful separation of newly created coins from transaction fees. First, the metric measures realized issuance, not the theoretical maximum subsidy schedule. Second, it depends on the accurate computation of transaction fees, because fees are included in the coinbase transaction output, but are not newly created bitcoins. Third, historical fee reconstruction may require a non-pruned node with the necessary block and undo data. Fourth, daily values vary with the number of blocks assigned to each UTC day, so deviations from the expected daily issuance can reflect normal variation in block production rather than changes in the protocol schedule. Fifth, the assignment of blocks to days depends on the block timestamp convention. In OBM, the default convention is to use the UTC date derived from the block timestamp returned by Bitcoin Core. This convention is transparent and reproducible, but alternative conventions, such as median time past, could produce small differences near daily boundaries.

Despite these limitations, \texttt{obm\_issuance\_btc\_daily} is one of the central series in the OBM dataset. It provides a transparent measure of realized Bitcoin monetary issuance and serves as a building block for cumulative supply, miner-revenue metrics, subsidy-era analysis, and empirical studies of Bitcoin's monetary schedule.
\subsection{\texttt{obm\_liveliness\_ratio\_daily}: Daily Liveliness Ratio}
\label{Daily Liveliness Ratio}


\paragraph{Definition.}
The daily liveliness ratio is a cumulative coin-age utilization measure. It compares the amount of coin age destroyed by spending with the amount of coin age created by the passage of time. In the OBM implementation, liveliness is computed from two already generated daily series: \texttt{obm\_cdd\_btcxdays\_daily} and \texttt{obm\_supply\_btc\_daily}.

Let \(\mathrm{CDD}_d\) denote daily Bitcoin Days Destroyed on UTC date \(d\), measured in BTC-days. Let \(\mathrm{SupplyBTC}_d\) denote the daily Bitcoin supply, measured in BTC. The cumulative CDD up to date \(d\) is:
\[
\mathrm{CumCDD}_d =
\sum_{\tau \leq d}
\mathrm{CDD}_{\tau}.
\]

The cumulative coin-days created up to date \(d\) is approximated as:
\[
\mathrm{CumCoinDaysCreated}_d =
\sum_{\tau \leq d}
\mathrm{SupplyBTC}_{\tau}.
\]

The liveliness ratio is then defined as:
\[
\mathrm{Liveliness}_d =
\frac{
\mathrm{CumCDD}_d
}{
\mathrm{CumCoinDaysCreated}_d
},
\qquad
\mathrm{CumCoinDaysCreated}_d > 0.
\]

If \(\mathrm{CumCoinDaysCreated}_d = 0\), the ratio is undefined and the output value is recorded as \texttt{NaN}. The resulting unit is dimensionless, because both numerator and denominator are measured in BTC-days. The output unit is therefore recorded as \texttt{ratio}.

This OBM definition uses daily supply as a transparent approximation to daily coin-day creation. A more exact continuous-time version would require tracking coin-day creation at the UTXO or block-interval level. The daily approximation is adopted here because it is auditable, reproducible, and directly computable from existing OBM series.

\paragraph{Economic interpretation.}
The liveliness ratio summarizes how much of the historically created coin age has been destroyed through spending. It is a cumulative indicator of holding and spending behavior. When coin days are destroyed rapidly relative to the creation of new coin days, liveliness rises. When coins remain inactive, and supply continues to accumulate coin days, liveliness tends to stabilize or decline more slowly.

For economic research, this metric is useful as a low-frequency, cumulative counterpart to daily CDD and dormancy. It provides a compact summary of long-run coin-age utilization and can help distinguish periods dominated by accumulation from periods characterized by substantial old-coin movement. However, it should not be interpreted as a daily activity measure. It is a cumulative ratio, and therefore reacts gradually except when daily CDD is unusually large.

The metric should also be interpreted cautiously. It does not adjust for lost coins, dormant but inaccessible coins, self-transfers, exchange activity, custodial wallet management, or entity-level behavior. It is a raw protocol-derived cumulative coin-age ratio, not an entity-adjusted measure of economic spending.

\paragraph{Similar metrics publicly available.}
The OBM series \texttt{obm\_liveliness\_ratio\_daily} is comparable to public metrics usually labeled \emph{Liveliness}. In OBM, the metric is defined as a derived cumulative ratio between cumulative Bitcoin Days Destroyed and cumulative coin-days created. Let \(\mathrm{CDD}_d\) denote daily Bitcoin Days Destroyed on UTC date \(d\), and let \(\mathrm{SupplyBTC}_d\) denote the Bitcoin supply on the same date. OBM computes
\[
\mathrm{Liveliness}_d
=
\frac{\sum_{\tau \leq d}\mathrm{CDD}_{\tau}}
{\sum_{\tau \leq d}\mathrm{SupplyBTC}_{\tau}}.
\]
The denominator uses the daily Bitcoin supply as the base for daily coin-day creation, so the OBM implementation is a transparent daily approximation of cumulative coin-day creation. The series is dimensionless, is reported as a ratio, and should be interpreted as a cumulative coin-age utilization measure rather than as a daily flow. A higher value indicates that a larger fraction of historically accumulated coin age has been destroyed through spending; a lower or declining value indicates that coin age is accumulating faster than it is being destroyed.

The closest public comparator is Glassnode's \emph{Bitcoin Liveliness} metric.\footnote{\url{https://studio.glassnode.com/charts/indicators.Liveliness?a=BTC}} Glassnode defines Liveliness as the ratio of the sum of Coin Days Destroyed to the sum of all coin days ever created, and states that the metric increases when long-term holders liquidate positions and decreases when they accumulate. Glassnode's methodological guide gives the same core interpretation, defining Liveliness as cumulative CDD divided by the total number of coin days ever created. This is conceptually the closest public equivalent to \texttt{obm\_liveliness\_ratio\_daily}. However, exact numerical equivalence should not be assumed without checking the denominator construction. OBM uses the cumulative sum of daily supply as a transparent daily approximation to cumulative coin-day creation, whereas a provider could compute coin-day creation at a finer block-level or UTXO-level resolution. Additional differences may arise from CDD conventions, fractional-day treatment, timestamp assignment, entity-adjustment status, lost-coin treatment, and chain-reorganization policy. 

ChainExposed provides a \emph{Liveliness} chart.\footnote{\url{https://chainexposed.com/Liveliness.html}} The page defines Liveliness as the sum of all Bitcoin Days ever destroyed divided by the sum of all Bitcoin Days ever created, and emphasizes that each data point incorporates Bitcoin history up to that date. This is highly aligned with the OBM definition. However, the page is a charting and interpretation source rather than a reproducible data descriptor. It does not provide a full algorithm specifying the exact CDD convention, whether coin-day creation is computed daily, block-by-block, or UTXO-by-UTXO, nor how timestamp conventions and historical edge cases are handled. 

Newhedge provides a \emph{Bitcoin Liveliness} chart.\footnote{\url{https://newhedge.io/bitcoin/liveliness}} The page describes Liveliness as an on-chain metric designed to measure the relative activity of the Bitcoin network over its entire lifespan by examining the age and movement of coins. This is consistent with the interpretation of \texttt{obm\_liveliness\_ratio\_daily} as a cumulative holding-versus-spending indicator. However, the public page is primarily explanatory and chart-oriented. It does not disclose a complete calculation algorithm, and full data access appears to be tied to the provider's platform. It is therefore useful as a public visual comparator, but not as a fully auditable methodological benchmark. 

CoinGlass provides a \emph{Bitcoin Liveliness} indicator.\footnote{\url{https://www.coinglass.com/pro/i/bitcoin-liveliness}} CoinGlass describes Bitcoin Liveliness as a long-term on-chain metric calculated as the ratio of Coin Days Destroyed to total coin days created. This is a close conceptual match to OBM. However, the public page does not provide a full reconstruction procedure, source-series details, or low-level choices regarding CDD, supply, timestamp convention, fractional days, and historical revisions. It should therefore be treated as a chart-level public comparator rather than as a reproducible algorithmic reference. 

BGeometrics provides a \emph{Bitcoin Liveliness} chart.\footnote{\url{https://charts.bgeometrics.com/bitcoin\_liveliness\_g.html}} The chart is publicly accessible and useful as an additional visual comparator, but its public page provides limited methodological information. Unless further provider documentation is available, it should be treated as a secondary charting reference rather than as an auditable implementation of the metric. 

Coin Metrics does not appear to expose a directly named public \emph{Liveliness} metric equivalent to \texttt{obm\_liveliness\_ratio\_daily}. However, it provides relevant component series. Its transferred-days-destroyed metric, \texttt{TxTfrValDayDst}, can serve as a CDD-like numerator, while supply metrics can be used to approximate the coin-days-created denominator. A Coin Metrics-based liveness ratio could therefore be constructed externally, but it would inherit Coin Metrics' transfer-value, full-day, and timestamp conventions. Such a construction should be described as a component-based approximation rather than as a directly published equivalent.

CryptoQuant similarly provides CDD-related metrics, including \emph{Coin Days Destroyed}, but does not appear to publish a directly named public Liveliness series. Its CDD documentation is useful for the numerator of a liveliness calculation because it defines CDD as the sum over spent UTXOs of value multiplied by lifespan. A CryptoQuant-based liveliness proxy would require combining cumulative CDD with a compatible cumulative coin-days-created denominator, and the result would depend on the provider's supply, timestamp, and age conventions. 

Several related metrics should be distinguished from Liveliness. Aggregate CDD, Dormancy, Binary CDD, Reserve Risk, CVDD, HODL Waves, and supply-age distributions all use coin-age information, but they are not direct equivalents. CDD is a destruction-flow measure rather than a ratio against cumulative coin-day creation. Dormancy divides CDD by spent or transferred value. Binary CDD transforms supply-adjusted CDD into a threshold indicator. Reserve Risk and CVDD introduce price or valuation components. HODL Waves and UTXO age bands describe the age distribution of currently unspent supply rather than the cumulative share of created coin-days that has been destroyed.

Overall, the strongest public comparator for \texttt{obm\_liveliness\_ratio\_daily} is Glassnode's \texttt{indicators.Liveliness}, because it explicitly defines the metric as cumulative Coin Days Destroyed divided by cumulative coin-days created and provides a named Bitcoin chart. ChainExposed and CoinGlass state essentially the same conceptual formula, while Newhedge and BGeometrics provide useful visual references with thinner methodological disclosure. Coin Metrics and CryptoQuant provide relevant CDD or supply components, but not a directly named liveliness ratio. The distinctive contribution of OBM is that it defines the ratio from auditable OBM source series, documents the daily-supply approximation used for coin-day creation, treats missing CDD flow dates as zero while requiring complete supply observations, inherits a consistent UTC date convention from the source series, and provides a reproducible CSV-to-CSV computation rather than relying on provider-specific black-box conventions.

\paragraph{Data source and input requirements.}
The metric is computed from two existing OBM CSV files rather than exported directly from the spent-output indexer database. The first input file must contain \texttt{obm\_cdd\_btcxdays\_daily}, which reports daily Bitcoin Days Destroyed in BTC-days. The second input file must contain \texttt{obm\_supply\_btc\_daily}, which reports daily Bitcoin supply in BTC.

Both input files must follow the standard OBM scalar schema:
\[
\texttt{date},\quad
\texttt{series\_id},\quad
\texttt{value},\quad
\texttt{unit},\quad
\texttt{frequency},\quad
\texttt{release\_version}.
\]

The script verifies that the CDD input file has series identifier \texttt{obm\_cdd\_btcxdays\_daily}, unit \texttt{BTC-days}, and frequency \texttt{daily}. It also verifies that the supply input file has series identifier \texttt{obm\_supply\_btc\_daily}, unit \texttt{BTC}, and frequency \texttt{daily}. Source values must be present, numeric, and non-negative.

The selected interval can be supplied explicitly through \texttt{\symbol{45}\symbol{45}start\_date} and \texttt{\symbol{45}\symbol{45}end\_date}. If either boundary is omitted, the script uses the first or last date available in the supply file. This convention reflects the role of supply as the denominator base for daily coin-day creation. The supply file must contain one observation for every selected date. The CDD file may be sparse: because CDD is a daily flow, a missing selected date in the CDD file is interpreted as zero daily CDD.

The metric does not require a running Bitcoin Core node, direct JSON-RPC access, address extraction, user clustering, entity identification, external price data, third-party APIs, or direct SQLite indexer access at computation time. However, it inherits the definitions, timestamp convention, and release metadata of the CDD and supply source series.

\paragraph{Algorithm.}
The script \texttt{compute\_obm\_liveliness\_ratio\_daily.py} implements the following procedure:

\begin{enumerate}
    \item Read the CDD input file supplied as the first positional argument.

    \item Read the supply input file supplied as the second positional argument.

    \item Verify that both input files exist, are non-empty, and contain the standard OBM columns:
    \[
    \texttt{date},\quad
    \texttt{series\_id},\quad
    \texttt{value},\quad
    \texttt{unit},\quad
    \texttt{frequency},\quad
    \texttt{release\_version}.
    \]

    \item Parse all dates using the \texttt{YYYY-MM-DD} format and interpret them as UTC dates.

    \item Parse all source \texttt{value} fields as decimal numbers. Missing, invalid, \texttt{NaN}, or negative source values cause the script to abort.

    \item Verify that neither source file contains duplicate dates.

    \item Validate that the first source file corresponds to \texttt{obm\_cdd\_btcxdays\_daily}, with unit \texttt{BTC-days} and frequency \texttt{daily}.

    \item Validate that the second source file corresponds to \texttt{obm\_supply\_btc\_daily}, with unit \texttt{BTC} and frequency \texttt{daily}.

    \item Infer or validate the selected date interval. If \texttt{\symbol{45}\symbol{45}start\_date} or \texttt{\symbol{45}\symbol{45}end\_date} is omitted, the script uses the first or last date available in the supply file.

    \item Verify that the supply file contains one observation for every date in the selected interval.

    \item Treat missing CDD dates in the selected interval as zero daily CDD. This convention is appropriate because CDD is a daily flow variable.

    \item Verify that the selected interval uses one common \texttt{release\_version}, considering all supply observations and the CDD observations that are present.

    \item For each date \(d\), update cumulative CDD:
    \[
    \mathrm{CumCDD}_d =
    \mathrm{CumCDD}_{d-1} + \mathrm{CDD}_d.
    \]

    \item For each date \(d\), update cumulative coin-days created:
    \[
    \mathrm{CumCoinDaysCreated}_d =
    \mathrm{CumCoinDaysCreated}_{d-1} + \mathrm{SupplyBTC}_d.
    \]

    \item Compute:
    \[
    \mathrm{Liveliness}_d =
    \frac{
    \mathrm{CumCDD}_d
    }{
    \mathrm{CumCoinDaysCreated}_d
    },
    \]
    whenever the denominator is positive.

    \item If \(\mathrm{CumCoinDaysCreated}_d = 0\), write \texttt{NaN} for that date.

    \item Write the resulting series to a CSV file using the standardized OBM schema.

    \item Optionally generate a plot when the \texttt{\symbol{45}\symbol{45}plot} flag is used.
\end{enumerate}

The script is therefore a deterministic transformation of two OBM source series. It does not query Bitcoin Core, does not maintain an outpoint state, and does not reconstruct spent outputs directly.

\paragraph{Metric-specific input parameters.}
The input parameters specific to this script are:

\begin{itemize}
    \item \texttt{cdd\_csv}: path to the \texttt{obm\_cdd\_btcxdays\_daily} CSV file. This file may be sparse over the selected interval. Missing selected dates are interpreted as zero daily CDD.

    \item \texttt{supply\_csv}: path to the \texttt{obm\_supply\_btc\_daily} CSV file. This file must contain one observation for every selected date.

    \item \texttt{\symbol{45}\symbol{45}start\_date}: optional starting date of the selected interval, inclusive, in \texttt{YYYY-MM-DD} format. If omitted, the first date in the supply file is used.

    \item \texttt{\symbol{45}\symbol{45}end\_date}: optional ending date of the selected interval, inclusive, in \texttt{YYYY-MM-DD} format. If omitted, the last date in the supply file is used.

    \item \texttt{\symbol{45}\symbol{45}output}: path of the output CSV file to be written. The default output CSV file is \texttt{obm\_liveliness\_ratio\_daily.csv}.

    \item \texttt{\symbol{45}\symbol{45}plot}: optional flag that instructs the script to generate a plot of the resulting series.

    \item \texttt{\symbol{45}\symbol{45}plot\_output}: optional path for the generated plot. The default generated plot file is \texttt{obm\_liveliness\_ratio\_daily.png}.
\end{itemize}

The script does not accept Bitcoin Core RPC parameters, \texttt{\symbol{45}\symbol{45}state\_db}, \texttt{\symbol{45}\symbol{45}height\_margin}, \texttt{\symbol{45}\symbol{45}min\_confirmations}, \texttt{\symbol{45}\symbol{45}commit\_every}, or \texttt{\symbol{45}\symbol{45}reset\_state\_db}. Those parameters belong to blockchain-scanning or indexer-backed exporters, whereas this metric is computed from existing OBM CSV files.

\paragraph{Aggregation rule.}
This metric is not aggregated as a daily sum. It is a cumulative ratio. For each date \(d\), the numerator is cumulative CDD:
\[
\mathrm{CumCDD}_d =
\sum_{\tau \leq d}
\mathrm{CDD}_{\tau},
\]
and the denominator is cumulative coin-days created, approximated as cumulative daily supply:
\[
\mathrm{CumCoinDaysCreated}_d =
\sum_{\tau \leq d}
\mathrm{SupplyBTC}_{\tau}.
\]

The daily liveliness ratio is:
\[
\mathrm{Liveliness}_d =
\frac{
\sum_{\tau \leq d}
\mathrm{CDD}_{\tau}
}{
\sum_{\tau \leq d}
\mathrm{SupplyBTC}_{\tau}
}.
\]

Because the metric is cumulative, monthly or annual versions should not be computed by summing daily liveliness values. If lower-frequency observations are needed, they should be obtained by sampling the daily series at the desired period endpoint, or by recomputing the cumulative numerator and denominator through the corresponding endpoint.

\paragraph{Output format.}
The output file contains one observation per UTC date. Each row has the following fields:

\begin{center}
\begin{tabular}{llp{0.40\textwidth}}
\toprule
\textbf{Column} & \textbf{Example} & \textbf{Description} \\
\midrule
\texttt{date} & \texttt{2024-01-01} & UTC calendar date \\
\texttt{series\_id} & \texttt{obm\_liveliness\_ratio\_daily} & Stable OBM series identifier \\
\texttt{value} & \texttt{0.612345678901} & Cumulative CDD divided by cumulative coin-days created \\
\texttt{unit} & \texttt{ratio} & Dimensionless unit \\
\texttt{frequency} & \texttt{daily} & Observation frequency \\
\texttt{release\_version} & \texttt{OBM v0.1.0} & Dataset release version inferred from source files \\
\bottomrule
\end{tabular}
\end{center}

Defined values are written with twelve decimal places. If the cumulative denominator is zero, the value field is written as \texttt{NaN}.

\paragraph{Technical validation.}
Several internal checks are used to validate this metric at computation time. First, the script verifies that both input files exist, are non-empty, and contain the required OBM schema columns. Second, it verifies that all source values are present, parseable as decimal numbers, non-\texttt{NaN}, and non-negative. Third, it verifies that neither source file contains duplicate dates. Fourth, it checks that the CDD input file has series identifier \texttt{obm\_cdd\_btcxdays\_daily}, unit \texttt{BTC-days}, and frequency \texttt{daily}. Fifth, it checks that the supply input file has series identifier \texttt{obm\_supply\_btc\_daily}, unit \texttt{BTC}, and frequency \texttt{daily}. Sixth, it verifies that the selected date interval is valid and that \texttt{\symbol{45}\symbol{45}start\_date} is not later than \texttt{\symbol{45}\symbol{45}end\_date}. Seventh, it verifies that the supply file contains complete observations for every date in the selected interval. Eighth, it treats missing CDD observations as zero daily CDD. Ninth, it verifies that the selected interval uses one common \texttt{release\_version}, considering all supply observations and all CDD observations present in the selected interval.

For each date, the script computes cumulative CDD and cumulative coin-days created. If the cumulative denominator is zero, the output value is written as \texttt{NaN}. Otherwise, the script writes the ratio:
\[
\mathrm{Liveliness}_d =
\frac{
\mathrm{CumCDD}_d
}{
\mathrm{CumCoinDaysCreated}_d
}.
\]

Additional validation checks should be performed before release. Defined values should be non-negative and, under consistent CDD and supply conventions, should normally lie between zero and one:
\[
0 \leq
\mathrm{Liveliness}_d
\leq 1.
\]
Values above one would indicate a definitional inconsistency between the CDD numerator and the coin-day-creation denominator. The cumulative CDD series and the cumulative coin-days-created denominator should both be non-decreasing. Large upward movements in liveliness should coincide with unusually large values of \texttt{obm\_cdd\_btcxdays\_daily}, and should be interpretable using the CDD age-band and spent-value age-band tables.

Comparisons with external liveliness series are useful as diagnostics but should not be interpreted as strict equality tests. Providers may differ in supply convention, timestamp convention, CDD definition, fractional-day treatment, treatment of lost coins, entity adjustment, and historical edge-case handling.

\paragraph{Known limitations.}
The daily liveliness ratio is useful but definition-sensitive. First, the \texttt{obm\_liveliness\_ratio\_daily} metric is a derived metric and therefore depends on the correctness, completeness, and compatibility of the source CDD and supply files. Second, it treats missing CDD dates as zero daily CDD, which is appropriate only if omitted CDD observations represent zero flow rather than missing data. Third, it requires a complete daily supply series across the selected interval. Fourth, it uses daily supply as an approximation to daily coin-day creation. A more exact version would require tracking coin-day creation at the UTXO or block-interval level. Fifth, it does not adjust for lost coins. Sixth, it is not entity-adjusted and therefore inherits the limitations of raw CDD. Seventh, it is cumulative and slow-moving, so it should not be interpreted as a high-frequency activity measure. Eighth, if either source series definition or release version changes, the liveliness series must be recomputed.

Despite these limitations, \texttt{obm\_liveliness\_ratio\_daily} is a useful OBM derived series. It summarizes cumulative coin-age destruction relative to cumulative coin-age creation and complements flow metrics such as CDD, dormancy, spent value, and age-band decompositions.
\subsection{\texttt{obm\_miner\_revenue\_btc\_daily}: Daily Miner Revenue in BTC}
\label{Daily Miner Revenue in BTC}


\paragraph{Definition.}
The daily miner-revenue series measures the total amount of BTC paid to miners through coinbase transactions in blocks assigned to a given UTC calendar day. Let \(B_d\) denote the set of blocks assigned to day \(d\). For each block \(b \in B_d\), let \(C_b\) denote the total output value of the coinbase transaction in that block. The block-level BTC-denominated miner revenue is defined as $\mathrm{MinerRevenueBTC}_b = C_b$. The daily miner-revenue series is then defined as:
\[
\mathrm{MinerRevenueBTC}_d =
\sum_{b \in B_d}
C_b.
\]

A block \(b\) is assigned to day \(d\) according to the UTC calendar date derived from its block timestamp \(t_b\): $d(b)=\mathrm{UTCDate}(t_b)$. The resulting series therefore reports the total BTC-denominated miner compensation paid through coinbase transactions in the Bitcoin main chain per UTC day. The coinbase transaction pays miners both newly issued BTC and transaction fees. Consequently, miner revenue can also be decomposed as:
\[
\mathrm{MinerRevenueBTC}_d =
\mathrm{IssuanceBTC}_d + \mathrm{FeesBTC}_d.
\]

\paragraph{Economic interpretation.}
Daily miner revenue in BTC is a flow variable that measures miner compensation before costs. It captures the BTC paid to miners for producing blocks, including both the newly issued component and the transaction-fee component.

For economic research, this metric is useful in several ways. First, it provides a direct measure of Bitcoin's BTC-denominated security budget, understood as the compensation paid to miners through the protocol. Second, it is useful for studying miner incentives across subsidy eras. Third, it helps analyze halving events, because subsidy reductions mechanically reduce the issuance component of miner revenue unless compensated by higher fees. Fourth, it provides the denominator for the fee-share-of-miner-revenue metric. Fifth, it helps distinguish changes in miner compensation caused by issuance dynamics from changes caused by the fee market.

The metric should not be interpreted as miner profit. It does not account for electricity costs, hardware depreciation, pool fees, financing costs, taxes, or other operating expenses. It is also not fiat-denominated miner income, because it reports miner revenue in BTC rather than in USD, EUR, or any other fiat currency. Finally, it does not identify the miner or mining pool that produced each block.

\paragraph{Similar metrics publicly available.}
The OBM series \texttt{obm\_miner\_revenue\_btc\_daily} is comparable to public metrics usually labeled \emph{miner revenue}, \emph{total miner revenue}, \emph{block reward revenue}, \emph{coinbase reward}, or \emph{miner income}. In OBM, the metric is defined as the BTC-denominated gross compensation assigned to miners in blocks included in a given UTC calendar day. For each block \(b\), let \(C_b\) denote the total output value of the coinbase transaction. The block-level miner revenue is $\mathrm{MinerRevenueBTC}_b = C_b$, and the daily value is
\[
\mathrm{MinerRevenueBTC}_d =
\sum_{b:\mathrm{UTCDate}(t_b)=d} C_b.
\]
Equivalently, since the coinbase transaction pays both newly issued bitcoins and transaction fees, the same daily value can be written as
\[
\mathrm{MinerRevenueBTC}_d =
\mathrm{Issuance}_d + \mathrm{FeesBTC}_d.
\]
This definition measures gross miner compensation in BTC. It does not measure miner profit, fiat-denominated miner income, hashprice, mining profitability, or production cost.

The closest public comparator is Glassnode's \emph{Bitcoin Miner Revenue (Total)}, exposed as \texttt{mining.RevenueSum}\footnote{\url{https://studio.glassnode.com/charts/mining.RevenueSum?a=BTC}} (Advanced plan required). Glassnode defines this metric as total miner revenue, that is, fees plus newly minted coins. This is conceptually very close to the metric \texttt{obm\_miner\_revenue\_btc\_daily}. The public chart page also allows the series to be viewed in native BTC units and in USD terms, and indicates access through CSV, JSON, API, Excel, and MCP interfaces. However, the publicly visible documentation does not provide a full node-level reconstruction algorithm. In particular, it does not fully specify whether the value is read directly from coinbase outputs, reconstructed as issuance plus fees, adjusted for underclaimed rewards, or assigned to days using a particular timestamp and reorganization policy.

Coin Metrics provides a close component-based benchmark through its native-unit issuance and fee metrics. Its \texttt{IssTotNtv} metric measures total issuance in native units\footnote{\url{https://api.coinmetrics.io/v4/timeseries/asset-metrics?assets=btc\&metrics=IssTotNtv}}, while \texttt{FeeTotNtv} measures total fees in native units\footnote{\url{https://api.coinmetrics.io/v4/timeseries/asset-metrics?assets=btc\&metrics=IssTotNtv}} (login required). Therefore, a Coin Metrics analog to \texttt{obm\_miner\_revenue\_btc\_daily} can be constructed as \texttt{IssTotNtv} + \texttt{FeeTotNtv}. This construction is conceptually aligned with OBM's validation identity. Nevertheless, it should be described as a derived comparator rather than as a direct replication unless the component conventions are harmonized. Potential differences include timestamp assignment, chain-reorganization treatment, underclaimed rewards, fee reconstruction, and historical edge-case handling. Coin Metrics also provides mining-entity flow metrics, but those are not equivalent to gross coinbase compensation because they track flows involving identified mining entities rather than the block-level sum of issuance and fees.

Blockchain.com provides a \emph{Miners Revenue (USD)} chart.\footnote{\url{https://www.blockchain.com/charts/miners-revenue}} The page defines the series as the total USD value of coinbase block rewards and transaction fees paid to miners. This is conceptually related to \texttt{obm\_miner\_revenue\_btc\_daily}, but it is not a direct equivalent because it is denominated in USD. A BTC-denominated analogue would require removing the price conversion or reconstructing the native coinbase value from block-level data. Blockchain.com is therefore useful for validating fiat-denominated miner-income narratives, but less suitable as a direct benchmark for an OBM series whose unit is BTC.

YCharts republishes a \emph{Bitcoin Miners Revenue Per Day} series.\footnote{\url{https://ycharts.com/indicators/bitcoin\_miners\_revenue\_per\_day}} The page identifies Blockchain.com as the source and reports the series as a financial indicator. It is therefore useful as an accessible secondary source for the USD-denominated Blockchain.com miner-revenue series. However, it does not provide an independent blockchain-level reconstruction algorithm, nor is it a direct native-unit comparator for \texttt{obm\_miner\_revenue\_btc\_daily}.

Newhedge provides a \emph{Bitcoin Daily Miner Revenue} chart.\footnote{\url{https://newhedge.io/bitcoin/miner-revenue}} The page defines Bitcoin Daily Miner Revenue as the daily amount of income earned by miners for participation in the mining process. It also provides image and API access options, although full access appears to require registration. The public definition is relevant, but the page does not disclose enough information to determine exact equivalence with OBM. In particular, it is not fully clear from the public page which views are BTC-denominated, which are USD-denominated, whether the series is based on coinbase outputs, and how daily boundaries are assigned.

BitInfoCharts reports current miner-compensation-related statistics.\footnote{\url{https://bitinfocharts.com/bitcoin/}} These include \emph{Reward Per Block} and \emph{Reward (last 24h)}, decomposed into subsidy plus fees. This is conceptually very close to OBM's definition, because miner compensation is represented as subsidy plus transaction fees. However, BitInfoCharts is mainly useful here as a current or short-window diagnostic source rather than as a clearly documented downloadable daily historical series equivalent to \texttt{obm\_miner\_revenue\_btc\_daily}. The public page does not provide a stable metric identifier or a full reproducible algorithm.

Bitcoin Magazine Pro provides a \emph{Bitcoin: Miner Revenue (Total)} chart.\footnote{\url{https://www.bitcoinmagazinepro.com/charts/bitcoin-miner-revenue-total/}} The page explains that miners generate revenue from two sources, transaction fees and mining rewards, and that the chart combines both sources to show total revenue earned by Bitcoin miners over time. This is a strong conceptual match to the OBM metric. Nevertheless, the page is primarily interpretive and chart-oriented. It does not disclose all calculation conventions, including whether the displayed series is denominated in BTC or USD under each view, whether it is based on actual coinbase outputs, whether it uses theoretical rewards, or how daily boundary rules are handled.

MacroMicro provides a \emph{Bitcoin - Miner Revenue} chart.\footnote{\url{https://en.macromicro.me/charts/143602/bitcoin-miner-revenue}} The page describes total miner revenue as coming from block rewards and transaction fees. This is conceptually aligned with \texttt{obm\_miner\_revenue\_btc\_daily}. However, the public documentation is explanatory rather than algorithmic, and the page does not provide enough detail to determine exact equivalence with OBM. It should therefore be used as a public charting reference rather than as a reproducible methodological benchmark.

Bitbo provides a \emph{Bitcoin Miner Monthly Revenue} chart.\footnote{\url{https://charts.bitbo.io/miner-monthly-revenue/}} The page states that the BTC amount includes both the block subsidy and transaction fees. This is conceptually aligned with OBM's decomposition of miner revenue into issuance plus fees. However, the Bitbo metric is monthly rather than daily, and the page is chart-oriented rather than algorithmic. It is therefore a useful secondary comparator for longer-horizon behavior, but not a direct daily benchmark for \texttt{obm\_miner\_revenue\_btc\_daily}.

Blockchair does not appear to provide a clearly named public daily \emph{miner revenue in BTC} metric as direct as Glassnode's \texttt{mining.RevenueSum}. Its Bitcoin charts catalog includes related series such as total transaction fees and fee-to-reward ratios, and its block-level data can be used to reconstruct the OBM concept by summing coinbase transaction outputs over blocks assigned to each UTC day.\footnote{\url{https://blockchair.com/bitcoin/blocks}}. Thus, Blockchair is best treated as a reconstruction source rather than a named daily miner-revenue metric unless the grouping and summation conventions are explicitly documented.

Overall, the strongest public comparator for \texttt{obm\_miner\_revenue\_btc\_daily} is Glassnode \texttt{mining.RevenueSum}, because it explicitly defines total miner revenue as fees plus newly minted coins and exposes the metric in BTC. A close Coin Metrics benchmark can be constructed from \texttt{IssTotNtv} plus \texttt{FeeTotNtv}, but this is a derived construction rather than a single, directly named gross miner-revenue metric. Blockchain.com and YCharts mainly provide USD-denominated miner revenue, while Newhedge, BitInfoCharts, Bitcoin Magazine Pro, MacroMicro, Bitbo, and Blockchair provide useful secondary references. The distinctive contribution of OBM is that it computes BTC-denominated miner revenue directly from Coinbase transaction outputs, assigns blocks to UTC calendar days using an explicit timestamp convention, and provides validation by linking miner revenue to daily issuance and daily transaction fees.

\paragraph{Data source and input requirements.}
The metric is exported from the persistent OBM spent-output indexer database, rather than computed by an independent metric-specific blockchain scan. The indexer, described in Sect.~\ref{sec:The Spent-Output Indexer}, is built from a running Bitcoin Core full node and stores reusable daily aggregates in a SQLite database. For \texttt{obm\_miner\_revenue\_btc\_daily}, the relevant source table is \texttt{daily\_aggregates}, and the relevant field is \texttt{coinbase\_output\_sats}.

The indexer computes miner revenue during its sequential scan of the blockchain. For each block, it reads the total output value of the coinbase transaction. This amount is the BTC-denominated compensation paid to the miner of the block, including both newly issued BTC and transaction fees. Internally, this quantity is stored in satoshis:
\[
\mathrm{coinbase\_output\_sats}_d =
\sum_{b \in B_d}
C^{\mathrm{sats}}_b.
\]

The exporter converts the stored satoshi-denominated aggregate into BTC by dividing by \(100{,}000{,}000\):
\[
\mathrm{MinerRevenueBTC}_d =
\frac{\mathrm{coinbase\_output\_sats}_d}{100{,}000{,}000}.
\]

The exporter script does not query Bitcoin Core directly. It requires an existing SQLite database generated by \texttt{obm\_spent\_output\_indexer.py}, with metadata identifying it as an OBM spent-output indexer database and with processed-date metadata covering the requested output interval. The script checks that the requested ending date is not beyond the maximum date processed by the indexer. If a requested date has no row in \texttt{daily\_aggregates}, the exporter writes a zero value for that date. This convention is appropriate for miner revenue because an absent aggregate row indicates that no block-level miner-revenue contribution was assigned to that UTC date under the indexer's timestamp convention.

This metric does not require address extraction, user clustering, entity identification, external price data, third-party APIs, or a direct Bitcoin Core connection at export time. However, it depends on the prior successful execution of the spent-output indexer, which itself requires a synchronized, non-pruned Bitcoin Core full node. The exported series inherits the indexer's UTC block-timestamp convention, chain-consistency checks, release metadata, and handling of historical edge cases.

\paragraph{Algorithm.}
The script \texttt{export\_obm\_miner\_revenue\_btc\_daily.py} implements the following procedure:

\begin{enumerate}
    \item Parse the user-provided date interval, \texttt{\symbol{45}\symbol{45}start\_date} and \texttt{\symbol{45}\symbol{45}end\_date}, using the format \texttt{YYYY-MM-DD}. Both dates are interpreted as UTC dates and both are included in the output.

    \item Open the persistent SQLite database generated by \texttt{obm\_spent\_output\_indexer.py}. The path to this database is provided through the \texttt{\symbol{45}\symbol{45}state\_db} argument.

    \item Validate that the SQLite database corresponds to the OBM spent-output indexer. The script checks the indexer metadata, including the expected \texttt{indexer\_id} and the presence of processing metadata such as \texttt{last\_processed\_height}, \texttt{last\_processed\_date}, and \texttt{max\_processed\_date}.

    \item Verify that the requested \texttt{\symbol{45}\symbol{45}end\_date} is not later than the maximum UTC date processed by the indexer. If the requested interval extends beyond the indexed range, the script aborts and requires the indexer to be run further before export.

    \item Infer the dataset release version from the indexer metadata. If the database does not contain a release-version field, the script uses the fallback value provided through the \texttt{\symbol{45}\symbol{45}release\_version} argument.

    \item For each UTC date \(d\) in the requested interval, query the \texttt{daily\_aggregates} table for the field \texttt{coinbase\_output\_sats}. If a row exists for date \(d\), the script reads the stored satoshi-denominated coinbase-output value. If no row exists, the script assigns zero miner revenue to that date.

    \item Convert each daily value from satoshis to BTC:
    \[
    \mathrm{MinerRevenueBTC}_d =
    \frac{\mathrm{coinbase\_output\_sats}_d}{100{,}000{,}000}.
    \]

    \item Write the resulting time series to a CSV file using the standardized OBM schema:
    \[
    \texttt{date},\quad
    \texttt{series\_id},\quad
    \texttt{value},\quad
    \texttt{unit},\quad
    \texttt{frequency},\quad
    \texttt{release\_version}.
    \]

    \item Optionally generate a plot of the exported series when the plotting flag is activated. If \texttt{\symbol{45}\symbol{45}plot\_output} is not provided, the plot is saved next to the CSV file with a \texttt{.png} extension. The plot title includes the series description and the selected date interval.
\end{enumerate}

This exporter does not query Bitcoin Core, does not maintain an outpoint state, and does not reconstruct miner revenue at export time. The coinbase-output aggregation has already been performed by the spent-output indexer, which stores the daily aggregate in \texttt{daily\_aggregates}. The export script is therefore a lightweight, deterministic transformation from the indexed database to the OBM CSV format.

\paragraph{Metric-specific input parameters.}
The input parameters specific to this exporter are:

\begin{itemize}
    \item \texttt{\symbol{45}\symbol{45}state\_db}: path to the persistent SQLite database generated by the Python script \texttt{obm\_spent\_output\_indexer.py}. This database must contain the \texttt{daily\_aggregates} table and the metadata required to verify that it corresponds to the OBM spent-output indexer.

    \item \texttt{\symbol{45}\symbol{45}start\_date}: starting date of the selected interval, inclusive, in \texttt{YYYY-MM-DD} format. The date is interpreted as a UTC calendar date.

    \item \texttt{\symbol{45}\symbol{45}end\_date}: ending date of the selected interval, inclusive, in \texttt{YYYY-MM-DD} format. The script verifies that this date is not later than the maximum date processed by the indexer.

    \item \texttt{\symbol{45}\symbol{45}release\_version}: fallback dataset release version used only if the indexer database does not contain a \texttt{release\_version} metadata field.

    \item \texttt{\symbol{45}\symbol{45}output}: path of the output CSV file to be written.

    \item \texttt{\symbol{45}\symbol{45}plot}: optional flag that instructs the script to generate a plot of the exported series.

    \item \texttt{\symbol{45}\symbol{45}plot\_output}: optional path for the generated plot. If this argument is omitted while \texttt{\symbol{45}\symbol{45}plot} is used, the plot is saved next to the CSV file using the same base name and a \texttt{.png} extension.
\end{itemize}

The exporter does not accept Bitcoin Core RPC parameters, \texttt{\symbol{45}\symbol{45}height\_margin}, \texttt{\symbol{45}\symbol{45}commit\_every}, \texttt{\symbol{45}\symbol{45}min\_confirmations}, or \texttt{\symbol{45}\symbol{45}reset\_state\_db}. These parameters belong to the spent-output indexer, not to the miner-revenue exporter. The exporter assumes that the indexer database has already been built and updated through the requested ending date. Its role is only to read \texttt{coinbase\_output\_sats} from \texttt{daily\_aggregates}, convert the stored satoshi values into BTC, and write the standardized OBM output file.

\paragraph{Aggregation rule.}
The daily value is computed as the sum of coinbase transaction output values for all blocks assigned to the same UTC calendar date:
\[
\mathrm{MinerRevenueBTC}_d =
\sum_{b \in B_d}
C_b.
\]

Equivalently, using the fee and issuance decomposition:
\[
\mathrm{MinerRevenueBTC}_d =
\mathrm{IssuanceBTC}_d +
\mathrm{FeesBTC}_d.
\]

The aggregation rule is therefore a daily sum of BTC-denominated miner compensation. Monthly versions of this metric, if distributed, should also be computed as sums of the corresponding daily values:
\[
\mathrm{MinerRevenueBTC}_m =
\sum_{d \in m}
\mathrm{MinerRevenueBTC}_d.
\]

This convention preserves the interpretation of the metric as the total BTC paid to miners over the corresponding period.

\paragraph{Output format.}
The output file contains one observation per UTC date. Each row has the following fields:

\begin{center}
\begin{tabular}{llp{0.45\textwidth}}
\toprule
\textbf{Column} & \textbf{Example} & \textbf{Description} \\
\midrule
\texttt{date} & \texttt{2024-01-01} & UTC calendar date \\
\texttt{series\_id} & \texttt{obm\_miner\_revenue\_btc\_daily} & Stable OBM series identifier \\
\texttt{value} & \texttt{931.13492017} & Total BTC paid to miners on that date \\
\texttt{unit} & \texttt{BTC} & Measurement unit \\
\texttt{frequency} & \texttt{daily} & Observation frequency \\
\texttt{release\_version} & \texttt{OBM v0.1.0} & Dataset release version \\
\bottomrule
\end{tabular}
\end{center}

\paragraph{Technical validation.}
The exporter implements several checks, and additional consistency checks are recommended to validate this metric. First, the script verifies that the requested date range is valid and that \texttt{\symbol{45}\symbol{45}start\_date} is not later than \texttt{\symbol{45}\symbol{45}end\_date}. Second, it checks that the SQLite state database exists, opens it, and sets the connection to query-only mode. Third, it verifies that the database metadata identify it as an OBM spent-output indexer database, by checking the expected \texttt{indexer\_id}. Fourth, it checks that the indexer database contains processing metadata, including \texttt{last\_processed\_height}, \texttt{last\_processed\_date}, and \texttt{max\_processed\_date}. Fifth, it verifies that the table \texttt{daily\_aggregates} is present in the database. Sixth, it verifies that the requested ending date is not later than the maximum date processed by the indexer. If the requested interval extends beyond the indexed range, the script aborts and requires the indexer to be updated before the export is attempted.

For each date in the requested interval, the exporter reads \texttt{coinbase\_output\_sats} from \texttt{daily\_aggregates}. If the date is absent from the aggregate table, the script exports zero miner revenue for that date. This convention is appropriate for \texttt{obm\_miner\_revenue\_btc\_daily} because miner revenue is a flow variable: if no block-level miner-revenue contribution is assigned to a UTC date under the indexer's timestamp convention, the daily value is zero rather than missing. Each retrieved value is converted from satoshis to BTC by dividing by \(100{,}000{,}000\), and the resulting file is written using the standard OBM schema:
\[
\texttt{date},\quad
\texttt{series\_id},\quad
\texttt{value},\quad
\texttt{unit},\quad
\texttt{frequency},\quad
\texttt{release\_version}.
\]

The OBM spent-output indexer is designed to be built from the Bitcoin genesis block onward. Accordingly, the exporter assumes that the indexer database contains a complete historical pass from block height 0 through the maximum processed height recorded in the database metadata. Under this invariant, a missing row in \texttt{daily\_aggregates} for a requested UTC date does not indicate an unprocessed date; it indicates that no block-level aggregate activity was assigned to that date under the indexer's timestamp convention.

The script also propagates the release version stored in the indexer metadata. If the metadata field is absent, it uses the fallback value supplied through \texttt{\symbol{45}\symbol{45}release\_version}. The number of rows written must equal the number of calendar days in the requested interval, because the exporter iterates explicitly over all dates between \texttt{\symbol{45}\symbol{45}start\_date} and \texttt{\symbol{45}\symbol{45}end\_date}, inclusive. Exported values should be non-negative, since coinbase output values are non-negative.

Additional validation checks can be performed using related OBM series exported from the same indexer database. In particular,
\[
\mathrm{MinerRevenueBTC}_d =
\mathrm{IssuanceBTC}_d + \mathrm{FeesBTC}_d
\]
should hold for every date \(d\), up to decimal representation. Equivalently,
\[
\mathrm{FeesBTC}_d =
\mathrm{MinerRevenueBTC}_d -
\mathrm{IssuanceBTC}_d.
\]
These identities should hold when \texttt{obm\_miner\_revenue\_btc\_daily}, \texttt{obm\_issuance\_btc\_daily}, and \texttt{obm\_fees\_btc\_daily} are exported from the same indexer database using the same timestamp convention and release version. Comparisons with external miner-revenue or block-reward series are useful as diagnostics but should not be interpreted as strict equality tests, because providers may differ in timestamp conventions, confirmation policies, reorganization handling, treatment of underclaimed issuance, historical edge-case treatment, and whether values are reported in BTC or fiat currency.

\paragraph{Known limitations.}
The daily miner-revenue series is conceptually simple, but it inherits several limitations from the indexer-based reconstruction pipeline. First, \texttt{obm\_miner\_revenue\_btc\_daily} is exported from the OBM spent-output indexer database and therefore depends on the correctness and completeness of the indexer run. Second, the metric depends on the block timestamp convention used to assign blocks to calendar days. Third, the series reports miner revenue in BTC, not in fiat currency, so it does not capture fiat-denominated miner income. Fourth, it does not measure miner profit, because it does not account for electricity costs, hardware depreciation, pool fees, financing costs, taxes, or other operating expenses. Fifth, it does not identify the miner, mining pool, or entity that produced each block. Sixth, the simple reorganization policy of the indexer detects inconsistencies and aborts, but it does not automatically roll back the state database.

Despite these limitations, \texttt{obm\_miner\_revenue\_btc\_daily} is a core OBM series. It provides a transparent BTC-denominated measure of daily miner compensation and serves as a foundation for studying miner incentives, halving dynamics, fee-market development, fee share of miner revenue, and Bitcoin's long-run security budget.
\subsection{\texttt{obm\_raw\_output\_value\_btc\_daily}: Daily Raw Output Value in BTC}
\label{Daily Raw Output Value in BTC}

\paragraph{Definition.}
The daily raw output value series measures the total BTC value of outputs created by non-coinbase transactions in blocks assigned to a given UTC calendar day. Let \(B_d\) denote the set of blocks assigned to day \(d\). For each non-coinbase transaction \(j\) included in a block \(b \in B_d\), let \(O_j\) denote the set of outputs created by that transaction, and let \(v_o\) denote the BTC value of output \(o \in O_j\). The daily raw output value is defined as:
\[
\mathrm{RawOutputValueBTC}_d =
\sum_{b \in B_d}
\sum_{j \in T_b^{\mathrm{noncb}}}
\sum_{o \in O_j}
v_o,
\]
where \(T_b^{\mathrm{noncb}}\) denotes the set of non-coinbase transactions included in block \(b\).

A block \(b\) is assigned to day \(d\) according to the UTC calendar date derived from its block timestamp \(t_b\): $d(b)=\mathrm{UTCDate}(t_b)$.

Coinbase transaction outputs are excluded from this metric. This exclusion is important because Coinbase outputs represent miner revenue, including newly issued BTC and transaction fees, and are already covered by the miner-revenue and issuance-related OBM metrics. The resulting series, therefore, reports raw output-side value created by ordinary, non-coinbase transactions.

\paragraph{Economic interpretation.}
The daily raw output value series is a raw measure of BTC-denominated output-side blockchain activity. It captures the total value assigned to transaction outputs created by non-coinbase transactions on each UTC day.

For economic research, this metric is useful as a proxy for transparent activity. It complements transaction count by measuring value rather than transaction frequency. It also complements the spent output value by looking at the output side of non-coinbase transactions rather than the input side. In empirical work, raw transaction-output value can be used as a reproducible proxy for on-chain transaction-value intensity, especially when the objective is to avoid opaque entity-adjustment heuristics.

However, the metric must be interpreted carefully. It is not an estimate of economically meaningful payment volume. Bitcoin transactions frequently create change outputs, and raw output value includes those change outputs. The metric may also include self-transfers, wallet consolidation, batching, exchange activity, custodial wallet reorganization, and other internal movements. Therefore, the series is intentionally named \texttt{obm\_raw\_output\_value\_btc\_daily}, rather than transaction volume, payment volume, or economic transfer volume.

The metric is most useful when interpreted as a transparent lower-level blockchain accounting series. Researchers who need entity-adjusted transfer volume should treat this OBM series as a reproducible raw baseline rather than as a substitute for heuristic-adjusted commercial metrics.

\paragraph{Similar metrics publicly available.}
The OBM series \texttt{obm\_raw\_output\_value\_btc\_daily} is comparable to public metrics usually labeled \emph{Output Value Per Day}, \emph{output volume}, \emph{transaction output value}, \emph{on-chain volume}, or \emph{transfer volume}. In OBM, the metric is defined as the total BTC value of outputs created by non-coinbase transactions in blocks assigned to a given UTC calendar day. Let \(B_d\) denote the set of blocks assigned to UTC date \(d\). For each non-coinbase transaction \(j\) in block \(b \in B_d\), let \(O_j\) denote the set of outputs created by that transaction, and let \(v_o\) denote the BTC value of output \(o\). The daily value is
\[
\mathrm{RawOutputValueBTC}_d
=
\sum_{b \in B_d}
\sum_{j \in T_b^{\mathrm{noncb}}}
\sum_{o \in O_j}
v_o .
\]
Coinbase transaction outputs are excluded because they represent miner revenue, which is covered by issuance and miner-revenue metrics. The metric is therefore a raw non-coinbase transaction-output value series. It includes change outputs, self-transfers, batching, exchange operations, custodial wallet management, and wallet reorganizations. It should not be interpreted as clean economic payment volume.

The closest public comparator is Blockchain.com's \emph{Output Value Per Day} chart.\footnote{\url{https://www.blockchain.com/charts/output-volume}} Blockchain.com defines the series as the total value of all transaction outputs per day and explicitly states that it includes coins returned to the sender as change. This is conceptually very close to \texttt{obm\_raw\_output\_value\_btc\_daily}, because both are raw output-side value measures rather than change-adjusted economic-transfer estimates. However, exact equivalence should not be assumed. The public page does not clearly specify whether Coinbase outputs are included or excluded in the chart, whereas OBM explicitly excludes them. Differences may also arise from timestamp conventions, chain reorganization handling, historical revisions, and daily boundary rules.

Blockchain.com's broader chart catalog also lists \emph{Output Value Per Day} under network activity and repeats that it is the total value of all transaction outputs per day, including change. The same catalog includes \emph{Estimated Transaction Value (BTC)}, which is related but conceptually different, as it estimates transfer value after accounting for change. For comparison with OBM, \emph{Output Value Per Day} is the closer raw-output comparator, while \emph{Estimated Transaction Value (BTC)} should be treated as an adjusted economic-transfer metric rather than as a direct equivalent. 

YCharts republishes Blockchain.com's \emph{Bitcoin Total Output Value Per Day} series.\footnote{\url{https://ycharts.com/indicators/bitcoin\_total\_output\_value\_per\_day}} This is useful as an accessible financial-data interface for the Blockchain.com output-volume series. Nevertheless, it should be treated as a secondary source rather than as an independent reconstruction. It inherits the same conceptual and methodological issues as the underlying Blockchain.com series, including the need to verify Coinbase treatment and timestamp conventions.

Blockchair provides a related \emph{on-chain volume} chart and chart-constructor interface.\footnote{\url{https://blockchair.com/bitcoin/charts/transaction-volume}} Its chart setup indicates a construction based on \emph{Output total - Blocks (BTC)} with a sum aggregation, and the general Bitcoin charts catalog lists \emph{On-chain volume (BTC)} and \emph{On-chain volume (USD)}. This is closely related to raw output value because it is output-side and block-aggregated. However, the public chart terminology differs from OBM's definition, and the page does not provide a full, reproducible algorithm specifying coinbase exclusion, timestamp assignment, change treatment, and reorganization handling. Blockchair also provides output-level explorer data that could support an independent reconstruction of the OBM series by summing non-coinbase outputs across blocks assigned to UTC dates. 

Glassnode provides \emph{Bitcoin Transfer Volume (Total)} (\texttt{transactions.TransfersVolumeSum}).\footnote{\url{https://studio.glassnode.com/charts/transactions.TransfersVolumeSum?a=BTC}} Glassnode describes this metric as the total amount of coins transferred on-chain, counting only successful transfers. This is related to \texttt{obm\_raw\_output\_value\_btc\_daily}, but it is not necessarily equivalent. Glassnode frames the measure as transfer volume rather than raw non-coinbase output value, and Glassnode also offers entity-adjusted and other filtered transfer-volume variants. As a result, Glassnode's metric is best treated as a transfer-volume comparator rather than as a strict raw-output-value benchmark, unless its underlying output and change conventions are aligned with OBM. 

Glassnode's public tutorial on on-chain activity further describes transfer volume as the total amount of BTC value sent across the Bitcoin network each day. This supports its use as a broad activity comparator, but it also highlights the conceptual difference: transfer volume is intended to capture value sent, whereas OBM's raw output value mechanically sums all non-coinbase transaction outputs, including change and self-transfer artifacts. 

Coin Metrics provides a related \emph{Xfer'd Val (native units)} metric, with MetricID \texttt{TxTfrValNtv}. It measures native units transferred during the interval.\footnote{\url{https://api.coinmetrics.io/v4/timeseries/asset-metrics?assets=btc\&metrics=TxTfrValNtv}} This is useful for comparing the general magnitude of value movement on-chain, but it is not a direct equivalent of our metric \texttt{obm\_raw\_output\_value\_btc\_daily}. Coin Metrics frames the metric as transferred value, and its adjusted variants apply heuristics to reduce noise from self-churn, cold-wallet shuffles, and other artifacts. OBM deliberately does not apply such adjustments and instead reports the raw output-side value. Therefore, Coin Metrics transfer-value metrics should be treated as related transfer-volume benchmarks rather than as raw output-value replications.

Newhedge provides related output-volume and transaction-volume pages\footnote{\url{https://newhedge.io/bitcoin/output-volume-per-transaction}} and transaction-volume indicators for Bitcoin. The output-volume-per-transaction page describes the metric as the value of Bitcoin transferred within each individual transaction. These indicators are relevant because they summarize output-side or transfer-side activity. However, the public pages do not provide enough detail to determine whether the underlying calculation includes change outputs, excludes coinbase outputs, uses raw outputs, or applies provider-specific filtering. Full data access may also require registration. Newhedge should therefore be treated as a chart-level comparator rather than as a fully auditable benchmark. 

BitInfoCharts reports indicators such as \emph{Bitcoins sent last 24h} and related sent-value statistics.\footnote{\url{https://bitinfocharts.com/bitcoin/}} These indicators are useful as broad diagnostics for on-chain value movement, but they are not clearly documented as raw non-coinbase output-value series. The public page does not specify whether the values are based on transaction outputs, transaction inputs, adjusted transfer estimates, or another convention. It is therefore not a direct methodological comparator for \texttt{obm\_raw\_output\_value\_btc\_daily}.

Blockchain.com's \emph{Confirmed Payments Per Day} chart\footnote{\url{https://www.blockchain.com/charts/n-payments}} is also related but should not be confused with this metric. That metric estimates the number of payments by treating transaction outputs as payments, except for one output assumed to be change when a transaction has at least two outputs. It is output-based, but it counts estimated payments rather than summing BTC value. It is therefore relevant for understanding output-side heuristics, but it is not a value-series equivalent. 

Overall, the closest public comparator for \texttt{obm\_raw\_output\_value\_btc\_daily} is Blockchain.com's \emph{Output Value Per Day}, because it explicitly reports the total value of transaction outputs and includes change. Blockchair's output-total-based on-chain volume chart is another close secondary comparator, and YCharts republishes the Blockchain.com output-volume series in a financial-data interface. Glassnode's \texttt{transactions.TransfersVolumeSum}, Coin Metrics' \texttt{TxTfrValNtv}, Newhedge's transaction-volume indicators, and BitInfoCharts' sent-value statistics are useful related transfer-volume references, but they are less direct because they may use transfer-value conventions, adjusted heuristics, or undocumented provider filters. The distinctive contribution of OBM is that it defines the raw output-value object explicitly, excludes coinbase outputs by construction, preserves raw output-side accounting, including change and self-transfers, assigns blocks to UTC calendar days using a documented Bitcoin Core timestamp convention, and computes the series reproducibly from decoded Bitcoin Core blocks.

\paragraph{Data source and input requirements.}

The metric is exported from the persistent OBM spent-output indexer database, rather than computed by an independent metric-specific blockchain scan. The indexer, described in Sect. \ref{sec:The Spent-Output Indexer}, is built from a running Bitcoin Core full node and stores reusable daily aggregates in a SQLite database. For \texttt{obm\_raw\_output\_value\_btc\_daily}, the relevant source table is \texttt{daily\_aggregates}, and the relevant fields are \texttt{spent\_value\_sats} and \texttt{fees\_sats}, or equivalent satoshi-denominated fields explicitly selected by the exporter.

The exporter relies on the transaction accounting identity that, for non-coinbase transactions, transaction fees equal input value minus output value. Therefore, non-coinbase output value can be recovered as input value minus fees. Aggregated by UTC date, the exporter computes:

\[
RawOutputValueSats_d = SpentValueSats_d - FeesSats_d .
\]

The resulting satoshi-denominated value is then converted into BTC by dividing by 100,000,000:

\[
RawOutputValueBTC_d = \frac{RawOutputValueSats_d}{100{,}000{,}000}.
\]

The exporter script does not query Bitcoin Core directly and does not retrieve decoded blocks at export time. It requires an existing SQLite database generated by \texttt{obm\_spent\_output\_indexer.py}. The path to this database is provided through the \texttt{\symbol{45}\symbol{45}state\_db} argument. The exporter identifies the daily aggregate table and the relevant date, spent-value, and fee columns either automatically or through explicit command-line arguments. Satoshi-denominated columns are preferred because they preserve exact integer accounting; BTC-denominated columns may be used as a fallback when explicitly selected or auto-detected.

If a requested date has no row in the daily aggregate table, the exporter typically treats it as a potential coverage issue and stops with an error. When the \texttt{\symbol{45}\symbol{45}missing\_dates\_as\_zero} flag is supplied, missing dates are exported as zero. This convention is appropriate for this flow variable only when the indexer database is known to fully cover the requested interval. Under that condition, an absent aggregate row indicates that no relevant raw-output contribution was assigned to that UTC date under the indexer's timestamp convention.

This metric does not require address extraction, user clustering, entity identification, external price data, third-party APIs, or a direct Bitcoin Core connection at export time. However, it depends on the prior successful execution of the spent-output indexer, which itself requires a synchronized, non-pruned Bitcoin Core full node and access to transaction-level data sufficient to reconstruct previous-output relationships and transaction fees. The exported series, therefore, inherits the indexer's UTC block-timestamp convention, previous-output reconstruction procedure, chain-consistency checks, release metadata, and handling of historical duplicate outpoint edge cases.

\paragraph{Algorithm.}
The script \texttt{export\_obm\_raw\_output\_value\_btc\_daily.py} implements the following procedure:

\begin{enumerate}
    \item Parse the user-provided date interval, \texttt{\symbol{45}\symbol{45}start\_date} and \texttt{\symbol{45}\symbol{45}end\_date}, using the format \texttt{YYYY-MM-DD}. Both dates are interpreted as UTC dates and included in the output.

    \item Verify that the starting date is not later than the ending date.

    \item Open the SQLite database generated by \texttt{obm\_spent\_output\_indexer.py}. The database path is provided through the \texttt{\symbol{45}\symbol{45}state\_db} argument. If the database file does not exist, the exporter stops with an error.

    \item Identify the daily aggregate table. If the user provides \texttt{\symbol{45}\symbol{45}table}, the exporter verifies that the table exists. Otherwise, it attempts to auto-detect a suitable daily aggregate table from a predefined list of common OBM table names, such as \texttt{daily\_aggregates}.

    \item Identify the date column. If the user provides \texttt{\symbol{45}\symbol{45}date\_column}, the exporter verifies that the column exists. Otherwise, it attempts to auto-detect a date column using common names such as \texttt{date}, \texttt{day}, or \texttt{utc\_date}.

    \item Identify the spent-value and fee columns. If the user provides \texttt{\symbol{45}\symbol{45}spent\_value\_column} and \texttt{\symbol{45}\symbol{45}fees\_column}, the exporter verifies that both columns exist. Otherwise, it attempts to auto-detect them. When \texttt{\symbol{45}\symbol{45}value\_mode auto} is used, the exporter first looks for satoshi-denominated columns, such as \texttt{spent\_value\_sats} and \texttt{fees\_sats}. If these are not found, it attempts to use BTC-denominated columns.

    \item Initialize the requested output interval with one entry per UTC date. At this stage, dates are marked as undefined until a matching aggregate row is read from the database.

    \item Query the selected daily aggregate table for all rows whose date lies between \texttt{\symbol{45}\symbol{45}start\_date} and \texttt{\symbol{45}\symbol{45}end\_date}, inclusive. The query retrieves the selected date, spent-value, and fee columns from the previous steps.

    \item For each returned date, verify that the date has not already appeared in the result set. Duplicate dates are treated as an error.

    \item For satoshi-denominated columns, parse the spent-value and fee fields as integer satoshi values. The exporter rejects non-integer values in this mode. For BTC-denominated columns, parse the fields using decimal arithmetic.

    \item Compute daily raw output value using the transaction-accounting identity:
    \[
    RawOutputValue_d = SpentValue_d - Fees_d.
    \]
    When satoshi-denominated columns are used, the calculation is performed as:
    \[
    RawOutputValueSats_d =
        SpentValueSats_d - FeesSats_d.
    \]

    \item Reject any observation for which the computed raw output value is negative, since this would indicate an inconsistent database state or incompatible column selection.

    \item Convert satoshi-denominated results to BTC:
    \[
    RawOutputValueBTC_d =
        \frac{RawOutputValueSats_d}{100{,}000{,}000}.
    \]

    \item Check whether any requested dates remain without a corresponding aggregate row. If \texttt{\symbol{45}\symbol{45}missing\_dates\_as\_zero} is not supplied, the exporter stops with an error. If \texttt{\symbol{45}\symbol{45}missing\_dates\_as\_zero} is supplied, missing dates are written as zero. This convention is appropriate only when the indexer database is known to fully cover the requested interval.

    \item Write the resulting time series to a CSV file using the standardized OBM schema:
    \[
    \texttt{date, series\_id, value, unit, frequency, release\_version}.
    \]
    Values are written in BTC with eight decimal places.

    \item Optionally generate a plot of the exported series when the \texttt{\symbol{45}\symbol{45}plot} flag is activated. If \texttt{\symbol{45}\symbol{45}plot\_output} is not provided, the plot is saved next to the CSV file using the same filename stem and the \texttt{.png} extension.

    \item Print diagnostic information unless \texttt{\symbol{45}\symbol{45}quiet} is supplied. The diagnostics include the output path, selected date range, database path, selected table, selected columns, resolved value mode, number of rows read, and number of missing dates.
\end{enumerate}

The exporter does not query Bitcoin Core or scan decoded blocks at export time. The computationally intensive blockchain processing, previous-output reconstruction, spent-value calculation, and fee calculation are performed in advance by the persistent OBM spent-output indexer. The raw-output-value exporter is therefore a deterministic transformation from the indexer database to the standard OBM CSV format.
   
\paragraph{Metric-specific input parameters.}
The input parameters specific to this exporter are:

\begin{itemize}
    \item \texttt{\symbol{45}\symbol{45}state\_db}: path to the persistent SQLite database generated by
    \texttt{obm\_spent\_output\_indexer.py}. The default path is
    \texttt{cache/obm\_spent\_output\_indexer.sqlite}.

    \item \texttt{\symbol{45}\symbol{45}table}: optional name of the daily aggregate table. If omitted, the exporter
    attempts to auto-detect a suitable daily aggregate table, such as
    \texttt{daily\_aggregates}.

    \item \texttt{\symbol{45}\symbol{45}date\_column}: optional name of the date column. If omitted, the exporter
    attempts to auto-detect a date column using common names such as \texttt{date},
    \texttt{day}, or \texttt{utc\_date}.

    \item \texttt{\symbol{45}\symbol{45}spent\_value\_column}: optional name of the column containing daily spent
    output value. Satoshi-denominated columns are preferred because they preserve exact
    integer accounting.

    \item \texttt{\symbol{45}\symbol{45}fees\_column}: optional name of the column containing daily transaction
    fees. Satoshi-denominated columns are preferred for the same reason.

    \item \texttt{\symbol{45}\symbol{45}value\_mode}: interpretation of the spent-value and fee columns. The
    accepted values are \texttt{auto}, \texttt{sats}, and \texttt{btc}. In \texttt{auto} mode,
    the exporter first tries to use satoshi-denominated columns and falls back to
    BTC-denominated columns if no suitable satoshi columns are found.

    \item \texttt{\symbol{45}\symbol{45}missing\_dates\_as\_zero}: optional flag indicating that requested dates
    absent from the daily aggregate table should be exported as zero. This option should
    be used only when the indexer database is known to fully cover the requested interval.
    Without this flag, missing aggregate rows are treated as a potential coverage or schema
    problem and cause the exporter to stop.
\end{itemize}

The script also accepts the standard output, release-version, plotting, and diagnostic
parameters: \texttt{\symbol{45}\symbol{45}start\_date}, \texttt{\symbol{45}\symbol{45}end\_date}, \texttt{\symbol{45}\symbol{45}output},
\texttt{\symbol{45}\symbol{45}release\_version}, \texttt{\symbol{45}\symbol{45}plot}, \texttt{\symbol{45}\symbol{45}plot\_output}, and
\texttt{\symbol{45}\symbol{45}quiet}. It does not accept Bitcoin Core RPC parameters or
\texttt{\symbol{45}\symbol{45}height\_margin}, because this metric is not reconstructed by scanning blocks at
export time. The required blockchain processing is performed beforehand by the persistent
spent-output indexer.

\paragraph{Aggregation rule.}
The daily value is computed as the sum of all non-coinbase transaction output values assigned to the same UTC calendar date:
\[
\mathrm{RawOutputValueBTC}_d =
\sum_{b: \mathrm{UTCDate}(t_b)=d}
\sum_{j \in T_b^{\mathrm{noncb}}}
\sum_{o \in O_j}
v_o.
\]

The aggregation rule is therefore a daily sum. Monthly versions of this metric, if distributed, should also be computed as sums of the corresponding daily values:
\[
\mathrm{RawOutputValueBTC}_m =
\sum_{d \in m}
\mathrm{RawOutputValueBTC}_d.
\]

This convention preserves the interpretation of the metric as the total raw non-coinbase output value created over the corresponding period.

\paragraph{Output format.}
The output file contains one observation per UTC date. Each row has the following fields:

\begin{center}
\begin{tabular}{llp{0.45\textwidth}}
\toprule
\textbf{Column} & \textbf{Example} & \textbf{Description} \\
\midrule
\texttt{date} & \texttt{2024-01-01} & UTC calendar date \\
\texttt{series\_id} & \texttt{obm\_raw\_output\_value\_btc\_daily} & Stable OBM series identifier \\
\texttt{value} & \texttt{587321.12345678} & Raw non-coinbase transaction output value \\
\texttt{unit} & \texttt{BTC} & Measurement unit \\
\texttt{frequency} & \texttt{daily} & Observation frequency \\
\texttt{release\_version} & \texttt{OBM v0.1.0} & Dataset release version \\
\bottomrule
\end{tabular}
\end{center}

Values are written with eight decimal places, preserving standard Bitcoin monetary precision.

\paragraph{Technical validation.}

Several internal checks are used to validate this metric during execution. First, the exporter verifies that the requested date range is valid and that \texttt{\symbol{45}\symbol{45}start\_date} is not later than \texttt{\symbol{45}\symbol{45}end\_date}. Second, it verifies that the SQLite state database specified through \texttt{\symbol{45}\symbol{45}state\_db} exists. Third, it verifies that the selected daily aggregate table exists, either because it was explicitly supplied through \texttt{\symbol{45}\symbol{45}table} or because it was successfully auto-detected by the exporter. Fourth, it verifies that the selected date, spent-value, and fee columns exist in the selected table. Fifth, when satoshi-denominated columns are used, it checks that the corresponding values are integer quantities. Sixth, it parses all numeric values using decimal arithmetic, avoiding binary floating-point artifacts. Seventh, it verifies that each date returned by the database query appears at most once. Eighth, it computes the raw output value using
\[
RawOutputValue_d = SpentValue_d - Fees_d,
\]
and rejects any observation for which the resulting value is negative. Ninth, it checks whether any requested dates are missing from the selected aggregate table. If \texttt{\symbol{45}\symbol{45}missing\_dates\_as\_zero} is not supplied, missing dates cause the exporter to stop, because they may indicate incomplete indexer coverage or an unexpected schema problem. If \texttt{\symbol{45}\symbol{45}missing\_dates\_as\_zero} is supplied, missing dates are exported as zero. Tenth, the exporter writes one row per requested UTC date using the standard OBM schema and reports diagnostic information, including the selected table, selected columns, the resolved value mode, the number of rows read, and the number of missing dates.

Additional consistency checks can be performed using related OBM metrics. For non-coinbase transactions, the transaction-level accounting identity is:
\[
InputValue_j = OutputValue_j + Fee_j .
\]
Therefore, subject to common block assignment, release version, and transaction coverage, the following daily relationship should hold:
\[
SpentValueBTC_d =
RawOutputValueBTC_d + FeesBTC_d .
\]
This identity links three metrics: \texttt{obm\_raw\_output\_value\_btc\_daily}, \texttt{obm\_spent\_value\_btc\_daily}, and \texttt{obm\_fees\_btc\_daily}. Deviations should be investigated because they may reveal differences in indexer coverage, timestamp convention, column selection, release version, coinbase treatment, missing-date handling, or historical edge-case handling.

For release validation, the indexer database should also be checked independently. In particular, the database should be verified to have been built from the Bitcoin genesis block onward, to cover the full requested interval, and to contain metadata consistent with the expected OBM spent-output indexer version. Selected dates can also be compared with the older direct block-scanning implementation as a diagnostic. External comparisons with blockchain explorers or public transaction-volume indicators should be interpreted cautiously, because many providers report adjusted transfer volume rather than raw output value, and may use change-output heuristics, entity clustering, alternative timestamp conventions, or different coinbase-treatment rules.

\paragraph{Known limitations.}

The daily raw output value series is transparent and reproducible, but it has important limitations. First, it is exported from the OBM spent-output indexer database and therefore depends on the indexer run's correctness, completeness, timestamp convention, and release version. Second, the exporter derives the metric from the accounting identity
\[
RawOutputValueBTC_d = SpentValueBTC_d - FeesBTC_d,
\]
rather than by independently summing transaction outputs at export time. This is exact under the OBM indexer convention when spent value and fees are computed from the same fully covered database, but it also means that the metric inherits any indexer-level errors, incomplete coverage, column-selection errors, or historical edge-case conventions affecting those components. Third, if \texttt{\symbol{45}\symbol{45}missing\_dates\_as\_zero} is used, missing aggregate rows are interpreted as zero-flow observations; this is appropriate only after verifying that the indexer database fully covers the requested interval.

Fourth, the metric measures raw output value rather than economically adjusted payment volume. It includes change outputs, self-transfers, batching, exchange operations, custodial wallet management, wallet consolidation, and other non-payment activities. Fifth, it excludes Coinbase outputs, so it should not be interpreted as the total value created in all transaction outputs, including miner payments. Sixth, it does not identify addresses, users, entities, exchanges, custodians, change outputs, or payment purposes. Seventh, it depends on the UTC block-timestamp convention used by the indexer to assign blocks and their aggregates to calendar dates. Eighth, it reports BTC-denominated value, not fiat-denominated value. Ninth, it may differ from commercial transaction-volume metrics that remove change outputs, apply entity-adjustment heuristics, classify internal transfers, use different timestamp conventions, or report adjusted transfer volume rather than raw output value.

Despite these limitations, \texttt{obm\_raw\_output\_value\_btc\_daily} is a useful OBM activity series. It provides a transparent, reproducible, output-side value measure derived from the same spent-output indexer used for spent value and transaction fees. Its main value is not that it estimates economic payment volume, but that it provides a raw baseline against which more sophisticated or heuristic-adjusted transaction-volume measures can be compared.
```

\subsection{\texttt{obm\_spent\_output\_count\_daily}: Daily Spent Output Count}
\label{Daily Spent Output Count}

\paragraph{Definition.}
The daily spent-output-count series measures the number of previous outputs consumed by non-coinbase transaction inputs in blocks assigned to a given UTC calendar day. Let \(B_d\) denote the set of blocks assigned to day \(d\). For each block \(b \in B_d\), let \(I_b\) denote the set of non-coinbase transaction inputs included in that block. Each input in \(I_b\) spends one previous output. The daily spent-output count is defined as:
\[
\mathrm{SpentOutputCount}_d =
\sum_{b \in B_d}
|I_b|,
\]
where \(|I_b|\) denotes the number of non-coinbase transaction inputs included in block \(b\).

A block \(b\) is assigned to day \(d\) according to the UTC calendar date derived from its block timestamp \(t_b\), $d(b)=\mathrm{UTCDate}(t_b)$. The resulting series therefore reports the number of previous outputs spent in the Bitcoin main chain per UTC day. It is an input-side activity metric: it counts consumed outputs, not transactions, users, entities, addresses, or payments.

\paragraph{Economic interpretation.}
Daily spent output count measures the number of UTXOs consumed on-chain each day. It is a raw indicator of input-side blockchain activity.

For economic research, this metric is useful in several ways. First, it complements transaction count, because a single Bitcoin transaction can spend one input or many inputs. Second, it helps interpret changes in spent value, CDD, dormancy, and age-threshold metrics. Third, it can help identify wallet-consolidation episodes, as consolidation transactions often consume many inputs. Fourth, it provides a denominator for derived indicators such as the average spent value per spent output. Fifth, it is useful for studying UTXO-flow dynamics and the structure of transaction input activity.

The metric should not be interpreted as a user count, an entity count, a payment count, or an economically meaningful transfer count. It does not distinguish ordinary payments from self-transfers, batching, wallet consolidation, exchange activity, custodial reorganization, or change management. It is a raw count of previous outputs consumed by transaction inputs.

\paragraph{Similar metrics publicly available.}
The OBM series \texttt{obm\_spent\_output\_count\_daily} is comparable to public metrics usually labeled \emph{spent outputs}, \emph{spent output count}, \emph{spent output age bands}, or \emph{UTXO spending activity}. In OBM, the metric is defined as the daily number of previous transaction outputs consumed by non-coinbase transaction inputs in blocks assigned to a given UTC calendar day. Coinbase transactions are excluded because they do not consume previous outputs. The metric is therefore a raw UTXO-level activity count: it measures how many previously unspent outputs were consumed on each day, not how many transactions occurred, how much BTC was transferred, or how many users were active. It is not entity-adjusted and does not attempt to identify exchanges, custodians, self-transfers, or change outputs.

The closest public comparators are Glassnode's \emph{Spent Outputs} age-band metrics. Glassnode exposes several charts that report the total number of spent outputs created within a given age interval, such as the \emph{Bitcoin Spent Outputs 1d-1w} chart.\footnote{\url{https://studio.glassnode.com/charts/indicators.Sol1D1W?a=BTC}} Glassnode describes this metric as the total number of spent outputs that were created between one day and one week ago. Other related Glassnode charts report spent-output counts for different age bands, such as \texttt{1h-24h}, \texttt{3y-5y}, and other lifespan intervals. These metrics are conceptually close to \texttt{obm\_spent\_output\_count\_daily}, because they count spent outputs rather than transactions or BTC volume. However, they are age-band-specific rather than a single raw total. A total spent-output-count comparator could, in principle, be approximated by summing all non-overlapping Glassnode spent-output age bands, provided that the full set of bands is available, mutually exclusive, exhaustive, and reported in absolute counts. The public pages do not provide a full open-source reconstruction algorithm, nor do they specify all low-level conventions needed for exact equivalence with OBM, such as timestamp assignment, fractional-age boundaries, chain reorganization handling, and whether any entity adjustment or filtering is applied.

CryptoQuant provides a closely related \emph{Spent Output Age Bands} framework.\footnote{\url{https://cryptoquant.com/asset/btc/chart/network-indicator/spent-output-age-bands}} CryptoQuant describes Spent Output Age Bands as a set of spent outputs created within specified age bands, with each colored band representing the total value of spent outputs whose lifespans fall within the corresponding interval. This is related to \texttt{obm\_spent\_output\_count\_daily}, but it is not a direct equivalent because the published banded metric is usually value-weighted rather than a simple count of spent outputs. CryptoQuant's methodology is nevertheless useful because it confirms the same underlying UTXO logic: spent outputs can be grouped according to the age of the consumed previous output. A count-based version would require replacing the value-weighted summation with an indicator count over spent outputs. The public documentation does not provide a complete node-level reconstruction script for such a raw count.

Glassnode's \emph{Spent Output Profit Ratio} family is also related, but not equivalent.\footnote{\url{https://studio.glassnode.com/charts/indicators.Sopr?a=BTC}} Glassnode defines SOPR as the realized value divided by the value at creation for spent outputs. Its documentation explains the concept of a spent output and uses spent outputs as the unit of analysis. However, SOPR is a profitability ratio, not a count. It is therefore useful for confirming that public providers use spent-output-level accounting, but it should not be treated as a direct comparator for \texttt{obm\_spent\_output\_count\_daily}.

CryptoQuant's SOPR-related metrics are similarly related but not direct equivalents. For example, its Long-Term Holder SOPR page defines the metric as a ratio of spent outputs older than a specified threshold, whereas the general SOPR framework evaluates whether moved coins are spent at a profit or loss. These metrics rely on spent outputs, but their outputs are ratios rather than counts. They should therefore be treated as methodological neighbors, not as substitutes for a raw daily spent-output-count series.

Coin Metrics does not appear to expose a named public metric exactly equivalent to the OBM metric \texttt{obm\_spent\_output\_count\_daily}. Its public transaction and transfer-value metrics, such as transaction count and native transfer value, are related to on-chain activity but do not directly count consumed UTXOs. Coin Metrics' transfer-value methodology for UTXO chains also applies heuristics in adjusted variants, such as excluding early spends or filtering likely self-churn. These metrics answer different questions from OBM's raw spent-output count. A Coin Metrics-based comparison would therefore require access to a spent-output-level count series or a custom reconstruction from lower-level blockchain data.

Blockchair provides block, transaction, input, and output-level explorer data, including Bitcoin output data.\footnote{\url{https://blockchair.com/bitcoin/outputs}}. This makes Blockchair a useful source for reconstruction. A researcher could reconstruct a daily spent-output-count series by identifying non-coinbase transaction inputs, resolving the previous outputs they consume, and counting those consumed outputs by the UTC date of the spending block. However, Blockchair does not appear to publish a clearly named public daily metric equivalent to \texttt{obm\_spent\_output\_count\_daily}. It should therefore be treated as a data source for independent reconstruction rather than as a named public comparator.

Blockchain.com provides transaction-count and output-volume charts, but these are not direct equivalents. For example, its \emph{Output Value Per Day} chart reports the total value of all transaction outputs per day, including change.\footnote{\url{https://www.blockchain.com/charts/output-volume}}. This is an output-value metric over newly created outputs, not a count of previously created outputs consumed as inputs. Similarly, transaction-count charts count transactions, not the number of spent UTXOs. These series may be useful for contextualizing network activity, but they do not measure the same object as \texttt{obm\_spent\_output\_count\_daily}.

Overall, no reviewed public provider appears to expose a simple named daily series exactly equivalent to the OBM metric \texttt{obm\_spent\_output\_count\_daily}. The closest public comparators are Glassnode's spent-output age-band count charts, because they count spent outputs within specified lifespan intervals. CryptoQuant's spent-output age-band framework is also relevant, although it is primarily value-based rather than count-based. SOPR metrics from Glassnode and CryptoQuant use spent outputs as their basic unit, but they report profitability ratios rather than counts. Blockchair can support independent reconstruction from lower-level data, while Blockchain.com and Coin Metrics provide related activity metrics that are not direct equivalents. The distinctive contribution of OBM is that it publishes the raw daily spent-output count directly, keeps the definition unadjusted and UTXO-based, excludes coinbase transactions by construction, assigns spending blocks to UTC calendar days using an explicit timestamp convention, and makes the computation auditable from full-node data.

\paragraph{Data source and input requirements.}
The metric is exported from the persistent OBM spent-output indexer database, rather than computed by an independent metric-specific blockchain scan. The indexer, described in Sect.~\ref{sec:The Spent-Output Indexer}, is built from a running Bitcoin Core full node and stores reusable daily aggregates in a SQLite database. For \texttt{obm\_spent\_output\_count\_daily}, the relevant source table is \texttt{daily\_aggregates}, and the relevant field is \texttt{spent\_output\_count}.

The indexer computes this quantity during its sequential scan of the blockchain. For each non-coinbase transaction input, it resolves the input to the previous output it spends and increments the daily spent-output count. Internally, the indexer stores:
\[
\mathrm{spent\_output\_count}_d =
\sum_{b \in B_d}
|I_b|.
\]

The exporter script does not query Bitcoin Core directly. It requires an existing SQLite database generated by \texttt{obm\_spent\_output\_indexer.py}, with metadata identifying it as an OBM spent-output indexer database and with processed-date metadata covering the requested output interval. The script checks that the requested ending date is not beyond the indexer's maximum date. If a requested date has no row in \texttt{daily\_aggregates}, the exporter writes a zero value for that date. This convention is appropriate because the metric is a daily count variable: if no spent-output aggregate is assigned to a UTC date under the indexer's timestamp convention, the spent-output count for that date is zero.

This metric does not require address extraction, user clustering, entity identification, external price data, third-party APIs, or a direct Bitcoin Core connection at export time. However, it depends on the prior successful execution of the spent-output indexer, which itself requires a synchronized, non-pruned Bitcoin Core full node and access to transaction-level data sufficient to reconstruct previous-output relationships. The exported series therefore inherits the indexer's UTC block-timestamp convention, previous-output reconstruction procedure, chain-consistency checks, release metadata, and handling of historical duplicate outpoint edge cases.

\paragraph{Algorithm.}
The script \texttt{export\_obm\_spent\_output\_count\_daily.py} implements the following procedure:

\begin{enumerate}
    \item Parse the user-provided date interval, \texttt{\symbol{45}\symbol{45}start\_date} and \texttt{\symbol{45}\symbol{45}end\_date}, using the format \texttt{YYYY-MM-DD}. Both dates are interpreted as UTC dates and both are included in the output.

    \item Open the persistent SQLite database generated by \texttt{obm\_spent\_output\_indexer.py}. The path to this database is provided through the \texttt{\symbol{45}\symbol{45}state\_db} argument.

    \item Validate that the SQLite database corresponds to the OBM spent-output indexer. The script checks the indexer metadata, including the expected \texttt{indexer\_id} and the presence of processing metadata such as \texttt{last\_processed\_height}, \texttt{last\_processed\_date}, and \texttt{max\_processed\_date}.

    \item Verify that the requested \texttt{\symbol{45}\symbol{45}end\_date} is not later than the maximum UTC date processed by the indexer. If the requested interval extends beyond the indexed range, the script aborts and requires the indexer to be run further before export.

    \item Infer the dataset release version from the indexer metadata. If the database does not contain a release-version field, the script uses the fallback value provided through the \texttt{\symbol{45}\symbol{45}release\_version} argument.

    \item For each UTC date \(d\) in the requested interval, query the \texttt{daily\_aggregates} table for the field \texttt{spent\_output\_count}. If a row exists for date \(d\), the script reads the stored count. If no row exists, the script assigns zero to that date.

    \item Write the resulting time series to a CSV file using the standardized OBM schema:
    \[
    \texttt{date},\quad
    \texttt{series\_id},\quad
    \texttt{value},\quad
    \texttt{unit},\quad
    \texttt{frequency},\quad
    \texttt{release\_version}.
    \]

    \item Optionally generate a plot of the exported series when the plotting flag is activated. If \texttt{\symbol{45}\symbol{45}plot\_output} is not provided, the plot is saved next to the CSV file with a \texttt{.png} extension. The plot title includes the series description and the selected date interval.
\end{enumerate}

This exporter does not query Bitcoin Core, does not maintain an outpoint state, and does not reconstruct spent outputs at export time. The computationally expensive previous-output resolution has already been performed by the spent-output indexer, which stores the daily spent-output count in \texttt{daily\_aggregates}. The export script is therefore a lightweight, deterministic transformation from the indexed database to the OBM CSV format.

\paragraph{Metric-specific input parameters.}
The input parameters specific to this exporter are:

\begin{itemize}
    \item \texttt{\symbol{45}\symbol{45}state\_db}: path to the persistent SQLite database generated by \texttt{obm\_spent\_output\_indexer.py}. This database must contain the \texttt{daily\_aggregates} table and the metadata required to verify that it corresponds to the OBM spent-output indexer.

    \item \texttt{\symbol{45}\symbol{45}start\_date}: starting date of the selected interval, inclusive, in \texttt{YYYY-MM-DD} format. The date is interpreted as a UTC calendar date.

    \item \texttt{\symbol{45}\symbol{45}end\_date}: ending date of the selected interval, inclusive, in \texttt{YYYY-MM-DD} format. The script verifies that this date is not later than the maximum date processed by the indexer.

    \item \texttt{\symbol{45}\symbol{45}release\_version}: fallback dataset release version used only if the indexer database does not contain a \texttt{release\_version} metadata field.

    \item \texttt{\symbol{45}\symbol{45}output}: path of the output CSV file to be written.

    \item \texttt{\symbol{45}\symbol{45}plot}: optional flag that instructs the script to generate a plot of the exported series.

    \item \texttt{\symbol{45}\symbol{45}plot\_output}: optional path for the generated plot. If this argument is omitted while \texttt{\symbol{45}\symbol{45}plot} is used, the plot is saved next to the CSV file using the same base name and a \texttt{.png} extension.
\end{itemize}

The exporter does not accept Bitcoin Core RPC parameters, \texttt{\symbol{45}\symbol{45}height\_margin}, \texttt{\symbol{45}\symbol{45}commit\_every}, \texttt{\symbol{45}\symbol{45}min\_confirmations}, or \texttt{\symbol{45}\symbol{45}reset\_state\_db}. These parameters belong to the spent-output indexer, not to the spent-output-count exporter. The exporter assumes that the indexer database has already been built and updated through the requested ending date. Its role is only to read \texttt{spent\_output\_count} from \texttt{daily\_aggregates} and write the standardized OBM output file.

\paragraph{Aggregation rule.}
The daily value is computed as the number of previous outputs spent by non-coinbase transaction inputs included in blocks assigned to the same UTC calendar date:
\[
\mathrm{SpentOutputCount}_d =
\sum_{b \in B_d}
|I_b|.
\]

The aggregation rule is therefore a daily count of spent previous outputs. Monthly versions of this metric, if distributed, should also be computed as sums of the corresponding daily values:
\[
\mathrm{SpentOutputCount}_m =
\sum_{d \in m}
\mathrm{SpentOutputCount}_d.
\]

This convention preserves the interpretation of the metric as the number of previous outputs consumed over the corresponding period.

\paragraph{Output format.}
The output file contains one observation per UTC date. Each row has the following fields:

\begin{center}
\begin{tabular}{llp{0.35\textwidth}}
\toprule
\textbf{Column} & \textbf{Example} & \textbf{Description} \\
\midrule
\texttt{date} & \texttt{2024-01-01} & UTC calendar date \\
\texttt{series\_id} & \texttt{obm\_spent\_output\_count\_daily} & Stable OBM series identifier \\
\texttt{value} & \texttt{752184} & Number of spent previous outputs \\
\texttt{unit} & \texttt{outputs} & Measurement unit \\
\texttt{frequency} & \texttt{daily} & Observation frequency \\
\texttt{release\_version} & \texttt{OBM v0.1.0} & Dataset release version \\
\bottomrule
\end{tabular}
\end{center}

\paragraph{Technical validation.}
Several internal checks are used to validate this metric at export time. First, the script verifies that the requested date range is valid and that \texttt{\symbol{45}\symbol{45}start\_date} is not later than \texttt{\symbol{45}\symbol{45}end\_date}. Second, it checks that the SQLite state database exists, opens it, and sets the connection to query-only mode. Third, it verifies that the database metadata identify it as an OBM spent-output indexer database, by checking the expected \texttt{indexer\_id}. Fourth, it checks that the indexer database contains processing metadata, including \texttt{last\_processed\_height}, \texttt{last\_processed\_date}, and \texttt{max\_processed\_date}. Fifth, it verifies that the table \texttt{daily\_aggregates} is present in the database. Sixth, it verifies that the requested ending date is not later than the maximum date processed by the indexer. If the requested interval extends beyond the indexed range, the script aborts and requires the indexer to be updated before the export is attempted.

For each date in the requested interval, the exporter reads \texttt{spent\_output\_count} from \texttt{daily\_aggregates}. If the date is absent from the aggregate table, the script exports zero for that date. This convention is appropriate for \texttt{obm\_spent\_output\_count\_daily} because it is a count variable: if no spent-output activity is assigned to a UTC date under the indexer's timestamp convention, the daily count is zero rather than missing. The resulting file is written using the standard OBM schema:
\[
\texttt{date},\quad
\texttt{series\_id},\quad
\texttt{value},\quad
\texttt{unit},\quad
\texttt{frequency},\quad
\texttt{release\_version}.
\]

The script also propagates the release version stored in the indexer metadata. If the metadata field is absent, it uses the fallback value supplied through \texttt{\symbol{45}\symbol{45}release\_version}. The number of rows written must equal the number of calendar days in the requested interval, because the exporter iterates explicitly over all dates between \texttt{\symbol{45}\symbol{45}start\_date} and \texttt{\symbol{45}\symbol{45}end\_date}, inclusive. Exported values should be non-negative integers.

Additional consistency checks can be performed using related OBM series exported from the same indexer database. In particular, if $\mathrm{SpentValueBTC}_d = 0$, then $\mathrm{SpentOutputCount}_d = 0$ should hold when both \texttt{obm\_spent\_value\_btc\_daily} and \texttt{obm\_spent\_output\_count\_daily} are exported from the same indexer database using the same timestamp convention and release version. Positive spent value should normally coincide with positive spent-output count. The converse should be interpreted cautiously only in edge cases involving zero-value outputs.

The metric can also be checked against transaction-count series. Since a non-coinbase transaction must have at least one input, the spent-output count should normally be at least as large as the number of non-coinbase transactions included on active dates. However, exact comparisons require a transaction-count series that separates coinbase and non-coinbase transactions. Comparisons with external input-count or spent-output-count series are useful as diagnostics but should not be interpreted as strict equality tests, because providers may differ in timestamp conventions, input filtering, treatment of zero-value outputs, chain-reorganization handling, and historical edge-case treatment.

\paragraph{Known limitations.}
The daily spent-output-count series is useful but should be interpreted carefully. First, \texttt{obm\_spent\_output\_count\_daily} is a raw input-side count, not an entity-adjusted measure of economic activity. It does not identify users, entities, custodians, exchanges, self-transfers, or change outputs. Second, the metric counts previous outputs consumed by transaction inputs, not transactions. Third, it depends on the block timestamp convention used to assign spending blocks to calendar days. Fourth, it can be affected by wallet consolidation, batching, exchange operations, custodial wallet management, and self-transfers. Fifth, it does not measure BTC value, fiat value, transaction fees, or coin age. Sixth, it requires the persistent spent-output indexer database and therefore depends on the correctness and completeness of the indexer run. Seventh, the simple reorganization policy of the indexer detects inconsistencies and aborts, but it does not automatically roll back the state database.

Despite these limitations, \texttt{obm\_spent\_output\_count\_daily} is a useful OBM activity series. It provides a transparent raw count of outputs consumed on-chain and complements transaction count, spent value, CDD, dormancy, and age-threshold metrics in the study of UTXO-flow dynamics.
\subsection{\texttt{obm\_spent\_value\_age\_band\_btc\_daily}: Daily Spent Output Value by Age Band in BTC}
\label{Daily Spent Output Value by Age Band in BTC}

\paragraph{Definition.}
The daily spent-output-value-by-age-band table decomposes the total daily spent output value according to the age of the outputs at the moment they are spent. Unlike the scalar metrics described above, this metric is a wide, vector-valued daily table: each row corresponds to one UTC calendar date, and each age-band column reports the BTC value of outputs spent on that date whose age falls within the corresponding band.

Let \(B_d\) denote the set of blocks assigned to day \(d\). For each non-coinbase transaction input \(i\) included in a transaction in block \(b \in B_d\), let \(v_i\) denote the BTC value of the previous output spent by that input, and let \(a_i\) denote the age of that previous output, measured in days. The age of the spent output is defined as:
\[
a_i =
\max\left(
0,
\frac{t^{\mathrm{spent}}_i - t^{\mathrm{created}}_i}{86400}
\right),
\]
where \(t^{\mathrm{created}}_i\) is the timestamp of the block in which the spent output was created, and \(t^{\mathrm{spent}}_i\) is the timestamp of the block in which it is spent.

For each age band \(k\), the daily age-band spent-value component is defined as:
\[
\mathrm{SpentValueBand}_{d,k} =
\sum_{b \in B_d}
\sum_{i \in I_b}
v_i \mathbf{1}\{a_i \in k\},
\]
where \(I_b\) denotes the set of non-coinbase transaction inputs included in block \(b\), and \(\mathbf{1}\{\cdot\}\) is an indicator function equal to one when the spent output age belongs to band \(k\), and zero otherwise.

A block \(b\) is assigned to day \(d\) according to the UTC calendar date derived from its block timestamp \(t_b\):
\[
d(b)=\mathrm{UTCDate}(t_b).
\]

The output table uses the following age bands: 0d--1d, 1d--1w, 1w--1m, 1m--3m, 3m--6m, 6m--1y, 1y--2y, 2y--3y, 3y--5y, 5y--7y, 7y--10y, and 10y+. The sum across all age-band columns should recover the total spent output value:
\[
\mathrm{SpentValueBTC}_d =
\sum_k
\mathrm{SpentValueBand}_{d,k}.
\]

\paragraph{Economic interpretation.}
The daily spent-output-value-by-age-band table describes the age distribution of BTC-denominated spent output. It indicates whether the daily spent value is dominated by recently created outputs, medium-aged outputs, or outputs that have remained inactive for several years.

For economic research, this table is useful in several ways. First, it decomposes the total spent value by output age, providing richer information than a single scalar spent-value series. Second, it helps distinguish young-output turnover from dormant-supply activation. Third, it allows researchers to examine whether spikes in spent value are driven by short-lived UTXO churn, wallet consolidation, or the movement of older coins. Fourth, it supports validation and interpretation of threshold metrics such as \texttt{obm\_cdd\_155d\_btc\_daily} and \texttt{obm\_cdd\_365d\_btc\_daily}. Fifth, it provides the basis for constructing age-band shares, such as the fraction of daily spent value associated with outputs older than one year.

The metric should not be interpreted as entity-adjusted payment volume, user activity, or economically meaningful settlement value. It is a raw spent-output decomposition. It does not identify users, entities, exchanges, custodians, change outputs, or self-transfers. As with other raw spent-output metrics, wallet consolidation, batching, custodial reorganization, and internal transfers may affect the observed age distribution.

\paragraph{Similar metrics publicly available.}
The OBM table \texttt{obm\_spent\_value\_age\_band\_btc\_daily} is comparable to public metrics usually labeled \emph{Spent Volume by Age}, \emph{Spent Volume Age Bands}, \emph{Spent Output Age Bands}, or \emph{SVAB}. Unlike scalar OBM series, this metric is a wide daily table: each row corresponds to one UTC calendar date, and each age-band column reports the BTC value of spent outputs assigned to that band. Let \(S_d\) denote the set of previous outputs consumed by non-coinbase transaction inputs in blocks assigned to UTC day \(d\), let \(v_i\) denote the BTC value of spent output \(i\), and let \(a_i\) denote its age in days. For each age band \(k=[\ell_k,u_k)\), the table reports
\[
\mathrm{SpentValueAgeBandBTC}_{d,k}
=
\sum_{i \in S_d}
v_i \mathbf{1}\{\ell_k \leq a_i < u_k\}.
\]
The last band is open-ended. In OBM, the fixed age-band columns correspond to each age band. The metric is raw and spend-output-based: it does not identify entities, exchanges, custodians, self-transfers, change outputs, or payment purposes. Its unit is BTC, not BTC-days.

The closest public comparator is Glassnode's \emph{Spent Volume by Age} chart.\footnote{\url{https://studio.glassnode.com/charts/breakdowns.SpentVolumeSumByAge?a=BTC}} Glassnode describes spent volume as the total volume of digital assets sold or spent, and the age-breakdown version as a categorization of that spent volume by the age of the coins being spent. This is conceptually very close to \texttt{obm\_spent\_value\_age\_band\_btc\_daily}, because both decompose spent value across coin-age cohorts. However, exact equivalence should not be assumed. Public documentation does not provide a full open-source reconstruction algorithm, and differences may arise from age-band boundaries, timestamp conventions, entity-adjustment status, smoothing, transfer-volume definitions, and chain-reorganization handling.

Glassnode also provides \emph{Spent Volume Age Bands} (SVAB).\footnote{\url{https://studio.glassnode.com/charts/indicators.Svab?a=BTC}} Glassnode defines SVAB as a separation of on-chain transfer volume based on coin age, where each band represents the percentage of spent volume previously moved within the time period shown in the legend. SVAB is therefore closely related, but it is usually expressed as a percentage distribution rather than as one BTC-denominated column per age band. It is useful for comparing the relative composition of spent value by age, whereas OBM provides the absolute BTC value in each band. Glassnode also provides an entity-adjusted SVAB variant.\footnote{\url{https://studio.glassnode.com/charts/indicators.SvabEntityAdjusted?a=BTC}} That version is less directly comparable to OBM because it discards transactions between addresses attributed to the same entity, while OBM remains raw and non-entity-adjusted.

Glassnode's \emph{Spent Output Age Bands} (SOAB) family is another related comparator.\footnote{\url{https://studio.glassnode.com/charts/indicators.Soab?a=BTC}} Glassnode's documentation describes SOAB as a metric that bundles spent coins into categories depending on their age and presents the bands as a proportion of total coins moved. The underlying classification by spent-output age is closely related to the OBM age-band table. However, SOAB is not a direct equivalent when reported as a percentage of spent outputs or a proportion of moved coins, rather than as an absolute BTC value per age band. It is therefore best understood as a related age-distribution indicator rather than as a substitute for \texttt{obm\_spent\_value\_age\_band\_btc\_daily}.

CryptoQuant provides a directly related \emph{Spent Output Age Bands} metric.\footnote{\url{https://cryptoquant.com/asset/btc/chart/network-indicator/spent-output-age-bands}} CryptoQuant describes the metric as a set of all spent outputs created within specified age bands, where each colored band represents the total value of spent outputs whose lifespan falls within the corresponding interval. This is one of the closest public formula-level analogs to OBM. Nevertheless, CryptoQuant publishes the information through provider-defined age bands and chart/API conventions, and the public documentation does not provide a full node-level reconstruction script. Exact comparison, therefore, requires checking whether the band boundaries, value units, timestamp conventions, fractional-age treatment, and reorganization policy match those used in OBM.

CryptoQuant also provides \emph{Exchange Inflow - Spent Output Age Bands}.\footnote{\url{https://cryptoquant.com/asset/btc/chart/flow-indicator/exchange-inflow-spent-output-age-bands}}. This metric is related but not equivalent. It restricts the age-band decomposition to outputs flowing into exchange wallets, whereas \texttt{obm\_spent\_value\_age\_band\_btc\_daily} includes all spent outputs regardless of destination. It is therefore useful for studying whether particular age cohorts are moving to exchanges, but it should not be used as a direct benchmark for the raw OBM age-band table.

Checkonchain provides \emph{Spent Volume Age Bands} and related revived-supply tools.\footnote{\url{https://charts.checkonchain.com/}} Its public chart catalog describes Spent Volume Age Bands as an on-chain spent-volume breakdown by coin age, available in absolute and relative forms. This makes it a useful public comparator for the economic interpretation of the OBM table. However, the public catalog does not provide a fully reproducible algorithm, nor does it specify the exact source data, band boundaries, entity filters, or timestamp conventions to be used. These should be verified before using it for numerical validation.

Glassnode also provides \emph{Spent Volume by Date Bands}.\footnote{\url{https://studio.glassnode.com/charts/indicators.SpentVolumeByDateBands?a=BTC}} This metric is similar to SVAB, but it groups spent volume by the absolute creation date of the UTXO rather than by floating age intervals. It is related to OBM because both characterize the spent value by the history of the consumed output. However, date-band grouping answers a different question from age-band grouping. OBM asks how old the spent outputs were on the spending date; date-band metrics ask when the spent outputs were originally created.

Coin Metrics does not appear to expose a named public table exactly equivalent to the OBM metric \texttt{obm\_spent\_value\_age\_band\_btc\_daily}. Its public transfer-value and transferred-days-destroyed metrics, such as \texttt{TxTfrValNtv} and \texttt{TxTfrValDayDst}, are related to value movement and coin-age destruction, but they do not publish the BTC value of spent outputs split across fixed age bands. A Coin Metrics-like age-band table would require thresholded or binned transfer-value data together with explicit alignment of transfer-value, timestamp, and output-age conventions.

Blockchair provides block-level, transaction-level, input-level, and output-level explorer data.\footnote{\url{https://blockchair.com/bitcoin/outputs}} This makes Blockchair useful as a reconstruction source. A researcher could resolve transaction inputs to previous outputs, compute the age of each consumed output, assign it to an age band, and aggregate the BTC value of spent outputs by UTC date and band. However, Blockchair does not appear to publish a named daily wide table equivalent to \texttt{obm\_spent\_value\_age\_band\_btc\_daily}. It should therefore be treated as a lower-level data source rather than as a direct public comparator.

HODL Waves and UTXO age-distribution charts are related but measure different objects. For example, Glassnode's HODL Waves and CryptoQuant's UTXO Age Bands describe the age distribution of currently unspent supply, not the age distribution of spent value.\footnote{\url{https://docs.glassnode.com/guides-and-tutorials/metric-guides/age-distribution/hodl-waves}} \footnote{\url{https://cryptoquant.com/asset/btc/chart/network-indicator/utxo-age-bands}} These stock metrics are useful for contextualizing the available supply in each age cohort, but they are not direct equivalents of \texttt{obm\_spent\_value\_age\_band\_btc\_daily}, which is a daily flow table.

Overall, the closest public comparators for the OBM metric \texttt{obm\_spent\_value\_age\_band\_btc\_daily} are Glassnode's \emph{Spent Volume by Age}, Glassnode's SVAB/SOAB families, CryptoQuant's \emph{Spent Output Age Bands}, and Checkonchain's \emph{Spent Volume Age Bands}. These sources show that age-band decompositions of spent coins are widely used in on-chain analysis. The distinctive contribution of OBM is that it provides a reproducible, full-node-derived, BTC-denominated wide daily table with explicit age-band columns, raw spent-output accounting, documented UTC timestamp convention, optional row-sum validation against total spent value, and direct compatibility with OBM threshold metrics such as \texttt{obm\_spent\_value\_lt155d\_btc\_daily}, \texttt{obm\_spent\_value\_ge155d\_btc\_daily}, and \texttt{obm\_spent\_value\_ge365d\_btc\_daily}.

\paragraph{Data source and input requirements.}
The metric is exported from the persistent OBM spent-output indexer database, rather than computed by an independent metric-specific blockchain scan. The indexer, described in Sect.~\ref{sec:The Spent-Output Indexer}, is built from a running Bitcoin Core full node and stores reusable daily age-band aggregates in a SQLite database. For \texttt{obm\_spent\_value\_age\_band\_btc\_daily}, the relevant source table is \texttt{daily\_age\_band\_aggregates}, and the relevant fields are \texttt{date}, \texttt{age\_band}, and \texttt{spent\_value\_sats}.

The indexer computes this table during its sequential scan of the blockchain. For each non-coinbase transaction input, it resolves the input to the previous output it spends, retrieves the previous output value and creation timestamp from the live outpoint state, computes the elapsed age in days, assigns the spent output to one of the fixed age bands, and accumulates the spent value in the corresponding daily age-band aggregate. Internally, the indexer stores:
\[
\mathrm{spent\_value\_sats}_{d,k} =
\sum_{b \in B_d}
\sum_{i \in I_b}
v^{\mathrm{sats}}_i \mathbf{1}\{a_i \in k\}.
\]

The exporter converts each stored satoshi-denominated aggregate into BTC by dividing by \(100{,}000{,}000\):
\[
\mathrm{SpentValueBand}_{d,k} =
\frac{\mathrm{spent\_value\_sats}_{d,k}}{100{,}000{,}000}.
\]

The exporter script does not query Bitcoin Core directly. It requires an existing SQLite database generated by \texttt{obm\_spent\_output\_indexer.py}, with metadata identifying it as an OBM spent-output indexer database and with processed-date metadata covering the requested output interval. The script checks that the requested ending date is not beyond the maximum date processed by the indexer. If a requested date-band pair has no row in \texttt{daily\_age\_band\_aggregates}, the exporter writes zero for that age-band column. This convention is appropriate because each age-band component is a flow variable: if no qualifying spent-output value is assigned to a given age band on a UTC date, the daily value for that band is zero.

This metric does not require address extraction, user clustering, entity identification, external price data, third-party APIs, or a direct Bitcoin Core connection at export time. However, it depends on the prior successful execution of the spent-output indexer, which itself requires a synchronized, non-pruned Bitcoin Core full node and access to transaction-level data sufficient to reconstruct previous-output values and creation timestamps. The exported table therefore inherits the indexer's UTC block-timestamp convention, previous-output reconstruction procedure, age-band assignment convention, fractional-day age convention, treatment of negative apparent ages, chain-consistency checks, release metadata, and handling of historical duplicate outpoint edge cases.

\paragraph{Algorithm.}
The script \texttt{export\_obm\_spent\_value\_age\_band\_btc\_daily.py} implements the following procedure:

\begin{enumerate}
    \item Parse the user-provided date interval, \texttt{\symbol{45}\symbol{45}start\_date} and \texttt{\symbol{45}\symbol{45}end\_date}, using the format \texttt{YYYY-MM-DD}. Both dates are interpreted as UTC dates and both are included in the output.

    \item Open the persistent SQLite database generated by \texttt{obm\_spent\_output\_indexer.py}. The path to this database is provided through the \texttt{\symbol{45}\symbol{45}state\_db} argument.

    \item Validate that the SQLite database corresponds to the OBM spent-output indexer. The script checks the indexer metadata, including the expected \texttt{indexer\_id} and the presence of processing metadata such as \texttt{last\_processed\_height}, \texttt{last\_processed\_date}, and \texttt{max\_processed\_date}.

    \item Verify that the database contains the table \texttt{daily\_age\_band\_aggregates} and that this table contains the required fields \texttt{date}, \texttt{age\_band}, and \texttt{spent\_value\_sats}.

    \item Verify that the requested \texttt{\symbol{45}\symbol{45}end\_date} is not later than the maximum UTC date processed by the indexer. If the requested interval extends beyond the indexed range, the script aborts and requires the indexer to be run further before export.

    \item Infer the dataset release version from the indexer metadata. If the database does not contain a release-version field, the script uses the fallback value provided through the \texttt{\symbol{45}\symbol{45}release\_version} argument.

    \item Initialize one complete output row for each UTC date in the requested interval, setting all age-band columns to zero.

    \item Query \texttt{daily\_age\_band\_aggregates} for all date-band observations in the requested interval.

    \item For each retrieved row, verify that \texttt{age\_band} belongs to the expected set of age-band labels. Unexpected labels cause the script to abort.

    \item For each retrieved row, verify that the corresponding \texttt{spent\_value\_sats} value is non-negative and that there is no duplicate date-band aggregate.

    \item Convert each value from satoshis to BTC:
    \[
    \mathrm{SpentValueBand}_{d,k} =
    \frac{\mathrm{spent\_value\_sats}_{d,k}}{100{,}000{,}000}.
    \]

    \item Write the resulting wide table to a CSV file. Each row corresponds to one UTC date, and each age-band column reports BTC-denominated spent value for that band.

    \item Optionally validate the row sums when the \texttt{\symbol{45}\symbol{45}validate\_row\_sums} flag is activated. In that case, the script checks that the sum across all age-band columns equals \texttt{daily\_aggregates.spent\_value\_sats}, converted to BTC, within the configured tolerance.

    \item Optionally generate a stacked area plot when the \texttt{\symbol{45}\symbol{45}plot} flag is activated. The stacked areas represent the contribution of each age band to total daily spent output value.
\end{enumerate}

This exporter does not query Bitcoin Core, does not maintain an outpoint state, and does not reconstruct spent outputs at export time. The computationally expensive previous-output resolution, age calculation, and age-band assignment have already been performed by the spent-output indexer. The export script is therefore a lightweight, deterministic transformation from the indexed database to a wide OBM CSV table.

\paragraph{Metric-specific input parameters.}
The input parameters specific to this exporter are:

\begin{itemize}
    \item \texttt{\symbol{45}\symbol{45}state\_db}: path to the persistent SQLite database generated by \texttt{obm\_spent\_output\_indexer.py}. This database must contain the \texttt{daily\_age\_band\_aggregates} table and the metadata required to verify that it corresponds to the OBM spent-output indexer.

    \item \texttt{\symbol{45}\symbol{45}start\_date}: starting date of the selected interval, inclusive, in \texttt{YYYY-MM-DD} format. The date is interpreted as a UTC calendar date.

    \item \texttt{\symbol{45}\symbol{45}end\_date}: ending date of the selected interval, inclusive, in \texttt{YYYY-MM-DD} format. The script verifies that this date is not later than the maximum date processed by the indexer.

    \item \texttt{\symbol{45}\symbol{45}release\_version}: fallback dataset release version used only if the indexer database does not contain a \texttt{release\_version} metadata field.

    \item \texttt{\symbol{45}\symbol{45}output}: path of the output CSV file to be written.

    \item \texttt{\symbol{45}\symbol{45}validate\_row\_sums}: optional flag that instructs the script to verify that the sum across all age-band columns equals total spent value from \texttt{daily\_aggregates.spent\_value\_sats}, when that aggregate is available.

    \item \texttt{\symbol{45}\symbol{45}row\_sum\_tolerance}: tolerance, in BTC, used by \texttt{\symbol{45}\symbol{45}validate\_row\_sums}. The default is \(0.00000001\) BTC.

    \item \texttt{\symbol{45}\symbol{45}plot}: optional flag that instructs the script to generate a stacked area plot of all age-band time series.

    \item \texttt{\symbol{45}\symbol{45}plot\_output}: optional path for the generated plot. If this argument is omitted while \texttt{\symbol{45}\symbol{45}plot} is used, the plot is saved next to the CSV file using the same base name and a \texttt{.png} extension.
\end{itemize}

The exporter does not accept Bitcoin Core RPC parameters, \texttt{\symbol{45}\symbol{45}height\_margin}, \texttt{\symbol{45}\symbol{45}min\_confirmations}, \texttt{\symbol{45}\symbol{45}commit\_every}, or \texttt{\symbol{45}\symbol{45}reset\_state\_db}. These parameters belong to the spent-output indexer, not to the age-band exporter. The exporter assumes that the indexer database has already been built and updated through the requested ending date. Its role is only to read \texttt{daily\_age\_band\_aggregates}, pivot it into a wide daily table, convert satoshi values into BTC, and write the standardized output.

\paragraph{Aggregation rule.}
For each UTC date \(d\), the value in each age-band column is computed as the sum of the BTC values of all previous outputs spent on that date whose age belongs to the corresponding band:
\[
\mathrm{SpentValueBand}_{d,k} =
\sum_{b \in B_d}
\sum_{i \in I_b}
v_i \mathbf{1}\{a_i \in k\}.
\]

The row sum across all age bands recovers total daily spent output value:
\[
\mathrm{SpentValueBTC}_d =
\sum_k
\mathrm{SpentValueBand}_{d,k}.
\]

Monthly versions of this table, if distributed, should be computed by summing each age-band column over the corresponding daily observations:
\[
\mathrm{SpentValueBand}_{m,k} =
\sum_{d \in m}
\mathrm{SpentValueBand}_{d,k}.
\]

This convention preserves the interpretation of each column as a BTC-denominated spent-value flow associated with a specific age band.

\paragraph{Output format.}
The output file contains one observation per UTC date and one column per age band. It therefore departs from the scalar OBM schema with a single \texttt{value} column, but retains the usual metadata fields.

\begin{center}
\begin{tabular}{lp{0.25\textwidth}p{0.25\textwidth}}
\toprule
\textbf{Column} & \textbf{Example} & \textbf{Description} \\
\midrule
\texttt{date} & \texttt{2024-01-01} & UTC calendar date \\
\texttt{series\_id} & \texttt{obm\_spent\_value\_age \_band\_btc\_daily} & Stable OBM table identifier \\
\texttt{spent\_value\_0d\_1d\_btc} & \texttt{12500.12345678} & BTC value of spent outputs aged 0d to 1d \\
\texttt{spent\_value\_1d\_1w\_btc} & \texttt{27100.00000000} & BTC value of spent outputs aged 1d to 1w \\
\texttt{spent\_value\_1w\_1m\_btc} & \texttt{50400.25000000} & BTC value of spent outputs aged 1w to 1m \\
\texttt{spent\_value\_1m\_3m\_btc} & \texttt{48200.00000000} & BTC value of spent outputs aged 1m to 3m \\
\texttt{spent\_value\_3m\_6m\_btc} & \texttt{39100.75000000} & BTC value of spent outputs aged 3m to 6m \\
\texttt{spent\_value\_6m\_1y\_btc} & \texttt{24200.00000000} & BTC value of spent outputs aged 6m to 1y \\
\texttt{spent\_value\_1y\_2y\_btc} & \texttt{22000.00000000} & BTC value of spent outputs aged 1y to 2y \\
\texttt{spent\_value\_2y\_3y\_btc} & \texttt{18000.00000000} & BTC value of spent outputs aged 2y to 3y \\
\texttt{spent\_value\_3y\_5y\_btc} & \texttt{14000.00000000} & BTC value of spent outputs aged 3y to 5y \\
\texttt{spent\_value\_5y\_7y\_btc} & \texttt{9000.00000000} & BTC value of spent outputs aged 5y to 7y \\
\texttt{spent\_value\_7y\_10y\_btc} & \texttt{4000.00000000} & BTC value of spent outputs aged 7y to 10y \\
\texttt{spent\_value\_10y\_plus\_btc} & \texttt{2000.00000000} & BTC value of spent outputs aged 10y or more \\
\texttt{unit} & \texttt{BTC} & Shared unit for all age-band columns \\
\texttt{frequency} & \texttt{daily} & Observation frequency \\
\texttt{release\_version} & \texttt{OBM v0.1.0} & Dataset release version \\
\bottomrule
\end{tabular}
\end{center}

All age-band columns are measured in BTC and are written with eight decimal places.

\paragraph{Technical validation.}
Several internal checks are used to validate this metric at export time. First, the script verifies that the requested date range is valid and that \texttt{\symbol{45}\symbol{45}start\_date} is not later than \texttt{\symbol{45}\symbol{45}end\_date}. Second, it checks that the SQLite state database exists, opens it, and sets the connection to query-only mode. Third, it verifies that the database metadata identify it as an OBM spent-output indexer database, by checking the expected \texttt{indexer\_id}. Fourth, it checks that the indexer database contains processing metadata, including \texttt{last\_processed\_height}, \texttt{last\_processed\_date}, and \texttt{max\_processed\_date}. Fifth, the script verifies that the table \texttt{daily\_age\_band\_aggregates} is present in the database and that it contains the fields \texttt{date}, \texttt{age\_band}, and \texttt{spent\_value\_sats}. Sixth, it verifies that the requested ending date is not later than the maximum date processed by the indexer. If the requested interval extends beyond the indexed range, the script aborts and requires the indexer to be updated before the export is attempted.

For each requested date, the exporter initializes all expected age-band columns to zero. For each row retrieved from \texttt{daily\_age\_band\_aggregates}, the script checks that the age-band label belongs to the expected set of labels, that the corresponding value is non-negative, and that the date-band pair is not duplicated. Each retrieved value is converted from satoshis to BTC by dividing by \(100{,}000{,}000\). The resulting file contains one complete wide row per calendar date in the requested interval.

When the optional \texttt{\symbol{45}\symbol{45}validate\_row\_sums} flag is used, the script also compares the sum across all age-band columns with total spent value from \texttt{daily\_aggregates.spent\_value\_sats}, converted to BTC. The validation identity is:
\[
\sum_k
\mathrm{SpentValueBand}_{d,k}
=
\mathrm{SpentValueBTC}_d.
\]
The check is performed within the configured \texttt{\symbol{45}\symbol{45}row\_sum\_tolerance}. This is a strong internal consistency check when both \texttt{daily\_age\_band\_aggregates} and \texttt{daily\_aggregates} are generated by the same indexer run.

Additional consistency checks can be performed using related OBM series. In particular, the row sum across all age-band columns should equal \texttt{obm\_spent\_value\_btc\_daily} for every date when both outputs are generated from the same indexer database and release version. Age-threshold metrics such as \texttt{obm\_cdd\_365d\_btc\_daily} should also be interpretable in relation to the sum of age-band columns beginning at the corresponding threshold, provided that the threshold boundary and the age-band definitions align exactly. Comparisons with external spent-output age-distribution metrics are useful as diagnostics but should not be interpreted as strict equality tests, because providers may differ in timestamp conventions, age-band boundaries, entity adjustment, change-output treatment, transfer-value definition, and historical edge-case handling.

\paragraph{Known limitations.}
The daily spent-output-value-by-age-band table is useful but definition-sensitive. First, \texttt{obm\_spent\_value\_age\_band\_btc\_daily} is a raw spent-output table, not an entity-adjusted age-distribution metric. It does not identify users, entities, custodians, exchanges, self-transfers, or change outputs. Second, the table depends on the block timestamp convention used both to compute output age and to assign spending blocks to calendar days. Third, the age-band boundaries are fixed, so small differences around a boundary can move an output from one column to another. Fourth, the table is wide and vector-valued rather than a scalar \texttt{value} series, so software expecting the standard scalar OBM schema must treat it separately. Fifth, the metric can be affected by self-transfers, wallet consolidation, batching, exchange operations, and custodial wallet management. Sixth, it reports BTC value, not fiat-denominated value. Seventh, it requires the persistent spent-output indexer database and therefore depends on the correctness and completeness of the indexer run. Eighth, the simple reorganization policy of the indexer detects inconsistencies and aborts, but it does not automatically roll back the state database.

Despite these limitations, \texttt{obm\_spent\_value\_age\_band\_btc\_daily} is a useful OBM age-distribution table. It provides a transparent decomposition of daily spent output value by output age and complements scalar spent-value, CDD, dormancy, threshold-based spent-value, and UTXO-flow metrics.
\subsection{\texttt{obm\_spent\_value\_btc\_daily}: Daily Spent Output Value in BTC}
\label{Daily Spent Output Value in BTC}

\paragraph{Definition.}
The daily spent-output-value series measures the total BTC value of previous outputs consumed by non-coinbase transaction inputs in blocks assigned to a given UTC calendar day. Let \(B_d\) denote the set of blocks assigned to day \(d\). For each non-coinbase transaction input \(i\) included in a transaction in block \(b \in B_d\), let \(v_i\) denote the value, in BTC, of the previous output spent by that input. The daily spent-output value is defined as:
\[
\mathrm{SpentValueBTC}_d =
\sum_{b \in B_d}
\sum_{i \in I_b}
v_i,
\]
where \(I_b\) denotes the set of non-coinbase transaction inputs included in block \(b\).

A block \(b\) is assigned to day \(d\) according to the UTC calendar date derived from its block timestamp \(t_b\): $d(b)=\mathrm{UTCDate}(t_b)$. The resulting series therefore reports the total BTC value of outputs spent in the Bitcoin main chain per UTC day. It is a raw spent-output measure: each input contributes the value of the previous output it consumes, regardless of whether the transaction represents a payment, a self-transfer, wallet consolidation, batching, change management, or custodial reorganization.

\paragraph{Economic interpretation.}
Daily spent output value is a BTC-denominated flow variable that measures the gross value of UTXOs consumed on-chain. It captures how much previously unspent BTC value is being turned over through transaction inputs on a given day.

For economic research, this metric is useful in several ways. First, it provides a raw measure of BTC-denominated spent-output activity. Second, it is the denominator for dormancy, where Bitcoin Days Destroyed is divided by spent value to obtain the average value-weighted age of spent coins. Third, it helps distinguish between transaction-count activity and value-weighted spent-output activity. Fourth, it is useful for studying UTXO turnover, consolidation episodes, wallet-management behavior, and periods of unusually high movement of previously unspent outputs. Fifth, it provides a foundation for age-band spent-value metrics and long-term-holder spent-value indicators.

The metric should not be interpreted as entity-adjusted payment volume, economically meaningful settlement value, or exchange-adjusted transfer volume. Bitcoin transactions often include change outputs, batching, self-transfers, exchange operations, custodial flows, and wallet consolidations. Consequently, raw spent-output value can be much larger than the value that would be obtained after entity adjustment or change-output filtering.

\paragraph{Similar metrics publicly available.}
The OBM series \texttt{obm\_spent\_value\_btc\_daily} is comparable to public metrics usually labeled \emph{spent value}, \emph{spent volume}, \emph{transfer volume}, \emph{output value}, or \emph{sent coins}. In OBM, the metric is defined as the daily sum of the BTC value of transaction outputs that are spent by transaction inputs in blocks assigned to a given UTC calendar day. Let \(S_d\) denote the set of previous outputs consumed by non-coinbase transaction inputs in blocks assigned to day \(d\), and let \(v_i\) denote the BTC value of spent output \(i\). The daily series is
\[
\mathrm{SpentValueBTC}_d =
\sum_{i \in S_d} v_i .
\]
Coinbase inputs are excluded because Coinbase transactions do not spend previous outputs. This definition is deliberately mechanical and UTXO-based: it measures the raw value of consumed outputs, without attempting to identify entities, remove self-transfers, detect change outputs, or infer economic payment volume. It should therefore be distinguished from adjusted transfer volume metrics and from output-volume metrics based on newly created outputs.

The closest public conceptual analog is Glassnode's family of \emph{Spent Volume} metrics. For example, Glassnode's \emph{Spent Volume by Age} chart is exposed as \texttt{breakdowns.SpentVolumeSumByAge}.\footnote{\url{https://studio.glassnode.com/charts/breakdowns.SpentVolumeSumByAge?a=BTC}} Glassnode describes spent volume as a metric that calculates the total volume of digital assets sold or spent, with the age-band version categorizing that volume by the age of the spent coins. This is conceptually close to \texttt{obm\_spent\_value\_btc\_daily}, because both are based on coins that are spent during a given interval. However, Glassnode's public page focuses on a breakdown metric rather than a simple raw daily total, and it does not provide a full reproducible algorithm specifying UTXO-level reconstruction, timestamp assignment, change treatment, reorganization policy, or whether the displayed value is raw, entity-adjusted, or address-based under each chart configuration. Glassnode also provides related transfer-volume metrics, such as \texttt{transactions.TransfersVolumeSum}, but those measure successful on-chain transfers rather than the raw value of consumed UTXOs. 

CryptoQuant provides a closely related \emph{Spent Output Value Bands} metric.\footnote{\url{https://cryptoquant.com/asset/btc/chart/network-indicator/spent-output-value-bands}} Its methodological documentation states that Spent Output Value Bands show the distribution of all spent outputs according to their value, and defines each band as the sum of the value of spent outputs falling within a given value interval:
\[
\sum_{o \in \mathrm{spent\ outputs}} \mathrm{value}_o \cdot
\mathbf{1}\{a_i \leq \mathrm{value}_o < b_i\}.
\]
This is one of the clearest public formula-level descriptions related to \texttt{obm\_spent\_value\_btc\_daily}. If all value bands were summed, the result would be conceptually close to OBM's raw spent-output value. Nevertheless, CryptoQuant publishes the metric primarily as a distribution by value bands, not as a single raw daily aggregate, and the public documentation does not provide an open-source reconstruction script or all low-level conventions needed for exact replication. 

Coin Metrics provides transfer-value metrics that are related but not identical. Its \emph{Xfer'd Val (native units)}, with MetricID \texttt{TxTfrValNtv}, is defined as the sum of native units transferred during an interval, and its adjusted version, \texttt{TxTfrValAdjNtv}, removes noise and artifacts using UTXO heuristics such as excluding early spends, discounting self-churn, and removing cold-wallet shuffles or other artifacts. The corresponding API endpoint for the unadjusted native transfer-value metric is available.\footnote{\url{https://api.coinmetrics.io/v4/timeseries/asset-metrics?assets=btc\&metrics=TxTfrValNtv}} These metrics are useful comparators for the general magnitude of value movement on-chain, but they are not exact equivalents of \texttt{obm\_spent\_value\_btc\_daily}. OBM sums the value of previous outputs consumed by transaction inputs, whereas Coin Metrics' transfer-value family is framed around transfers of native units, and its adjusted version deliberately applies heuristics to remove non-economic activity. Therefore, Coin Metrics is useful as a benchmark for transfer-volume behavior, but not as a raw spent-output-value replication unless its transfer construction is explicitly aligned with the OBM spent-output definition. 

Blockchain.com provides a related but different \emph{Output Value Per Day} chart.\footnote{\url{https://www.blockchain.com/charts/output-volume}} The page defines the series as the total value of all transaction outputs per day and explicitly states that it includes coins returned to the sender as change. This metric is not the same as \texttt{obm\_spent\_value\_btc\_daily}, because Blockchain.com's chart sums newly created transaction outputs, whereas OBM sums previous outputs consumed as inputs. In a UTXO system, these two quantities are mechanically related but not identical on a daily basis because transaction fees, coinbase outputs, boundary timing, and the distinction between inputs and outputs matter. Blockchain.com also provides an \emph{Estimated Transaction Value (BTC)} chart that excludes coins returned as change, but it is an adjusted economic-transfer estimate rather than a raw spent-output-value series. 

YCharts republishes Blockchain.com's \emph{Bitcoin Total Output Value Per Day} series.\footnote{\url{https://ycharts.com/indicators/bitcoin\_total\_output\_value\_per\_day}} This is useful as an accessible financial-data reference for Blockchain.com's output-volume series, but it inherits the same limitation: it reports total created output value per day, not the raw BTC value of spent previous outputs. It also does not provide an independent blockchain-level reconstruction algorithm. 

Newhedge provides a \emph{Bitcoin Transaction Volume} chart.\footnote{\url{https://newhedge.io/bitcoin/transaction-volume}} The page defines the metric as daily transaction volume transferred on the Bitcoin network in BTC, sats, and USD. It also provides a related \emph{Output Volume per Transaction} chart.\footnote{\url{https://newhedge.io/bitcoin/output-volume-per-transaction}} These series are relevant because they describe BTC value movement on-chain, but the public pages do not disclose a full algorithm and do not make clear whether the underlying quantity is raw input value, raw output value, transfer value, or adjusted transfer value. Access to full data also appears to require registration. Newhedge should therefore be treated as a useful chart-level comparator rather than as a reproducible methodological benchmark. 

BitInfoCharts reports current and historical indicators related to sent coins. Its Bitcoin statistics page reports \emph{Bitcoins sent last 24h} and \emph{Bitcoins sent avg. per hour}, and its comparison pages include \emph{Sent coins in USD per day}.\footnote{\url{https://bitinfocharts.com/bitcoin/}}. These indicators are useful as broad diagnostic comparators for on-chain value movement, but they are not clearly documented as raw spent-output-value series. The public pages do not provide a reproducible algorithm, nor do they specify whether the series is based on transaction inputs, transaction outputs, transfer amounts, change-adjusted estimates, or another convention. 

Blockchair provides block, transaction, and output-level explorer data, including a Bitcoin outputs explorer.\footnote{\url{https://blockchair.com/bitcoin/outputs}} It also provides a Bitcoin charts catalog. .\footnote{\url{https://blockchair.com/bitcoin/charts}} Blockchair is therefore useful as a reconstruction source: a researcher could, in principle, reconstruct a series similar to \texttt{obm\_spent\_value\_btc\_daily} by resolving transaction inputs to previous outputs and summing the BTC value of those previous outputs by UTC day. However, Blockchair does not appear to provide a clearly named public daily \emph{spent output value in BTC} chart equivalent to OBM. Its public charts and explorer data should therefore be treated as reconstruction inputs or secondary validation sources rather than as a named methodological equivalent. 

Overall, no reviewed public provider appears to expose a named daily series that exactly matches \texttt{obm\_spent\_value\_btc\_daily} as a raw full-node-derived sum of consumed previous-output values. The closest public analogs are Glassnode's spent-volume family and CryptoQuant's Spent Output Value Bands. Coin Metrics' \texttt{TxTfrValNtv} and \texttt{TxTfrValAdjNtv}, Blockchain.com's output-volume and estimated-transaction-value charts, Newhedge's transaction-volume charts, BitInfoCharts' sent-coins indicators, YCharts' output-volume series, and Blockchair's output-level data are useful secondary references, but they correspond to related concepts rather than to the exact OBM definition. The distinctive contribution of OBM is that it defines the series explicitly at the spent-output level, keeps it unadjusted, reports it in BTC, assigns spending blocks to UTC calendar days using a documented timestamp convention, and makes the reconstruction auditable from full-node data.

\paragraph{Data source and input requirements.}
The metric is exported from the persistent OBM spent-output indexer database, rather than computed by an independent metric-specific blockchain scan. The indexer, described in Sect.~\ref{sec:The Spent-Output Indexer}, is built from a running Bitcoin Core full node and stores reusable daily aggregates in a SQLite database. For \texttt{obm\_spent\_value\_btc\_daily}, the relevant source table is \texttt{daily\_aggregates}, and the relevant field is \texttt{spent\_value\_sats}.

The indexer computes spent output value during its sequential scan of the blockchain. For each non-coinbase transaction input, it resolves the input to the previous output it spends, retrieves the previous output value from the live outpoint state, and adds that value to the daily spent-value aggregate. Internally, this quantity is stored in satoshis:
\[
\mathrm{spent\_value\_sats}_d =
\sum_{b \in B_d}
\sum_{i \in I_b}
v^{\mathrm{sats}}_i.
\]

The exporter converts the stored satoshi-denominated aggregate into BTC by dividing by \(100{,}000{,}000\):
\[
\mathrm{SpentValueBTC}_d =
\frac{\mathrm{spent\_value\_sats}_d}{100{,}000{,}000}.
\]

The exporter script does not query Bitcoin Core directly. It requires an existing SQLite database generated by \texttt{obm\_spent\_output\_indexer.py}, with metadata identifying it as an OBM spent-output indexer database and with processed-date metadata covering the requested output interval. The script checks that the requested ending date is not beyond the maximum date processed by the indexer. If a requested date has no row in \texttt{daily\_aggregates}, the exporter writes a zero value for that date. This convention is appropriate for spent value because an absent aggregate row indicates that no spent-output value was assigned to that UTC date under the indexer's timestamp convention.

This metric does not require address extraction, user clustering, entity identification, external price data, third-party APIs, or a direct Bitcoin Core connection at export time. However, it depends on the prior successful execution of the spent-output indexer, which itself requires a synchronized, non-pruned Bitcoin Core full node and access to transaction-level data sufficient to reconstruct previous-output values. The exported series therefore inherits the indexer's UTC block-timestamp convention, previous-output reconstruction procedure, chain-consistency checks, release metadata, and handling of historical duplicate outpoint edge cases.

\paragraph{Algorithm.}
The script \texttt{export\_obm\_spent\_value\_btc\_daily.py} implements the following procedure:

\begin{enumerate}
    \item Parse the user-provided date interval, \texttt{\symbol{45}\symbol{45}start\_date} and \texttt{\symbol{45}\symbol{45}end\_date}, using the format \texttt{YYYY-MM-DD}. Both dates are interpreted as UTC dates and both are included in the output.

    \item Open the persistent SQLite database generated by \texttt{obm\_spent\_output\_indexer.py}. The path to this database is provided through the \texttt{\symbol{45}\symbol{45}state\_db} argument.

    \item Validate that the SQLite database corresponds to the OBM spent-output indexer. The script checks the indexer metadata, including the expected \texttt{indexer\_id} and the presence of processing metadata such as \texttt{last\_processed\_height}, \texttt{last\_processed\_date}, and \texttt{max\_processed\_date}.

    \item Verify that the requested \texttt{\symbol{45}\symbol{45}end\_date} is not later than the maximum UTC date processed by the indexer. If the requested interval extends beyond the indexed range, the script aborts and requires the indexer to be run further before export.

    \item Infer the dataset release version from the indexer metadata. If the database does not contain a release-version field, the script uses the fallback value provided through the \texttt{\symbol{45}\symbol{45}release\_version} argument.

    \item For each UTC date \(d\) in the requested interval, query the \texttt{daily\_aggregates} table for the field \texttt{spent\_value\_sats}. If a row exists for date \(d\), the script reads the stored satoshi-denominated spent-value aggregate. If no row exists, the script assigns zero spent value to that date.

    \item Convert each daily value from satoshis to BTC:
    \[
    \mathrm{SpentValueBTC}_d =
    \frac{\mathrm{spent\_value\_sats}_d}{100{,}000{,}000}.
    \]

    \item Write the resulting time series to a CSV file using the standardized OBM schema:
    \[
    \texttt{date},\quad
    \texttt{series\_id},\quad
    \texttt{value},\quad
    \texttt{unit},\quad
    \texttt{frequency},\quad
    \texttt{release\_version}.
    \]

    \item Optionally generate a plot of the exported series when the plotting flag is activated. If \texttt{\symbol{45}\symbol{45}plot\_output} is not provided, the plot is saved next to the CSV file with a \texttt{.png} extension. The plot title includes the series description and the selected date interval.
\end{enumerate}

This exporter does not query Bitcoin Core, does not maintain an outpoint state, and does not reconstruct spent outputs at export time. The computationally expensive previous-output resolution has already been performed by the spent-output indexer, which stores the daily spent-value aggregate in \texttt{daily\_aggregates}. The export script is therefore a lightweight, deterministic transformation from the indexed database to the OBM CSV format.

\paragraph{Metric-specific input parameters.}
The input parameters specific to this exporter are:

\begin{itemize}
    \item \texttt{\symbol{45}\symbol{45}state\_db}: path to the persistent SQLite database generated by the OBM Python script \texttt{obm\_spent\_output\_indexer.py}. This database must contain the \texttt{daily\_aggregates} table and the metadata required to verify that it corresponds to the OBM spent-output indexer.

    \item \texttt{\symbol{45}\symbol{45}start\_date}: starting date of the selected interval, inclusive, in \texttt{YYYY-MM-DD} format. The date is interpreted as a UTC calendar date.

    \item \texttt{\symbol{45}\symbol{45}end\_date}: ending date of the selected interval, inclusive, in \texttt{YYYY-MM-DD} format. The script verifies that this date is not later than the maximum date processed by the indexer.

    \item \texttt{\symbol{45}\symbol{45}release\_version}: fallback dataset release version used only if the indexer database does not contain a \texttt{release\_version} metadata field.

    \item \texttt{\symbol{45}\symbol{45}output}: path of the output CSV file to be written.

    \item \texttt{\symbol{45}\symbol{45}plot}: optional flag that instructs the script to generate a plot of the exported series.

    \item \texttt{\symbol{45}\symbol{45}plot\_output}: optional path for the generated plot. If this argument is omitted while \texttt{\symbol{45}\symbol{45}plot} is used, the plot is saved next to the CSV file using the same base name and a \texttt{.png} extension.
\end{itemize}

The exporter does not accept Bitcoin Core RPC parameters, \texttt{\symbol{45}\symbol{45}height\_margin}, \texttt{\symbol{45}\symbol{45}commit\_every}, \texttt{\symbol{45}\symbol{45}min\_confirmations}, or \texttt{\symbol{45}\symbol{45}reset\_state\_db}. These parameters belong to the spent-output indexer, not to the spent-value exporter. The exporter assumes that the indexer database has already been built and updated through the requested ending date. Its role is only to read \texttt{spent\_value\_sats} from \texttt{daily\_aggregates}, convert the stored satoshi values into BTC, and write the standardized OBM output file.

\paragraph{Aggregation rule.}
The daily value is computed as the sum of the values of all previous outputs spent by non-coinbase transaction inputs included in blocks assigned to the same UTC calendar date:
\[
\mathrm{SpentValueBTC}_d =
\sum_{b \in B_d}
\sum_{i \in I_b}
v_i.
\]

The aggregation rule is therefore a daily sum of BTC-denominated spent-output value. Monthly versions of this metric, if distributed, should also be computed as sums of the corresponding daily values:
\[
\mathrm{SpentValueBTC}_m =
\sum_{d \in m}
\mathrm{SpentValueBTC}_d.
\]

This convention preserves the interpretation of the metric as the total BTC value of outputs consumed over the corresponding period.

\paragraph{Output format.}
The output file contains one observation per UTC date. Each row has the following fields:

\begin{center}
\begin{tabular}{llp{0.45\textwidth}}
\toprule
\textbf{Column} & \textbf{Example} & \textbf{Description} \\
\midrule
\texttt{date} & \texttt{2024-01-01} & UTC calendar date \\
\texttt{series\_id} & \texttt{obm\_spent\_value\_btc\_daily} & Stable OBM series identifier \\
\texttt{value} & \texttt{428318.49281735} & Total BTC value of spent previous outputs \\
\texttt{unit} & \texttt{BTC} & Measurement unit \\
\texttt{frequency} & \texttt{daily} & Observation frequency \\
\texttt{release\_version} & \texttt{OBM v0.1.0} & Dataset release version \\
\bottomrule
\end{tabular}
\end{center}

\paragraph{Technical validation.}
Several internal checks are used to validate this metric at export time. First, the script verifies that the requested date range is valid and that \texttt{\symbol{45}\symbol{45}start\_date} is not later than \texttt{\symbol{45}\symbol{45}end\_date}. Second, it checks that the SQLite state database exists, opens it, and sets the connection to query-only mode. Third, it verifies that the database metadata identify it as an OBM spent-output indexer database, by checking the expected \texttt{indexer\_id}. Fourth, it checks that the indexer database contains processing metadata, including \texttt{last\_processed\_height}, \texttt{last\_processed\_date}, and \texttt{max\_processed\_date}. Fifth, it verifies that the \texttt{daily\_aggregates} table is present in the database. Sixth, it verifies that the requested ending date is not later than the maximum date processed by the indexer. If the requested interval extends beyond the indexed range, the script aborts and requires the indexer to be updated before the export is attempted.

For each date in the requested interval, the exporter reads \texttt{spent\_value\_sats} from the \texttt{daily\_aggregates} table. If the date is absent from the aggregate table, the script exports zero spent value for that date. This convention is appropriate for \texttt{obm\_spent\_value\_btc\_daily} because spent value is a flow variable: if no spent-output value is assigned to a UTC date under the indexer's timestamp convention, the daily value is zero rather than missing. Each retrieved value is converted from satoshis to BTC by dividing by \(100{,}000{,}000\), and the resulting file is written using the standard OBM schema:
\[
\texttt{date},\quad
\texttt{series\_id},\quad
\texttt{value},\quad
\texttt{unit},\quad
\texttt{frequency},\quad
\texttt{release\_version}.
\]

The script also propagates the release version stored in the indexer metadata. If the metadata field is absent, it uses the fallback value supplied through \texttt{\symbol{45}\symbol{45}release\_version}. The number of rows written must equal the number of calendar days in the requested interval, because the exporter iterates explicitly over all dates between \texttt{\symbol{45}\symbol{45}start\_date} and \texttt{\symbol{45}\symbol{45}end\_date}, inclusive. Exported values should be non-negative, since spent-output values are non-negative.

Additional consistency checks can be performed using related OBM series exported from the same indexer database. In particular, \texttt{obm\_dormancy\_days\_daily} should satisfy
\[
\mathrm{Dormancy}_d =
\frac{\mathrm{CDD}_d}{\mathrm{SpentValueBTC}_d}
\]
for every date \(d\) with positive spent value. Equivalently,
\[
\mathrm{CDD}_d =
\mathrm{Dormancy}_d
\times
\mathrm{SpentValueBTC}_d.
\]
These identities should hold when \texttt{obm\_dormancy\_days\_daily}, \texttt{obm\_cdd\_btcxdays\_daily}, and \texttt{obm\_spent\_value\_btc\_daily} are generated from the same source series or indexer release. Comparisons with external transfer-value or spent-value series are useful as diagnostics but should not be interpreted as strict equality tests, because providers may differ in timestamp conventions, transfer-value definitions, change-output treatment, entity adjustment, historical edge-case treatment, and whether the metric is based on raw spent outputs or economically adjusted transfers.

\paragraph{Known limitations.}
The daily spent-output-value series is useful but definition-sensitive. First, \texttt{obm\_spent\_value\_btc\_daily} is a raw spent-output metric, not an entity-adjusted transfer-volume metric. It does not identify users, entities, custodians, exchanges, self-transfers, or change outputs. Second, the metric depends on the block timestamp convention used to assign spending blocks to calendar days. Third, the series reports BTC value, not fiat-denominated value. Fourth, the metric can be inflated by wallet consolidation, batching, exchange operations, and self-transfers. Fifth, it requires the persistent spent-output indexer database and therefore depends on the correctness and completeness of the indexer run. Sixth, the simple reorganization policy of the indexer detects inconsistencies and aborts, but it does not automatically roll back the state database.

Despite these limitations, \texttt{obm\_spent\_value\_btc\_daily} is a core OBM series. It provides a transparent BTC-denominated measure of raw spent-output activity and serves as a foundation for dormancy, age-band spent-value metrics, long-term-holder spent-value indicators, and broader UTXO-flow analysis.
\subsection{\texttt{obm\_spent\_value\_ge155d\_btc\_daily}: Spent Output Value whose age is at least 155 Days in BTC}
\label{Spent Output Value whose age is at least 155 Days in BTC}

\paragraph{Definition.}
The 155-day CDD-threshold series measures the total BTC value of outputs spent in blocks assigned to a given UTC calendar day whose age is at least 155 days. Let \(B_d\) denote the set of blocks assigned to day \(d\). For each non-coinbase transaction input \(i\) included in a transaction in block \(b \in B_d\), let \(v_i\) denote the value, in BTC, of the previous output spent by that input, and let \(a_i\) denote the age of that previous output, measured in days. The age of the spent output is defined as:
\[
a_i =
\max\left(
0,
\frac{t^{\mathrm{spent}}_i - t^{\mathrm{created}}_i}{86400}
\right),
\]
where \(t^{\mathrm{created}}_i\) is the timestamp of the block in which the spent output was created, and \(t^{\mathrm{spent}}_i\) is the timestamp of the block in which it is spent.

The daily 155-day CDD-threshold spent-value series is then defined as:
\[
\mathrm{CDD155BTC}_d =
\sum_{b \in B_d}
\sum_{i \in I_b}
v_i \mathbf{1}\{a_i \geq 155\},
\]
where \(I_b\) denotes the set of non-coinbase transaction inputs included in block \(b\), and \(\mathbf{1}\{\cdot\}\) is an indicator function equal to one when the condition is satisfied and zero otherwise.

A block \(b\) is assigned to day \(d\) according to the UTC calendar date derived from its block timestamp \(t_b\):
\[
d(b)=\mathrm{UTCDate}(t_b).
\]

The resulting series reports the BTC value of spent outputs that had remained unspent for at least 155 days before being consumed. Although the metric belongs to the CDD family of coin-age indicators, it is not measured in BTC-days. It uses output age only as a binary threshold. Therefore, its unit is BTC.

\paragraph{Economic interpretation.}
The 155-day CDD-threshold spent-value series measures the movement of BTC value that had remained inactive for at least 155 days. It can be interpreted as a raw long-term holder spent-value indicator under a 155-day inactivity threshold.

For economic research, this metric is useful in several ways. First, it identifies days on which relatively old coins are spent. Second, it helps distinguish the movement of longer-held supply from the ordinary turnover of recently active coins. Third, it complements raw Bitcoin Days Destroyed, because raw CDD weights each output by its full age, whereas this metric applies a fixed age threshold and then sums BTC value. Fourth, it can be used to construct ratios measuring the share of spent value associated with outputs aged at least 155 days. Fifth, it provides a transparent raw measure for long-term holder activity, dormant-supply activation, and threshold-based UTXO flow analysis.

The metric should not be interpreted as conventional Bitcoin Days Destroyed, because conventional CDD multiplies value by age and is measured in BTC-days. It should also not be interpreted as entity-adjusted long-term holder selling, because it does not identify users, entities, exchanges, custodians, changes in outputs, or self-transfers. It is a raw spent-output threshold metric.

\paragraph{Relationship with the 365-day threshold metric.}
This metric is the 155-day counterpart of \texttt{obm\_cdd\_365d\_btc\_daily}. Both series use the same threshold-spent-value logic, but with different age cutoffs:
\[
\mathrm{CDD155BTC}_d =
\sum_{b \in B_d}
\sum_{i \in I_b}
v_i \mathbf{1}\{a_i \geq 155\},
\]
and:
\[
\mathrm{CDD365BTC}_d =
\sum_{b \in B_d}
\sum_{i \in I_b}
v_i \mathbf{1}\{a_i \geq 365\}.
\]

Since every output aged at least 365 days is also aged at least 155 days, the 365-day threshold series is a subset of the 155-day threshold series. Therefore, when both series are exported from the same indexer database and release version, the following relation should hold:
\[
0 \leq
\mathrm{CDD365BTC}_d
\leq
\mathrm{CDD155BTC}_d
\leq
\mathrm{SpentValueBTC}_d.
\]

The difference between the 155-day and 365-day series isolates the BTC value of outputs spent with age at least 155 days but less than 365 days:
\[
\mathrm{CDD155BTC}_d -
\mathrm{CDD365BTC}_d.
\]

This difference can be interpreted as a medium-old spent-value component under the OBM threshold convention.

\paragraph{Similar metrics publicly available.}
The OBM series \texttt{obm\_spent\_value\_ge155d\_btc\_daily} is comparable to public metrics usually labelled \emph{Spent Volume by LTH/STH}, \emph{Spent Volume by Age}, \emph{Spent Output Age Bands}, \emph{Long-Term Holder spending}, or \emph{long-term-holder spent output metrics}. In OBM, the metric is defined as the daily BTC value of spent outputs that are at least 155 days old. Let \(S_d\) denote the set of previous outputs consumed by non-coinbase transaction inputs in blocks assigned to UTC day \(d\), let \(v_i\) denote the BTC value of spent output \(i\), and let \(a_i\) denote its age in days. The daily value is
\[
\mathrm{CDD155ValueBTC}_d =
\sum_{i \in S_d} v_i \mathbf{1}\{a_i \geq 155\}.
\]
Despite the reference to CDD in the identifier, the unit of this series is BTC rather than BTC-days. The metric does not sum coin days destroyed; instead, it uses output age as a threshold and sums the BTC value of qualifying spent outputs. It therefore measures how much BTC from outputs aged at least 155 days was spent on each day. The series is raw and spend-output-based: it does not identify users, exchanges, custodians, self-transfers, or changes to outputs.

The most relevant public conceptual comparator is Glassnode's \emph{Spent Volume by LTH/STH} chart.\footnote{\url{https://studio.glassnode.com/charts/breakdowns.SpentVolumeSumByLthSth?a=BTC}} Glassnode defines Long-Term Holder and Short-Term Holder supply using an entity-level average purchasing date, with weights given by a logistic function centered at an age of 155 days and a transition width of 10 days. The spent-volume-by-LTH/STH chart is therefore closely related to \texttt{obm\_spent\_value\_ge155d\_btc\_daily}, because both aim to isolate spending by older, long-term-held coins. However, the two metrics are not identical. OBM applies a hard spent-output age threshold, \(a_i \geq 155\), at the UTXO level. Glassnode's LTH/STH methodology is entity-based and uses a logistic transition around the 155-day boundary. Consequently, the Glassnode series is a useful public comparator for long-term-holder spending, but not a direct replication of the OBM thresholded spent-output value.

Glassnode's \emph{Spent Volume by Age} family is another close comparator.\footnote{\url{https://studio.glassnode.com/charts/breakdowns.SpentVolumeSumByAge?a=BTC}} Glassnode describes spent volume as the total volume of digital assets sold or spent, and the age-band version as a categorization of that spent volume by the age of the coins being spent. This is conceptually close to \texttt{obm\_spent\_value\_ge155d\_btc\_daily}, because both classify spent value by the age of the consumed coins. A rough public approximation to OBM could be obtained by summing all Glassnode age cohorts older than approximately 155 days, provided the bands are available in absolute BTC units, and the cohort boundaries align closely with the 155-day threshold. However, exact equivalence should not be assumed because Glassnode's public page does not provide a full open-source reconstruction algorithm, and because band boundaries, timestamp rules, entity-adjustment status, smoothing, and reorganization policies may differ from OBM.

CryptoQuant provides a directly related \emph{Spent Output Age Bands} metric.\footnote{\url{https://cryptoquant.com/asset/btc/chart/network-indicator/spent-output-age-bands}} CryptoQuant describes Spent Output Age Bands as the set of all spent outputs created within specified age bands, with each colored band representing the total value of spent outputs whose lifespan falls within the corresponding interval. This is very close to the OBM construction in spirit. The OBM metric can be interpreted as a single-threshold aggregate across all spent outputs with a lifespan of at least 155 days, whereas CryptoQuant presents the information as age bands. A 155-day-threshold comparator could therefore be approximated by summing CryptoQuant bands above the 155-day boundary if the relevant bands are available and sufficiently granular. Nevertheless, CryptoQuant does not appear to publish a named daily series exactly equivalent to \texttt{obm\_spent\_value\_ge155d\_btc\_daily}, and its public documentation does not provide a full node-level reconstruction script or all low-level conventions required for exact replication.

CryptoQuant also provides \emph{Long Term Holder SOPR}, which uses a 155-day age condition.\footnote{\url{https://cryptoquant.com/asset/btc/chart/market-indicator/long-term-holder-sopr}} Its documentation defines Long-Term Holder SOPR as a ratio based on spent outputs that lived more than 155 days. This confirms the importance of the 155-day threshold in public on-chain analytics. However, LTH-SOPR is not a spent-value metric. It is a profitability ratio comparing the realized value at spending with the value at creation. It should therefore be treated as a related long-term-holder spent-output metric, but not as a direct comparator for \texttt{obm\_spent\_value\_ge155d\_btc\_daily}.

Glassnode provides analogous long-term holder profitability and supply metrics, including charts based on the same 155-day long-term holder heuristic.\footnote{\url{https://studio.glassnode.com/charts/supply.LthSum?a=BTC}} These metrics are useful for contextualizing the long-term-holder threshold, but they are not direct equivalents. Long-Term Holder Supply is a stock metric that measures supply classified as long-term held, whereas \texttt{obm\_spent\_value\_ge155d\_btc\_daily} is a flow metric that measures the BTC value of old outputs actually spent on a given day.

Bitcoin Magazine Pro provides a \emph{Bitcoin: Long-Term Holder Supply} chart.\footnote{\url{https://www.bitcoinmagazinepro.com/charts/long-term-holder-supply/}} The page defines long-term-holder supply as bitcoins held for 155 days or more. This is useful as an accessible public reference for the 155-day convention. However, it is again a stock metric, not a daily spent-value flow. It should therefore be used only as contextual evidence for the threshold and not as a benchmark for \texttt{obm\_spent\_value\_ge155d\_btc\_daily}.

Coin Metrics does not appear to expose a named public metric exactly equivalent to our OBM metric \texttt{obm\_spent\_value\_ge155d\_btc\_daily}. It provides transferred-days-destroyed and transfer-value metrics, such as \texttt{TxTfrValDayDst} and \texttt{TxTfrValNtv}, but these do not isolate the BTC value of spent outputs whose age exceeds a fixed 155-day threshold. A Coin Metrics-like version would require age-band or thresholded transfer-value data, together with explicit alignment of transfer-value, timestamp, and age conventions.

Blockchair provides block-level, transaction-level, input-level, and output-level explorer data, including Bitcoin output data.\footnote{\url{https://blockchair.com/bitcoin/outputs}} This makes Blockchair useful as a reconstruction source. A researcher could resolve transaction inputs to previous outputs, compute the age of each consumed output, and sum the BTC value of those with \(a_i \geq 155\) by the UTC date of the spending block. However, Blockchair does not appear to publish a named daily metric equivalent to \texttt{obm\_spent\_value\_ge155d\_btc\_daily}. It should therefore be treated as a source of lower-level data rather than as a direct public comparator.

Overall, no reviewed provider appears to expose a named public daily series exactly equivalent to \texttt{obm\_spent\_value\_ge155d\_btc\_daily}. The closest comparators are Glassnode's \emph{Spent Volume by LTH/STH}, Glassnode's \emph{Spent Volume by Age}, and CryptoQuant's \emph{Spent Output Age Bands}. LTH-SOPR and long-term-holder supply metrics from Glassnode, CryptoQuant, and Bitcoin Magazine Pro are useful related indicators because they use or explain the 155-day long-term-holder convention, but they measure profitability ratios or supply stocks rather than thresholded spent value. The distinctive contribution of OBM is that it publishes the 155-day thresholded spent-value series directly, defines the cutoff explicitly as \(a_i \geq 155\), reports the result in BTC, keeps the calculation raw and non-entity-adjusted, assigns spending blocks to UTC calendar days using a documented timestamp convention, and makes the reconstruction auditable from full-node data.

\paragraph{Data source and input requirements.}
The metric is exported from the persistent OBM spent-output indexer database, rather than computed by an independent metric-specific blockchain scan. The indexer, described in Sect.~\ref{sec:The Spent-Output Indexer}, is built from a running Bitcoin Core full node and stores reusable daily aggregates in a SQLite database. For \texttt{obm\_spent\_value\_ge155d\_btc\_daily}, the relevant source table is \texttt{daily\_aggregates}, and the relevant field is \texttt{spent\_value\_155d\_sats}.

The indexer computes this quantity during its sequential scan of the blockchain. For each non-coinbase transaction input, it resolves the input to the previous output it spends, retrieves the previous output value and creation timestamp from the live outpoint state, computes the elapsed age in days, and adds the output value to the 155-day aggregate if the age is at least 155 days. Internally, the indexer stores:
\[
\mathrm{spent\_value\_155d\_sats}_d =
\sum_{b \in B_d}
\sum_{i \in I_b}
v^{\mathrm{sats}}_i
\mathbf{1}\{a_i \geq 155\}.
\]

The exporter converts the stored satoshi-denominated aggregate into BTC by dividing by \(100{,}000{,}000\):
\[
\mathrm{CDD155BTC}_d =
\frac{\mathrm{spent\_value\_155d\_sats}_d}{100{,}000{,}000}.
\]

The exporter script does not query Bitcoin Core directly. It requires an existing SQLite database generated by \texttt{obm\_spent\_output\_indexer.py}, with metadata identifying it as an OBM spent-output indexer database and with processed-date metadata covering the requested output interval. The script checks that the requested ending date is not beyond the maximum date processed by the indexer. If a requested date has no row in \texttt{daily\_aggregates}, the exporter writes a zero value for that date. This convention is appropriate because the metric is a flow variable: if no qualifying spent-output value is assigned to a UTC date under the indexer's timestamp convention, the daily value is zero.

This metric does not require address extraction, user clustering, entity identification, external price data, third-party APIs, or a direct Bitcoin Core connection at export time. However, it depends on the prior successful execution of the spent-output indexer, which itself requires a synchronized, non-pruned Bitcoin Core full node and access to transaction-level data sufficient to reconstruct previous-output values and creation timestamps. The exported series therefore inherits the indexer's UTC block-timestamp convention, previous-output reconstruction procedure, fractional-day age convention, treatment of negative apparent ages, chain-consistency checks, release metadata, and handling of historical duplicate outpoint edge cases.

\paragraph{Algorithm.}
The script \texttt{export\_obm\_spent\_value\_ge155d\_btc\_daily.py} implements the following procedure:

\begin{enumerate}
    \item Parse the user-provided date interval, \texttt{\symbol{45}\symbol{45}start\_date} and \texttt{\symbol{45}\symbol{45}end\_date}, using the format \texttt{YYYY-MM-DD}. Both dates are interpreted as UTC dates and both are included in the output.

    \item Open the persistent SQLite database generated by \texttt{obm\_spent\_output\_indexer.py}. The path to this database is provided through the \texttt{\symbol{45}\symbol{45}state\_db} argument.

    \item Validate that the SQLite database corresponds to the OBM spent-output indexer. The script checks the indexer metadata, including the expected \texttt{indexer\_id} and the presence of processing metadata such as \texttt{last\_processed\_height}, \texttt{last\_processed\_date}, and \texttt{max\_processed\_date}.

    \item Verify that the requested \texttt{\symbol{45}\symbol{45}end\_date} is not later than the maximum UTC date processed by the indexer. If the requested interval extends beyond the indexed range, the script aborts and requires the indexer to be run further before export.

    \item Infer the dataset release version from the indexer metadata. If the database does not contain a release-version field, the script uses the fallback value provided through the \texttt{\symbol{45}\symbol{45}release\_version} argument.

    \item For each UTC date \(d\) in the requested interval, query the \texttt{daily\_aggregates} table for the field \texttt{spent\_value\_155d\_sats}. If a row exists for date \(d\), the script reads the stored satoshi-denominated value. If no row exists, the script assigns zero to that date.

    \item Convert each daily value from satoshis to BTC:
    \[
    \mathrm{CDD155BTC}_d =
    \frac{\mathrm{spent\_value\_155d\_sats}_d}{100{,}000{,}000}.
    \]

    \item Write the resulting time series to a CSV file using the standardized OBM schema:
    \[
    \texttt{date},\quad
    \texttt{series\_id},\quad
    \texttt{value},\quad
    \texttt{unit},\quad
    \texttt{frequency},\quad
    \texttt{release\_version}.
    \]

    \item Optionally generate a plot of the exported series when the plotting flag is activated. If \texttt{\symbol{45}\symbol{45}plot\_output} is not provided, the plot is saved next to the CSV file with a \texttt{.png} extension. The plot title includes the series description and the selected date interval.
\end{enumerate}

This exporter does not query Bitcoin Core, does not maintain an outpoint state, and does not reconstruct spent outputs at export time. The computationally expensive previous-output resolution and age-threshold classification have already been performed by the spent-output indexer, which stores the daily 155-day threshold aggregate in \texttt{daily\_aggregates}. The export script is therefore a lightweight, deterministic transformation from the indexed database to the OBM CSV format.

\paragraph{Metric-specific input parameters.}
The input parameters specific to this exporter are:

\begin{itemize}
    \item \texttt{\symbol{45}\symbol{45}state\_db}: path to the persistent SQLite database generated by the OBM auxiliary script \texttt{obm\_spent\_output\_indexer.py}. This database must contain the \texttt{daily\_aggregates} table and the metadata required to verify that it corresponds to the OBM spent-output indexer.

    \item \texttt{\symbol{45}\symbol{45}start\_date}: starting date of the selected interval, inclusive, in \texttt{YYYY-MM-DD} format. The date is interpreted as a UTC calendar date.

    \item \texttt{\symbol{45}\symbol{45}end\_date}: ending date of the selected interval, inclusive, in \texttt{YYYY-MM-DD} format. The script verifies that this date is not later than the maximum date processed by the indexer.

    \item \texttt{\symbol{45}\symbol{45}release\_version}: fallback dataset release version used only if the indexer database does not contain a \texttt{release\_version} metadata field.

    \item \texttt{\symbol{45}\symbol{45}output}: path of the output CSV file to be written.

    \item \texttt{\symbol{45}\symbol{45}plot}: optional flag that instructs the script to generate a plot of the exported series.

    \item \texttt{\symbol{45}\symbol{45}plot\_output}: optional path for the generated plot. If this argument is omitted while \texttt{\symbol{45}\symbol{45}plot} is used, the plot is saved next to the CSV file using the same base name and a \texttt{.png} extension.
\end{itemize}

The exporter does not accept Bitcoin Core RPC parameters, \texttt{\symbol{45}\symbol{45}height\_margin},  \texttt{\symbol{45}\symbol{45}commit\_every}, \texttt{\symbol{45}\symbol{45}min\_confirmations}, or \texttt{\symbol{45}\symbol{45}reset\_state\_db}. These parameters belong to the spent-output indexer, not to the 155-day threshold exporter. The exporter assumes that the indexer database has already been built and updated through the requested ending date. Its role is only to read \texttt{spent\_value\_155d\_sats} from \texttt{daily\_aggregates}, convert the stored satoshi values into BTC, and write the standardized OBM output file.

\paragraph{Aggregation rule.}
The daily value is computed as the sum of the values of all previous outputs spent by non-coinbase transaction inputs included in blocks assigned to the same UTC calendar date, restricted to outputs whose age is at least 155 days:
\[
\mathrm{CDD155BTC}_d =
\sum_{b \in B_d}
\sum_{i \in I_b}
v_i \mathbf{1}\{a_i \geq 155\}.
\]

The aggregation rule is therefore a daily sum of BTC-denominated spent-output value subject to a 155-day age threshold. Monthly versions of this metric, if distributed, should also be computed as sums of the corresponding daily values:
\[
\mathrm{CDD155BTC}_m =
\sum_{d \in m}
\mathrm{CDD155BTC}_d.
\]

This convention preserves the interpretation of the metric as the total BTC value of outputs aged at least 155 days that were spent over the corresponding period.

\paragraph{Output format.}
The output file contains one observation per UTC date. Each row has the following fields:

\begin{center}
\begin{tabular}{llp{0.35\textwidth}}
\toprule
\textbf{Column} & \textbf{Example} & \textbf{Description} \\
\midrule
\texttt{date} & \texttt{2024-01-01} & UTC calendar date \\
\texttt{series\_id} & \texttt{obm\_spent\_value\_ge155d\_btc\_daily} & Stable OBM series identifier \\
\texttt{value} & \texttt{58291.49281735} & BTC value of spent outputs aged at least 155 days \\
\texttt{unit} & \texttt{BTC} & Measurement unit \\
\texttt{frequency} & \texttt{daily} & Observation frequency \\
\texttt{release\_version} & \texttt{OBM v0.1.0} & Dataset release version \\
\bottomrule
\end{tabular}
\end{center}

\paragraph{Technical validation.}
Several internal checks are used to validate this metric at export time. First, the script verifies that the requested date range is valid and that \texttt{\symbol{45}\symbol{45}start\_date} is not later than \texttt{\symbol{45}\symbol{45}end\_date}. Second, it checks that the SQLite state database exists, opens it, and sets the connection to query-only mode. Third, it verifies that the database metadata identify it as an OBM spent-output indexer database, by checking the expected \texttt{indexer\_id}. Fourth, it checks that the indexer database contains processing metadata, including \texttt{last\_processed\_height}, \texttt{last\_processed\_date}, and \texttt{max\_processed\_date}. Fifth, it verifies that the table \texttt{daily\_aggregates} is present in the database. Sixth, it verifies that the requested ending date is not later than the maximum date processed by the indexer. If the requested interval extends beyond the indexed range, the script aborts and requires the indexer to be updated before the export is attempted.

For each date in the requested interval, the exporter reads \texttt{spent\_value\_155d\_sats} from \texttt{daily\_aggregates}. If the date is absent from the aggregate table, the script exports zero for that date. This convention is appropriate for \texttt{obm\_spent\_value\_ge155d\_btc\_daily} because it is a flow variable: if no qualifying spent-output value is assigned to a UTC date under the indexer's timestamp convention, the daily value is zero rather than missing. Each retrieved value is converted from satoshis to BTC by dividing by \(100{,}000{,}000\), and the resulting file is written using the standard OBM schema:
\[
\texttt{date},\quad
\texttt{series\_id},\quad
\texttt{value},\quad
\texttt{unit},\quad
\texttt{frequency},\quad
\texttt{release\_version}.
\]

The script also propagates the release version stored in the indexer metadata. If the metadata field is absent, it uses the fallback value supplied through \texttt{\symbol{45}\symbol{45}release\_version}. The number of rows written must equal the number of calendar days in the requested interval, because the exporter iterates explicitly over all dates between \texttt{\symbol{45}\symbol{45}start\_date} and \texttt{\symbol{45}\symbol{45}end\_date}, inclusive. Exported values should be non-negative, since spent-output values are non-negative.

A basic internal consistency check is:
\[
0 \leq
\mathrm{CDD155BTC}_d
\leq
\mathrm{SpentValueBTC}_d.
\]
The upper bound should hold because the 155-day threshold series is a subset of total spent value. Therefore, \texttt{obm\_spent\_value\_ge155d\_btc\_daily} should never exceed \texttt{obm\_spent\_value\_btc\_daily} when both series are exported from the same indexer database, using the same timestamp convention and release version.

A second consistency check links this metric to its 365-day counterpart:
\[
0 \leq
\mathrm{CDD365BTC}_d
\leq
\mathrm{CDD155BTC}_d.
\]
This relation should hold because the 365-day threshold set is contained in the 155-day threshold set.

Additional consistency checks can be performed by comparing this series with raw CDD and dormancy. Such comparisons are conceptual rather than exact identities, because raw CDD is measured in BTC-days and this metric is measured in BTC. Comparisons with external long-term-holder spent-value, old-coin movement, or age-threshold metrics are useful as diagnostics but should not be interpreted as strict equality tests, because providers may differ in timestamp conventions, output-age thresholding, entity adjustment, change-output treatment, transfer-value definition, long-term-holder threshold definition, and historical edge-case handling.

\paragraph{Known limitations.}
The 155-day CDD-threshold spent-value series is useful but definition-sensitive. First, \texttt{obm\_spent\_value\_ge155d\_btc\_daily} is a raw spent-output metric, not an entity-adjusted long-term-holder metric. It does not identify users, entities, custodians, exchanges, self-transfers, or change outputs. Second, the metric depends on the block timestamp convention used both to compute output age and to assign spending blocks to calendar days. Third, the 155-day threshold is sharp: an output aged slightly below 155 days does not contribute, while an output aged 155 days or more does. Fourth, the metric is measured in BTC, not BTC-days, despite belonging to the CDD-related family of indicators. Fifth, it can be affected by self-transfers, wallet consolidation, batching, exchange operations, and custodial wallet management. Sixth, it requires the persistent spent-output indexer database and therefore depends on the correctness and completeness of the indexer run. Seventh, the simple reorganization policy of the indexer detects inconsistencies and aborts, but it does not automatically roll back the state database.

Despite these limitations, \texttt{obm\_spent\_value\_ge155d\_btc\_daily} is a useful OBM coin-age threshold series. It provides a transparent BTC-denominated measure of outputs spent after at least 155 days of inactivity and complements raw CDD, dormancy, spent-value, 365-day threshold metrics, and age-band metrics in the study of dormant supply, long-term-holder behavior, UTXO turnover, and on-chain activity.
\subsection{\texttt{obm\_spent\_value\_ge365d\_btc\_daily}: Spent Output Value whose age is at least 365 Days in BTC}
\label{Spent Output Value whose age is at least 365 Days in BTC}

\paragraph{Definition.}
The 365-day CDD-threshold series measures the total BTC value of outputs spent in blocks assigned to a given UTC calendar day whose age is at least 365 days. Let \(B_d\) denote the set of blocks assigned to day \(d\). For each non-coinbase transaction input \(i\) included in a transaction in block \(b \in B_d\), let \(v_i\) denote the value, in BTC, of the previous output spent by that input, and let \(a_i\) denote the age of that previous output, measured in days. The age of the spent output is defined as:
\[
a_i =
\max\left(
0,
\frac{t^{\mathrm{spent}}_i - t^{\mathrm{created}}_i}{86400}
\right),
\]
where \(t^{\mathrm{created}}_i\) is the timestamp of the block in which the spent output was created, and \(t^{\mathrm{spent}}_i\) is the timestamp of the block in which it is spent.

The daily 365-day CDD-threshold spent-value series is then defined as:
\[
\mathrm{CDD365BTC}_d =
\sum_{b \in B_d}
\sum_{i \in I_b}
v_i \mathbf{1}\{a_i \geq 365\},
\]
where \(I_b\) denotes the set of non-coinbase transaction inputs included in block \(b\), and \(\mathbf{1}\{\cdot\}\) is an indicator function equal to one when the condition is satisfied and zero otherwise.

A block \(b\) is assigned to day \(d\) according to the UTC calendar date derived from its block timestamp \(t_b\): $d(b)=\mathrm{UTCDate}(t_b)$. The resulting series reports the BTC value of spent outputs that had remained unspent for at least 365 days before being consumed. Although the metric belongs to the CDD family of coin-age indicators, it is not measured in BTC-days. It uses output age only as a binary threshold. Therefore, its unit is BTC.

\paragraph{Economic interpretation.}
The 365-day CDD-threshold spent-value series measures the movement of BTC value that had remained inactive for at least one year. It can be interpreted as a raw, long-dormant, spent-value indicator.

For economic research, this metric is useful in several ways. First, it identifies days on which long-inactive coins are spent. Second, it helps distinguish the movement of older supply from the ordinary turnover of recently active coins. Third, it complements raw Bitcoin Days Destroyed, because raw CDD weights each output by its full age, whereas this metric only applies a one-year activity threshold and then sums BTC value. Fourth, it can be used to construct ratios measuring the share of spent value associated with outputs aged at least 365 days. Fifth, it provides a transparent raw measure for long-term holder activity, dormant-supply activation, and age-threshold UTXO flow analysis.

The metric should not be interpreted as conventional Bitcoin Days Destroyed, because conventional CDD multiplies value by age and is measured in BTC-days. It should also not be interpreted as entity-adjusted long-term holder selling, because it does not identify users, entities, exchanges, custodians, changes in outputs, or self-transfers. It is a raw spent-output threshold metric.

\paragraph{Similar metrics publicly available.}
The OBM series \texttt{obm\_spent\_value\_ge365d\_btc\_daily} is comparable to public metrics usually labeled \emph{Spent Volume by Age}, \emph{Spent Output Age Bands}, \emph{old-coin spent volume}, or \emph{long-term-holder spent volume}. In OBM, the metric is defined as the daily BTC value of spent outputs that are at least 365 days old. Let \(S_d\) denote the set of previous outputs consumed by non-coinbase transaction inputs in blocks assigned to UTC day \(d\), let \(v_i\) denote the BTC value of spent output \(i\), and let \(a_i\) denote its age in days. The daily value is
\[
\mathrm{CDD365ValueBTC}_d =
\sum_{i \in S_d} v_i \mathbf{1}\{a_i \geq 365\}.
\]
Despite the reference to CDD in the identifier, the unit of this series is BTC rather than BTC-days. The metric does not sum coin days destroyed; instead, it uses output age as a threshold and sums the BTC value of qualifying spent outputs. It therefore measures how many BTC from coins that have been dormant for at least 1 year were spent each day. The series is raw and spend-output-based: it does not identify users, exchanges, custodians, self-transfers, or changes to outputs.

The closest public comparator is Glassnode's \emph{Spent Volume by Age} family, exposed through different charts.\footnote{\url{https://studio.glassnode.com/charts/breakdowns.SpentVolumeSumByAge?a=BTC}} Glassnode describes spent volume as the total volume of digital assets sold or spent, and the age-band version as a categorization of that spent volume by the age of the coins being spent. The chart reports age cohorts including bands such as \texttt{6m-12m}, \texttt{1y-2y}, \texttt{2y-3y}, and older groups. This is conceptually close to \texttt{obm\_spent\_value\_ge365d\_btc\_daily}: summing Glassnode's age cohorts from \texttt{1y-2y} upward would approximate a public ``spent value older than 365 days'' benchmark. However, it would remain an approximation unless all cohort boundaries, timestamp conventions, entity-adjustment status, smoothing choices, and reorganization policies are harmonized. Glassnode provides data access through its platform, but the public chart page does not offer a full open-source reconstruction algorithm comparable to OBM.

Glassnode also provides \emph{Spent Output Age Bands}.\footnote{\url{https://studio.glassnode.com/charts/indicators.Soab?a=BTC}} Its documentation describes SOAB as a metric that bundles spent coins into categories according to their age, with each band representing the percentage of spent outputs created within the time period denoted in the legend. This is closely related, but not identical, because SOAB is often expressed as a relative distribution rather than as a BTC-denominated absolute spent-value total. It is useful for validating the age-threshold logic of \texttt{obm\_spent\_value\_ge365d\_btc\_daily}, but it is not a direct substitute unless absolute BTC values for the relevant age bands are available and summed.

CryptoQuant provides a directly related \emph{Spent Output Age Bands} metric.\footnote{\url{https://cryptoquant.com/asset/btc/chart/network-indicator/spent-output-age-bands}} Its methodological documentation states that Spent Output Age Bands are the set of all spent outputs created within specified age bands, and that each colored band represents the total value of spent outputs whose lifespan falls inside the denoted interval. In formula form, the metric can be written as
\[
\sum_{o \in \mathrm{spent\ outputs}} \mathrm{value}_o
\mathbf{1}\{a_i \leq \mathrm{lifespan}_o < b_i\}.
\]
This is very close to the OBM construction. In particular, \texttt{obm\_spent\_value\_ge365d\_btc\_daily} can be interpreted as the sum of all CryptoQuant-style spent-output age bands with lower bound greater than or equal to 365 days. Nevertheless, CryptoQuant publishes the measure primarily as a banded indicator, and the public documentation does not provide a full-node-level reconstruction script or all the low-level conventions required for exact replication.

CryptoQuant also provides \emph{Exchange Inflow - Spent Output Age Bands}.\footnote{\url{https://cryptoquant.com/asset/btc/chart/flow-indicator/exchange-inflow-spent-output-age-bands}} This metric is related but not equivalent. It restricts spent outputs to those flowing into exchange wallets, whereas OBM counts all spent outputs that satisfy the 365-day age threshold, regardless of destination. It is therefore useful for studying whether old coins are moving into exchanges, but it should not be used as a direct benchmark for \texttt{obm\_spent\_value\_ge365d\_btc\_daily}.

Coin Metrics does not appear to expose a named public metric exactly equivalent to our metric \texttt{obm\_spent\_value\_ge365d\_btc\_daily}. It provides \texttt{TxTfrValDayDst}, a transferred-days-destroyed metric, and \texttt{TxTfrValNtv}, a native transfer-value metric, but these do not isolate the BTC value of spent outputs whose age exceeds a fixed 365-day threshold. A related measure could, in principle, be constructed if age-band transfer-value data were available, but the standard public Coin Metrics transfer-value and CDD metrics are not direct equivalents.

Blockchair provides block-level, transaction-level, and output-level explorer data.\footnote{\url{https://blockchair.com/bitcoin/outputs}} This makes Blockchair useful as a reconstruction source: a researcher could resolve spent inputs to previous outputs, compute each output's age, and sum the BTC value of outputs whose age is at least 365 days. However, Blockchair does not appear to publish a named daily metric equivalent to \texttt{obm\_spent\_value\_ge365d\_btc\_daily}. It should therefore be treated as a source of raw data for independent reconstruction rather than as a direct public comparator.

HODL-wave and UTXO-age-band charts are related but measure different objects. For example, Glassnode's HODL Waves and CryptoQuant's UTXO Age Bands describe the age distribution of currently unspent supply.\footnote{\url{https://docs.glassnode.com/guides-and-tutorials/metric-guides/age-distribution/hodl-waves}} \footnote{\url{https://cryptoquant.com/asset/btc/chart/network-indicator/utxo-age-bands}} These are stock metrics, not flow metrics. They measure how much supply remains unspent in each age band, whereas \texttt{obm\_spent\_value\_ge365d\_btc\_daily} measures the BTC value of old outputs spent on a given day. They are useful for contextualizing the pool of old coins that could potentially move, but they are not direct equivalents.

Overall, no reviewed public provider appears to expose a named daily series exactly equivalent to \texttt{obm\_spent\_value\_ge365d\_btc\_daily}. The closest comparators are Glassnode's \emph{Spent Volume by Age} and CryptoQuant's \emph{Spent Output Age Bands}, because both classify spent value by output age. A 365-day-threshold series can be approximated by summing all age bands that are one year or older, provided the provider's bands are sufficiently granular and available in absolute BTC units. The distinctive contribution of OBM is that it publishes the thresholded series directly, defines the age cutoff explicitly as \(a_i \geq 365\), reports the result in BTC, keeps the calculation raw and non-entity-adjusted, assigns spending blocks to UTC calendar days using a documented timestamp convention, and makes the reconstruction auditable from full-node data.

\paragraph{Data source and input requirements.}
The metric is exported from the persistent OBM spent-output indexer database, rather than computed by an independent metric-specific blockchain scan. The indexer, described in Sect.~\ref{sec:The Spent-Output Indexer}, is built from a running Bitcoin Core full node and stores reusable daily aggregates in a SQLite database. For \texttt{obm\_spent\_value\_ge365d\_btc\_daily}, the relevant source table is \texttt{daily\_aggregates}, and the relevant field is \texttt{spent\_value\_365d\_sats}.

The indexer computes this quantity during its sequential scan of the blockchain. For each non-coinbase transaction input, it resolves the input to the previous output it spends, retrieves the previous output value and creation timestamp from the live outpoint state, computes the elapsed age in days, and adds the output value to the 365-day aggregate if the age is at least 365 days. Internally, the indexer stores:
\[
\mathrm{spent\_value\_365d\_sats}_d =
\sum_{b \in B_d}
\sum_{i \in I_b}
v^{\mathrm{sats}}_i
\mathbf{1}\{a_i \geq 365\}.
\]

The exporter converts the stored satoshi-denominated aggregate into BTC by dividing by \(100{,}000{,}000\):
\[
\mathrm{CDD365BTC}_d =
\frac{\mathrm{spent\_value\_365d\_sats}_d}{100{,}000{,}000}.
\]

The exporter script does not query Bitcoin Core directly. It requires an existing SQLite database generated by \texttt{obm\_spent\_output\_indexer.py}, with metadata identifying it as an OBM spent-output indexer database and with processed-date metadata covering the requested output interval. The script checks that the requested ending date is not beyond the maximum date processed by the indexer. If a requested date has no row in \texttt{daily\_aggregates}, the exporter writes a zero value for that date. This convention is appropriate because the metric is a flow variable: if no qualifying spent-output value is assigned to a UTC date under the indexer's timestamp convention, the daily value is zero.

This metric does not require address extraction, user clustering, entity identification, external price data, third-party APIs, or a direct Bitcoin Core connection at export time. However, it depends on the prior successful execution of the spent-output indexer, which itself requires a synchronized, non-pruned Bitcoin Core full node and access to transaction-level data sufficient to reconstruct previous-output values and creation timestamps. The exported series therefore inherits the indexer's UTC block-timestamp convention, previous-output reconstruction procedure, fractional-day age convention, treatment of negative apparent ages, chain-consistency checks, release metadata, and handling of historical duplicate outpoint edge cases.

\paragraph{Algorithm.}
The script \texttt{export\_obm\_spent\_value\_ge365d\_btc\_daily.py} implements the following procedure:

\begin{enumerate}
    \item Parse the user-provided date interval, \texttt{\symbol{45}\symbol{45}start\_date} and \texttt{\symbol{45}\symbol{45}end\_date}, using the format \texttt{YYYY-MM-DD}. Both dates are interpreted as UTC dates and both are included in the output.

    \item Open the persistent SQLite database generated by \texttt{obm\_spent\_output\_indexer.py}. The path to this database is provided through the \texttt{\symbol{45}\symbol{45}state\_db} argument.

    \item Validate that the SQLite database corresponds to the OBM spent-output indexer. The script checks the indexer metadata, including the expected \texttt{indexer\_id} and the presence of processing metadata such as \texttt{last\_processed\_height}, \texttt{last\_processed\_date}, and \texttt{max\_processed\_date}.

    \item Verify that the requested \texttt{\symbol{45}\symbol{45}end\_date} is not later than the maximum UTC date processed by the indexer. If the requested interval extends beyond the indexed range, the script aborts and requires the indexer to be run further before export.

    \item Infer the dataset release version from the indexer metadata. If the database does not contain a release-version field, the script uses the fallback value provided through the \texttt{\symbol{45}\symbol{45}release\_version} argument.

    \item For each UTC date \(d\) in the requested interval, query the \texttt{daily\_aggregates} table for the field \texttt{spent\_value\_365d\_sats}. If a row exists for date \(d\), the script reads the stored satoshi-denominated value. If no row exists, the script assigns zero to that date.

    \item Convert each daily value from satoshis to BTC:
    \[
    \mathrm{CDD365BTC}_d =
    \frac{\mathrm{spent\_value\_365d\_sats}_d}{100{,}000{,}000}.
    \]

    \item Write the resulting time series to a CSV file using the standardized OBM schema:
    \[
    \texttt{date},\quad
    \texttt{series\_id},\quad
    \texttt{value},\quad
    \texttt{unit},\quad
    \texttt{frequency},\quad
    \texttt{release\_version}.
    \]

    \item Optionally generate a plot of the exported series when the plotting flag is activated. If \texttt{\symbol{45}\symbol{45}plot\_output} is not provided, the plot is saved next to the CSV file with a \texttt{.png} extension. The plot title includes the series description and the selected date interval.
\end{enumerate}

This exporter does not query Bitcoin Core, does not maintain an outpoint state, and does not reconstruct spent outputs at export time. The computationally expensive previous-output resolution and age-threshold classification have already been performed by the spent-output indexer, which stores the daily 365-day threshold aggregate in \texttt{daily\_aggregates}. The export script is therefore a lightweight, deterministic transformation from the indexed database to the OBM CSV format.

\paragraph{Metric-specific input parameters.}
The input parameters specific to this exporter are:

\begin{itemize}
    \item \texttt{\symbol{45}\symbol{45}state\_db}: path to the persistent SQLite database generated by theo OBM auxiliary script \texttt{obm\_spent\_output\_indexer.py}. This database must contain the \texttt{daily\_aggregates} table and the metadata required to verify that it corresponds to the OBM spent-output indexer.

    \item \texttt{\symbol{45}\symbol{45}start\_date}: starting date of the selected interval, inclusive, in \texttt{YYYY-MM-DD} format. The date is interpreted as a UTC calendar date.

    \item \texttt{\symbol{45}\symbol{45}end\_date}: ending date of the selected interval, inclusive, in \texttt{YYYY-MM-DD} format. The script verifies that this date is not later than the maximum date processed by the indexer.

    \item \texttt{\symbol{45}\symbol{45}release\_version}: fallback dataset release version used only if the indexer database does not contain a \texttt{release\_version} metadata field.

    \item \texttt{\symbol{45}\symbol{45}output}: path of the output CSV file to be written.

    \item \texttt{\symbol{45}\symbol{45}plot}: optional flag that instructs the script to generate a plot of the exported series.

    \item \texttt{\symbol{45}\symbol{45}plot\_output}: optional path for the generated plot. If this argument is omitted while \texttt{\symbol{45}\symbol{45}plot} is used, the plot is saved next to the CSV file using the same base name and a \texttt{.png} extension.
\end{itemize}

The exporter does not accept Bitcoin Core RPC parameters, \texttt{\symbol{45}\symbol{45}height\_margin}, \texttt{\symbol{45}\symbol{45}commit\_every}, \texttt{\symbol{45}\symbol{45}min\_confirmations}, or \texttt{\symbol{45}\symbol{45}reset\_state\_db}. These parameters belong to the spent-output indexer, not to the 365-day threshold exporter. The exporter assumes that the indexer database has already been built and updated through the requested ending date. Its role is only to read \texttt{spent\_value\_365d\_sats} from \texttt{daily\_aggregates}, convert the stored satoshi values into BTC, and write the standardized OBM output file.

\paragraph{Aggregation rule.}
The daily value is computed as the sum of the values of all previous outputs spent by non-coinbase transaction inputs included in blocks assigned to the same UTC calendar date, restricted to outputs whose age is at least 365 days:
\[
\mathrm{CDD365BTC}_d =
\sum_{b \in B_d}
\sum_{i \in I_b}
v_i \mathbf{1}\{a_i \geq 365\}.
\]

The aggregation rule is therefore a daily sum of BTC-denominated spent-output value subject to a 365-day age threshold. Monthly versions of this metric, if distributed, should also be computed as sums of the corresponding daily values:
\[
\mathrm{CDD365BTC}_m =
\sum_{d \in m}
\mathrm{CDD365BTC}_d.
\]

This convention preserves the interpretation of the metric as the total BTC value of outputs aged at least 365 days that were spent over the corresponding period.

\paragraph{Output format.}
The output file contains one observation per UTC date. Each row has the following fields:

\begin{center}
\begin{tabular}{llp{0.30\textwidth}}
\toprule
\textbf{Column} & \textbf{Example} & \textbf{Description} \\
\midrule
\texttt{date} & \texttt{2024-01-01} & UTC calendar date \\
\texttt{series\_id} & \texttt{obm\_spent\_value\_ge365d\_btc\_daily} & Stable OBM series identifier \\
\texttt{value} & \texttt{34182.49281735} & BTC value of spent outputs aged at least 365 days \\
\texttt{unit} & \texttt{BTC} & Measurement unit \\
\texttt{frequency} & \texttt{daily} & Observation frequency \\
\texttt{release\_version} & \texttt{OBM v0.1.0} & Dataset release version \\
\bottomrule
\end{tabular}
\end{center}

\paragraph{Technical validation.}
Several internal checks are used to validate this metric at export time. First, the script verifies that the requested date range is valid and that \texttt{\symbol{45}\symbol{45}start\_date} is not later than \texttt{\symbol{45}\symbol{45}end\_date}. Second, it checks that the SQLite state database exists, opens it, and sets the connection to query-only mode. Third, it verifies that the database metadata identify it as an OBM spent-output indexer database, by checking the expected \texttt{indexer\_id}. Fourth, it checks that the indexer database contains processing metadata, including \texttt{last\_processed\_height}, \texttt{last\_processed\_date}, and \texttt{max\_processed\_date}. Fifth, it verifies that the table \texttt{daily\_aggregates} is present in the database. Sixth, it verifies that the requested ending date is not later than the maximum date processed by the indexer. If the requested interval extends beyond the indexed range, the script aborts and requires the indexer to be updated before the export is attempted.

For each date in the requested interval, the exporter reads \texttt{spent\_value\_365d\_sats} from \texttt{daily\_aggregates}. If the date is absent from the aggregate table, the script exports zero for that date. This convention is appropriate for \texttt{obm\_spent\_value\_ge365d\_btc\_daily} because it is a flow variable: if no qualifying spent-output value is assigned to a UTC date under the indexer's timestamp convention, the daily value is zero rather than missing. Each retrieved value is converted from satoshis to BTC by dividing by \(100{,}000{,}000\), and the resulting file is written using the standard OBM schema:
\[
\texttt{date},\quad
\texttt{series\_id},\quad
\texttt{value},\quad
\texttt{unit},\quad
\texttt{frequency},\quad
\texttt{release\_version}.
\]

The script also propagates the release version stored in the indexer metadata. If the metadata field is absent, it uses the fallback value supplied through \texttt{\symbol{45}\symbol{45}release\_version}. The number of rows written must equal the number of calendar days in the requested interval, because the exporter iterates explicitly over all dates between \texttt{\symbol{45}\symbol{45}start\_date} and \texttt{\symbol{45}\symbol{45}end\_date}, inclusive. Exported values should be non-negative, since spent-output values are non-negative.

A basic internal consistency check is:
\[
0 \leq
\mathrm{CDD365BTC}_d
\leq
\mathrm{SpentValueBTC}_d.
\]
The upper bound should hold because the 365-day threshold series is a subset of total spent value. Therefore, \texttt{obm\_spent\_value\_ge365d\_btc\_daily} should never exceed \texttt{obm\_spent\_value\_btc\_daily} when both series are exported from the same indexer database, using the same timestamp convention and release version.

Additional consistency checks can be performed by comparing this series with raw CDD and dormancy. Such comparisons are conceptual rather than exact identities, because raw CDD is measured in BTC-days and this metric is measured in BTC. Comparisons with external long-term-holder spent-value, old-coin movement, or age-threshold metrics are useful as diagnostics but should not be interpreted as strict equality tests, because providers may differ in timestamp conventions, output-age thresholding, entity adjustment, change-output treatment, transfer-value definition, and historical edge-case handling.

\paragraph{Known limitations.}
The 365-day CDD-threshold spent-value series is useful but definition-sensitive. First, \texttt{obm\_spent\_value\_ge365d\_btc\_daily} is a raw spent-output metric, not an entity-adjusted long-term-holder metric. It does not identify users, entities, custodians, exchanges, self-transfers, or change outputs. Second, the metric depends on the block timestamp convention used both to compute output age and to assign spending blocks to calendar days. Third, the 365-day threshold is sharp: an output aged 364.99 days does not contribute, while an output aged 365.00 days does. Fourth, the metric is measured in BTC, not BTC-days, despite belonging to the CDD-related family of indicators. Fifth, it can be affected by self-transfers, wallet consolidation, batching, exchange operations, and custodial wallet management. Sixth, it requires the persistent spent-output indexer database and therefore depends on the correctness and completeness of the indexer run. Seventh, the simple reorganization policy of the indexer detects inconsistencies and aborts, but it does not automatically roll back the state database.

Despite these limitations, \texttt{obm\_spent\_value\_ge365d\_btc\_daily} is a useful OBM coin-age threshold series. It provides a transparent BTC-denominated measure of outputs spent after at least one year of inactivity and complements raw CDD, dormancy, spent-value, and age-band metrics in the study of dormant supply, long-term-holder behavior, UTXO turnover, and on-chain activity.
\subsection{\texttt{obm\_spent\_value\_lt155d\_btc\_daily}: Spent Output Value Younger Than 155 Days in BTC}
\label{Spent Output Value Younger Than 155 Days in BTC}

\paragraph{Definition.}
The spent-output-value younger-than-155-days series measures the total BTC value of outputs spent in blocks assigned to a given UTC calendar day whose age is strictly less than 155 days. Unlike the threshold metrics exported directly from the spent-output indexer, this series is derived from two already generated OBM CSV files. Let \(\mathrm{SpentValueBTC}_d\) denote total daily spent output value, and let \(\mathrm{SpentValueGE155BTC}_d\) denote the daily spent output value associated with outputs aged at least 155 days. The younger-than-155-days series is defined as:
\[
\mathrm{SpentValueLT155BTC}_d =
\mathrm{SpentValueBTC}_d -
\mathrm{SpentValueGE155BTC}_d.
\]

Equivalently, if \(B_d\) denotes the set of blocks assigned to day \(d\), \(I_b\) denotes the set of non-coinbase transaction inputs included in block \(b\), \(v_i\) denotes the BTC value of the previous output spent by input \(i\), and \(a_i\) denotes the age of that previous output in days, then:
\[
\mathrm{SpentValueLT155BTC}_d =
\sum_{b \in B_d}
\sum_{i \in I_b}
v_i \mathbf{1}\{a_i < 155\}.
\]

A block \(b\) is assigned to day \(d\) according to the UTC calendar date derived from its block timestamp \(t_b\): $d(b)=\mathrm{UTCDate}(t_b)$. The resulting series reports the BTC value of spent outputs that had remained unspent for less than 155 days before being consumed. It is therefore the young-output complement of the 155-day threshold spent-value series. Its unit is BTC.

\paragraph{Economic interpretation.}
The younger-than-155-days spent-value series measures the movement of relatively young BTC value. It captures the portion of raw spent-output activity associated with outputs that had not yet reached the 155-day age threshold.

For economic research, this metric is useful in several ways. First, it provides the short-age counterpart to the 155-day threshold spent-value series. Second, it helps decompose total spent value into younger and older spent-output components. Third, it can be used to construct young-output and old-output spent-value shares. Fourth, it is useful for studying recently active coin turnover, short-term-holder behavior under an age-threshold convention, and changes in the age composition of on-chain activity. Fifth, it complements CDD, dormancy, and age-threshold metrics by separating whether spent value is dominated by younger or older outputs.

The metric should not be interpreted as entity-adjusted short-term-holder selling. It does not identify users, entities, exchanges, custodians, change outputs, or self-transfers. It is a raw spent-output metric derived from two source OBM series.

\paragraph{Relationship with the 155-day threshold metric.}
This metric is constructed as the complement of the spent-value series for outputs aged at least 155 days. For each UTC date \(d\), the intended decomposition is:
\[
\mathrm{SpentValueBTC}_d =
\mathrm{SpentValueLT155BTC}_d +
\mathrm{SpentValueGE155BTC}_d.
\]

This relationship should hold when the relevant source series are generated using the same threshold definition, timestamp convention, and release version.

\paragraph{Similar metrics publicly available.}
The OBM series \texttt{obm\_spent\_value\_lt155d\_btc\_daily} is comparable to public metrics usually labeled \emph{Spent Volume by LTH/STH}, \emph{Spent Volume by Age}, \emph{Spent Output Age Bands}, \emph{Short-Term Holder spent volume}, or \emph{short-term-holder spending}. In OBM, the metric is defined as the daily BTC value of spent outputs whose age is strictly lower than 155 days. Let \(S_d\) denote the set of previous outputs consumed by non-coinbase transaction inputs in blocks assigned to UTC day \(d\), let \(v_i\) denote the BTC value of spent output \(i\), and let \(a_i\) denote its age in days. The daily value is
\[
\mathrm{SpentValueLT155dBTC}_d = 
\sum_{i \in S_d} v_i \mathbf{1}\{a_i < 155\}.
\]
The metric is measured in BTC, not in BTC-days. It therefore measures the BTC value of spent outputs satisfying an age condition, rather than the amount of coin age destroyed. It is also the exact complement of \texttt{obm\_spent\_value\_ge155d\_btc\_daily} within the total spent-output-value series:
\[
\texttt{obm\_spent\_value\_btc\_daily} 
=
\]
\[
\texttt{obm\_spent\_value\_lt155d\_btc\_daily}
+
\texttt{obm\_spent\_value\_ge155d\_btc\_daily}.
\]
The series is raw and spend-output-based: it does not identify users, exchanges, custodians, self-transfers, or changes to outputs.

The most relevant public conceptual comparator is Glassnode's \emph{Spent Volume by LTH/STH} chart.\footnote{\url{https://studio.glassnode.com/charts/breakdowns.SpentVolumeSumByLthSth?a=BTC}} Glassnode describes spent volume as the total volume of digital assets sold or spent, and the LTH/STH breakdown as a categorization of this spent volume into two cohorts based on holding duration. Glassnode also documents the 155-day convention, stating that coins younger than 155 days are statistically more likely to be re-spent and are considered the more liquid, short-term holder portion of the supply. This makes the Glassnode LTH/STH spent-volume breakdown the closest public analog to \texttt{obm\_spent\_value\_lt155d\_btc\_daily}. However, the two are not identical. OBM applies a hard threshold at the spent-output level, \(a_i < 155\). Glassnode's broader LTH/STH methodology is entity-oriented and, in some contexts, uses a probabilistic or smoothed transition around the 155-day boundary. Therefore, Glassnode's series is a useful public comparator for short-term holder spending, but not a direct replication of the OBM thresholded spent-output value.

Glassnode's \emph{Spent Volume by Age} family is another close comparator.\footnote{\url{https://studio.glassnode.com/charts/breakdowns.SpentVolumeSumByAge?a=BTC}} Glassnode describes this metric as spent volume categorized into different age cohorts, from recently acquired coins to older dormant coins. A rough public approximation to \texttt{obm\_spent\_value\_lt155d\_btc\_daily} could be obtained by summing all absolute BTC spent-volume age cohorts younger than 155 days, provided that the relevant bands are available, mutually exclusive, exhaustive below the threshold, and reported in compatible units. Exact equivalence should not be assumed because public age bands may not align exactly with the 155-day boundary, and because the page does not provide a full open-source reconstruction algorithm specifying timestamp assignment, entity adjustment, smoothing, and reorganization policy.

Glassnode also provides \emph{Spent Volume Age Bands} (SVAB).\footnote{\url{https://studio.glassnode.com/charts/indicators.Svab?a=BTC}} SVAB separates on-chain transfer volume according to coin age, with each band representing the percentage of spent volume that last moved within the interval denoted in the legend. This is related to our OBM metric \texttt{obm\_spent\_value\_lt155d\_btc\_daily}, but it is not a direct equivalent when expressed as percentages rather than BTC-denominated absolute values. Entity-adjusted SVAB is even less directly comparable, because it discards transactions between addresses attributed to the same entity, while OBM remains raw and non-entity-adjusted.

CryptoQuant provides a directly related \emph{Spent Output Age Bands} metric.\footnote{\url{https://cryptoquant.com/asset/btc/chart/network-indicator/spent-output-age-bands}} CryptoQuant describes Spent Output Age Bands as the set of all spent outputs created within specified age bands, with each colored band representing the total value of spent outputs whose lifespan falls within the corresponding interval. This is very close to the OBM construction in spirit. The OBM metric can be interpreted as a single-threshold aggregate across all spent outputs with a lifespan below 155 days, whereas CryptoQuant presents the information as age bands. A 155-day short-term spent-value comparator could therefore be approximated by summing CryptoQuant bands below the 155-day boundary if the relevant bands are available and sufficiently granular. Nevertheless, CryptoQuant does not appear to publish a named daily series exactly equivalent to \texttt{obm\_spent\_value\_lt155d\_btc\_daily}, and its public documentation does not provide a full node-level reconstruction script or all low-level conventions required for exact replication. 

CryptoQuant's \emph{Short Term Holder SOPR} is also related.\footnote{\url{https://cryptoquant.com/asset/btc/chart/market-indicator/short-term-holder-sopr}} Its documentation defines STH-SOPR over spent outputs that lived more than one hour and less than 155 days. This confirms the public use of the 155-day short-term-holder boundary. However, STH-SOPR is a profitability ratio, not a spent-value metric. It compares the USD value of spent outputs at the time of spending with their value at the time of creation. It should therefore be treated as a related short-term-holder spent-output indicator, but not as a direct comparator for \texttt{obm\_spent\_value\_lt155d\_btc\_daily}. 

Glassnode provides analogous short-term holder and long-term holder supply and profitability metrics based on the same 155-day convention. These include short-term holder supply, long-term holder supply, and SOPR variants. They are useful for contextualizing the threshold, but they are not direct equivalents. Supply metrics are stocks, whereas our metric \texttt{obm\_spent\_value\_lt155d\_btc\_daily} is a daily flow of spent BTC. SOPR metrics are profitability ratios, whereas OBM reports the BTC value of spent outputs.

Coin Metrics does not appear to expose a named public metric exactly equivalent to our OBM metric \texttt{obm\_spent\_value\_lt155d\_btc\_daily}. It provides transfer-value metrics such as \texttt{TxTfrValNtv} and transferred-days-destroyed metrics such as \texttt{TxTfrValDayDst}, but these do not isolate the BTC value of spent outputs whose age is below a fixed 155-day threshold. A Coin Metrics-like version would require age-band or thresholded transfer-value data, together with explicit alignment of transfer-value, timestamp, and age conventions.

Blockchair provides block-level, transaction-level, input-level, and output-level explorer data, including Bitcoin output data.\footnote{\url{https://blockchair.com/bitcoin/outputs}} This makes Blockchair useful as a reconstruction source. A researcher could resolve transaction inputs to previous outputs, compute the age of each consumed output, and sum the BTC value of those with \(a_i < 155\) by the UTC date of the spending block. However, Blockchair does not appear to publish a named daily metric equivalent to \texttt{obm\_spent\_value\_lt155d\_btc\_daily}. It should therefore be treated as a source of lower-level data rather than as a direct public comparator.

Overall, no reviewed provider appears to expose a named public daily series exactly equivalent to \texttt{obm\_spent\_value\_lt155d\_btc\_daily}. The closest comparators are Glassnode's \emph{Spent Volume by LTH/STH}, Glassnode's \emph{Spent Volume by Age}, Glassnode's SVAB family, and CryptoQuant's \emph{Spent Output Age Bands}. CryptoQuant's STH-SOPR and related Glassnode STH/LTH metrics are useful related indicators because they use or explain the 155-day short-term-holder convention, but they measure profitability ratios, cohort stocks, or age-band percentages rather than raw thresholded spent value. The distinctive contribution of OBM is that it publishes the short-term thresholded spent-value series directly, defines the cutoff explicitly as \(a_i < 155\), reports the result in BTC, keeps the calculation raw and non-entity-adjusted, assigns spending blocks to UTC calendar days using a documented timestamp convention, and makes the reconstruction auditable from full-node-derived source series.

\paragraph{Data source and input requirements.}
The metric is computed from two existing OBM CSV files, rather than exported directly from the spent-output indexer database. The first input file must contain:
\[
\texttt{obm\_spent\_value\_btc\_daily},
\]
which reports total daily spent output value. The second input file must contain:
\[
\texttt{obm\_spent\_value\_ge155d\_btc\_daily},
\]
which reports daily spent output value for outputs aged at least 155 days.

Both input files must follow the standard OBM schema:
\[
\texttt{date},\quad
\texttt{series\_id},\quad
\texttt{value},\quad
\texttt{unit},\quad
\texttt{frequency},\quad
\texttt{release\_version}.
\]

The script verifies that the first input file has series identifier \(\texttt{obm\_spent\_value\_btc\_daily}\), unit \(\texttt{BTC}\), and frequency \(\texttt{daily}\). It also verifies that the second input file has series identifier \(\texttt{obm\_spent\_value\_ge155d\_btc\_daily}\), unit \(\texttt{BTC}\), and frequency \(\texttt{daily}\). Source values must be present, numeric, and non-negative.

The selected date interval can be supplied explicitly through \texttt{\symbol{45}\symbol{45}start\_date} and \texttt{\symbol{45}\symbol{45}end\_date}. If either boundary is omitted, the script infers the interval from the common dates available in both input files. The script requires both input files to contain complete observations for every date in the selected interval. It also verifies that the selected interval uses one common \texttt{release\_version} across both source files.

This metric does not require address extraction, user clustering, entity identification, external price data, third-party APIs, a direct Bitcoin Core connection, or direct SQLite indexer access at computation time. However, it inherits the definitions, timestamp convention, and release metadata of the two source series.

\paragraph{Algorithm.}
The script \texttt{compute\_obm\_spent\_value\_lt155d\_btc\_daily.py} implements the following procedure:

\begin{enumerate}
    \item Read the total spent-value input file supplied as the first positional argument.

    \item Read the spent-value aged at least 155 days input file supplied as the second positional argument.

    \item Verify that both input files exist, are non-empty, and contain the standard OBM columns:
    \[
    \texttt{date},\quad
    \texttt{series\_id},\quad
    \texttt{value},\quad
    \texttt{unit},\quad
    \texttt{frequency},\quad
    \texttt{release\_version}.
    \]

    \item Parse all dates using the \texttt{YYYY-MM-DD} format and interpret them as UTC dates.

    \item Parse all \texttt{value} fields as decimal numbers. Missing, invalid, or negative source values cause the script to abort.

    \item Verify that neither source file contains duplicate dates.

    \item Validate that the first source file corresponds to \texttt{obm\_spent\_value\_btc\_daily}, with unit \texttt{BTC} and frequency \texttt{daily}.

    \item Validate that the second source file corresponds to \texttt{obm\_spent\_value\_ge155d\_btc\_daily}, with unit \texttt{BTC} and frequency \texttt{daily}.

    \item Infer or validate the selected date interval. If \texttt{\symbol{45}\symbol{45}start\_date} or \texttt{\symbol{45}\symbol{45}end\_date} is omitted, the script uses the first or last common date available in both input files.

    \item Verify that both source files contain one observation for every date in the selected interval.

    \item Verify that all observations in the selected interval share one common \texttt{release\_version} across both source files.

    \item For each date \(d\), compute:
    \[
    \mathrm{SpentValueLT155BTC}_d =
    \mathrm{SpentValueBTC}_d -
    \mathrm{SpentValueGE155BTC}_d.
    \]

    \item If the computed value is slightly negative but its absolute value is no larger than the configured \texttt{\symbol{45}\symbol{45}negative\_tolerance}, set it to zero. This handles negligible decimal-rounding effects.

    \item If the computed value is negative beyond the configured tolerance, abort the computation, because this indicates inconsistent source files or incompatible definitions.

    \item Write the resulting series to a CSV file using the standardized OBM schema.

    \item Optionally generate a plot when the \texttt{\symbol{45}\symbol{45}plot} flag is used.
\end{enumerate}

The script is therefore a deterministic transformation of two OBM source series. It does not query Bitcoin Core, does not maintain an outpoint state, and does not reconstruct spent outputs directly.

\paragraph{Metric-specific input parameters.}
The input parameters specific to this script are:

\begin{itemize}
    \item \texttt{spent\_value\_csv}: path to the \texttt{obm\_spent\_value\_btc\_daily} CSV file.

    \item \texttt{spent\_value\_ge155d\_csv}: path to the \texttt{obm\_spent\_value\_ge155d\_btc\_daily} CSV file.

    \item \texttt{\symbol{45}\symbol{45}start\_date}: optional starting date of the selected interval, inclusive, in \texttt{YYYY-MM-DD} format. If omitted, the first common date in both input files is used.

    \item \texttt{\symbol{45}\symbol{45}end\_date}: optional ending date of the selected interval, inclusive, in \texttt{YYYY-MM-DD} format. If omitted, the last common date in both input files is used.

    \item \texttt{\symbol{45}\symbol{45}output}: path of the output CSV file to be written. The default name of the file is \texttt{obm\_spent\_value\_lt155d\_btc\_daily.csv}.

    \item \texttt{\symbol{45}\symbol{45}plot}: optional flag that instructs the script to generate a plot of the resulting series.

    \item \texttt{\symbol{45}\symbol{45}plot\_output}: optional path for the generated plot. The default name of the file is \texttt{obm\_spent\_value\_lt155d\_btc\_daily.png}.

    \item \texttt{\symbol{45}\symbol{45}negative\_tolerance}: tolerance for tiny negative values caused by decimal rounding. Values whose absolute magnitude is no greater than this threshold are set to zero. The default is \(0.00000001\) BTC.
\end{itemize}

The script does not accept Bitcoin Core RPC parameters, \texttt{\symbol{45}\symbol{45}state\_db}, \texttt{\symbol{45}\symbol{45}height\_margin}, \texttt{\symbol{45}\symbol{45}min\_confirmations}, \texttt{\symbol{45}\symbol{45}commit\_every}, or \texttt{\symbol{45}\symbol{45}reset\_state\_db}. Those parameters belong to blockchain-scanning or indexer-backed exporters, whereas this metric is computed from existing OBM CSV files.

\paragraph{Aggregation rule.}
The daily value is computed as the difference between two daily spent-value aggregates:
\[
\mathrm{SpentValueLT155BTC}_d =
\mathrm{SpentValueBTC}_d -
\mathrm{SpentValueGE155BTC}_d.
\]

Equivalently, it is the sum of the BTC values of all spent outputs whose age is strictly less than 155 days:
\[
\mathrm{SpentValueLT155BTC}_d =
\sum_{b \in B_d}
\sum_{i \in I_b}
v_i \mathbf{1}\{a_i < 155\}.
\]

Monthly versions of this metric, if distributed, should be computed as sums of the corresponding daily values:
\[
\mathrm{SpentValueLT155BTC}_m =
\sum_{d \in m}
\mathrm{SpentValueLT155BTC}_d.
\]

Equivalently, a monthly version may be computed by subtracting the monthly summed \(\geq 155\)-day spent value from the monthly summed total spent value:
\[
\mathrm{SpentValueLT155BTC}_m =
\mathrm{SpentValueBTC}_m -
\mathrm{SpentValueGE155BTC}_m.
\]

This preserves the interpretation of the metric as a flow of BTC value over the corresponding period.

\paragraph{Output format.}
The output file contains one observation per UTC date. Each row has the following fields:

\begin{center}
\begin{tabular}{llp{0.30\textwidth}}
\toprule
\textbf{Column} & \textbf{Example} & \textbf{Description} \\
\midrule
\texttt{date} & \texttt{2024-01-01} & UTC calendar date \\
\texttt{series\_id} & \texttt{obm\_spent\_value\_lt155d\_btc\_daily} & Stable OBM series identifier \\
\texttt{value} & \texttt{370027.00000000} & BTC value of spent outputs younger than 155 days \\
\texttt{unit} & \texttt{BTC} & Measurement unit \\
\texttt{frequency} & \texttt{daily} & Observation frequency \\
\texttt{release\_version} & \texttt{OBM v0.1.0} & Dataset release version inferred from source files \\
\bottomrule
\end{tabular}
\end{center}

\paragraph{Technical validation.}
Several internal checks are used to validate this metric at computation time. First, the script verifies that both input files exist, are non-empty, and contain the required OBM schema columns. Second, it verifies that all source values are present, parseable as decimal numbers, and non-negative. Third, it verifies that neither source file contains duplicate dates. Fourth, it checks that the first input file has series identifier \texttt{obm\_spent\_value\_btc\_daily}, unit \texttt{BTC}, and frequency \texttt{daily}. Fifth, it checks that the second input file has series identifier \texttt{obm\_spent\_value\_ge155d\_btc\_daily}, unit \texttt{BTC}, and frequency \texttt{daily}. Sixth, it verifies that the two input files have at least one overlapping date. Seventh, it verifies that \texttt{\symbol{45}\symbol{45}start\_date} is not later than \texttt{\symbol{45}\symbol{45}end\_date}. Eighth, it verifies that both input files contain complete observations for every date in the selected interval. Ninth, it verifies that the selected interval uses one common \texttt{release\_version} across both source files.

For each date in the selected interval, the script computes:
\[
\mathrm{SpentValueLT155BTC}_d =
\mathrm{SpentValueBTC}_d -
\mathrm{SpentValueGE155BTC}_d.
\]
If the result is negative only within the configured \texttt{\symbol{45}\symbol{45}negative\_tolerance}, the value is set to zero. If the result is negative beyond that tolerance, the script aborts. This prevents a silent combination of inconsistent source files or incompatible definitions.

The primary validation identity is:
\[
\mathrm{SpentValueBTC}_d =
\mathrm{SpentValueLT155BTC}_d +
\mathrm{SpentValueGE155BTC}_d.
\]

This identity should hold for every date when all involved series share the same timestamp convention, threshold definition, and release version. Comparisons with external short-term-holder or young-coin spent-value metrics are useful as diagnostics but should not be interpreted as strict equality tests, because providers may differ in timestamp conventions, age-threshold definitions, entity adjustment, change-output treatment, transfer-value definition, and historical edge-case handling.

\paragraph{Known limitations.}
The younger-than-155-days spent-value series is useful but definition-sensitive. First, \texttt{obm\_spent\_value\_lt155d\_btc\_daily} is a derived metric and therefore depends on the correctness, completeness, and compatibility of both source CSV files. Second, it inherits the UTC block-timestamp convention of the source series. Third, it inherits the 155-day threshold convention used by the \(\geq 155\)-day source series. Fourth, it is a raw, spent-output metric, not an entity-adjusted short-term holder activity measure. It does not identify users, entities, custodians, exchanges, self-transfers, or change outputs. Fifth, it can be affected by wallet consolidation, batching, exchange operations, custodial wallet management, and self-transfers. Sixth, it reports BTC value, not fiat-denominated value. Seventh, if either the source series definition or the release version changes, this derived series must be recomputed.

Despite these limitations, \texttt{obm\_spent\_value\_lt155d\_btc\_daily} is a useful OBM-derived series. It provides the young-output counterpart to the 155-day threshold spent-value series and supports decomposition of daily spent value into younger and older spent-output components.
\subsection{\texttt{obm\_supply\_btc\_daily}: Bitcoin Supply}
\label{Bitcoin Supply}


\paragraph{Definition.}
The Bitcoin supply series measures the cumulative amount of newly issued BTC from the beginning of the OBM daily calendar, \texttt{2009-01-01}, to a user-specified ending date. This convention provides a complete UTC daily series from the start of Bitcoin's calendar history; dates before the genesis block contribute zero issuance when the source issuance series is complete. It is derived directly from the daily realized issuance series \texttt{obm\_issuance\_btc\_daily}. Let \(\mathrm{Issuance}_d\) denote the realized Bitcoin issuance on UTC calendar day \(d\), as defined in the previous subsection. From \(s=\mathrm{2009-01-01}\) to the selected ending date \(e\), with both dates included, supply at date \(d\) is defined as:

\[
\mathrm{Supply}_{d;s}
=
\sum_{k=s}^{d} \mathrm{Issuance}_k,
\qquad s \leq d \leq e.
\]

The subscript \(s\) emphasizes that this is an interval-specific cumulative series. The cumulative value is reset at \texttt{2009-01-01}. Therefore, $\mathrm{Supply}_{s;s} = \mathrm{Issuance}_s$.

Since this metric intends to reflect the total circulating Bitcoin supply, the selected starting date is fixed to \texttt{2009-01-01}. In OBM, \texttt{obm\_supply\_btc\_daily} is computed as a cumulative sum of realized daily issuance and is reported as an end-of-day supply stock when the accumulation starts from the beginning of the OBM daily calendar.

\paragraph{Economic interpretation.}
Accumulated Bitcoin issuance measures how many new bitcoins have been created over a selected research window. It is useful when the researcher is interested not in the daily flow of newly issued BTC, but in the total quantity issued since the genesis block.

For economic research, this metric is useful in several settings. First, it allows researchers to measure realized issuance over specific monetary regimes, such as pre-halving and post-halving windows. Second, it supports comparisons of cumulative issuance across subperiods with different block subsidy levels. Third, it can be used to construct interval-specific supply-flow variables or to describe the cumulative monetary expansion associated with a selected sample period. Fourth, it provides a simple bridge between the daily issuance flow and stock-like monetary variables.

The metric should be interpreted as a cumulative transformation of \texttt{obm\_issuance\_btc\_daily}. It does not introduce new information beyond the daily issuance series, but it makes explicit a transformation that is frequently needed in descriptive and econometric applications.

\paragraph{Similar metrics publicly available.}
The OBM series \texttt{obm\_supply\_btc\_daily} is comparable to public metrics usually labeled \emph{current supply}, \emph{circulating supply}, \emph{total bitcoins}, or \emph{total issued supply}. However, the comparison requires care. In OBM, supply is a cumulative-flow variable derived from the daily realized issuance series \texttt{obm\_issuance\_btc\_daily}. The cumulative value is set at the Bitcoin genesis block date, so this metric should be interpreted as total circulating supply. The main validation identity is
\[
\mathrm{Supply}_{d;s}
-
\mathrm{Supply}_{d-1;s}
=
\mathrm{Issuance}_d,
\qquad d>s.
\]

The closest public comparator is Coin Metrics' \emph{Current Supply}, with MetricID \texttt{SplyCur}.\footnote{\url{https://community-api.coinmetrics.io/v4/timeseries/asset-metrics?assets=btc\&metrics=SplyCur}} Coin Metrics describes current supply as the sum of all unspent output values for UTXO chains and notes that the metric can also be characterized as ``total issued supply'', because it captures the sum of all native units visible in the ledger up to the metric calculation point. 

Since \texttt{obm\_supply\_btc\_daily} accumulation starts at the beginning of Bitcoin history and uses a daily realized issuance series that captures all coinbase-created supply, \texttt{obm\_supply\_btc\_daily} should approximate a supply-stock series such as \texttt{SplyCur}. Differences may arise from timestamp conventions, treatment of reorganizations, underclaimed coinbase rewards, unspendable outputs, and the distinction between summing realized issuance and summing currently unspent output values.

Glassnode provides a \emph{Bitcoin Circulating Supply} metric, exposed as \texttt{supply.Current}\footnote{\url{https://studio.glassnode.com/charts/supply.Current?a=BTC}}, that is directly related to ours. Glassnode defines it as the total amount of all coins ever created or issued, that is, the circulating supply. This makes it a close public reference for full-history accumulated issuance. However, it does not fully specify whether the value is reconstructed from actual coinbase outputs, inferred from the protocol schedule, adjusted for underclaimed rewards, or treated under a specific timestamp and reorganization convention. Consequently, it is a useful benchmark for full-history accumulation, but might not be an exact equivalent of \texttt{obm\_supply\_btc\_daily}.

Blockchain.com provides a \emph{Total Circulating Bitcoin} chart.\footnote{\url{https://www.blockchain.com/charts/total-bitcoins}} The chart shows how many bitcoins have been mined or put in circulation. Its methodology states that the number of bitcoins in circulation is calculated from the theoretical reward defined by the Bitcoin protocol. This is a valuable public reference for the theoretical supply path, but it is not equivalent to \texttt{obm\_supply\_btc\_daily}. Our metric accumulates the \textit{realized} issuance flow derived from Coinbase outputs after subtracting transaction fees, whereas Blockchain.com's methodology is explicitly based on the theoretical reward schedule. Therefore, Blockchain.com is useful for broad validation against the protocol-implied supply path, but not as a strict benchmark for realized supply.

YCharts republishes a \emph{Bitcoin Supply} series sourced from Blockchain.com.\footnote{\url{https://ycharts.com/indicators/bitcoin\_supply}} This is a stock series rather than an accumulated-flow series. It can be used as a secondary comparator, but it inherits the methodological limitations of the underlying Blockchain.com supply series. It should therefore not be treated as an independent reproducible implementation of accumulated realized issuance.

Newhedge provides a \emph{Bitcoin Circulating Supply} page.\footnote{\url{https://newhedge.io/bitcoin/circulating-supply}} The page reports circulating supply, percentage issued, left to mine, total mined blocks, and daily issuance. These quantities are related to accumulated issuance because circulating supply can be interpreted as a full-history cumulative issuance stock under a given convention. However, the public page does not disclose a complete algorithm for reconstructing the cumulative quantity from block-level coinbase data, and some API features appear to require a paid plan. It is therefore useful as a public diagnostic reference, but not as a fully auditable comparator.

Blockchair provides a \emph{Bitcoin circulation} chart.\footnote{\url{https://blockchair.com/bitcoin/charts/circulation}} This is a supply-stock series rather than a named accumulated-issuance flow. Blockchair also provides block-level and transaction-level data, so a researcher could, in principle, reconstruct a cumulative issuance series by processing Coinbase transactions and fees. Nevertheless, the public circulation chart itself should be treated as a stock comparator rather than as an explicit implementation of \texttt{obm\_supply\_btc\_daily}.

MacroMicro provides a daily issuance chart\footnote{\url{https://en.macromicro.me/charts/29069/bitcoin-issuance}} (sign-in required). This is not a cumulative-issuance metric, but it can be used as a public reference. A cumulative version could be produced by summing the MacroMicro daily issuance series over a selected interval. However, the public documentation is chart-oriented and does not specify whether the daily values are based on realized coinbase outputs, block count multiplied by subsidy, or another provider-specific convention. Any accumulated series derived from it would therefore inherit those undocumented choices.

Overall, the closest public comparators for \texttt{obm\_supply\_btc\_daily} are Coin Metrics \texttt{SplyCur} and Glassnode \texttt{supply.Current}, because both represent supply stocks that correspond conceptually to full-history supply. Blockchain.com, YCharts, Newhedge, and Blockchair provide additional supply-stock references, while MacroMicro provides a daily issuance flow that could be accumulated externally. The main distinction is that \texttt{obm\_supply\_btc\_daily} defines supply as a transparent cumulative sum of realized daily issuance, with inherited release version, and a validation identity based on first differences. Public supply charts are useful comparators, but they may rely on theoretical reward schedules or undocumented provider conventions rather than on an auditable accumulation of realized issuance.

\paragraph{Data source and input requirements.}
The metric is derived from the OBM daily realized issuance file \texttt{obm\_issuance\_btc\_daily.csv}. Unlike \texttt{obm\_issuance\_btc\_daily}, this series does not require direct access to Bitcoin Core, block metadata, coinbase transactions, or fee reconstruction. It only requires a valid OBM-compatible input CSV file containing the daily issuance series.

The input file must follow the standard OBM schema:

\[
\texttt{date},\quad
\texttt{series\_id},\quad
\texttt{value},\quad
\texttt{unit},\quad
\texttt{frequency},\quad
\texttt{release\_version}.
\]

The script verifies that the input series identifier is \texttt{obm\_issuance\_btc\_daily}, that the unit is BTC, and that the frequency is daily. It also ensures that all required dates for the selected interval are present.

\paragraph{Algorithm.}
The script \texttt{compute\_obm\_supply\_btc\_daily.py} implements the following procedure:

\begin{enumerate}
    \item Read the input CSV file provided as the mandatory positional argument. This file is expected to contain the daily realized issuance series \texttt{obm\_issuance\_btc\_daily}.

    \item Parse the user-provided date ending date, \texttt{\symbol{45}\symbol{45}end\_date}, using the format \texttt{YYYY-MM-DD}. Both dates are interpreted as UTC and included in the output.

    \item Validate the input schema. The script checks that the input file contains the fields:
    \[
    \texttt{date},\quad
    \texttt{series\_id},\quad
    \texttt{value},\quad
    \texttt{unit},\quad
    \texttt{frequency},\quad
    \texttt{release\_version}.
    \]

    \item Validate that the input series identifier is \texttt{obm\_issuance\_btc\_daily}, that the unit is \texttt{BTC}, and that the frequency is \texttt{daily}.

    \item Select all observations between \texttt{2009-01-01} and \texttt{\symbol{45}\symbol{45}end\_date}, inclusive.

    \item Verify that every date in the requested interval has exactly one corresponding daily issuance observation. If any date is missing, the script aborts.

    \item Initialize the cumulative value at zero before the starting date (\texttt{2009-01-01}).

    \item Iterate through the selected dates in chronological order. For each date \(d\), add the daily issuance value \(\mathrm{Issuance}_d\) to the running cumulative total: $\mathrm{Supply}_{d;s} = \mathrm{Supply}_{d-1;s} +\mathrm{Issuance}_d$, with the convention that the running total before date \(s\) is zero.

    \item Infer the \texttt{release\_version} from the source observations in the selected interval. If multiple release versions are present in the selected interval, the script aborts to avoid mixing releases.

    \item Write the resulting time series to a CSV file using the standardized OBM schema:
    \[
    \texttt{date},\quad
    \texttt{series\_id},\quad
    \texttt{value},\quad
    \texttt{unit},\quad
    \texttt{frequency},\quad
    \texttt{release\_version}.
    \]

    \item Optionally generate a plot of the accumulated series when the plotting flag is activated. The plot title includes the series description and the selected date interval.
\end{enumerate}

\paragraph{Metric-specific input parameters.}
The input parameter specific to this metric is:

\begin{itemize}
     \item \texttt{\symbol{45}\symbol{45}end\_date}: ending date of the accumulation interval, inclusive.
\end{itemize}

The script also accepts \texttt{\symbol{45}\symbol{45}output}, \texttt{\symbol{45}\symbol{45}plot}, and \texttt{\symbol{45}\symbol{45}plot\_output}, following the same output and plotting conventions used by the other OBM scripts. The source file \texttt{obm\_issuance\_btc\_daily.csv} is provided as a mandatory positional argument.

In Bitcoin Core, RPC parameters are not required because this metric is not reconstructed directly from the blockchain. It is a deterministic transformation of an already generated OBM time series.

\paragraph{Aggregation rule.}
The daily value is computed as the running cumulative sum of daily realized issuance from $s=$\texttt{2009-01-01}:

\[
\mathrm{Supply}_{d;s}
=
\sum_{k=s}^{d} \mathrm{Issuance}_k.
\]

Equivalently, the recursive form is $\mathrm{Supply}_{d;s} = \mathrm{Supply}_{d-1;s} + \mathrm{Issuance}_d,
\qquad d > s$, with $\mathrm{Supply}_{s;s} = \mathrm{Issuance}_s$. The aggregation rule is therefore an interval-specific cumulative sum. If monthly versions are distributed, they should be defined carefully. One possibility is to compute the daily accumulated series first and then report the end-of-month accumulated value. Another possibility is to compute monthly issuance flows and accumulate them from the selected starting month. The chosen convention should be documented explicitly.

\paragraph{Relationship with \texttt{obm\_issuance\_btc\_daily}.}
The series \texttt{obm\_supply\_btc\_daily} is directly derived from \texttt{obm\_issuance\_btc\_daily}. The latter measures the daily flow of newly created BTC, whereas the former measures the cumulative sum of those daily flows over a selected interval.

This relationship implies the following consistency condition: $\mathrm{Supply}_{d;s} - \mathrm{Supply}_{d-1;s} = \mathrm{Issuance}_d$, with $d > s$. Therefore, first differences of \texttt{obm\_supply\_btc\_daily} must reproduce the corresponding observations of \texttt{obm\_issuance\_btc\_daily}. This identity is the main internal validation check for the series.

\paragraph{Output format.}
The output file contains one observation per UTC date in the selected interval. Each row has the following fields:

\begin{center}
\begin{tabular}{llp{0.45\textwidth}}
\toprule
\textbf{Column} & \textbf{Example} & \textbf{Description} \\
\midrule
\texttt{date} & \texttt{2024-01-01} & UTC calendar date \\
\texttt{series\_id} & \texttt{obm\_supply\_btc\_daily} & Stable OBM series identifier \\
\texttt{value} & \texttt{918.75000000} & Accumulated issuance since \texttt{2009-01-01} \\
\texttt{unit} & \texttt{BTC} & Measurement unit \\
\texttt{frequency} & \texttt{daily} & Observation frequency \\
\texttt{release\_version} & \texttt{OBM v0.1.0} & Dataset release version inherited from the source interval \\
\bottomrule
\end{tabular}
\end{center}

\paragraph{Technical validation.}
Several checks are either implemented by the script or recommended for validating this metric. First, the script verifies that the requested date range is valid and that the ending date is no sooner than \texttt{2009-01-01}. Second, it checks that the input file follows the standard OBM schema. Third, it verifies that the input file corresponds to \texttt{obm\_issuance\_btc\_daily}, with unit \texttt{BTC} and frequency \texttt{daily}. Fourth, it checks that every date in the requested interval is present exactly once. Fifth, it verifies that all observations in the selected interval belong to a single release version. Sixth, the script rejects negative daily issuance values, which ensures that the resulting accumulated series is non-decreasing.

The primary validation identity is $\Delta \mathrm{Supply}_{d;s} = \mathrm{Issuance}_d$, for all dates \(d > s\). This identity can be checked by differencing the accumulated series and comparing the result with the source daily issuance series.

\paragraph{Known limitations.}
The supply series is simple and reproducible, but it has several limitations. First, it is a derived series and not an independent full-node reconstruction. Second, it inherits all definitional choices and potential revisions of \texttt{obm\_issuance\_btc\_daily}. Third, it resets at \texttt{2009-01-01}, so its value depends on this fixed starting date. Fourth, it should not be interpreted as the total circulating supply unless the source series covers the full issuance history. Fifth, if the underlying daily issuance file is revised in a later OBM release, the accumulated series should be regenerated from the revised source file.

Despite these limitations, \texttt{obm\_supply\_btc\_daily} is a useful derived OBM series. It provides a convenient measure of cumulative realized issuance over arbitrary research intervals and supports monetary-supply, halving-period, and supply-growth analysis.

\subsection{\texttt{obm\_tx\_count\_daily}: Daily Transaction Count}
\label{Daily Transaction Count}


\paragraph{Definition.}
The daily transaction count measures the number of Bitcoin transactions confirmed in blocks whose timestamps fall within a given UTC calendar day. Let \(B_d\) denote the set of blocks assigned to day \(d\), and let \(N_b\) denote the number of transactions included in block \(b\). The daily transaction count is defined as:

\[
\mathrm{TxCount}_d =
\sum_{b \in B_d} N_b.
\]

A block \(b\) is assigned to day \(d\) according to the UTC calendar date derived from its block timestamp \(t_b\): $d(b) = \mathrm{UTCDate}(t_b)$.

The resulting series therefore reports the total number of confirmed transactions per UTC day. Coinbase transactions are included in the count because they are valid transactions contained in blocks and are part of the block-level transaction total reported by Bitcoin Core. This convention also makes the metric directly reproducible from block metadata.

\paragraph{Economic interpretation.}
The transaction count is a basic proxy for Bitcoin network activity. It captures the number of transactions settled on-chain each day, regardless of transaction size, economic value, fees paid, or the number of inputs and outputs. For economic research, this metric is useful as a broad indicator of transaction demand, settlement activity, and network usage. It is especially useful when combined with other series, such as total fees, block weight, spent output value, or Bitcoin Days Destroyed, because each captures a different dimension of on-chain activity.

The metric should not be interpreted as a direct count of users or payments. A single transaction may aggregate multiple inputs, create multiple outputs, represent self-transfer activity, or correspond to the internal operations of exchanges, custodians, or other intermediaries. Conversely, many economic payments can also be represented off-chain and therefore not appear as separate Bitcoin transactions. The series is therefore best interpreted as a count of confirmed on-chain transactions, not as a count of users, purchases, or economically distinct transfers.

\paragraph{Similar metrics publicly available.}
The OBM series \texttt{obm\_tx\_count\_daily} is comparable to public metrics usually labeled \emph{transaction count}, \emph{transactions per day}, \emph{confirmed transactions per day}, or \emph{number of transactions in blockchain per day}. In OBM, the metric is defined as the number of Bitcoin transactions confirmed in blocks assigned to a given UTC calendar day. Coinbase transactions are included because they are valid transactions contained in blocks and are part of the block-level transaction count reported by Bitcoin Core. This convention makes the metric directly reproducible from block metadata, using the \texttt{nTx} field when available or, equivalently, the length of the decoded block transaction array.

The closest public comparator is Coin Metrics' \emph{Tx Cnt}, with MetricID \texttt{TxCnt}.\footnote{\url{https://community-api.coinmetrics.io/v4/timeseries/asset-metrics?assets=btc\&metrics=TxCnt}} Coin Metrics defines this metric as the sum count of transactions during the interval and makes it available as a native transaction-count series. 

This is a strong benchmark for \texttt{obm\_tx\_count\_daily}, because it is a named transaction-count metric with explicit interval-based aggregation. However, exact equality should not be assumed without checking the treatment of coinbase transactions, timestamp assignment, chain reorganizations, and data revision policy. Coin Metrics also provides related adjusted or category-specific transaction metrics, such as exchange transaction count, but these should not be confused with the raw transaction-count series.

Glassnode provides transaction-count metrics for Bitcoin, including daily confirmed transaction-count series and entity-adjusted variants. Its documentation describes transaction counts as measuring the number of confirmed transactions each day, including all transaction types, such as payments, exchange deposits and withdrawals, custodial operations, and Lightning channel openings or closings. This raw confirmed-transaction concept is close to \texttt{obm\_tx\_count\_daily}. Glassnode also provides an entity-adjusted transaction-count metric, exposed with the name of \texttt{transactions.EntityAdjustedCount}, which estimates the number of transactions between different entities by excluding transactions within addresses controlled by the same entity\footnote{\url{https://studio.glassnode.com/charts/transactions.EntityAdjustedCount?a=BTC}} (Professional plan required). The entity-adjusted series is economically interesting, but it is not a direct comparator for \texttt{obm\_tx\_count\_daily} because it relies on proprietary clustering heuristics. Glassnode's raw transaction-count series is a closer conceptual benchmark, whereas the entity-adjusted version should be treated as a derived, interpretation-oriented alternative.

Blockchain.com provides a \emph{Confirmed Transactions Per Day} chart.\footnote{\url{https://www.blockchain.com/charts/n-transactions}} The page describes the series as the number of daily confirmed transactions and highlights its use as a measure of Bitcoin network activity. Blockchain.com also provides a Charts and Statistics API.\footnote{\url{https://www.blockchain.com/api/charts\_api}} The API documentation states that date parameters are represented in UTC and that chart data can be requested programmatically. This makes Blockchain.com one of the most useful public comparators for the daily transaction count. However, the public chart page does not fully specify low-level implementation conventions, such as coinbase inclusion, reorganization handling, or whether all historical revisions are preserved.

Blockchair provides a \emph{Bitcoin transaction count} chart.\footnote{\url{https://blockchair.com/bitcoin/charts/transaction-count}} The page describes the metric as the number of confirmed Bitcoin transactions. Blockchair also provides a cumulative transaction-count chart.\footnote{\url{https://blockchair.com/bitcoin/charts/total-transaction-count}} The daily chart is conceptually close to \texttt{obm\_tx\_count\_daily}, while the cumulative chart is related but not equivalent. Blockchair is also useful because it exposes block-level and transaction-level explorer data, so \texttt{obm\_tx\_count\_daily} could, in principle, be reconstructed externally by summing block-level transaction counts by date. Nevertheless, the public chart description is concise and does not provide a full, reproducible algorithm that specifies timestamp conventions, coinbase treatment, or reorganization policy.

Bitbo provides a \emph{Bitcoin Transactions Per Day} chart.\footnote{\url{https://charts.bitbo.io/tx-per-day/}} The page states that the chart displays the daily number of Bitcoin transactions processed on the network. It also notes that the default chart shows a 30-day moving average, while raw daily transaction data can also be viewed. This distinction matters for comparison with \texttt{obm\_tx\_count\_daily}: the raw daily view is the relevant comparator, whereas the moving-average display is a smoothed derivative. Bitbo is useful for visual inspection and broad cross-checking, but the public page is not a full algorithmic specification.

Newhedge provides a \emph{Bitcoin Transactions per Day} chart.\footnote{\url{https://newhedge.io/bitcoin/transactions-per-day}} The page defines the metric as the daily number of confirmed transactions on the Bitcoin network. It also offers image and API access, although full access appears to require registration. The definition is directly aligned with \texttt{obm\_tx\_count\_daily}, but the public page does not disclose a full reconstruction algorithm, nor does it specify the treatment of coinbase transactions, block timestamp conventions, or data revisions.

BitInfoCharts provides a \emph{Bitcoin Transactions} historical chart.\footnote{\url{https://bitinfocharts.com/comparison/bitcoin-transactions.html}} The chart is described as the number of transactions in the blockchain per day. BitInfoCharts also reports current 24-hour transaction statistics.\footnote{\url{https://bitinfocharts.com/bitcoin/}} These series are close public comparators for \texttt{obm\_tx\_count\_daily}, but their documentation is largely descriptive. The public pages do not provide a reproducible algorithm or clarify low-level choices such as coinbase inclusion, daily boundary assignment, or reorganization handling.

YCharts provides a \emph{Bitcoin Transactions Per Day} series.\footnote{\url{https://ycharts.com/indicators/bitcoin\_transactions\_per\_day}} YCharts reports current, previous-day, and historical values, and its indicator page identifies the series as a daily Bitcoin transaction-count measure. It is useful as an accessible secondary source, especially because it provides a long historical chart. However, it should be treated mainly as a republished financial data indicator rather than as a transparent blockchain reconstruction methodology. The public page does not provide the low-level algorithm used to compute the transaction counts.

Token Terminal provides a \emph{Bitcoin Transaction Count} metric.\footnote{\url{https://tokenterminal.com/explorer/projects/bitcoin/metrics/transaction-count}} The page defines the Bitcoin transaction-count metric as the total number of unique transactions processed by the Bitcoin blockchain on a daily basis. This is conceptually related to \texttt{obm\_tx\_count\_daily}. However, Token Terminal's cross-chain analytics framing and the use of the phrase ``unique transactions'' make it less directly comparable unless its Bitcoin-specific methodology is examined in detail. The public description is not sufficient to determine exact equivalence with \texttt{obm\_tx\_count\_daily}.

Overall, the strongest public comparators for \texttt{obm\_tx\_count\_daily} are Coin Metrics \texttt{TxCnt}, Blockchain.com's \emph{Confirmed Transactions Per Day}, Glassnode's raw transaction-count series, and Blockchair's \emph{transaction count} chart. Bitbo, Newhedge, BitInfoCharts, YCharts, and Token Terminal provide useful secondary references. The main distinction is that \texttt{obm\_tx\_count\_daily} provides an explicitly reproducible full-node implementation: it counts transactions directly from block metadata, includes coinbase transactions by definition, assigns blocks to UTC days using the block timestamp returned by Bitcoin Core, and documents the height-margin procedure used to avoid missing boundary blocks caused by non-monotonic block timestamps.

\paragraph{Data source and input requirements.}
The metric is obtained from a running Bitcoin Core full node through the JSON-RPC interface. For each block, the script queries Bitcoin Core to obtain the block hash and then retrieves the decoded block object. The number of transactions in the block is read from the \texttt{nTx} field in the block metadata when available. If this field is not available, the script falls back to counting the number of elements in the block's \texttt{tx} array.

This metric does not require reconstructing previous transaction outputs, accessing the UTXO set, extracting addresses, parsing transaction values, or external price data. Consequently, it is among the simplest and most robust OBM metrics. It can be computed from block-level information alone.

\paragraph{Algorithm.}
The script \texttt{compute\_obm\_tx\_count\_daily.py} implements the following procedure:

\begin{enumerate}
    \item Parse the user-provided date interval, \texttt{\symbol{45}\symbol{45}start\_date} and \texttt{\symbol{45}\symbol{45}end\_date}, using the format \texttt{YYYY-MM-DD}. Both dates are interpreted as UTC dates, and both are included in the output.

    \item Convert the starting date into the timestamp corresponding to 00:00:00 UTC of that day, and the ending date into the timestamp corresponding to 23:59:59 UTC of that day.

    \item Query the local Bitcoin Core node using \texttt{getblockchaininfo} to determine the current best chain height.

    \item Use \texttt{getblockhash} and \texttt{getblock} to retrieve block timestamps and locate an approximate height interval covering the requested date range. This is done through a binary search over block heights.

    \item Expand the approximate height interval using the metric-specific safety parameter, called \texttt{\symbol{45}\symbol{45}height\_margin}. This margin is added before the estimated starting height and after the estimated ending height.

    \item Scan all blocks in the expanded height interval. For each height \(h\), the script:
    \begin{enumerate}
        \item obtains the corresponding block hash using \texttt{getblockhash};
        \item retrieves the decoded block object using \texttt{getblock};
        \item extracts the block timestamp \(t_b\);
        \item assigns the block to a UTC calendar date \(d(b)\);
        \item extracts the number of transactions \(N_b\), preferably from \texttt{nTx};
        \item adds \(N_b\) to the daily total if \(d(b)\) lies between the requested starting and ending dates.
    \end{enumerate}

    \item Initialize all dates in the requested interval with a value of zero before scanning. This ensures that the output file contains one row for every calendar day in the requested interval, even if no block was assigned to a particular day.

    \item Write the resulting time series to a CSV file using the standardized OBM schema:
    \[
    \texttt{date},\quad
    \texttt{series\_id},\quad
    \texttt{value},\quad
    \texttt{unit},\quad
    \texttt{frequency},\quad
    \texttt{release\_version}.
    \]

    \item Optionally generate a plot of the resulting series when the plotting flag is activated. The plot title includes the series description and the selected date interval.
\end{enumerate}

\paragraph{Metric-specific input parameter.}
The only input parameter specific to this metric is:

\begin{itemize}
    \item \texttt{\symbol{45}\symbol{45}height\_margin}: number of extra blocks scanned before and after the approximate height interval associated with the requested date range.
\end{itemize}

This parameter is needed because Bitcoin block timestamps are not strictly monotonic with respect to block height. Although block heights are strictly ordered by chain position, miners' timestamps can occasionally shift slightly backward or forward relative to neighboring blocks. As a result, a binary search based only on block timestamps could identify a height interval that is too narrow and accidentally exclude blocks whose timestamps fall inside the requested UTC date interval.

To avoid this problem, the script first obtains an approximate height interval and then scans an expanded interval $[h_{\min}^{\mathrm{scan}}, h_{\max}^{\mathrm{scan}}]$, where:

\[
h_{\min}^{\mathrm{scan}}
=
\max(0, h_{\min}^{\mathrm{approx}} - m),
\]
\[
h_{\max}^{\mathrm{scan}}
=
\min(h_{\mathrm{tip}}, h_{\max}^{\mathrm{approx}} + m),
\]

and \(m\) is the value of \texttt{\symbol{45}\symbol{45}height\_margin}. The default value is 288 blocks, approximately two days of Bitcoin block production at the expected rate of 144 blocks per day. This choice is deliberately conservative. The margin does not change the dates written to the output file; it only widens the internal block scan. During the scan, the script still counts only blocks whose UTC dates fall between \texttt{\symbol{45}\symbol{45}start\_date} and \texttt{\symbol{45}\symbol{45}end\_date}, inclusive.

A larger value of \texttt{\symbol{45}\symbol{45}height\_margin} increases robustness at the cost of scanning more blocks. A smaller value improves speed but increases the risk of missing boundary blocks in unusual timestamp configurations. For ordinary daily updates, the default value is sufficient. For full historical reconstruction or formal archival releases, a larger margin can be used as an additional safeguard. OBM repository uses the default value.

\paragraph{Aggregation rule.}
The daily value is computed as the sum of the number of transactions in all blocks assigned to the same UTC calendar date:

\[
\mathrm{TxCount}_d =
\sum_{b: \mathrm{UTCDate}(t_b)=d} N_b.
\]

The aggregation rule is therefore a daily sum. Monthly versions of this metric should also be computed as sums of the corresponding daily values:

\[
\mathrm{TxCount}_m =
\sum_{d \in m} \mathrm{TxCount}_d.
\]

\paragraph{Output format.}
The output file contains one observation per UTC date. Each row has the following fields:

\begin{center}
\begin{tabular}{llp{0.4\textwidth}}
\toprule
\textbf{Column} & \textbf{Example} & \textbf{Description} \\
\midrule
\texttt{date} & \texttt{2024-01-01} & UTC calendar date \\
\texttt{series\_id} & \texttt{obm\_tx\_count\_daily} & Stable OBM series identifier \\
\texttt{value} & \texttt{731566} & Number of confirmed transactions \\
\texttt{unit} & \texttt{transactions} & Measurement unit \\
\texttt{frequency} & \texttt{daily} & Observation frequency \\
\texttt{release\_version} & \texttt{OBM v0.1.0} & Dataset release version \\
\bottomrule
\end{tabular}
\end{center}

\paragraph{Technical validation.}
Several internal checks are used to validate this metric. First, the script verifies that the requested date range is valid and that the starting date is not later than the ending date. Second, it checks that the requested interval is not entirely beyond the current chain tip. Third, the output is initialized with all calendar dates in the requested interval, which helps detect missing observations. Fourth, the daily values can be independently checked by summing the block-level transaction counts of all blocks assigned to the same UTC day. Finally, the number of scanned blocks and the effective height interval are printed during execution, allowing the user to verify that the requested period has been covered.

For additional validation, the daily transaction count can be compared with external Bitcoin data providers over overlapping periods. Such comparisons should be interpreted as diagnostic rather than definitive, because providers may differ in timestamp conventions, treatment of coinbase transactions, reorganization handling, or data revision policies.

\paragraph{Known limitations.}
The transaction count is simple and reproducible, but it has several limitations. First, it measures transactions, not users or economically distinct payments. Second, it does not distinguish between small and large transactions. Third, it does not account for batching, change outputs, self-transfers, exchange operations, or off-chain settlement. Fourth, the assignment of blocks to days depends on the block timestamp convention. In OBM, the default convention is to use the UTC date derived from the block timestamp returned by Bitcoin Core. This convention is transparent and reproducible, but alternative conventions, such as median time past, could produce small differences near daily boundaries.

Despite these limitations, \texttt{obm\_tx\_count\_daily} is one of the most useful baseline series in the OBM dataset. It provides a transparent measure of confirmed on-chain activity and serves as a natural reference variable for interpreting more complex metrics, such as fees, block-space utilization, spent output value, dormancy, and Bitcoin Days Destroyed.

\subsection{\texttt{obm\_utxo\_eod\_count\_daily}: End-of-Day UTXO Count}
\label{End-of-Day UTXO Count}

\paragraph{Definition.}
The end-of-day UTXO count series reports the number of spendable unspent transaction outputs after processing the highest-height block assigned to a given UTC calendar day. This is a protocol-state metric, not a daily flow. Let \(B_d\) denote the set of blocks assigned to day \(d\). For each block \(b \in B_d\), let \(h_b\) denote its height, and let \(\mathrm{UTXOCountAfter}_b\) denote the number of spendable UTXOs after processing block \(b\). The end-of-day UTXO count is defined as:
\[
\mathrm{UTXOCountEOD}_d =
\mathrm{UTXOCountAfter}_{b^\ast},
\]
where
\[
b^\ast =
\arg\max_{b \in B_d} h_b.
\]
That is, \(b^\ast\) is the highest-height block among all blocks assigned to UTC date \(d\).

A block \(b\) is assigned to day \(d\) according to the UTC calendar date derived from its block timestamp \(t_b\):
\[
d(b)=\mathrm{UTCDate}(t_b).
\]

If no block is assigned to date \(d\), the series is undefined for that date and the output value is recorded as \texttt{NaN}. This convention avoids assigning a misleading value to a state variable for a date with no end-of-day block under the selected timestamp convention.

The unit of measurement is \texttt{outputs}. The series identifier places \texttt{count} after \texttt{eod} because the metric measures the count of UTXOs under an end-of-day observation convention:
\[
\texttt{obm\_utxo\_eod\_count\_daily}.
\]

\paragraph{Economic interpretation.}
The UTXO set is the collection of unspent outputs that define the spendable state of the Bitcoin ledger. Its size is relevant for scalability, node resource requirements, and the long-run evolution of Bitcoin's state burden. A growing UTXO set means that full nodes must maintain a larger set of spendable outputs, whereas a declining UTXO count indicates that spending activity is consolidating more outputs than it creates.

For economic and empirical research, this metric is useful as a network-state variable. It complements transaction count, spent-output count, block weight, and raw output value by capturing the cumulative effect of output creation and spending. Periods of rapid UTXO growth may indicate output splitting, increased address and wallet fragmentation, a high number of small outputs, or activity patterns that increase the state burden. Periods of declining UTXO count may indicate consolidation episodes.

The metric should not be interpreted as a count of users, addresses, wallets, entities, or economically meaningful agents. A single user can control many UTXOs, and a single entity can create or spend a very large number of outputs. The metric also does not measure the BTC value held in the UTXO set. It measures the number of unspent outputs, not their monetary value.

\paragraph{Similar metrics publicly available.}
The OBM series \texttt{obm\_utxo\_eod\_count\_daily} is comparable to public metrics usually labelled \emph{UTXO count}, \emph{UTXO set count}, \emph{Unspent Transaction Outputs}, or \emph{UTXO set size}. In OBM, the metric is defined as the number of spendable unspent transaction outputs after processing the highest-height block assigned to a given UTC calendar day. Let \(B_d\) denote the set of blocks assigned to UTC date \(d\), and let \(\mathrm{UTXOCountAfter}_b\) denote the number of UTXOs after block \(b\) has been processed. OBM defines
\[
b^\ast = \arg\max_{b \in B_d} h_b,
\]
and
\[
\mathrm{UTXOCountEOD}_d = \mathrm{UTXOCountAfter}_{b^\ast}.
\]
If no block is assigned to date \(d\), the value is undefined and is written as \texttt{NaN}. The metric is therefore an end-of-day protocol-state variable, not a daily flow. The script updates the count by adding non-provably-unspendable outputs created in each block and subtracting one UTXO for each non-coinbase transaction input. Coinbase outputs are included when they are not provably unspendable, even before they mature, because they are members of the UTXO set. Standard \texttt{OP\_RETURN} or \texttt{nulldata} outputs are excluded because they are provably unspendable.

The closest public comparator is Blockchain.com's \emph{Unspent Transaction Outputs} chart:
\[
\texttt{https://www.blockchain.com/charts/utxo-count}.
\]
Blockchain.com defines the series as the total number of valid unspent transaction outputs and explicitly states that it excludes invalid UTXOs with opcode \texttt{OP\_RETURN}. This is conceptually very close to \texttt{obm\_utxo\_eod\_count\_daily}, because both count valid unspent outputs and exclude provably unspendable outputs. However, exact equivalence should not be assumed. The public page does not fully specify whether observations are end-of-day, forward-filled, sampled at a fixed time, or based on another snapshot convention. It also does not fully document block-timestamp convention, reorganization policy, treatment of dates with no blocks, or whether all forms of provably unspendable outputs are treated exactly as in OBM.

Glassnode provides a closely related \emph{Bitcoin: UTXO Set Growth} chart:
\[
\texttt{https://studio.glassnode.com/charts/btc-utxo-set-growth?a=BTC}.
\]
Glassnode describes UTXOs as the fundamental accounting system of Bitcoin and states that the metric presents both the total number of UTXOs in the set and the 7-day exponential moving average of daily change. This is a strong conceptual comparator for \texttt{obm\_utxo\_eod\_count\_daily}. The total UTXO-set count is the relevant comparison; the 7-day EMA of daily change is a derived flow-like diagnostic and should not be confused with the end-of-day stock count. The public Glassnode page provides a clear interpretation, but it does not expose a full reconstruction algorithm specifying the precise snapshot timing, treatment of unspendable outputs, immature coinbase outputs, no-block dates, and chain reorganizations.

CryptoQuant provides a \emph{Bitcoin: UTXO Count} chart:
\[
\texttt{https://cryptoquant.com/asset/btc/chart/network-stats/utxo-count}.
\]
CryptoQuant describes the metric as the total number of unspent transaction outputs existing at a specified point. This is closely aligned with the OBM concept of an end-of-day UTXO-set count. Nevertheless, the public chart description does not fully specify the snapshot convention used for daily observations, the exact treatment of provably unspendable outputs, the handling of immature coinbase outputs, or the chain-state and reorganization policy. It is therefore a useful public comparator, but not a fully auditable methodological benchmark.

Newhedge provides a \emph{Bitcoin UTXO Count} chart:
\[
\texttt{https://newhedge.io/bitcoin/utxo-count}.
\]
The page defines Bitcoin UTXO Count as the total number of unspent transaction outputs on the Bitcoin network. It also provides chart controls and API-related access options. This is conceptually aligned with \texttt{obm\_utxo\_eod\_count\_daily}. However, the public page does not provide a full algorithmic specification. In particular, it does not clarify whether the daily value is an end-of-day observation, a sampled snapshot, a forward-filled state, or an average; nor does it specify how unspendable outputs, no-block dates, timestamp boundaries, and reorganizations are treated.

Bitbo provides a related \emph{Bitcoin Unspent Outputs By Year} chart:
\[
\texttt{https://charts.bitbo.io/unspent-outputs/}.
\]
The page describes the chart as showing the number of unspent transaction outputs per year since the genesis block. This is related to UTXO-set size, but it is less directly comparable to OBM because the public chart is organized at a yearly rather than daily frequency. It is useful for broad long-run visualization of UTXO-set growth, but it is not a direct daily end-of-day benchmark for \texttt{obm\_utxo\_eod\_count\_daily}.

ChainMetrics exposes a broad set of UTXO metrics, including \emph{UTXO count (Total)}, \emph{UTXO Created Count}, \emph{UTXO Spent Count}, and UTXO age-count breakdowns:
\[
\texttt{https://chainmetrics.com/metric/m23\_utxo\_age\_24h}.
\]
These metrics are conceptually relevant because they distinguish total UTXO stock from created and spent UTXO flows. However, public access and documentation appear to be more limited, and the linked pages do not provide a full reproducible algorithm for the total UTXO count. They are therefore useful as secondary references, especially for age-distribution and UTXO-flow analysis, but not as primary auditable comparators.

Coin Metrics does not appear to expose a directly named public \emph{UTXO count} metric equivalent to \texttt{obm\_utxo\_eod\_count\_daily} in the standard public metric pages reviewed here. It provides many transaction, transfer, supply, and network-usage metrics, but a direct UTXO-set-count stock series is not as clearly documented as Blockchain.com's UTXO-count chart or Glassnode's UTXO-set-growth chart. If a Coin Metrics UTXO-count metric is available through a broader or licensed catalogue, its comparability would still depend on the snapshot convention, unspendable-output policy, and chain-state treatment.

Block explorers and lower-level data sources can support independent reconstruction. A UTXO count can be reconstructed by scanning the blockchain in height order, adding newly created spendable outputs and subtracting consumed outputs. Bitcoin Core itself can also report current UTXO-set information through node-state queries such as \texttt{gettxoutsetinfo}, which is useful for checking the count near the chain tip. However, such sources are not necessarily named daily public time series. They should be treated as reconstruction or validation tools unless the provider publishes a documented daily UTXO-count series.

Related metrics should be distinguished from \texttt{obm\_utxo\_eod\_count\_daily}. UTXO value bands, UTXO age bands, HODL waves, UTXOs in profit or loss, UTXO created count, and UTXO spent count all use the UTXO model, but they measure different objects. Value-band and age-band metrics decompose the UTXO set by value or age. Profit/loss metrics compare UTXO creation prices with current prices. Created and spent counts are daily flows. OBM's \texttt{obm\_utxo\_eod\_count\_daily} is instead the end-of-day stock count of UTXOs under an explicit counting convention.

Overall, the strongest public comparators for \texttt{obm\_utxo\_eod\_count\_daily} are Blockchain.com's \emph{Unspent Transaction Outputs}, Glassnode's \emph{UTXO Set Growth}, CryptoQuant's \emph{UTXO Count}, and Newhedge's \emph{Bitcoin UTXO Count}. Bitbo and ChainMetrics provide useful secondary references, while block explorers and Bitcoin Core node-state queries can support independent reconstruction or validation. The distinctive contribution of OBM is that it defines the daily observation precisely as the UTXO count after the highest-height block assigned to the UTC date, includes immature coinbase outputs when they are not provably unspendable, excludes standard \texttt{OP\_RETURN}/\texttt{nulldata} outputs, writes \texttt{NaN} when no block is assigned to a date, and documents both genesis-scan and checkpoint-based computation modes for reproducibility.

\paragraph{Data source and input requirements.}
The metric is obtained from a running Bitcoin Core full node through the JSON-RPC interface. For each block in the scan interval, the script retrieves the decoded block object using:
\[
\texttt{getblock <block\_hash> 2}.
\]
The decoded block contains the transactions, their inputs, their outputs, and the decoded \texttt{scriptPubKey} metadata needed to identify provably unspendable outputs.

This metric does not require reconstructing previous transaction outputs. It does not require the spent-output indexer database, address extraction, user clustering, entity identification, external price data, or third-party APIs. It also does not require \texttt{txindex=1}, because each non-coinbase input spends exactly one UTXO and the count can therefore be updated without resolving the input to its previous transaction output.

The metric requires access to a synchronized Bitcoin Core full node and JSON-RPC credentials or cookie authentication. The script uses the locally verified main chain reported by the node. As with other directly scanned OBM metrics, the block timestamp returned by Bitcoin Core is used to assign each block to a UTC calendar date.

\paragraph{Algorithm.}
The script \texttt{compute\_obm\_utxo\_eod\_count\_daily\_v2.py} implements the following procedure:

\begin{enumerate}
    \item Parse the user-provided output date interval, \texttt{\symbol{45}\symbol{45}start\_date} and \texttt{\symbol{45}\symbol{45}end\_date}, using the format \texttt{YYYY-MM-DD}. Both dates are interpreted as UTC dates and both are included in the output.

    \item Determine whether the script runs in genesis mode or checkpoint mode. Genesis mode is used when \texttt{\symbol{45}\symbol{45}start\_date} is \texttt{2009-01-03} and no checkpoint count is supplied. Checkpoint mode is used when \texttt{\symbol{45}\symbol{45}start\_date\_eod\_utxo\_count} is supplied.

    \item If \texttt{\symbol{45}\symbol{45}start\_date} is different from \texttt{2009-01-03}, require the user to supply the parameter \texttt{\symbol{45}\symbol{45}start\_date\_eod\_utxo\_count}. This value is interpreted as the trusted UTXO count at the end of \texttt{\symbol{45}\symbol{45}start\_date}, under the same counting convention used by the script.

    \item Connect to the local Bitcoin Core node through JSON-RPC, using either explicit RPC credentials, environment variables, or cookie authentication.

    \item Query the local node using \texttt{getblockchaininfo} to determine the current best-chain height.

    \item Locate an approximate ending height covering \texttt{\symbol{45}\symbol{45}end\_date}. This is done by using block timestamps and binary search over block heights.

    \item Expand the approximate ending height using the safety parameter \texttt{\symbol{45}\symbol{45}height\_margin}. This reduces the risk of missing boundary blocks because Bitcoin block timestamps are not strictly monotonic with respect to height.

    \item In genesis mode, initialize $\mathrm{utxo\_count}=0$ and start scanning from block height 0.

    \item In checkpoint mode, initialize: $\mathrm{utxo\_count} = \texttt{\symbol{45}\symbol{45}start\_date\_eod\_utxo\_count}$. The script then locates the highest-height block assigned to \texttt{\symbol{45}\symbol{45}start\_date} or earlier, writes the supplied checkpoint value for \texttt{\symbol{45}\symbol{45}start\_date}, and starts block-by-block computation from the following chain position.

    \item Scan blocks in chain-height order. For each processed block \(b\), retrieve the decoded block using \texttt{getblock} with verbosity level 2.

    \item For every transaction output in the block, determine whether the output is provably unspendable. Standard \texttt{OP\_RETURN} outputs are excluded when Bitcoin Core reports \texttt{scriptPubKey.type = nulldata}. The script also defensively excludes scripts whose hexadecimal representation begins with \texttt{6a}, corresponding to \texttt{OP\_RETURN}.

    \item Count every non-provably-unspendable output as a newly created spendable output. This includes non-provably-unspendable coinbase outputs. Coinbase maturity is not used as an exclusion because immature coinbase outputs are still part of the UTXO set.

    \item Count one spent output for every input of every non-coinbase transaction. Coinbase inputs do not spend previous outputs and therefore do not reduce the UTXO count.

    \item Update the running count after each block:
    \[
    \mathrm{UTXOCountAfter}_b =
    \mathrm{UTXOCountBefore}_b
    +
    \mathrm{CreatedSpendableOutputs}_b
    -
    \mathrm{SpentOutputs}_b.
    \]

    \item Verify that the running UTXO count never becomes negative. A negative value would indicate an inconsistent checkpoint or counting rule.

    \item Assign the block to a UTC date using its block timestamp.

    \item For each selected date, store the UTXO count after the highest-height block assigned to that date.

    \item Write \texttt{NaN} for selected dates with no assigned block.

    \item Write the resulting time series to a CSV file using the standardized OBM schema:
    \[
    \texttt{date},\quad
    \texttt{series\_id},\quad
    \texttt{value},\quad
    \texttt{unit},\quad
    \texttt{frequency},\quad
    \texttt{release\_version}.
    \]

    \item Optionally generate a plot of the resulting series when the plotting flag is activated.
\end{enumerate}

The script therefore computes the metric directly from decoded block data. It does not query the spent-output indexer, does not reconstruct previous outputs, and does not need to store the full UTXO set. It maintains only a running integer count.

\paragraph{Metric-specific input parameters.}
The metric-specific input parameters are:

\begin{itemize}
    \item \texttt{\symbol{45}\symbol{45}start\_date\_eod\_utxo\_count}: trusted UTXO count at the end of \texttt{\symbol{45}\symbol{45}start\_date}, under the same counting convention used by the script. This parameter is mandatory whenever \texttt{\symbol{45}\symbol{45}start\_date} is different from \texttt{2009-01-03}. If supplied, the script writes this value for \texttt{\symbol{45}\symbol{45}start\_date} and starts block-by-block computation from the following chain position.

    \item \texttt{\symbol{45}\symbol{45}height\_margin}: number of extra blocks scanned around the checkpoint boundary and after the approximate ending height. The default is 288 blocks.
\end{itemize}

The checkpoint parameter is needed because the UTXO count is cumulative. Without either a genesis scan or a trusted checkpoint, the script cannot determine the UTXO count at an arbitrary later date from that date's blocks alone.

The \texttt{\symbol{45}\symbol{45}height\_margin} parameter is needed because Bitcoin block timestamps are not strictly monotonic with respect to block height. A timestamp-based binary search provides only an approximate height boundary. To avoid missing boundary blocks whose timestamps fall inside the relevant UTC date, the script expands the boundary scan by \(m\) blocks:
\[
h^{\mathrm{scan}}_{\max} =
\min(h_{\mathrm{tip}}, h^{\mathrm{approx}}_{\max} + m).
\]
In checkpoint mode, the same margin is used to locate the checkpoint boundary around the end of \texttt{\symbol{45}\symbol{45}start\_date}.

The script also accepts standard RPC, output, release-version, and plotting parameters, such as
\texttt{\symbol{45}\symbol{45}rpc\_url}, \texttt{\symbol{45}\symbol{45}rpc\_user}, \texttt{\symbol{45}\symbol{45}rpc\_password}, \texttt{\symbol{45}\symbol{45}cookie\_path}, \texttt{\symbol{45}\symbol{45}use\_default\_cookie}, \texttt{\symbol{45}\symbol{45}rpc\_timeout}, \texttt{\symbol{45}\symbol{45}output}, \texttt{\symbol{45}\symbol{45}release\_version}, \texttt{\symbol{45}\symbol{45}plot}, \texttt{\symbol{45}\symbol{45}plot\_output}, and \texttt{\symbol{45}\symbol{45}quiet}.

\paragraph{Aggregation rule.}
This metric is not aggregated as a daily sum. It is an end-of-day state variable. The running state evolves block by block according to:
\[
\mathrm{UTXOCountAfter}_b =
\mathrm{UTXOCountBefore}_b
+
\mathrm{CreatedSpendableOutputs}_b
-
\mathrm{SpentOutputs}_b.
\]

For each UTC date \(d\), the reported value is:
\[
\mathrm{UTXOCountEOD}_d =
\mathrm{UTXOCountAfter}_{b^\ast},
\qquad
b^\ast =
\arg\max_{b \in B_d} h_b.
\]

Monthly versions of this metric, if distributed, should not be computed by summing daily UTXO counts. Since the metric is a state variable, a monthly version should preferably be computed as an end-of-month observation:
\[
\mathrm{UTXOCountEOM}_m =
\mathrm{UTXOCountEOD}_{d^\ast},
\]
where \(d^\ast\) is the last date in month \(m\) for which a UTXO-count observation is defined.

\paragraph{Output format.}
The output file contains one observation per UTC date. Each row has the following fields:

\begin{center}
\begin{tabular}{llp{0.35\textwidth}}
\toprule
\textbf{Column} & \textbf{Example} & \textbf{Description} \\
\midrule
\texttt{date} & \texttt{2024-01-01} & UTC calendar date \\
\texttt{series\_id} & \texttt{obm\_utxo\_eod\_count\_daily} & Stable OBM series identifier \\
\texttt{value} & \texttt{160123456} & End-of-day UTXO count \\
\texttt{unit} & \texttt{outputs} & Number of UTXOs \\
\texttt{frequency} & \texttt{daily} & Observation frequency \\
\texttt{release\_version} & \texttt{OBM v0.1.0} & Dataset release version \\
\bottomrule
\end{tabular}
\end{center}

Values are integer quantities. Dates with no assigned block are written as \texttt{NaN}. In checkpoint mode, the supplied \texttt{\symbol{45}\symbol{45}start\_date\_eod\_utxo\_count} is written as the value for \texttt{\symbol{45}\symbol{45}start\_date}.

\paragraph{Technical validation.}
Several internal checks are used to validate this metric during execution. First, the script verifies that the requested date range is valid and that \texttt{\symbol{45}\symbol{45}start\_date} is not later than \texttt{\symbol{45}\symbol{45}end\_date}. Second, it verifies that \texttt{\symbol{45}\symbol{45}height\_margin} is non-negative. Third, it verifies that \texttt{\symbol{45}\symbol{45}start\_date\_eod\_utxo\_count}, when provided, is non-negative. Fourth, it enforces that \texttt{\symbol{45}\symbol{45}start\_date\_eod\_utxo\_count} is provided whenever \texttt{\symbol{45}\symbol{45}start\_date} is different from \texttt{2009-01-03}. Fifth, it checks that RPC authentication is valid, including cookie-file existence and format when cookie authentication is used. Sixth, it connects to Bitcoin Core and retrieves the current chain tip using \texttt{getblockchaininfo}. Seventh, it locates and expands the ending height for the requested date interval. Eighth, in checkpoint mode, it locates the highest-height block assigned to \texttt{\symbol{45}\symbol{45}start\_date} or earlier and starts scanning after that checkpoint boundary. Ninth, it scans blocks in chain-height order and retrieves decoded blocks using \texttt{getblock} with verbosity level 2. Tenth, it excludes provably unspendable outputs at creation time. Eleventh, it counts non-provably-unspendable outputs created by both coinbase and non-coinbase transactions. Twelfth, it subtracts one UTXO for each non-coinbase transaction input. Thirteenth, it verifies that the running UTXO count never becomes negative. Fourteenth, it records the UTXO count after the highest-height block assigned to each selected date. Fifteenth, it writes one output row per selected UTC date.

Additional consistency checks should be performed before release. Defined values should be non-negative integers. Dates with positive \texttt{obm\_block\_count\_daily} under the same timestamp convention should normally have a defined UTXO count. Dates with no assigned block should be reported as \texttt{NaN}. Near the chain tip, the final UTXO count can be compared with Bitcoin Core's \texttt{gettxoutsetinfo}, allowing for differences in exact chain state, timestamp boundary, and output-treatment convention. In checkpoint mode, the supplied \texttt{\symbol{45}\symbol{45}start\_date\_eod\_utxo\_count} should be independently verified, and the first output row should equal that checkpoint value.

When corresponding daily flow variables are available, daily changes should satisfy:
\[
\Delta \mathrm{UTXOCount}_d =
\mathrm{CreatedSpendableOutputs}_d -
\mathrm{SpentOutputs}_d.
\]
Large positive changes may indicate output splitting or periods of high output creation. Large negative changes may indicate consolidation episodes. Comparisons with external UTXO-count series should be interpreted cautiously because providers may differ in timestamp convention, treatment of provably unspendable outputs, reorganization handling, checkpoint convention, or forward-filling rules.

\paragraph{Known limitations.}
The end-of-day UTXO count series is transparent and reproducible, but it has several limitations. First, it is a state variable, so it requires either a genesis scan or a trusted checkpoint value. Second, in checkpoint mode, the correctness of all subsequent values depends on the correctness of \texttt{\symbol{45}\symbol{45}start\_date\_eod\_utxo\_count}. Third, the metric depends on the block timestamp convention used to assign blocks to UTC dates. Fourth, dates with no assigned block are recorded as \texttt{NaN}, not forward-filled. Fifth, the script excludes provably unspendable outputs using the decoded script information available from Bitcoin Core, especially \texttt{nulldata} and \texttt{OP\_RETURN}-style scripts. Sixth, immature coinbase outputs are included because they are part of the UTXO set, even though they are not yet spendable under the coinbase maturity rule. Seventh, the metric does not identify addresses, users, wallets, entities, exchanges, custodians, or ownership. Eighth, it measures the count of UTXOs, not the BTC value held in the UTXO set. Ninth, it is computed directly from decoded block data and does not require the spent-output indexer.

Despite these limitations, \texttt{obm\_utxo\_eod\_count\_daily} is a useful OBM network-state series. It provides a simple, full-node-derived measure of UTXO set size and complements block count, transaction count, spent-output count, raw output value, block weight, and other activity metrics.




\bibliographystyle{plainnat}
\bibliography{references}

@misc{nakamoto2008bitcoin,
  author       = {Nakamoto, Satoshi},
  title        = {Bitcoin: A Peer-to-Peer Electronic Cash System},
  year         = {2008},
  url          = {https://bitcoin.org/bitcoin.pdf},
  note         = {Accessed: 2026-05-04}
}

@misc{bitcoincore2026,
  author       = {{Bitcoin Core Project}},
  title        = {Bitcoin Core},
  year         = {2026},
  url          = {https://bitcoincore.org/},
  note         = {Accessed: 2026-05-04}
}

@misc{coinmetrics2026community,
  author       = {{Coin Metrics}},
  title        = {Coin Metrics Community Data and API Documentation},
  year         = {2026},
  url          = {https://docs.coinmetrics.io/},
  note         = {Accessed: 2026-05-04}
}

@misc{glassnode2026pricing,
  author       = {{Glassnode}},
  title        = {Glassnode Pricing: Digital Asset Data and Analytics},
  year         = {2026},
  url          = {https://glassnode.com/pricing/studio},
  note         = {Accessed: 2026-05-04}
}

@misc{cryptoquant2026pricing,
  author       = {{CryptoQuant}},
  title        = {CryptoQuant Pricing},
  year         = {2026},
  url          = {https://cryptoquant.com/pricing},
  note         = {Accessed: 2026-05-04}
}

@misc{blockchaincom2026charts,
  author       = {{Blockchain.com}},
  title        = {Blockchain.com Charts: Total Transaction Fees BTC},
  year         = {2026},
  url          = {https://www.blockchain.com/charts/transaction-fees},
  note         = {Accessed: 2026-05-04}
}

@misc{blockchair2026api,
  author       = {{Blockchair}},
  title        = {Blockchair API},
  year         = {2026},
  url          = {https://blockchair.com/api},
  note         = {Accessed: 2026-05-04}
}

@misc{blockchair2026dumps,
  author       = {{Blockchair}},
  title        = {Blockchair Database Dumps},
  year         = {2026},
  url          = {https://blockchair.com/dumps},
  note         = {Accessed: 2026-05-04}
}

@misc{mempool2026api,
  author       = {{mempool.space}},
  title        = {mempool.space REST API Documentation},
  year         = {2026},
  url          = {https://mempool.space/docs/api/rest},
  note         = {Accessed: 2026-05-04}
}

@misc{bitbo2026pricing,
  author       = {{Bitbo}},
  title        = {Bitbo Subscription Pricing},
  year         = {2026},
  url          = {https://charts.bitbo.io/pricing/},
  note         = {Accessed: 2026-05-04}
}

@misc{newhedge2026api,
  author       = {{Newhedge}},
  title        = {Newhedge API},
  year         = {2026},
  url          = {https://newhedge.io/api},
  note         = {Accessed: 2026-05-04}
}

@misc{bitinfocharts2026bitcoin,
  author       = {{BitInfoCharts}},
  title        = {Bitcoin Statistics},
  year         = {2026},
  url          = {https://bitinfocharts.com/bitcoin/},
  note         = {Accessed: 2026-05-04}
}

@misc{ycharts2026fees,
  author       = {{YCharts}},
  title        = {Bitcoin Total Transaction Fees Per Day},
  year         = {2026},
  url          = {https://ycharts.com/indicators/bitcoin_total_transaction_fees_per_day_btc},
  note         = {Accessed: 2026-05-04}
}

@misc{bitcoinmagazinepro2026home,
  author       = {{Bitcoin Magazine Pro}},
  title        = {Bitcoin Magazine Pro},
  year         = {2026},
  url          = {https://www.bitcoinmagazinepro.com/},
  note         = {Accessed: 2026-05-04}
}

@misc{checkonchain2026charts,
  author       = {{Checkonchain}},
  title        = {Bitcoin On-Chain Analysis and Charts},
  year         = {2026},
  url          = {https://charts.checkonchain.com/},
  note         = {Accessed: 2026-05-04}
}

@misc{tradingdigits2026realizedprice,
  author       = {{Trading Digits}},
  title        = {Bitcoin Realized Price Chart},
  year         = {2026},
  url          = {https://www.tradingdigits.io/realized-price},
  note         = {Accessed: 2026-05-04}
}

\end{document}